\documentclass[english,3p]{elsarticle}
\usepackage[T1]{fontenc}
\usepackage[utf8]{inputenc}
\usepackage{array}
\usepackage{verbatim}
\usepackage{float}
\usepackage{units}
\usepackage{amsmath}
\usepackage{amsthm}
\usepackage{amssymb}
\usepackage{stmaryrd}
\usepackage{graphicx}

\makeatletter

\providecommand{\tabularnewline}{\\}
\floatstyle{ruled}
\newfloat{algorithm}{tbp}{loa}
\providecommand{\algorithmname}{Algorithm}
\floatname{algorithm}{\protect\algorithmname}

\numberwithin{equation}{section}
\numberwithin{figure}{section}

\usepackage{algpseudocode,algorithm,algorithmicx}
\usepackage{tikz}
\usetikzlibrary{shapes,arrows,calc,decorations.pathreplacing}

\@ifundefined{showcaptionsetup}{}{%
 \PassOptionsToPackage{caption=false}{subfig}}
\usepackage{subfig}
\makeatother

\usepackage{babel}
\begin{document}

\title{A Conservative Discontinuous Galerkin Discretization for the Chemically
Reacting Navier-Stokes Equations}

\author{Ryan F. Johnson and Andrew D. Kercher}
\address{Laboratories for Computational Physics and Fluid Dynamics,  U.S. Naval Research Laboratory, 4555 Overlook Ave SW, Washington, DC 20375}

\begin{abstract}
We present a detailed description and verification of a discontinuous
Galerkin finite element method (DG) for the multi-component chemically
reacting compressible Navier-Stokes equations that retains the desirable
properties of DG, namely discrete conservation and high-order accuracy
in smooth regions of the flow. Pressure equilibrium between adjacent
elements is maintained through the consistent evaluation of the thermodynamics
model and the resulting weak form, as well as the proper choice of
nodal basis. As such, the discretization does not generate unphysical
pressure oscillations in smooth regions of the flow or at material
interfaces where the temperature is continuous. Additionally, we present
an $hp$-adaptive DG method for solving systems of ordinary differential
equations, DGODE, which is used to resolve the temporal evolution
of the species concentrations due to stiff chemical reactions. The
coupled solver is applied to several challenging test problems including
multi-component shocked flows as well as chemically reacting detonations,
deflagrations, and shear flows with detailed kinetics. We demonstrate
that the discretization does not produce unphysical pressure oscillations
and, when applicable, we verify that it maintains discrete conservation.
The solver is also shown to reproduce the expected temperature and
species profiles throughout a detonation as well as the expected two-dimensional
cellular detonation structure. We also demonstrate that the solver
can produce accurate, high-order, approximations of temperature and
species profiles without artificial stabilization for the case of
a one-dimensional pre-mixed flame. Finally, high-order solutions of
two- and three-dimensional multi-component chemically reacting shear
flows, computed without any additional stabilization, are presented.
\end{abstract}
\begin{keyword}
High order finite elements; Discontinuous Galerkin method; Chemistry;
Combustion; 
\end{keyword}
\maketitle
\global\long\def\middlebar{\,\middle|\,}%
\global\long\def\average#1{\left\{  \!\!\left\{  #1\right\}  \!\!\right\}  }%
\global\long\def\expnumber#1#2{{#1}\mathrm{e}{#2}}%

\let\svthefootnote\thefootnote\let\thefootnote\relax\footnotetext{\\ \hspace*{120pt}Distribution A. Approved for public release: distribution unlimited.}\addtocounter{footnote}{-1}\let\thefootnote\svthefootnote

\section{Introduction\label{sec:Introduction}}

In this work we provide a detailed description and verification of
a conservative, high-order, method for the multi-component chemically
reacting Navier-Stokes flows~\citep{Joh19}. The method is based
on the discontinuous Galerkin finite element method (DG)~\citep{Ree73,Bas97,Bas97_2,Coc98,Coc00,Arn02,Har02,Fid05,Luo06,Luo07,Luo08,Per08,Har13},
which has become an increasingly popular approach for modeling a wide
range of fluid dynamics. This is due to the fact that the method is
fully conservative, able to achieve high-order accuracy on unstructured
grids, and it naturally supports local polynomial, $p$, adaptivity.
Furthermore, in contrast to the continuous Galerkin finite element
method, DG does not require additional stabilization for pure advection
problems. As such, DG has the potential to be a powerful tool for
simulating multi-component chemically reacting flows. However, previous
applications of DG to multi-component chemically reacting flows were
not capable of maintaining pressure equilibrium, which resulted in
unphysical pressure oscillations not only at material interfaces but
in smooth regions of the flow~\citep{Bil11,Lv15}. These unphysical
oscillations were suppressed by incorporating a nonconservative flux,
an approach previously developed in the context of finite volume methods,
where the discrete solution is inherently discontinuous. This approach
is known as the \emph{double flux} \emph{method}, and it has the undesirable
effect that the discretization no longer achieves the discrete conservation
of energy, which is critical for the reliable approximation of shock
locations and speeds, as well as the correct determination of heat
release in combustion processes.

In this work, we present a DG discretization for the multi-component
chemically reacting Navier-Stokes equations that does not generate
unphysical pressure oscillations in smooth regions of the flow and
across material interfaces when the temperature is continuous without
the use of additional stabilization. This is achieved by
\begin{itemize}
\item Evaluating the thermodynamics exactly, i.e., defining temperature
such that the internal energy of a discrete solution and the species
weighted polynomial representation for internal energy are equivalent.
\item Representing the discrete solution in terms of a nodal basis with
coefficients defined on the element interfaces so that pressure equilibrium
between adjacent elements is maintained in smooth regions of the flow
as well as defining the basis coefficients of the nonlinear flux,
which is used to numerically evaluate the resulting weak form, in
a manner that maintains pressure equilibrium.
\end{itemize}
By preserving pressure equilibrium between adjacent elements in smooth
regions of the flow at each stage of the approximation, i.e., evaluation
of the exact thermodynamics, through the representation of the discrete
solution, and approximation of the weak form, the discretization is
capable of simulating multi-component flows without generating unphysical
pressure oscillations and therefore maintains the desirable properties
of DG, namely discrete conservation and high order accuracy for smooth
flows. As with all numerical methods for convection dominated flows,
instabilities in the solution are generated at discontinuous interfaces,
e.g., shocks and detonation fronts, that are not grid aligned. In
this case, additional stabilization is required, which we implement
via residual based artificial viscosity of the form of~\citep{Har13}.

In addition to applying DG to the spatial formulation, we have developed
an $hp$-adaptive DG method to solve the ordinary differential equations
(ODEs) that describes the time split species evolution for chemically
reacting flows. The method, termed DGODE, removes the need for a third
party library since it is built on the existing DG infrastructure.
Accurate and efficient integration of the potentially stiff chemical
source term is ensured via local adaptive refinement of both the temporal
resolution, $h$, as well as the polynomial degree, $p$. The effectiveness
of local $hp$-adaptivity for resolving the disparate chemical times
scales is studied in the context of the GRI-3.0 mechanism~\citep{GRI30}
where the polynomial degree of the local approximation, $p$, is compared
to the stiffness associated with the Jacobian of the chemical source
term for both a homogeneous reactor and a one-dimensional $C_{2}H_{4}$
and air detonation wave.

Finally, the coupled reacting Navier-Stokes flow solver is applied
to several multi-component non-reacting and chemically reacting test
cases in one, two, and three dimensions. In particular, we study a
one-dimensional multi-component shock tube and two-dimensional shock-bubble
interaction to test the ability of the solver to compute high-order
solutions to multi-component high-speed flows. We study a detonation
wave in one and two dimensions and analyze the ability of DGODE to
locally adapt the polynomial degree in order to accurately and efficiently
integrate complex chemical source terms in the presence of non-trivial
fluid dynamics. We analyze the conservation error of the formulation
for problems with shocks and detonations and comment on the stability
for smooth unsteady reacting flows. Where applicable, we present these
results in comparison to previous experimental and computational work.
Finally, we solve a three-dimensional multi-component chemically reacting
shear flow in the presence of a splitter plate to test the ability
of the solver to compute high-order solutions of smooth reacting Navier-Stokes
flows without the need for additional stabilization.

\subsection{Background\label{sec:Background}}

Overcoming unphysical pressure oscillations generated at material
interfaces of multi-component flows has been one of the primary challenges
in the simulation of chemically reacting flows. The source of these
oscillations has been attributed to variations in the thermodynamic
properties of multi-component gases. Furthermore, it was previously
concluded that any fully conservative Godunov-type scheme would be
unable to maintain a pressure equilibrium across material fronts~\citep{Abg88}.
However, the original analysis assumed that the variable ratio of
specific heats, $\gamma$, was only a function of species concentrations
and was frozen at each time step~\citep{Kar94,Abg96}. Jenny et al.~\citep{Jen97}
also analyzed the conditions under which pressure oscillations are
generated and concluded that the pressure remains in equilibrium across
a material interface if
\begin{enumerate}
\item The interface is grid aligned throughout the time step.\label{enu:grid-aligned}
\item The ratio of specific heats, $\gamma$, is continuous across the interface.\label{enu:continuous-ratio-of-specific-heats}
\item The temperature is continuous across the interface, i.e., discrete
temperature equilibrium between adjacent elements is maintained.\label{enu:continuous-temperature}
\end{enumerate}
Satisfaction of item~\ref{enu:grid-aligned} is outside the scope
of this work as it would require an implicit shock-fitting approach~\citep{Zah18,Cor18}
or a Lagrangian method for chemically reacting flows~\citep{Sun20}.
Item~\ref{enu:continuous-ratio-of-specific-heats} represents a special
case and is not generally satisfied in multi-component flows. However,
item~\ref{enu:continuous-temperature} is generally satisfied in
smooth regions of the flow, therefore a numerical scheme, if applied
to multi-component compressible Navier-Stokes flows, should be capable
of satisfying this condition discretely while also maintaining pressure
equilibrium.

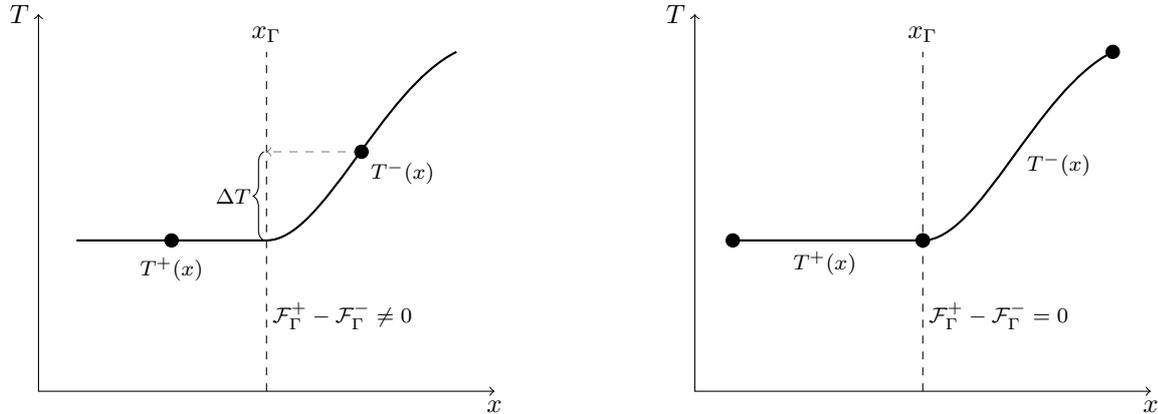
\begin{figure}
\begin{centering}
\begin{minipage}[c][1\totalheight][t]{0.475\textwidth}%
\begin{center}
\subfloat[\label{fig:discrete-temperature-equilibrium-FV}Loss of discrete temperature
equilibrium between adjacent elements in the case of low order finite
volume methods where piecewise constant, $p=0$, interpolation is
used. High-order reconstruction can reduce the error at the interface
but cannot, in general, eliminate it.]{\begin{centering}
\noindent\begin{minipage}[c][1\totalheight][t]{1\columnwidth}%
\begin{tikzpicture}[scale=5.0]
\coordinate (y) at (-0.1,1);
\coordinate (x) at (1.1,0);
\draw[<->] (y) node[left] {$T$} -- (-0.1,0) -- (x) node[below] {$x$};

\path
coordinate (start) at (0.5,0.4)
coordinate (c1) at +(0.65,0.4)
coordinate (c2) at +(0.8,0.8)
coordinate (end) at (1.0,0.9)
coordinate (p0) at (0.25,0.4)
coordinate (p1) at (0.75,0.635);

\draw[style={thick}] (0.0,0.4) -- (0.5,0.4);
\draw[style={dashed}] (0.5,0.0) -- (0.5,0.9) node[above] {$x_{\Gamma}$};

\draw[style={->,dashed,color=gray}] (p1) -- (0.5,0.635);

\draw[style={thick}] (start) .. controls (c1) and (c2) .. (end);
\filldraw [black] (p0) circle (0.5pt) node[below,yshift=-0.1cm, black] {\footnotesize $T^+(x)$};
\filldraw [black] (p1) circle (0.5pt) node[below right, black] {\footnotesize $T^-(x)$};

\draw[decorate,decoration={brace,amplitude=4pt},xshift=-0.2pt,yshift=0pt] (0.5,0.4) -- (0.5,0.635) node [black,midway,xshift=-0.4cm]  {\footnotesize $\Delta T$};

\node[] at (0.7,0.2) {\footnotesize $\mathcal{F}^+_{\Gamma}-\mathcal{F}^-_{\Gamma}\ne0$};

\end{tikzpicture}%
\end{minipage}
\par\end{centering}
\raggedright{}}
\par\end{center}%
\end{minipage}\hfill{}%
\begin{minipage}[c][1\totalheight][t]{0.475\textwidth}%
\begin{center}
\subfloat[\label{fig:discrete-temperature-equilibrium-FE}Satisfaction of discrete
temperature equilibrium in the case of finite element methods, with
$p>0$, where the solution is given in terms of, possibly piecewise,
continuous basis functions defined over the element.$\protect\phantom{\qquad\qquad\qquad\qquad\qquad\qquad\qquad\qquad\qquad\qquad}$$\protect\phantom{\qquad\qquad\qquad\qquad\qquad\qquad\qquad\qquad\qquad\qquad}$]{\begin{centering}
\noindent\begin{minipage}[c][1\totalheight][t]{1\columnwidth}%
\begin{tikzpicture}[scale=5.0]
\coordinate (y) at (-0.1,1);
\coordinate (x) at (1.1,0);
\draw[<->] (y) node[left] {$T$} -- (-0.1,0) -- (x) node[below] {$x$};

\path
coordinate (start) at (0.5,0.4)
coordinate (c1) at +(0.65,0.4)
coordinate (c2) at +(0.8,0.8)
coordinate (end) at (1.0,0.9)
coordinate (p0) at (0.5,0.4)
coordinate (p1) at (0.5,0.4);

\draw[style={thick}] (0.0,0.4) -- (0.5,0.4);
\draw[style={dashed}] (0.5,0.0) -- (0.5,0.9) node[above] {$x_{\Gamma}$};

\draw[style={thick}] (start) .. controls (c1) and (c2) .. (end);
\filldraw [black] (0,0.4) circle (0.5pt);
\filldraw [black] (p0) circle (0.5pt) node[left,xshift=-0.75cm,yshift=-0.3cm, black] {\footnotesize $T^+(x)$};
\filldraw [black] (p1) circle (0.5pt) node[right,xshift=+1.25cm,yshift=+1.0cm, black] {\footnotesize $T^-(x)$};
\filldraw [black] (end) circle (0.5pt);

\node[] at (0.7,0.2) {\footnotesize $\mathcal{F}^+_{\Gamma}-\mathcal{F}^-_{\Gamma}=0$};

\end{tikzpicture}%
\end{minipage}
\par\end{centering}
\raggedright{}}
\par\end{center}%
\end{minipage}
\par\end{centering}
\caption{\label{fig:discrete-temperature-equilibrium}The loss of temperature
equilibrium between adjacent elements leads to unphysical pressure
oscillations generated at the interface of adjacent elements where
the jump in the normal flux does not vanish, i.e., $\mathcal{F}^{+}-\mathcal{F}^{-}\protect\ne0$.}
\end{figure}

In the case of finite volume methods, discrete temperature equilibrium
is not, in general, easily maintained. For a piece-wise constant discrete
solution, i.e., DG$\left(p=0\right)$, only a constant temperature
profile will remain in equilibrium at the interfaces between adjacent
elements. In the case of a smoothly varying temperature profile, equilibrium
between adjacent elements will be lost, as shown in Figure~\ref{fig:discrete-temperature-equilibrium-FV}.
A more accurate reconstruction of the interface state, based on information
from the neighboring elements, may reduce this error, but cannot,
in general, eliminate it. Furthermore, the reconstruction would need
to maintain equilibrium between all adjacent elements in a fully multi-dimensional
setting, where as most reconstruction algorithms are dimensionally
split for efficiency. For the DG$\left(p>0\right)$ discretization
described in this work, the discrete solution is given in terms piecewise
continuous basis functions with basis coefficients located on the
element boundary as shown in Figure~\ref{fig:discrete-temperature-equilibrium-FE}.
As such, interpolation is not required to define the interface state,
and discrete equilibrium is automatically maintained between adjacent
elements regardless of the dimensional setting.

As with the DG method described in this work, continuous Galerkin
finite element methods (CG) also trivially maintain pressure equilibrium
between adjacent elements since the discrete solution is continuous
throughout the domain by design, see Figure~\ref{fig:discrete-temperature-equilibrium}.
However, in contrast to DG methods, CG methods require additional
stabilization for smooth convection dominated flows and are therefore
not considered in this work.

Alternatively, finite difference methods discretize the strong form
of the equations where the discrete solution is coupled through a
discrete difference operator applied directly to the, linear or nonlinear,
function corresponding to the underlying equation. In the case of
computational fluid dynamics, stability of the advection operator
is maintained by evaluating the flux at the state with components
defined in terms of weighted sums with adjacent elements. The weighted
sums are unique for each component of the state, which is problematic
for maintaining pressure equilibrium since any deviation in the relationship
defining pressure will result in the loss of equilibrium and the subsequent
generation of unphysical pressure oscillations.

Due to the difficulties described above, various techniques have been
developed for avoiding the generation of unphysical pressure oscillations.
A nonconservative approach, referred to as the \emph{double flux}
\emph{method}, is one popular option~\citep{Abg01,Bil03,Hou11,Bil11,Lv15}.
The method assumes consistent fluid thermal properties through a material
interface, thereby breaking energy conservation but maintaining pressure
equilibrium across interfaces. It has been successfully applied to
complex multidimensional reacting Navier-Stokes flows, including detonations,
in the context of both finite volume~\citep{Hou11} and DG discretizations~\citep{Bil11,Lv15}
with frozen thermodynamics.

Quasi-conservative methods based on the solution of an additional
transport equation for a given function of the ratio of specific heats
has been developed. Early attempts using this approach did not conserve
species mass concentrations~\citep{Abg96}, which lead to temperature
errors across material interfaces and subsequent unphysical heating
due to thermal diffusion. In response, a modified weighted essentially
non-oscillatory scheme was developed that solved for the mass fraction
in conservative form, thereby preventing temperature and species conservation
errors~\citep{Joh12}.

Methods based on the inclusion of additional transport equations have
also been purposed, as reviewed by~\citep{Tho18} in their presentation
of the five equation quasi-conservative model. This approach employs
a number fraction model to avoid pressure oscillations and is applicable
to Navier-Stokes flows since it includes the effects of species diffusion,
viscosity, and thermal conductivity. Although these methods have been
successful at preventing unphysical pressure oscillations, their applicability
to chemically reacting flows has yet to be demonstrated. In contrast,
the formulation presented in this work is fully conservative and does
not require the solution of additional transport equations while being
directly applicable to both inviscid and viscous multi-component chemically
reacting flows.

Conservative schemes based on exact thermodynamics have also been
successfully developed for compressible chemically reacting flows~\citep{Ora98,Ora00,Che00,Haw05,Tay09,Tay15_AIAA_ASM}.
However, this alone will not prevent the generation of unphysical
oscillations at material interfaces. Specifically, oscillations are
still generated in smooth regions of the flow if the discrete representation
of the solution is discontinuous, as is the case for finite volume
and finite difference methods. These oscillations are then suppressed
via limiting, artificial viscosity, or filtering. 

In the case of structured grids, conservative high-order finite difference
methods have been developed for compressible reacting flow problems
without shocks~\citep{Ken94,Che00,Haw05} %
. In this approach, a numerical filter is used to create artificial
viscosity and suppress unphysical oscillations, for details see~\citep{Ken94}.
Although high-order finite difference stencils are readily available
on the interior of the domain, ensuring the formal order of accuracy
of the method at the boundaries of the domain is not always obvious.
In some cases, the formal order of accuracy can be ensured through
the application of specially derived difference formulas at the boundary.
In general, this is not the case, especially when embedded surfaces
must be employed to represented non-cartesian aligned geometries.
The DG discretization we employ also supports arbitrarily high-order
approximations. However, unlike finite difference methods, it is capable
of achieving high-order accuracy on unstructured grids and it does
not require special treatment at the boundaries. Furthermore, the
DG discretization purposed in this work does not require additional
stabilization for well resolved smooth flows.

Additionally, finite volume schemes with detailed finite rate chemistry,
which derive the exact thermodynamic quantities from the energy have
been used to successfully simulate detonations and deflagrations.
Methods based on flux corrected transport (FCT) have been used to
study two-dimensional cellular detonation structures~\citep{Ora98,Tay15_AIAA_ASM}.
These results have been reproduced in two- and three-dimensional simulations
using an extended Roe solver for self-sustaining detonations where
unphysical oscillations did not corrupt the numerical results~\citep{Dei03}.
Simulations of detonations were also successfully performed using
a second order Godunov scheme and the Colella--Glaz Riemann solver~\citep{Tay09}.
In this work, we show that the our method is also capable of simulating
self-sustaining detonations, the results of which are reported in
Sections~\ref{subsec:One-dimensional-detonation-wave} and~\ref{subsec:Two-dimensional-detonation-wave}.

In addition, total variational diminishing type finite volume schemes
where the exact thermodynamics quantities were derived from the energy
have been used to carry out direct numerical simulations. Turbulent
three-dimensional flames have been simulated using a finite volume
formulation~\citep{Pol15}. The interfacial fluxes were computed
using an HLLC approximate Riemann solver where the reconstructed solutions
at the cell interfaces are limited using the piecewise-parabolic method
of Colella and Woodward~\citep{Col84}. %

In this work we also derive exact thermodynamics quantities from the
conserved energy, however, the formulation presented in this work
does not require stabilization or limiting in smooth regions of the
flow. The formulation is based on a DG discretization, and we show
that unphysical pressure oscillations exist as derived by ~\citep{Jen97},
and as detailed in~\ref{sec:Discontinuities}. These conclusions
are verified in Section~\ref{sec:discrete-pressure-equilibrium},
where we apply the conservative DG formulation to a series of problems
containing various types of material interfaces and analyze the magnitude
of pressure oscillations generated.

\section{Chemically Reacting Navier-Stokes Equations\label{subsec:Chemically-Reacting-Navier-Stokes}}

Let $\Omega\subset\mathbb{R}^{d}$ be a given $d$-dimensional domain.
We consider the nonlinear conservation law governing the unsteady
chemically reacting Navier-Stokes equations, in strong form, defined
for piecewise smooth, $\mathbb{R}^{m}$-valued functions $y$, and
gradient $\nabla y$, given as

\begin{align}
\frac{\partial y}{\partial t}+\nabla\cdot\mathcal{F}\left(y,\nabla y\right)-\mathcal{S}\left(y\right)=0 & \textup{ in }\Omega,\label{eq:conservation-strong-primal}
\end{align}
where $\mathcal{F}:\mathbb{R}^{m}\rightarrow\mathbb{R}^{m\times d}$
is a given flux function, $\mathcal{S}:\mathbb{R}^{m}\rightarrow\mathbb{R}^{m}$
is a given source term, and $t$ denotes time. The flux function 

\begin{equation}
F\left(y,\nabla y\right)=\left(\mathcal{F}^{c}\left(y\right)-\mathcal{F}^{v}\left(y,\nabla y\right)\right)\label{eq:flux_function}
\end{equation}
is defined in terms of the convective flux $\mathcal{F}^{c}\left(y\right)$,
which is only a function of the state $y$, and viscous flux $\mathcal{F}^{v}\left(y,\nabla y\right)$,
which is a function of the state and the gradient, $\nabla y$. The
chemically reacting Navier-Stokes flow state variable is given by

\begin{equation}
y=\left(\rho v_{1},\ldots,\rho v_{d},\rho e_{t},C_{i},\ldots,C_{n_{s}}\right),\label{eq:reacting-navier-stokes-state}
\end{equation}
where $m=d+n_{s}+1$, $n_{s}$ is the number of thermally perfect
species, $\rho:\mathbb{R}^{n_{s}}\rightarrow\mathbb{R}$ is density,
$\left(v_{1},\ldots,v_{d}\right):\mathbb{R}^{m}\rightarrow\mathbb{R}^{d}$
is velocity, $e_{t}:\mathbb{R}^{m}\rightarrow\mathbb{R}$ is the specific
total energy, and $C:\Omega\rightarrow\mathbb{R}^{n_{s}}$ are the
species concentrations. The density is calculated from the concentrations
as

\begin{equation}
\rho=\sum_{i=1}^{n_{s}}W_{i}C_{i},\label{eq:density_definition}
\end{equation}

\noindent where $W_{i}$ is the molecular weight of species $i$.

The $k$-th spatial convective flux component is given by
\begin{equation}
\mathcal{F}_{k}^{c}\left(y\right)=\left(\rho v_{k}v_{1}+p\delta_{k1},\ldots,\rho v_{k}v_{d}+p\delta_{kd},v_{k}\left(\rho e_{t}+p\right),v_{k}C_{1},\ldots,v_{k}C_{n_{s}}\right).\label{eq:reacting-navier-stokes-spatial-convective-flux-component}
\end{equation}
The pressure, $p:\mathbb{R}^{m}\rightarrow\mathbb{R}$, is calculated
from the equation of state,

\begin{equation}
p=R^{0}T\sum_{i=1}^{n_{s}}C_{i},\label{eq:EOS-1}
\end{equation}
where $T:\mathbb{R}^{m}\rightarrow\mathbb{R}$, is the temperature
and $R^{0}=8314.4621$ J/Kmol/K is the universal gas constant. The
total energy, $\rho e_{t}$, is given as the sum of the internal and
kinetic and energies as

\begin{equation}
\rho e_{t}=\rho u+\frac{1}{2}\sum_{k=1}^{d}\rho v_{k}v_{k},\label{eq:internal_energy_conserved_state}
\end{equation}
where $\rho u:\mathbb{R}^{m}\rightarrow\mathbb{R}$ is the internal
energy. The internal energy is also defined as the mass weighted sum
of thermally perfect species specific internal energies that are $n_{p}$-order
polynomials with respect to temperature, 

\begin{equation}
\rho u=\sum_{i=1}^{n_{s}}W_{i}C_{i}\sum_{k=0}^{n_{p}}a_{ik}T^{k}.\label{eq:internal_energy_polynomial}
\end{equation}
In this work, all thermodynamic polynomials are continuous refits
of the analytic form from NASA’s polynomial representations~\citep{Mcb02}.
The temperature is defined consistently to ensure equivalency between
the definition of internal energy, $\rho u$, given by Equation~(\ref{eq:internal_energy_conserved_state})
and the definition given by Equation~(\ref{eq:internal_energy_polynomial}),
i.e., find $T$ such that

\begin{eqnarray}
0 & = & \rho u-\sum_{i=1}^{n_{s}}W_{i}C_{i}\sum_{k=1}^{K}a_{ik}T^{k},\nonumber \\
 & = & \left(\rho e_{t}-\frac{1}{2}\sum_{k=1}^{n}\rho v_{k}v_{k}\right)-\sum_{i=1}^{n_{s}}W_{i}C_{i}\sum_{k=1}^{K}a_{ik}T^{k}.\label{eq:newton_eq}
\end{eqnarray}
The temperature is computed such that the following is satisfied to
machine precision for a given an initial temperature:

\begin{equation}
\delta T=\frac{\rho u-\sum_{i=1}^{n_{s}}W_{i}C_{i}\sum_{k=1}^{K}a_{ik}T^{k}}{\frac{\partial\rho u}{\partial T}},\label{eq:temperature-decrement}
\end{equation}
where $\delta T$ is the temperature decrement corresponding to Newton's
method and

\begin{equation}
\frac{\partial\rho u}{\partial T}=\sum_{i=1}^{n_{s}}W_{i}C_{i}\sum_{k=1}^{K}ka_{ik}T^{k-1}\label{eq:partial-u-partial-T}
\end{equation}
is the partial derivative of internal energy with respect to temperature,
see~\ref{sec:Source-term-Jacobian}. In practice, we observe that
the temperature converges within five nonlinear iterations with an
initial guess of $T=500$ K. In our experience, divergence of the
nonlinear solver very rarely occurs. In the case that the solution
does diverge, the cause is attributed to other numerical issues such
as negative concentrations. Once the temperature and state are known,
pressure is computed by evaluating Equation~(\ref{eq:EOS-1}) thereby
satisfying the thermodynamics state exactly. It is important to note
that we do not rely on the ratio of specific heats, $\gamma$, to
evaluate pressure to calculate the fluxes from the conserved state.

The $k$-th spatial component of the viscous flux is given by
\begin{equation}
\mathcal{F}_{k}^{v}\left(y,\nabla y\right)=\left(\tau_{1k},\ldots,\tau_{dk},\sum_{j=1}^{d}\tau_{kj}v_{j}-W_{i}C_{i}h_{i}V_{ik}-q_{k},C_{1}V_{1k},\ldots,C_{n_{s}}V_{n_{s}k}\right),\label{eq:navier-stokes-viscous-flux-spatial-component}
\end{equation}

\noindent where $q:\mathbb{R}^{m}\times\mathbb{R}^{m\times d}\rightarrow\mathbb{R}^{d}$
is the thermal heat flux, $\tau:\mathbb{R}^{m}\times\mathbb{R}^{m\times d}\rightarrow\mathbb{R}^{d\times d}$
is the viscous stress tensor, $\left(h_{1},\ldots,h_{n_{s}}\right):\mathbb{R}^{m}\rightarrow\mathbb{R}^{n_{s}}$
are the species specific enthalpies, and $\left(\left(V_{11},\dots,V_{1d}\right),\dots,\left(V_{n_{s}1},\dots,V_{n_{s}d}\right)\right):\mathbb{R}^{m}\times\mathbb{R}^{n_{s}\times d}\rightarrow\mathbb{R}^{n_{s}\times d}$
are the species diffusion velocities. The $k$-th spatial component
of the viscous stress tensor is given by

\noindent 
\begin{equation}
\tau_{k}\left(y,\nabla y\right)=\mu\left(\frac{\partial v_{1}}{\partial x_{k}}+\frac{\partial v_{k}}{\partial x_{1}}-\delta_{k1}\frac{2}{3}\sum_{j=1}^{d}\frac{\partial v_{j}}{\partial x_{j}},\ldots,\frac{\partial v_{d}}{\partial x_{k}}+\frac{\partial v_{k}}{\partial x_{d}}-\delta_{kd}\frac{2}{3}\sum_{j=1}^{d}\frac{\partial v_{j}}{\partial x_{j}}\right),\label{eq:reacting-navier-stokes-viscous-stress-tensor-component}
\end{equation}

\noindent where $\mu:\mathbb{R}^{m}\rightarrow\mathbb{R}$ is the
dynamic viscosity. The $k$-th spatial component of the heat flux
is given as
\begin{eqnarray}
q_{k}\left(y,\nabla y\right) & = & -\lambda\sum_{j=1}^{m}T_{y_{j}}\left(y\right)\frac{\partial y_{j}}{\partial x_{k}},\nonumber \\
 & = & -\lambda\frac{\partial T}{\partial x_{k}}.\label{eq:reacting-navier-stokes-heat-flux-component}
\end{eqnarray}
where $\lambda:\mathbb{R}^{m}\rightarrow\mathbb{R}$ is the thermal
conductivity and where $T_{y_{j}}$ is the partial derivatives of
$T$ with respect to state component $y_{j}$, see~\ref{sec:Source-term-Jacobian}.

The transport properties are calculated using mixture averaged properties.
The $k$-th spatial component of the diffusion velocity for the $i$-th
species is given as

\begin{equation}
V_{ik}=\frac{\bar{D}_{i}}{C_{i}}\frac{\partial C_{i}}{\partial x_{k}}-\frac{\bar{D}_{i}}{\rho}\frac{\partial\rho}{\partial x_{k}},\label{eq:diffusion_velocity}
\end{equation}
and the species mixture averaged diffusion coefficients $\left(\bar{D}_{1},\ldots,\bar{D}_{n_{s}}\right):\mathbb{R}^{m}\rightarrow\mathbb{R}^{n_{s}}$,
from~\citep{Kee89}, are defined for the $i$-th species as
\begin{equation}
\bar{D}_{i}=\frac{p_{atm}}{p\bar{W}}\frac{\sum_{j=1,j\ne i}^{n_{s}}X_{j}W_{j}}{\sum_{j=1,j\ne i}^{n_{s}}X_{j}/D_{ij}},\label{eq:diffusion}
\end{equation}
where $p_{atm}=101325$ Pa, $X_{j}$ is the mole fraction of species
$j$, $D_{ij}$ is the diffusion coefficient of species $i$ to species
$j$, and $\bar{W}:\mathbb{R}^{m}\rightarrow\mathbb{R}$ is the mixture
molecular weight, defined as
\begin{equation}
\bar{W}=\rho/\sum_{i=1}^{n_{s}}C_{i}.\label{eq:mixture_molecular_weight}
\end{equation}
and the mole fractions $\left(X_{1},\ldots,X_{n_{s}}\right):\mathbb{R}^{n_{s}}\rightarrow\mathbb{R}^{n_{s}}$
can be calculated directly from concentrations, 
\begin{equation}
X_{i}=C_{i}/\sum_{i=1}^{n_{s}}C_{i}.\label{eq:mole_fractions}
\end{equation}
The Wilke model~\citep{Wil50} is used to calculate viscosity

\begin{eqnarray}
\mu & = & \sum_{i=1}^{n_{s}}\frac{X_{i}\mu_{i}}{X_{i}+\sum_{i=1,i\ne j}^{n_{s}}\left(X_{j}\phi_{ij}\right)},\label{eq:viscosity}
\end{eqnarray}
where

\begin{eqnarray*}
\phi_{ij} & = & \frac{\left(1+\left(\frac{W_{j}}{W_{i}}\right)^{1/4}\sqrt{\left(\frac{\mu_{i}}{\mu_{j}}\right)}\right)^{2}}{\sqrt{8\left(1+\frac{W_{i}}{W_{j}}\right)}},
\end{eqnarray*}
and $\mu_{i}$ and $\mu_{j}$ are the species specific viscosities
for species $i$ and $j$, respectively. The Mathur model~\citep{Mat67}
is used to calculate conductivity,

\begin{equation}
\lambda=\frac{1}{2}\left(\sum_{i=1}^{n_{s}}X_{i}\lambda_{i}+\frac{1}{\sum_{i=1}^{n_{s}}\frac{X_{i}}{\lambda_{i}}}\right),\label{eq:conductivity}
\end{equation}
where $\lambda_{i}$ is the conductivity of species $i$.

Finally, the source term, which includes the detailed chemical kinetics,
is given by

\begin{equation}
\mathcal{S}\left(y\right)=\left(0,\ldots,0,0,\omega_{1},\ldots,\omega_{n_{s}}\right),\label{eq:reacting-navier-stokes-source-term}
\end{equation}
where $\omega_{i}$ is the production rate of species $i$, which
is the sum of the progress reaction rates from any arbitrary number
of reactions and reaction types, cf.~\citep{chemkin89}. 

\subsection{Additional thermodynamic relationships\label{subsec:Additional-thermodyanmic-relationships}}

It is often useful to extract from the aforementioned formulation
a relationship where internal energy is linearly related to pressure
at the current state, $\rho u=p/\left(\bar{\gamma}-1\right)$. The
relationship using $\bar{\gamma}$ can then be used to apply specific
conditions, e.g., characteristic boundary conditions, see~\ref{sec:Non-reflective-Inflow-Outflow},
that were developed in the context of calorically perfect gases~\citep{Poi92}.
Here we present the steps to formulate $\bar{\gamma}$ and comment
on where $\bar{\gamma}$ is equivalent to the ratio of specific heats

\begin{equation}
\gamma=\frac{c_{p}}{c_{v}}=\frac{c_{p}}{c_{p}-R},\label{eq:gamma_identity}
\end{equation}
where $c_{p}$ is the specific heat at constant pressure, $c_{v}$
is the specific heat at constant volume, and $R$ is the mixture gas
constant

\begin{equation}
R=\frac{R^{0}\sum_{i=1}^{n_{s}}C_{i}}{\rho}.\label{eq:R_mix}
\end{equation}

To find $\bar{\gamma}$ we use the definition of internal energy in
terms of enthalpy and pressure,

\begin{equation}
\rho u=\rho h-p.\label{eq:internal_energy_enthalpy_relationship}
\end{equation}
Here the enthalpy is

\begin{equation}
\rho h=\rho\sum_{i=1}^{n_{s}}Y_{i}\int_{0}^{T}c_{p,i}dT=\rho\sum_{i=1}^{n_{s}}Y_{i}h_{i},\label{eq:total_enthalpy_definition}
\end{equation}
where
\begin{equation}
Y_{i}=W_{i}C_{i}/\rho,\label{eq:mass-fractions}
\end{equation}
is the mass fraction of species $i$, $c_{p,i}$ is the specific heat
at constant pressure per unit mass of species $i$, and $h_{i}$ is
the species specific enthalpy polynomial of temperature that is degree
$n_{p}$, $h_{i}=\sum_{k=0}^{n_{p}}a_{ik}T^{k}+R^{0}T$. We reduce
the definition of internal energy to achieve the equivalent formulation
that contains the expression $\rho u=\frac{p}{\bar{\gamma}-1}$ by
introducing the mean value of $c_{p}$ and $c_{p,i}$ from reference
temperature, $T_{0}$, to current temperature, $T$, 

\begin{equation}
\bar{c}_{p,i}=\frac{1}{T-T_{0}}\int_{T_{0}}^{T}c_{p,i}dT=\frac{h_{i}-h_{i}^{0}}{T-T_{0}}\label{eq:cp_eff_enthalpy_specific}
\end{equation}
and

\begin{equation}
\bar{c}_{p}=\sum_{i=1}^{n_{s}}\frac{Y_{i}}{T-T_{0}}\int_{T_{0}}^{T}c_{p,i}dT=\frac{\sum_{i=1}^{n_{s}}Y_{i}\left(h_{i}-h_{i}^{0}\right)}{T-T_{0}},\label{eq:cp_eff_enthalpy}
\end{equation}
where $h_{i}^{0}$ is the species specific enthalpy at $T_{0}$. Using
the following definitions

\begin{eqnarray}
\bar{\gamma} & = & \frac{\bar{c}_{p}}{\bar{c}_{p}-R}=\frac{\sum_{i=1}^{n_{s}}\frac{Y_{i}\left(h_{i}-h_{i}^{0}\right)}{T-T_{0}}}{\sum_{i=1}^{n_{s}}\frac{Y_{i}\left(h_{i}-h_{i}^{0}\right)}{T-T_{0}}-R},\label{eq:cp_eff}\\
\rho u & = & \frac{p}{\bar{\gamma}-1}+\rho\sum_{i=1}^{n_{s}}Y_{i}\left(h_{i}^{0}-\bar{c}_{p,i}T_{0}\right),\label{eq:internal_energy_gamma_m_one}
\end{eqnarray}
the inviscid total energy conservation without reactions becomes

\begin{eqnarray}
\frac{\partial\left(\frac{p}{\bar{\gamma}-1}+\frac{1}{2}\sum_{k=1}^{d}\rho v_{k}v_{k}\right)}{\partial t}+\nabla\cdot\left(\left(\frac{p}{\bar{\gamma}-1}+\frac{1}{2}\sum_{k=1}^{d}\rho v_{k}v_{k}+p\right)\left(v_{1},\dots,v_{d}\right)\right) & = & 0,\label{eq:gamma_m_one_equivalent}
\end{eqnarray}
where the term $\rho\sum_{i=1}^{n_{s}}Y_{i}\left(h_{i}^{0}-\bar{c}_{p,i}T_{0}\right)$
in Equation~(\ref{eq:internal_energy_gamma_m_one}) is eliminated
from Equation~(\ref{eq:gamma_m_one_equivalent}) by fixing $T_{0}$
to 0 K and multiplying the non-reacting inviscid form of the species
conservation equations from Equation~(\ref{eq:conservation-law-strong-form})
by $W_{i}h_{i}^{0}$ and summing over all species conservation equations.
Equation~(\ref{eq:gamma_m_one_equivalent}) is equivalent to the
non-reacting inviscid form of the conservation of energy from Equation~(\ref{eq:conservation-strong-primal}).
Equation~(\ref{eq:gamma_m_one_equivalent}) has the same form of
the compressible Euler equations, and therefore is convenient for
evaluating characteristic boundary conditions and other functions
that rely on a linear relationship between internal energy in pressure.
However, those formulations may require the flow to be non-reacting
with constant thermodynamic properties as $\bar{\gamma}$ is assumed
to be constant in calorically perfect flows.

\begin{figure}[H]
\subfloat[\label{fig:Cp_difference}The difference between $\bar{c}_{p}$ and
$c_{p}$ with reference temperature $T_{0}=200$ K, for a varying
mixture of $CH_{4}$ and $O_{2}$ at two different temperatures, $T=300$
K and $T=1000$ K.]{\begin{centering}
\includegraphics[width=0.45\columnwidth]{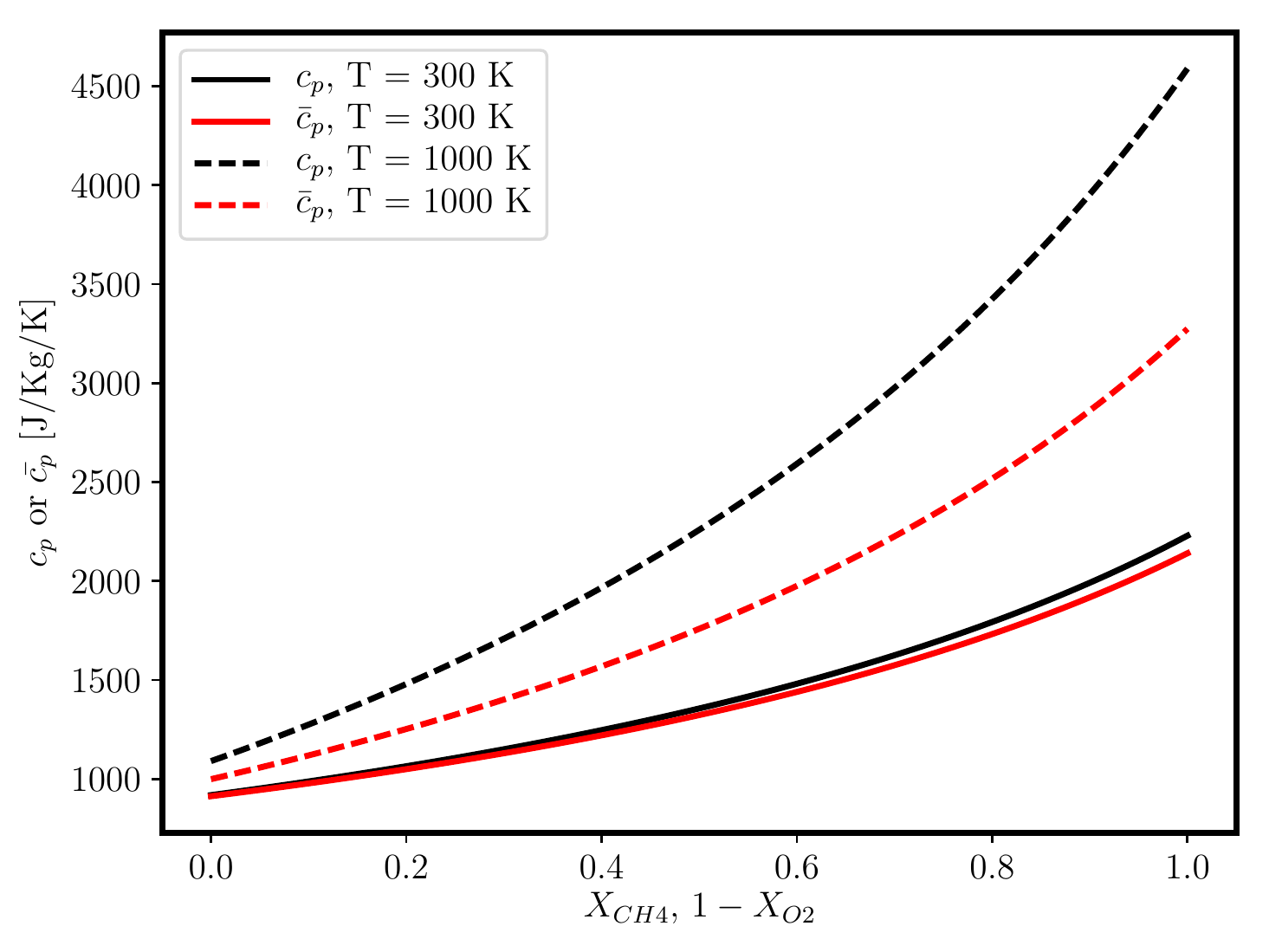}
\par\end{centering}
}\hfill{}
\begin{centering}
\subfloat[\label{fig:gamma_difference}The difference between $\bar{\gamma}$
and $\gamma=c_{p}/(c_{p}-R)$ with reference temperature $T_{0}=200$
K, for a varying mixture of $CH_{4}$ and $O_{2}$ at two different
temperatures, $T=300$ K and $T=1000$ K.]{\begin{centering}
\includegraphics[width=0.45\columnwidth]{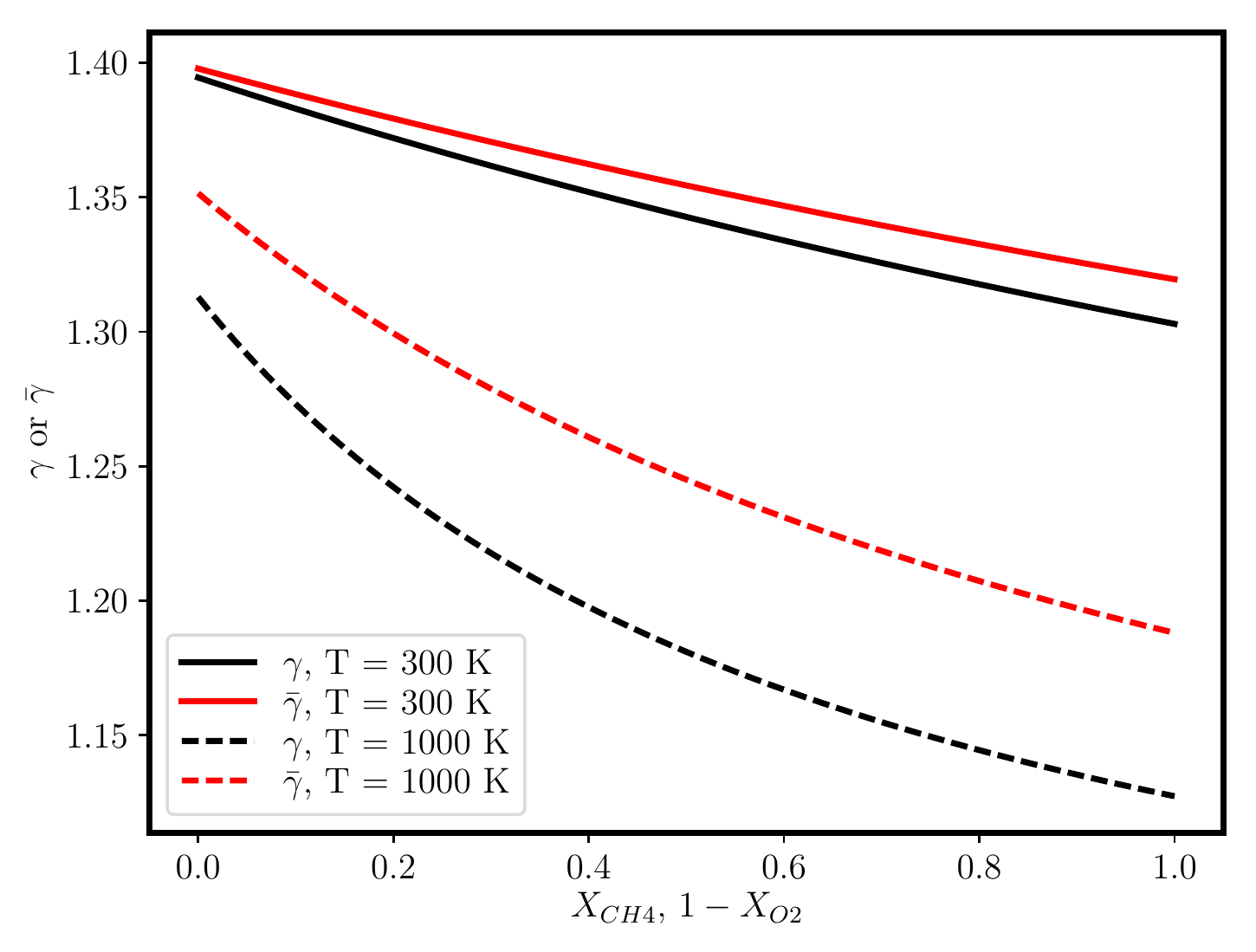}
\par\end{centering}
}
\par\end{centering}
\caption{The difference between $c_{p}$ evaluated from NASA polynomials, Equation~(\ref{eq:mixture_average_cp_polynomial}),
and $\bar{c}_{p}$ evaluated from Equation~(\ref{eq:cp_eff_enthalpy})
and the difference between $\gamma$ evaluated from NASA polynomials,
Equation~(\ref{eq:gamma_identity}), and $\bar{\gamma}$ evaluated
from Equation~(\ref{eq:cp_eff}).\label{fig:gamma_cp_differences}}
\end{figure}

It is also important to note that $\bar{c}_{p,i}$, given by Equation~(\ref{eq:cp_eff_enthalpy_specific}),
and $\bar{c}_{p}$, given by Equation~(\ref{eq:cp_eff_enthalpy}),
are not equivalent to the evaluations of $c_{p,i}$ and $c_{p}$ from
polynomial expressions, e.g., the analytic form of NASA’s polynomial
representations~\citep{Mcb02}. This is also true for $\gamma$,
Equation~(\ref{eq:gamma_identity}), and $\bar{\gamma}$, Equation~(\ref{eq:cp_eff}).
Rather, $\bar{c}_{p,i}$ and $\bar{c}_{p}$ can be viewed as the mean
value of $c_{p,i}$ and $c_{p}$, where $\bar{c}_{p}$ approaches
$c_{p}$ and $\bar{\gamma}$ approaches $\gamma$ as $T-T_{0}\rightarrow0$.
This difference can be shown mathematically by using the following
polynomial definitions of specific heat at constant pressure,

\begin{eqnarray}
c_{pi} & = & \sum_{k=0}^{n_{p}}b_{ik}T^{k}\label{eq:species_cp_polynomail}
\end{eqnarray}
and mixture averaged $c_{p}$, 

\begin{eqnarray}
c_{p} & = & \sum_{i=1}^{n_{s}}Y_{i}\sum_{k=0}^{n_{p}}b_{ik}T^{k}.\label{eq:mixture_average_cp_polynomial}
\end{eqnarray}
We arrive at the total enthalpy in polynomial form by integrating
of Equations~(\ref{eq:species_cp_polynomail})~and~(\ref{eq:mixture_average_cp_polynomial})
from $0$ to $T$ and substituting the result into Equation~(\ref{eq:total_enthalpy_definition}),

\begin{eqnarray}
\rho h & =\rho\sum_{i=1}^{n_{s}}Y_{i}\left(\int_{0}^{T}c_{p,i}dT\right)= & \rho\sum_{i=1}^{n_{s}}Y_{i}\left(\sum_{k=0}^{n_{p}}b_{ik}\frac{T^{k+1}}{k+1}\right).\label{eq:total_enthalpy_cp_polynomial}
\end{eqnarray}
By substituting Equation~(\ref{eq:total_enthalpy_cp_polynomial})
in Equation~(\ref{eq:cp_eff_enthalpy}), we arrive at $\bar{c_{p}}$
in terms of the $c_{p,i}$ polynomial coefficients and temperature,

\begin{equation}
\bar{c}_{p}=\frac{\sum_{i=1}^{n_{s}}Y_{i}\left(\sum_{k=0}^{n_{p}}b_{ik}\frac{T^{k+1}}{k+1}-\sum_{k=0}^{n_{p}}b_{ik}\frac{T_{0}^{k+1}}{k+1}\right)}{T-T_{0}}.\label{eq:specific-heat-constant-pressure-mean-value}
\end{equation}
Therefore, $\bar{c}_{p}$ and $c_{p}$ as well as $\bar{\gamma}$
and $\gamma$ are only equivalent if $c_{p}$ is constant with respect
to temperature, i.e., $n_{p}=0$. Figure~\ref{fig:Cp_difference}
shows the difference between $c_{p}$ evaluated from NASA polynomials
and $\bar{c}_{p}$ evaluated from Equation~(\ref{eq:cp_eff_enthalpy}).
Additionally, Figure~\ref{fig:gamma_difference} shows the difference
between $\gamma$ evaluated from Equation~(\ref{eq:gamma_identity})
with $c_{p}$ from Equation~(\ref{eq:mixture_average_cp_polynomial})
and $\bar{\gamma}$ evaluated from Equation~(\ref{eq:cp_eff}). The
displayed values were calculated using a reference temperature, $T_{0}$,
of $200$ K and mixture of methane, $CH_{4}$, and oxygen, $O_{2}$.
The thermodynamic properties were evaluated at a temperature, $T$,
of 300 K and 1000 K with the mixture varying from pure methane to
pure oxygen, $X_{CH4}=1-X_{O2}$. The values for $c_{p}$ and $\bar{c}_{p}$
as well as $\gamma$ and $\bar{\gamma}$ are significantly different
at higher temperatures due to the large difference between the actual
temperature and the reference temperature, $T-T_{0}$. This is indicative
the care that must be taken when using formulations or relationships
that rely on the ratio of specific heats with mixture varying thermodynamics
properties.

An additional source of confusion would be in the use of $\gamma$
or $\bar{\gamma}$ in the evaluation of the speed of sound. The speed
of sound is derived in the context of constant entropy and involves
the ratio of the derivatives of internal energy and enthalpy, as outlined
in Anderson~\citep{And20}. Using the same steps outlined in Anderson
for chemically reacting flows with variable thermodynamics, we came
to the conclusion that the formulation presented here does not require
any mean valued thermodynamics, such as $\bar{c}_{p}$, to evaluate
the speed of sound. Therefore the mixture speed of sound is 

\begin{equation}
c=\sqrt{\gamma RT},\label{eq:speed-of-sound}
\end{equation}
where $\gamma$ is evaluated using $c_{p}$ from Equation~(\ref{eq:mixture_average_cp_polynomial}).
In general, the formulation presented in this work does not rely on
the speed of sound. However, we do make for the evaluation of both
the numerical flux and characteristic boundary conditions.

\section{Formulation\label{sec:Formulation}}

In this section we present DG discretizations for nonlinear conservation
laws over a given $d$-dimensional domain $\Omega\subset\mathbb{R}^{d}$.
In particular, we consider the multi-dimensional unsteady chemically
reacting Navier-Stokes equations described in Section~\ref{subsec:Chemically-Reacting-Navier-Stokes}
and a one-dimensional system of ordinary differential equations (ODEs)
in Section~\ref{subsec:dg-ode}. We assume that $\Omega$ is partitioned
by $\mathcal{T}$, consisting of disjoint sub-domains or cells $\kappa$,
so that $\overline{\Omega}=\cup_{\kappa\in\mathcal{T}}\overline{\kappa}$,
with interfaces $\epsilon$ composing a set $\mathcal{E}$ so that
$\cup_{\epsilon\in\mathcal{E}}\epsilon=\cup_{\kappa\in\mathcal{T}}\partial\kappa$,
over which an oriented normal $n:\epsilon\rightarrow\mathbb{R}^{d}$
is defined. Furthermore, we assume that $\mathcal{E}$ consists of
two disjoint subsets: the interior interfaces
\begin{equation}
\epsilon_{\mathcal{I}}\in\mathcal{E_{I}}=\left\{ \epsilon_{\mathcal{I}}\in\mathcal{E}\middlebar\epsilon_{\mathcal{I}}\cap\partial\Omega=\emptyset\right\} \label{eq:interior-interfaces}
\end{equation}
and exterior interfaces 
\begin{equation}
\epsilon_{\partial}\in\mathcal{E}_{\partial}=\left\{ \epsilon_{\partial}\in\mathcal{E}\middlebar\epsilon_{\partial}\subset\partial\Omega\right\} ,\label{eq:boundary-interfaces}
\end{equation}
so that $\mathcal{E}=\mathcal{E_{I}}\cup\mathcal{E}_{\partial}$.
For interior interfaces, $\epsilon_{\mathcal{I}}\in\mathcal{E_{I}}$,
there exists $\kappa^{+},\kappa^{-}\in\mathcal{T}$ such that $\epsilon_{\mathcal{I}}=\partial\kappa^{+}\cap\partial\kappa^{-}$
and $n^{+},n^{-}$ denote the outward facing normal of $\kappa^{+},\kappa^{-}$
respectively, so that $n^{+}=-n^{-}$.

In order to discretize the nonlinear conservation laws considered
in this work, we assume that there is a continuous, invertible mapping
\begin{equation}
u:\hat{\Omega}\rightarrow\Omega,\label{eq:shape-mapping}
\end{equation}
from a reference domain $\hat{\Omega}\subset\mathbb{R}^{d}$ to the
physical domain $\Omega\subset\mathbb{R}^{d}$, see~\citep[Section 2.4.1]{Cor20_Ref,Ker20}
for additional details. We introduce a discrete (finite-dimensional)
subspace $V_{h}^{p}$ over $\mathcal{T}$ using standard piecewise
polynomials, cf.~\citep{Har13}. Let $\mathcal{P}_{p}$ denote the
space of polynomials spanned by the monomials $\boldsymbol{x}^{\alpha}$
with multi-index $\alpha\in\mathbb{N}_{0}^{d}$ , satisfying $\sum_{i=1}^{d}\alpha_{i}\leq p$.
In the case of a simplicial grid,
\begin{eqnarray}
V_{h}^{p} & = & \left\{ v\in\left[L^{2}\left(\Omega\right)\right]^{m}\middlebar\forall\kappa\in\mathcal{T},\left.v\right|_{\kappa}\circ u\in\left[\mathcal{P}_{p}\right]^{m}\right\} .\label{eq:discrete-subspace-simplex}
\end{eqnarray}
Let $\mathcal{Q}_{p}$ denote the tensor-product space of polynomials
spanned by the monomials $\boldsymbol{x}^{\alpha}$ with multi-index
$\alpha\in\mathbb{N}_{0}^{d}$ , satisfying $\alpha_{i}\leq p$ for
$i=1\ldots d$. In the case of a cuboid grid,

\begin{eqnarray}
V_{h}^{p} & = & \left\{ v\in\left[L^{2}\left(\Omega\right)\right]^{m}\middlebar\forall\kappa\in\mathcal{T},\left.v\right|_{\kappa}\circ u\in\left[\mathcal{Q}_{p}\right]^{m}\right\} .\label{eq:discrete-subspace-cubiod}
\end{eqnarray}

\subsection{Discretization\label{subsec:Discretization}}

We augment~(\ref{eq:conservation-strong-primal}) with initial and
boundary conditions as follows

\begin{align}
\frac{\partial y}{\partial t}+\nabla\cdot\mathcal{F}\left(y,\nabla y\right)-\mathcal{S}\left(y\right)=0 & \textup{ in }\kappa\qquad\forall\kappa\in\mathcal{T},\label{eq:conservation-law-strong-form}\\
y\left(\cdot,t_{0}\right)-y_{0}=0 & \textup{ in }\kappa\qquad\forall\kappa\in\mathcal{T},\label{eq:conservation-law-initial-condition}\\
n\cdot\mathcal{F}\left(y,\nabla y\right)-n\cdot\mathcal{F}_{\partial}\left(y,\nabla y\right)=0 & \textup{ on }\epsilon\qquad\forall\epsilon\in\mathcal{E}_{\partial},\label{eq:conservation-law-flux-boundary-condition}\\
G_{\partial}\left(y_{\partial}\right):\left(y^{+}-y_{\partial}\right)\otimes n=0 & \textup{ on }\epsilon\qquad\forall\epsilon\in\mathcal{E}_{\partial},\label{eq:conservation-law-state-boundary-condition}
\end{align}
where $G\left(y\right)$ is the partial linearization of the viscous,
or diffusive, flux, $\mathcal{F}^{v}$, with respect to gradient,
$\nabla y$, sometimes referred to as the homogeneity tensor~\citep{Har13}.
The initial conditions are given by $y_{0}$ in Equation~(\ref{eq:conservation-law-initial-condition}).

Following Hartmann and Leicht~\citep{Har13}, the boundary conditions
given by Equation~(\ref{eq:conservation-law-flux-boundary-condition})
are imposed through the boundary flux, $\mathcal{F}_{\partial}\left(y,\nabla y\right)=\mathcal{F}_{\partial}^{c}\left(y\right)-\mathcal{F}_{\partial}^{\nu}\left(y,\nabla y\right)$,
where $\mathcal{F}_{\partial}^{c}\left(y\right)$ and $\mathcal{F}_{\partial}^{\nu}\left(y,\nabla y\right)$
are the convective and viscous fluxes, respectively, at the boundary
and boundary. The condition on the state at the boundary given by
Equation~(\ref{eq:conservation-law-state-boundary-condition}) is
imposed through the boundary state $y_{\partial}$ and the boundary
modified homogeneity tensor, $G_{\partial}\left(y_{\partial}\right)$.

The DG (semi-)discretization of Equations~(\ref{eq:conservation-law-strong-form})-~(\ref{eq:conservation-law-flux-boundary-condition})
is given as: find $\frac{\partial y}{\partial t}\in V_{h}^{p}$ such
that

\begin{gather}
\sum_{\kappa\in\mathcal{T}}\left(\frac{\partial y}{\partial t},v\right)_{\kappa}-\sum_{\kappa\in\mathcal{T}}\left(\mathcal{F}\left(y,\nabla y\right),\nabla v\right)_{\kappa}+\sum_{\epsilon\in\mathcal{E}}\left(h\left(y,n\right),\left\llbracket v\right\rrbracket \right)_{\mathcal{E}}-\sum_{\epsilon\in\mathcal{E}}\left(\average{\mathcal{F}^{\nu}\left(y,\nabla y\right)}\cdot n-\delta\left(y,n\right),\left\llbracket v\right\rrbracket \right)_{\mathcal{E}}\nonumber \\
+\sum_{\kappa\in\mathcal{T}}\left(\left(\average y-y^{+}\right)\otimes n,\left(G^{\top}\left(y^{+}\right):\nabla v\right)\right)_{\partial\kappa}-\sum_{\kappa\in\mathcal{T}}\left(\mathcal{S}\left(y\right),v\right)_{\kappa}=0\qquad\forall v\in V_{h}^{p},\label{eq:semi-discretization}
\end{gather}
where $\left(\cdot,\cdot\right)$ denotes the inner product, $h\left(y,n\right)$
is the numerical flux, and $\delta\left(y,n\right)$ is a penalty
term that is required for stability, $\left\llbracket \cdot\right\rrbracket $
denotes the jump operator, $\average{\cdot}$ denotes the average
operator. In this work, the numerical flux chosen to be the HLLC approximate
Riemann~\citep{Tor13}, see also Appendix B of ~\citep{Lv15}. The
penalty term $\delta$ is implemented via the modified formulation
of Bassi and Rebay~\citep{Bas98,Bas00,Bas02}, commonly known as
BR2.

The DG space semi-discretization is integrated temporally with either
a second or third order strong-stability-preserving Runge-Kutta method~\citep{Got01,Spi02},
denoted SSP-RK2 and SSP-RK3 respectively. The time step is restricted
by the Courant-Friedrichs-Lewy number, CFL, defined as

\begin{equation}
\mathrm{CFL}=\frac{\Delta t}{\left(2p+1\right)\min\left(\Delta x\right)}\left(\left|v\right|+c\right),\label{eq:cfl}
\end{equation}
where $p$ is the polynomial degree. The optimal error estimates associated
with the RK2+DG method are $\mathcal{O}\left(h^{p+1}+\tau^{2}\right)$
where $h$ and $\tau$ are the length and time scales respectively~\citep{Zha04}.
This is verified for the formulation presented in this manuscript
in Section~\ref{subsec:thermal-bubble} where we study convergence
under grid refinement for the advection of a thermal bubble for approximations
corresponding to $p=3,4,5$.

The discretization of the convection and diffusion operators is decoupled
from the source term discretization via Strang operator splitting~\citep{Str68}.
As such, we have developed an $hp$-adaptive finite element method
for stiff ordinary differential equation (ODE) integration, which
is described in Section~\ref{subsec:dg-ode}, to facilitate integration
of the system that arises from this type of splitting.

The discrete solution is given by the coefficients of a nodal basis
defined over each element. In this work, the coefficients are defined
at the corresponding Gauss-Lobatto points of the element. In this
case, the trace of the solution is readily available, and does not
require interpolation, therefore maintaining pressure equilibrium
across element interfaces. The volume and surface terms of Equation~(\ref{eq:semi-discretization})
are numerically evaluated using a quadrature free approach~\citep{Atk96,Atk98}.
In the case of multi-component flows, the evaluation of the nonlinear
flux presents unique challenges, the details of which are discussed
in Section~\ref{subsec:nonlinear-flux-evaluation}.

On interior interfaces, $\mathcal{E_{I}}$, the jump, average, numerical
flux, and penalty term are defined in terms of the interior and exterior
traces, as follows
\begin{align}
\left\llbracket v\right\rrbracket =v^{+}-v^{-} & \textup{ on }\epsilon\qquad\forall\epsilon\in\mathcal{E_{I}},\label{eq:jump-interior}\\
\average y=\frac{1}{2}\left(y^{+}+y^{-}\right) & \textup{ on }\epsilon\qquad\forall\epsilon\in\mathcal{E_{I}},\label{eq:average-state-interior}\\
\average{\mathcal{F}^{\nu}\left(y,\nabla y\right)}=\frac{1}{2}\left(\mathcal{F}^{\nu}\left(y^{+},\nabla y^{+}\right)+\mathcal{F}^{\nu}\left(y^{-},\nabla y^{-}\right)\right) & \textup{ on }\epsilon\qquad\forall\epsilon\in\mathcal{E_{I}},\label{eq:average-flux-interior}\\
h\left(y,n\right)=h\left(y^{+},y^{-},n\right) & \textup{ on }\epsilon\qquad\forall\epsilon\in\mathcal{E_{I}},\label{eq:numerical-flux-interior}\\
\delta\left(y,n\right)=\delta\left(y^{+},y^{-},n\right) & \textup{ on }\epsilon\qquad\forall\epsilon\in\mathcal{E_{I}}.\label{eq:penalty-term-interior}
\end{align}

On the exterior interfaces, $\mathcal{E}_{\partial}$, we define the
following,

\begin{align}
\left\llbracket v\right\rrbracket =v^{+} & \textup{ on }\epsilon\qquad\forall\epsilon\in\mathcal{E}_{\partial},\label{eq:jump-exterior}\\
\average y=y_{\partial}\left(y^{+},n^{+}\right) & \textup{ on }\epsilon\qquad\forall\epsilon\in\mathcal{E}_{\partial},\label{eq:average-state-exterior}\\
\average{\mathcal{F}^{\nu}\left(y,\nabla y\right)}=\mathcal{F}_{\partial}^{\nu}\left(y_{\partial}\left(y^{+},n^{+}\right),\nabla y^{+}\right) & \textup{ on }\epsilon\qquad\forall\epsilon\in\mathcal{E}_{\partial},\label{eq:average-flux-exterior}\\
h\left(y,n\right)=h_{\partial}\left(y^{+},n^{+}\right) & \textup{ on }\epsilon\qquad\forall\epsilon\in\mathcal{E}_{\partial},\label{eq:numerical-flux-exterior}\\
\delta\left(y,n\right)=\delta_{\partial}\left(y^{+},n^{+}\right) & \textup{ on }\epsilon\qquad\forall\epsilon\in\mathcal{E}_{\partial},\label{eq:penalty-term-exterior}
\end{align}
where $y_{\partial}\left(y^{+},n^{+}\right)$ is the prescribed boundary
state, $h_{\partial}\left(y^{+},n^{+}\right)$ is the numerical boundary
flux, and $\mathcal{F}_{\partial}^{\nu}\left(y_{\partial}\left(y^{+}\right),\nabla y^{+},n^{+}\right)$
is the viscous boundary flux. The numerical and viscous fluxes are
defined consistently with the imposed boundary condition such that~(\ref{eq:conservation-law-flux-boundary-condition})
is satisfied, cf.~\citep{Har13}.

In this work we apply the following boundary conditions, where $k$-th
spatial component of the viscous boundary flux is denoted $\mathcal{F}_{\partial,k}^{\nu}\left(y_{\partial}\left(y^{+}\right),\nabla y^{+}\right)$
, the boundary stress tensor, $\tau_{\partial}=\tau\left(y_{\partial}\left(y^{+},n^{+}\right),\nabla y^{+}\right)$,
and boundary heat flux, $q_{\partial}=q\left(y_{\partial}\left(y^{+},n^{+}\right),\nabla y^{+}\right)$,
are both evaluated at the boundary state and the interior gradient.

\subsubsection*{Inflow:}

The inflow boundary condition is specified as

\begin{align}
y_{\partial}\left(y^{+},n^{+}\right)=y_{\infty} & \textup{ on }\epsilon\qquad\forall\epsilon\in\mathcal{E}_{\mathrm{in}},\label{eq:boundary-state-inflow}\\
\mathcal{F}_{\partial,k}^{\nu}\left(y_{\partial}\left(y^{+},n^{+}\right),\nabla y^{+}\right)=\mathcal{F}_{k}^{\nu}\left(y_{\infty},\nabla y^{+}\right) & \textup{ on }\epsilon\qquad\forall\epsilon\in\mathcal{E}_{\mathrm{in}},\label{eq:boundary-viscous-flux-inflow}\\
h_{\partial}\left(y^{+},n^{+}\right)=\mathcal{F}^{c}\left(y_{\infty}\right)\cdot n^{+} & \textup{ on }\epsilon\qquad\forall\epsilon\in\mathcal{E}_{\mathrm{in}},\label{eq:boundary-numerical-flux-inflow}\\
\delta_{\partial}\left(y^{+},n^{+}\right)=\delta\left(y^{+},y_{\infty}\right) & \textup{ on }\epsilon\qquad\forall\epsilon\in\mathcal{E}_{\mathrm{in}},\label{eq:boundary-penalty-term-inflow}
\end{align}
where $y_{\infty}:\Omega\rightarrow\mathbb{R}^{m}$ is a prescribed
state.

\subsubsection*{Outflow:}

We do not restrict the flow at an outflow boundary. The boundary condition
is therefore given by

\begin{align}
y_{\partial}\left(y^{+},n^{+}\right)=y^{+} & \textup{ on }\epsilon\qquad\forall\epsilon\in\mathcal{E}_{\mathrm{out}},\label{eq:boundary-state-outflow}\\
\mathcal{F}_{\partial,k}^{\nu}\left(y_{\partial}\left(y^{+},n^{+}\right),\nabla y^{+}\right)=\mathcal{F}_{k}^{\nu}\left(y^{+},\nabla y^{+}\right) & \textup{ on }\epsilon\qquad\forall\epsilon\in\mathcal{E}_{\mathrm{out}},\label{eq:boundary-viscous-flux-outflow}\\
h_{\partial}\left(y^{+},n^{+}\right)=\mathcal{F}^{c}\left(y^{+}\right)\cdot n^{+} & \textup{ on }\epsilon\qquad\forall\epsilon\in\mathcal{E}_{\mathrm{out}},\label{eq:boundary-numerical-flux-outflow}\\
\delta_{\partial}\left(y^{+},n^{+}\right)=0 & \textup{ on }\epsilon\qquad\forall\epsilon\in\mathcal{E}_{\mathrm{out}}.\label{eq:boundary-penalty-term-outflow}
\end{align}
Here the penalty term is evaluated at the interior state, $\delta\left(y^{+},y^{+}\right)$,
which is trivially zero.

\subsubsection*{Slip wall:}

At a slip wall we require the flow be parallel to the boundary. We
therefore define the boundary velocity, $\left(v_{\partial,1},\ldots,v_{\partial,d}\right):\mathbb{R}^{m}\times\mathbb{R}^{d}\rightarrow\mathbb{R}^{d}$
as
\begin{equation}
v_{\partial}\left(y^{+},n^{+}\right)=\left(v_{1}^{+}-\left(\sum_{k=1}^{d}v_{k}^{+}n_{k}^{+}\right)n_{1}^{+}\right),\cdots,\left(v_{d}^{+}-\left(\sum_{k=1}^{d}v_{k}^{+}n_{k}^{+}\right)n_{d}^{+}\right).\label{eq:boundary-velocity-slip-wall}
\end{equation}
where the normal component has been set to zero. The boundary condition
is specified as

\begin{align}
y_{\partial}\left(y^{+},n^{+}\right)=\left(\rho^{+}v_{\partial,1},\ldots,\rho^{+}v_{\partial,d},\left(\rho e_{t}\right)^{+},C_{i}^{+},\ldots,C_{n_{s}}^{+}\right) & \textup{ on }\epsilon\qquad\forall\epsilon\in\mathcal{E}_{\mathrm{slip}},\label{eq:boundary-state-slip-wall}\\
\mathcal{F}_{\partial,k}^{\nu}\left(y_{\partial}\left(y^{+},n^{+}\right),\nabla y^{+}\right)=\left(\tau_{\partial,1k},\ldots,\tau_{\partial,dk},{\textstyle \sum_{j=1}^{d}}\tau_{\partial,kj}v_{j},0,\ldots,0\right) & \textup{ on }\epsilon\qquad\forall\epsilon\in\mathcal{E}_{\mathrm{slip}},\label{eq:boundary-viscous-flux-slip-wall}\\
h_{\partial}\left(y^{+},n^{+}\right)=h\left(y^{+},2y_{\partial}\left(y^{+},n^{+}\right)-y^{+},n^{+}\right) & \textup{ on }\epsilon\qquad\forall\epsilon\in\mathcal{E}_{\mathrm{slip}},\label{eq:boundary-numerical-flux-slip-wall}\\
\delta_{\partial}\left(y^{+},n^{+}\right)=\delta\left(y^{+},y_{\partial}\left(y^{+},n^{+}\right)\right) & \textup{ on }\epsilon\qquad\forall\epsilon\in\mathcal{E}_{\mathrm{slip}},\label{eq:boundary-penalty-term-slip-wall}
\end{align}
where we have also required the species diffusion velocities and thermal
heat flux be zero. The viscous flux is computed from the boundary
state and the interior gradient. Following Hartmann and Leicht~\citep{Har13},
the numerical flux is evaluated at the interior state and the reflected
state, $2y_{\partial}\left(y^{+},n^{+}\right)-y^{+}$, which differs
from the boundary state given by~(\ref{eq:boundary-state-slip-wall}).

\subsubsection*{Adiabatic wall:}

At an adiabatic wall, the flow moves at a specified wall velocity,
while the species diffusion velocities and thermal heat flux are set
to zero. The boundary condition is given as

\begin{align}
y_{\partial}\left(y^{+},n^{+}\right)=\left(\rho^{+}v_{\partial,1},\ldots,\rho^{+}v_{\partial,d},\left(\rho e_{t}\right)^{+},C_{i}^{+},\ldots,C_{n_{s}}^{+}\right) & \textup{ on }\epsilon\qquad\forall\epsilon\in\mathcal{E}_{\mathrm{adi}},\label{eq:boundary-state-adiabatic-wall}\\
\mathcal{F}_{\partial,k}^{\nu}\left(y_{\partial}\left(y^{+},n^{+}\right),\nabla y^{+}\right)=\left(\tau_{\partial,1k},\ldots,\tau_{\partial,dk},{\textstyle \sum_{j=1}^{d}}\tau_{\partial,kj}v_{j},0,\ldots,0\right) & \textup{ on }\epsilon\qquad\forall\epsilon\in\mathcal{E}_{\mathrm{adi}},\label{eq:boundary-viscous-flux-adiabatic-wall}\\
h_{\partial}\left(y^{+},n^{+}\right)=\mathcal{F}^{c}\left(y_{\partial}\left(y^{+},n^{+}\right)\right)\cdot n^{+} & \textup{ on }\epsilon\qquad\forall\epsilon\in\mathcal{E}_{\mathrm{adi}},\label{eq:boundary-numerical-flux-adiabatic-wall}\\
\delta_{\partial}\left(y^{+},n^{+}\right)=\delta\left(y^{+},y_{\partial}\left(y^{+},n^{+}\right)\right) & \textup{ on }\epsilon\qquad\forall\epsilon\in\mathcal{E}_{\mathrm{adi}},\label{eq:boundary-penalty-term-adiabatic-wall}
\end{align}
where $\left(v_{\partial,1},\ldots,v_{\partial,d}\right):\Omega\rightarrow\mathbb{R}^{d}$
is the prescribed boundary velocity. 

\subsubsection*{Isothermal wall:}

At an isothermal wall, the flow moves at a specified wall velocity
and is actively heated, or cooled, to a specified temperature. Again,
the species diffusion velocities are set to zero, but the thermal
heat flux is unspecified. The boundary condition is given as

\begin{align}
y_{\partial}\left(y^{+},n^{+}\right)=\left(\rho_{\partial}v_{\partial,1},\ldots,\rho_{\partial}v_{\partial,d},\rho u_{\partial}+0.5\cdot{\textstyle \sum_{j=1}^{d}}\rho_{\partial}v_{\partial,k}v_{\partial,k},C_{\partial,i},\ldots,C_{\partial,n_{s}}\right) & \textup{ on }\epsilon\qquad\forall\epsilon\in\mathcal{E}_{\mathrm{iso}},\label{eq:boundary-state-isothermal-wall}\\
\mathcal{F}_{\partial,k}^{\nu}\left(y_{\partial}\left(y^{+},n^{+}\right),\nabla y^{+}\right)=\left(\tau_{\partial,1k},\ldots,\tau_{\partial,dk},{\textstyle \sum_{j=1}^{d}}\tau_{\partial,kj}v_{\partial,j}-q_{\partial,k},0,\ldots,0\right) & \textup{ on }\epsilon\qquad\forall\epsilon\in\mathcal{E}_{\mathrm{iso}},\label{eq:boundary-viscous-flux-isothermal-wall}\\
h_{\partial}\left(y^{+},n^{+}\right)=\mathcal{F}^{c}\left(y_{\partial}\left(y^{+},n^{+}\right)\right)\cdot n^{+} & \textup{ on }\epsilon\qquad\forall\epsilon\in\mathcal{E}_{\mathrm{iso}},\label{eq:boundary-numerical-flux-isothermal-wall}\\
\delta_{\partial}\left(y^{+},n^{+}\right)=\delta\left(y^{+},y_{\partial}\left(y^{+},n^{+}\right)\right) & \textup{ on }\epsilon\qquad\forall\epsilon\in\mathcal{E}_{\mathrm{iso}},\label{eq:boundary-penalty-term-isothermal-wall}
\end{align}
where $\left(v_{\partial,1},\ldots,v_{\partial,d}\right):\Omega\rightarrow\mathbb{R}^{d}$
is the prescribed boundary velocity, $T_{\partial}:\Omega\rightarrow\mathbb{R}$
is the prescribed boundary temperature, $\left(C_{\partial,i},\ldots,C_{\partial,n_{s}}\right):\mathbb{R}^{m}\rightarrow\mathbb{R}^{n_{s}}$
are the boundary concentrations, where the $i$-th component is given
as
\begin{equation}
C_{\partial,i}\left(y^{+}\right)=\frac{T^{+}}{T_{\partial}}C_{i}^{+}.\label{eq:boundary-concentrations-isothermal-wall}
\end{equation}
The boundary density is defined as
\begin{equation}
\rho_{\partial}=\sum_{i=1}^{n_{s}}W_{i}C_{\partial,i},\label{eq:density-isothermal-wall}
\end{equation}
and the boundary internal energy is given by
\begin{equation}
\rho u_{\partial}=\sum_{i=1}^{n_{s}}W_{i}C_{\partial,i}\sum_{k=0}^{n_{p}}a_{ik}T_{\partial}^{k}.\label{eq:internal_energy_polynomial-isothermal-wall}
\end{equation}

\subsubsection*{Characteristic:}

The conserved state at characteristic boundaries is determined based
on incoming and outgoing characteristics. The boundary condition is
given as

\begin{align}
y_{\partial}\left(y^{+},n^{+}\right)=y^{*}\left(y^{+},y_{\infty},n^{+}\right) & \textup{ on }\epsilon\qquad\forall\epsilon\in\mathcal{E}_{\mathrm{cha}},\label{eq:boundary-state-characteristic}\\
\mathcal{F}_{\partial,k}^{\nu}\left(y_{\partial}\left(y^{+},n^{+}\right),\nabla y^{+}\right)=\mathcal{F}_{k}^{\nu}\left(y_{\partial}\left(y^{+},n^{+}\right),\nabla y^{+}\right) & \textup{ on }\epsilon\qquad\forall\epsilon\in\mathcal{E}_{\mathrm{cha}},\label{eq:boundary-viscous-flux-characteristic}\\
h_{\partial}\left(y^{+},n^{+}\right)=h\left(y^{+},y_{\partial}\left(y^{+},n^{+}\right),n^{+}\right) & \textup{ on }\epsilon\qquad\forall\epsilon\in\mathcal{E}_{\mathrm{cha}},\label{eq:boundary-numerical-flux-characteristic}\\
\delta_{\partial}\left(y^{+},n^{+}\right)=\delta\left(y^{+},y_{\partial}\left(y^{+},n^{+}\right)\right) & \textup{ on }\epsilon\qquad\forall\epsilon\in\mathcal{E}_{\mathrm{cha}},\label{eq:boundary-penalty-term-characteristic}
\end{align}
where $y^{*}\left(y^{+},y_{\infty},n^{+}\right)$ is the characteristic
boundary value. The derivation $y^{*}\left(y^{+},y_{\infty},n^{+}\right)$
for non-reflecting inflow and outflow boundary conditions is given
in~\ref{sec:Non-reflective-Inflow-Outflow}.

\subsubsection{Nonlinear flux evaluation and consistent interpolation\label{subsec:nonlinear-flux-evaluation}}

In the case of single component flows, methods for computing the nodal
basis coefficients corresponding to the flux\footnote{We restrict the discussion to the nonlinear convective flux, where
the $k$-th spatial-component is given by~(\ref{eq:reacting-navier-stokes-spatial-convective-flux-component}),
as we have found this term to be the most sensitive to the loss in
pressure equilibrium, however, the viscous terms of~(\ref{eq:semi-discretization})
are also evaluated in this same manner.} include: interpolation, i.e., defining the high order basis coefficients
corresponding to the solution, $y$, in terms of a linear combination
of the low order basis function that are then used in the nonlinear
flux evaluation, and $L^{2}$ projection, which requires the solution
of an auxiliary problem but redistributes the error more evenly throughout
the element, see~\citep{Atk96,Atk98}. To analyze both approaches
in the case of multi-component flows, we must consider the trial functions
corresponding to the solution variables, i.e., the conserved state
variables, which interpolate onto the span of $\left\{ \varphi_{1},\ldots,\varphi_{s}\right\} $
\begin{align}
y_{_{i}} & =\sum_{s}y_{i}\left(x_{s}\right)\varphi_{s},\label{eq:state-interpolation}
\end{align}
as well as the flux projection, which interpolates onto the span of
$\left\{ \varphi_{1},\ldots,\varphi_{r}\right\} $ 
\begin{equation}
\left(\Pi\left(\mathcal{F}\left(y\right)\right)\right)_{ij}=\sum_{r}\mathcal{F}_{ij}\left(y\left(x_{r}\right)\right)\varphi_{r},\label{eq:flux-projection}
\end{equation}
where $r\ge s$.

\begin{figure}[H]
\subfloat[\label{fig:interpolated-state}The normalized exact and discrete solutions
corresponding to a piecewise linear approximation, i.e., $p=1$, with
basis coefficients defined at the points $x_{0}$ and $x_{1}$. In
this example, the error associated with the discrete approximation
of the species concentrations results in a loss of pressure equilibrium
when represented in a higher degree basis, e.g., $p=2$, with basis
coefficients defined at the points $x_{0},x_{1}$ and $x_{2}$. The
difference between the exact and discrete values of the species concentrations
are denoted $\Delta C_{i}$ for $i=1,2$. Equilibrium is maintained
at the points corresponding to the original linear basis as there
is no error associated with the discrete approximations in regards
to pressure equilibrium.]{\begin{centering}
\includegraphics[width=0.45\columnwidth]{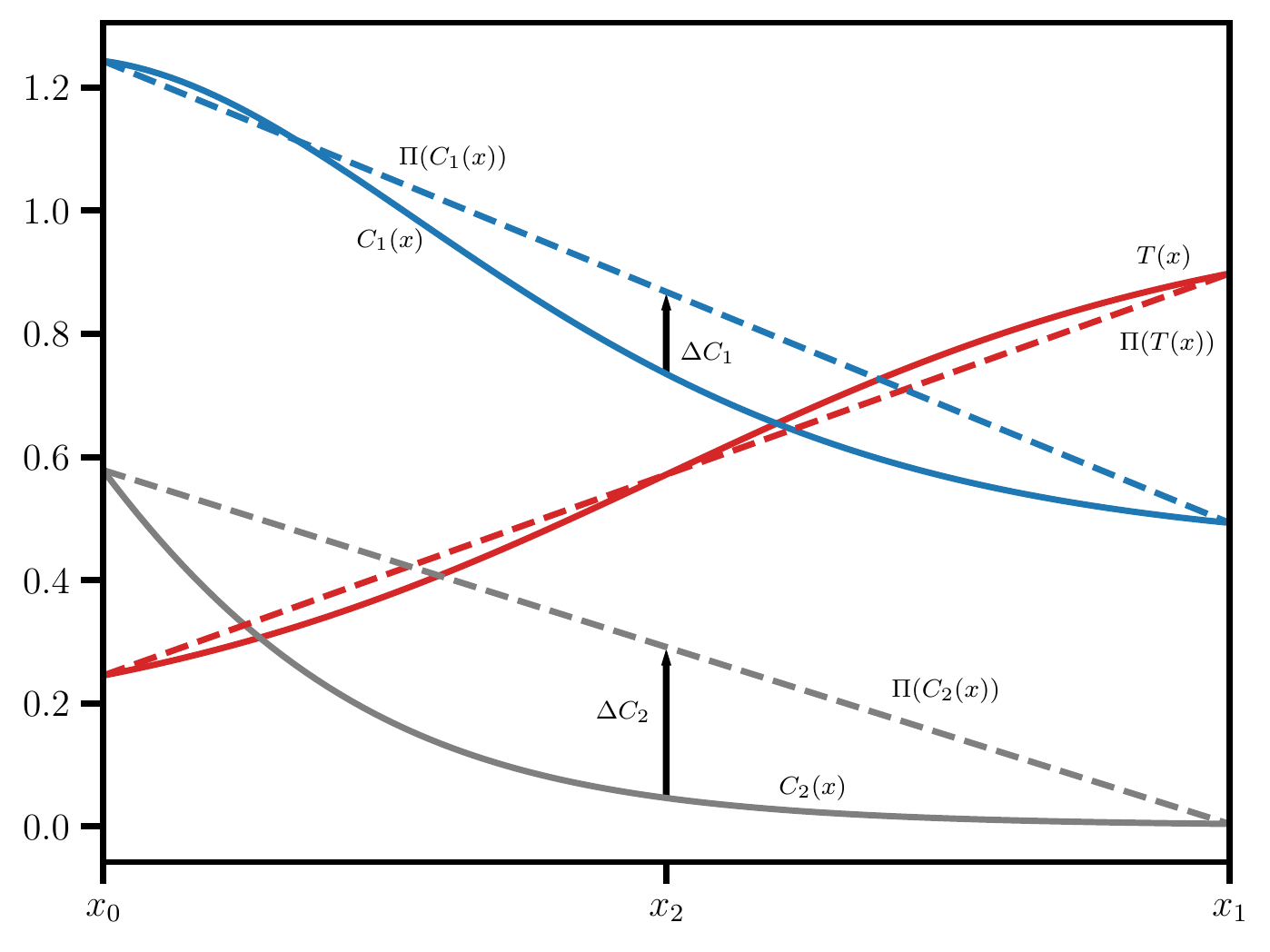}
\par\end{centering}
}\hfill{}
\begin{centering}
\subfloat[\label{fig:projected-pressure}The exact pressure, $p\left(x\right)$,
and the resulting discrete projection of the pressure, $\Pi\left(p\left(x\right)\right)$,
corresponding to a piecewise quadratic flux approximation, i.e., $p=2$,
defined in terms of both interpolation and the $L^{2}$ projection.
Defining the basis coefficients via interpolation maintains pressure
equilibrium between adjacent elements but it is lost within the element.
In contrast, if the basis coefficients defined by the $L^{2}$ projection
of $R^{0}T\left(x\right)\sum_{i}C_{i}\left(x\right)$, pressure equilibrium
is maintained within the element but lost between adjacent elements.
Equilibrium is only maintained if the pressure is projected onto the
same basis as the discrete solution.]{\begin{centering}
\includegraphics[width=0.45\columnwidth]{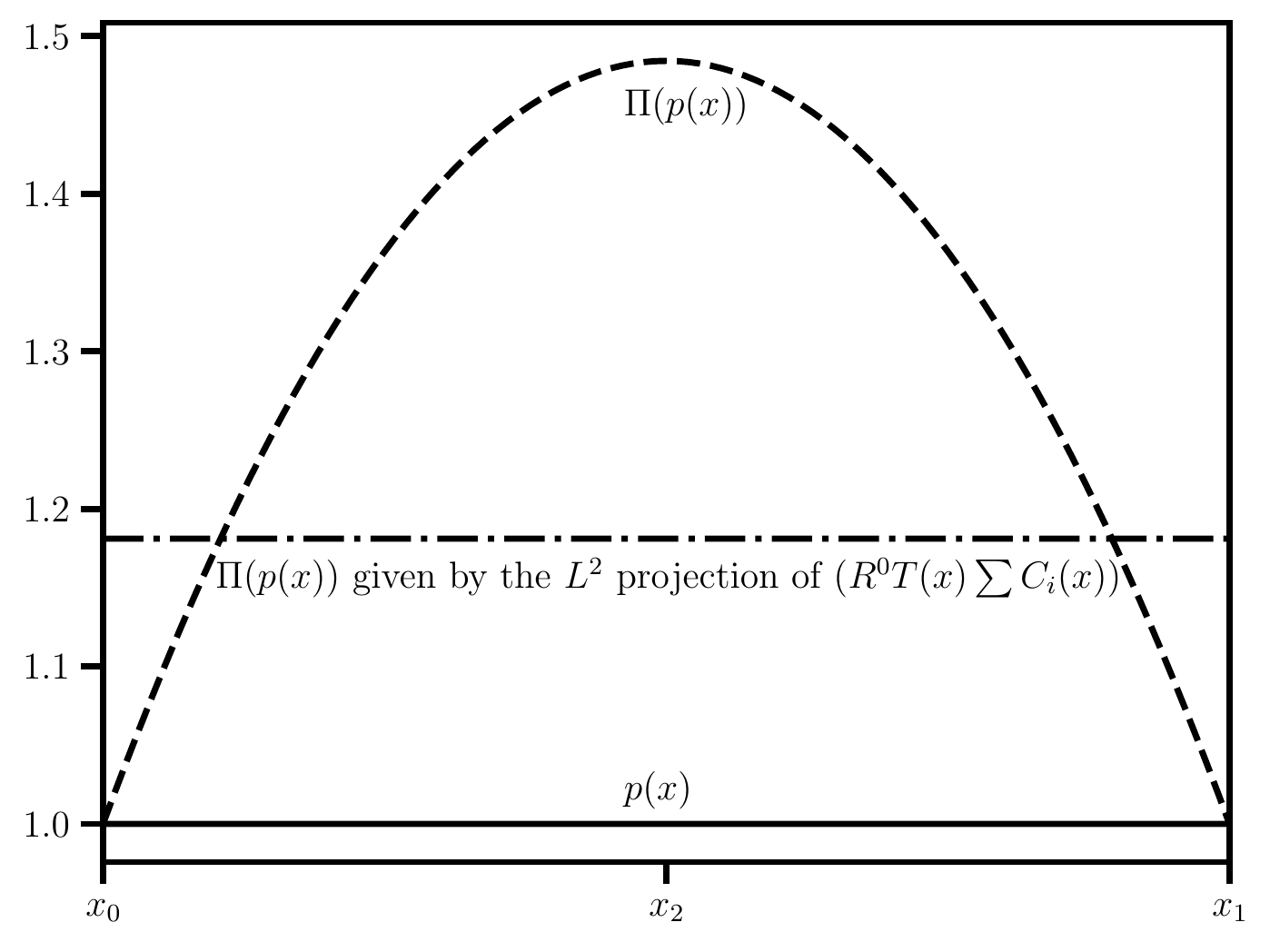}
\par\end{centering}
}
\par\end{centering}
\caption{\label{fig:pressure-equilibrium} The loss of pressure equilibrium
for the higher degree basis coefficients corresponding to the nonlinear
flux resulting from the error associated with the projection of the
discrete solution. As a simplification, the analysis is given in terms
of species concentrations and temperature, thereby avoiding the nonlinear
relationship between the total energy, $\rho e_{t}$, of the conserved
state, $y$, and the pressure. However, the analysis would not change
if we were to consider the more complicated case where temperature
was not substituted for total energy.}
\end{figure}

At this point, the basis coefficients of the flux, $\mathcal{F}_{ij}$,
can be directly evaluated in terms of the trial functions corresponding
to the conserved state~(\ref{eq:state-interpolation}). For multi-component
flows, this is only guaranteed to maintain pressure equilibrium for
$r=s$, as the basis coefficients corresponding to the nodal values
of pressure that are evaluated from the interpolated state, $y$,
which is not constrained to remain in equilibrium as shown in Figure~\ref{fig:pressure-equilibrium}.
This results in a loss of equilibrium within the element, while it
is maintained between adjacent elements since the nodal values of
the interpolated state correspond to the original space, define over
$s$. However, the error associated with the loss of equilibrium is
reduced as the discrete approximation is refined.

Alternatively, we can define an auxiliary problem and solve for the
coefficients $\mathcal{F}_{ij}$ via an $L^{2}$ projection. This
approach redistributes the error throughout the element, in contrast
to the approach based on interpolation of the trial space~\citep{Atk96,Atk98}.
In the case of multi-component flows, one might seek to minimize the
error associated with the pressure projection for the evaluation of
the flux coefficients. In this case, we must consider the temperature
projection, which interpolates onto the span $\left\{ \varphi_{1},\ldots,\varphi_{m}\right\} $
\begin{equation}
\Pi\left(T\right)=\sum_{m}T\left(y\left(x_{m}\right)\right)\varphi_{m},\label{eq:temperature-projection-1}
\end{equation}
and species concentrations, which interpolate onto the span of $\left\{ \varphi_{1},\ldots,\varphi_{s}\right\} $
\begin{equation}
C_{i}=\sum_{s}C_{i}\left(x_{s}\right)\varphi_{s},\label{eq:concentrations-projection-1}
\end{equation}
where $m\ge s$. The pressure, the projection of which is onto the
span $\left\{ \varphi_{1},\ldots,\varphi_{r}\right\} $
\begin{equation}
\Pi\left(p\right)=\sum_{r}p\left(y\left(x_{r}\right)\right)\varphi_{r},\label{eq:pressure-projection-1}
\end{equation}
is defined via~(\ref{eq:EOS-1}), and is given in terms of the sum
of the concentrations, which is computed as
\begin{eqnarray}
\sum_{i}C_{i} & = & \sum_{i}\sum_{s}C_{i}\left(x_{s}\right)\varphi_{s}\label{eq:temperature-projection-1-1-1}\\
 & = & \left(\sum_{s}C_{1}\left(x_{s}\right)\varphi_{s}+\cdots+\sum_{s}C_{ns}\left(x_{s}\right)\varphi_{s}\right)\\
 & = & \left(C_{1}\left(x_{1}\right)\varphi_{1}+\cdots+C_{1}\left(x_{s}\right)\varphi_{s}+\cdots+C_{ns}\left(x_{1}\right)\varphi_{1}+\cdots+C_{ns}\left(x_{s}\right)\varphi_{s}\right)\\
 & = & \left(C_{1}\left(x_{1}\right)\varphi_{1}+\cdots+C_{n_{s}}\left(x_{1}\right)\varphi_{1}+\cdots+C_{1}\left(x_{s}\right)\varphi_{s}+\cdots+C_{ns}\left(x_{s}\right)\varphi_{s}\right)\\
 & = & \left(\left(C_{1}\left(x_{1}\right)+\cdots+C_{n_{s}}\left(x_{1}\right)\right)\varphi_{1}+\cdots+\left(C_{1}\left(x_{s}\right)+\cdots+C_{ns}\left(x_{s}\right)\right)\varphi_{s}\right)\\
 & = & \sum_{s}\left(\sum_{i}C_{i}\left(x_{s}\right)\right)\varphi_{s}
\end{eqnarray}
 Furthermore, we introduce a scalar valued test function that interpolates
onto the span of $\left\{ \varphi_{1},\ldots,\varphi_{q}\right\} $
\begin{align}
v & =\sum_{q}v\left(x_{q}\right)\varphi_{q}.\label{eq:test-function-interpolation}
\end{align}
Finally, the following linear system can be solved in order to determine
the expansion coefficients of pressure, $p\left(y\left(x_{r}\right)\right)$,
\begin{equation}
\left(\sum_{r}p\left(y\left(x_{r}\right)\right)\varphi_{r},\sum_{q}v\left(x_{q}\right)\varphi_{q}\right)_{\Omega}=\left(R^{o}\sum_{m}T\left(y\left(x_{m}\right)\varphi_{m}\right)\sum_{s}\left(\sum_{i}C_{i}\left(x_{s}\right)\varphi_{s}\right),\sum_{q}v\left(x_{q}\right)\varphi_{q}\right)_{\Omega}.\label{eq:pressure-projection-linear-system}
\end{equation}
By factoring out the constants and rearranging the sums, we can write
the resulting system as
\begin{equation}
\sum_{q}v\left(x_{q}\right)\left(\sum_{r}p\left(y\left(x_{r}\right)\right)\left(\varphi_{r},\varphi_{q}\right)_{\Omega}\right)=\sum_{q}v\left(x_{q}\right)\left(R^{o}\sum_{m}T\left(y\left(x_{m}\right)\right)\sum_{s}\left(\sum_{i}C_{i}\left(x_{s}\right)\right)\left(\varphi_{m}\varphi_{s},\varphi_{q}\right)_{\Omega}\right).\label{eq:pressure-projection-linear-system-factored}
\end{equation}
This approach is attractive because it will maintain equilibrium within
an element. However, if the temperature or the species concentrations
vary within the element, the value of pressure can be modified away
from the equilibrium value as shown in Figure~\ref{fig:pressure-equilibrium}.
As such, pressure equilibrium between adjacent elements will be lost
and unphysical pressure oscillations will be generated. 

Since both approaches described above result in the loss of equilibrium,
either with an element or between adjacent elements, we must consider
alternative approaches. In the case of multi-component flows, we have
found the simplest approach to work best. We project the pressure
such that equilibrium is maintained, i.e., into the same space as
the conserved state, $y$. Therefore, the pressure projection interpolates
onto the span $\left\{ \varphi_{1},\ldots,\varphi_{s}\right\} $
\begin{equation}
\Pi\left(p\right)=\sum_{s}p\left(y\left(x_{s}\right)\right)\varphi_{s}.\label{eq:pressure-projection}
\end{equation}
Furthermore, the modified flux projection interpolates onto the span
of $\left\{ \varphi_{1},\ldots,\varphi_{r}\right\} $
\begin{equation}
\left(\Pi\left(\mathcal{F}\left(y\right)\right)\right)_{ij}=\sum_{r}\mathcal{F}_{ij}\left(\tilde{y}\left(x_{r}\right)\right)\varphi_{r},\label{eq:flux-projection-modified}
\end{equation}
where $r\ge s$, and it is now evaluated in terms of the modified
state $\tilde{y}:\mathbb{R}^{m}\times\mathbb{R}\rightarrow\mathbb{R}^{m}$,
which is defined as
\begin{equation}
\tilde{y}\left(y,p\right)=\left(\rho v_{1},\ldots,\rho v_{d},\tilde{\rho u}+\frac{1}{2}\sum_{k=1}^{d}\rho v_{k}v_{k},C_{i},\ldots,C_{n_{s}}\right).\label{eq:interpolated-state-modified}
\end{equation}
The internal energy, $\tilde{\rho u}$ , is given in terms of the
pressure and the unmodified concentrations as
\begin{equation}
\tilde{\rho u}=\sum_{i=1}^{n_{s}}W_{i}C_{i}\sum_{k=0}^{n_{p}}a_{ik}\left(\frac{p}{R^{o}\sum_{i=1}^{n_{s}}C_{i}}\right)^{k}.\label{eq:internal_energy_polynomial-projection}
\end{equation}
In this way, consistency between the volume and surface terms is preserved
as well as pressure equilibrium both internally and between adjacent
elements. A similar conclusion was reached in the context of reconstructed
finite volume methods where it was found necessary to reconstruct
the primitive variables, which include velocity and pressure, in order
to maintain equilibrium at an isolated material or contact discontinuity~\citep{Joh06,Joh11}.
However, in the case of the formulation presented in this work, the
above analysis applies to all regions of the flow, as the discussion
was not restricted to the case of an isolated discontinuity, where
equilibrium is trivially maintained since the primitive variables
are readily available at the interface, see Section~\ref{sec:discrete-pressure-equilibrium}.

\subsubsection{Artificial Viscosity\label{subsec:artificial-viscosity}}

In order to stabilize the solution at physical discontinuities, we
augment the nonlinear flux locally on each cell with a residual based
artificial viscosity similar to~\citep{Har13}, of the form
\begin{equation}
\mathcal{F}^{a\nu}\left(y,\nabla y\right)=\left(S_{av}\left(p\right)+\left(C_{av}\tilde{h}_{k}^{2}\left|\frac{\partial T}{\partial y}\cdot\frac{\mathcal{R}\left(y,\nabla y\right)}{T}\right|\right)\right)\nabla y,\label{eq:artifical-viscosity}
\end{equation}
here $\tilde{h}_{k}=\nicefrac{h_{k}}{\left(p+1\right)}$ is evaluated
using the characteristic length scale of the element, $h_{k}$, and
the polynomial degree, $p$, of the element, $\mathcal{R}\left(y,\nabla y\right)$
is the strong form of the residual~(\ref{eq:conservation-law-strong-form}),
$T'=\nicefrac{\partial T}{\partial y}$ is the Jacobian of temperature
with respect to the state variables, see~\ref{sec:Source-term-Jacobian},
$C_{av}$ is a user defined coefficient, and $S_{av}$ is a pressure
dependent shock sensor. The exact form of the $S_{av}$ used in this
work is given by Ching et al.~\citep{Chi19}, which is based on the
shock sensor originally described by Persson and Peraire~\citep{Per06}.

\subsection{DGODE: an $hp$-adaptive discontinuous Galerkin method for ordinary
differential equation integration\label{subsec:dg-ode}}

We present an $hp$-adaptive discontinuous Galerkin method for the
integration of potentially stiff ordinary differential equations,
termed DGODE. Consider the case of a one-dimensional nonlinear ODE
integration, given in strong form, defined for piecewise smooth, $\mathbb{R}^{m}$-valued,
functions $y$, as
\begin{align}
y'-\mathcal{S}\left(y\right)=0 & \textup{ in }\kappa\qquad\forall\kappa\in\mathcal{T},\label{eq:ode-strong-form}\\
y\left(t_{0}\right)-y_{0}=0 & \textup{ on }\epsilon\qquad\forall\epsilon\in\mathcal{E}_{0},\label{eq:ode-initial-condition}
\end{align}
where we have decomposed the exterior interfaces $\mathcal{E}_{\partial}$
into disjoint subsets of inflow and outflow interfaces $\mathcal{E}_{\partial}=\mathcal{E}_{\text{0}}\cup\mathcal{E}_{f}$
over which a boundary state, $y_{\partial}\left(y^{+}\right)$, which
may be a function of the interior state $y^{+}$, is defined. At the
inflow interface $\epsilon_{\text{0}}\in\mathcal{E}_{0}$ located
at $t=t_{0}$, the inflow boundary condition~(\ref{eq:ode-initial-condition})
is applied, i.e., $y_{\partial}\left(y^{+}\right)=y_{0}$. No boundary
condition is imposed at the outflow interface $\epsilon_{f}\in\mathcal{E}_{f}$
located at $t=t_{f}$, i.e., $y_{\partial}\left(y^{+}\right)=y^{+}$.

\subsubsection{Weak formulation \label{subsec:dg-ode-weak-form}}

Let $V$ be vector-valued Sobolev space,
\begin{eqnarray}
V & = & \left[H^{1}\left(\mathcal{T}\right)\right]^{m}=\left\{ y\in\left[L^{2}\left(\Omega\right)\right]^{m}\bigl|\forall\kappa\in\mathcal{T},\left.y\right|_{\kappa}\in\left[H^{1}\left(\kappa\right)\right]^{m}\right\} ,\label{eq:dg-ode-solution-space-vector}
\end{eqnarray}
defined over a mesh $\mathcal{T}$. The weak formulation is obtained
by integrating the conservation law~(\ref{eq:ode-strong-form}) against
a test function and integrating by parts: find $y\in V$ such that
\begin{gather}
N\left(y,v\right)=\sum_{\epsilon\in\mathcal{E}}\left(h\left(y,n\right),\left\llbracket v\right\rrbracket \right)_{\mathcal{E_{I}}}-\sum_{\kappa\in\mathcal{T}}\left(y,\nabla v\right)_{\kappa}-\sum_{\kappa\in\mathcal{T}}\left(\mathcal{S}\left(y\right),v\right)_{\kappa}=0\qquad\forall v\in V,\label{eq:dgode-weak-form-integrated-by-parts}
\end{gather}
where $h\left(y,n\right)$ is the numerical flux function and the
jump operator is defined by Equations~(\ref{eq:jump-interior}) and~(\ref{eq:jump-exterior})
on interior and exterior interfaces respectively.

On interior interfaces the numerical flux is defined as the upwind
numerical flux for linear advection based on the normal velocity $n\cdot v$,
\begin{align}
h\left(y,n\right)=h\left(y^{+},y^{-},n\right)=\begin{cases}
\left(n\cdot v\right)y^{-} & \textup{ if }n\cdot v<0\\
\left(n\cdot v\right)y^{+} & \textup{ if }n\cdot v\geq0
\end{cases}, & \textup{ on }\epsilon\qquad\forall\epsilon\in\mathcal{E_{I}},\label{eq:upwind-numerical-flux}
\end{align}
where the velocity is defined as $v=\left(1\right)$ and is constant
throughout the domain.

On exterior interfaces the numerical flux is defined as
\begin{align}
h\left(y,n\right)=h_{\partial}\left(y^{+},n\right)=\begin{cases}
\left(n\cdot v\right)y_{\partial}\left(y^{+}\right) & \textup{ if }n\cdot v<0\\
\left(n\cdot v\right)y^{+} & \textup{ if }n\cdot v\geq0
\end{cases}, & \textup{ on }\epsilon\qquad\forall\epsilon\in\mathcal{E}_{\partial}.\label{eq:upwind-numerical-flux-boundary}
\end{align}
where $y_{\partial}\left(y^{+}\right)$ is a prescribed boundary state,
which may or may not depend on $y^{+}$, the interior trace of $y$.

\subsubsection{Discretization \label{subsec:dg-ode-discretization}}

In order to discretize Equation~(\ref{eq:dgode-weak-form-integrated-by-parts}),
we restrict $V$ to a discrete (finite-dimensional) subspace $V_{h}^{p}$,
as defined by Equation~(\ref{eq:discrete-subspace-simplex}), so
that the discretized weak formulation is: find $y\in V_{h}^{p}$ such
that

\begin{gather}
N_{h}\left(y,v\right)=\sum_{\epsilon\in\mathcal{E}}\left(h\left(y,n\right),\left\llbracket v\right\rrbracket \right)_{\mathcal{E}}-\sum_{\kappa\in\mathcal{T}_{h}}\left(y,\nabla v\right)_{\kappa}-\sum_{\kappa\in\mathcal{T}_{h}}\left(\mathcal{S}\left(y\right),v\right)_{\kappa}=0\qquad\forall v\in V_{h}^{p}.\label{eq:dg-ode-weak-form-discretized}
\end{gather}

Furthermore, let $n=\dim V_{h}^{p}$ and $\left(\varphi_{1},\ldots,\varphi_{n}\right)$
be a basis for $V_{h}^{p}$, then the discrete residual $\left(\mathcal{R}_{1}\left(y\right),\ldots,\mathcal{R}_{n}\left(y\right)\right)\in\mathbb{R}^{n}$
is defined by
\begin{equation}
\mathcal{R}_{i}\left(y\right)=\sum_{\epsilon\in\mathcal{E}}\left(h\left(y,n\right),\left\llbracket \varphi_{i}\right\rrbracket \right)_{\mathcal{E}}-\sum_{\kappa\in\mathcal{T}}\left(\mathcal{F}\left(y\right),\nabla\varphi_{i}\right)_{\kappa}-\sum_{\kappa\in\mathcal{T}}\left(\mathcal{S}\left(y\right),\varphi_{i}\right)_{\kappa}\label{eq:dg-ode-residual}
\end{equation}
for $i=1,\ldots,n$. 

We can therefore, write~\ref{eq:dg-ode-weak-form-discretized} as
\begin{equation}
\mathcal{R}\left(y\right)=0,\label{eq:dg-ode-residual-discretized-weak-formulation}
\end{equation}
and we can solve (\ref{eq:dg-ode-residual-discretized-weak-formulation})
iteratively via Newton's method: starting with an initial guess $y^{0}$,
we can solve for $y^{1},y^{2},\ldots$ by solving the linear system
of equations 
\begin{equation}
\mathcal{R}'\left[y^{k}\right]\left(y^{k+1}-y^{k}\right)=-\mathcal{R}\left(y^{k}\right)\label{eq:dg-ode-newtons-method}
\end{equation}
until a certain convergence criterion is satisfied and the application
of the Jacobian of the discrete residual to a function $\left(\mathcal{R}_{1}'\left[y\right]\delta y,\ldots,\mathcal{R}_{n}'\left[y\right]\delta y\right)\in\mathbb{R}^{n}$
is then given as 
\begin{equation}
\mathcal{R}_{i}'\left[y\right]w=\sum_{\epsilon\in\mathcal{E}}\left(h_{y}\left(y,n\right)\delta y,\left\llbracket \varphi_{i}\right\rrbracket \right)_{\mathcal{E}}-\sum_{\kappa\in\mathcal{T}}\left(\delta y,\nabla\varphi_{i}\right)_{\kappa}-\sum_{\kappa\in\mathcal{T}}\left(\mathcal{S}'\left(y\right)\delta y,\varphi_{i}\right)_{\kappa}\label{eq:dg-ode-residual-jacobian}
\end{equation}
where the $\mathcal{S}'\left(y\right)\delta y$ is the Jacobian of
the chemical source term, the derivation of which is given in~\ref{sec:Source-term-Jacobian}.

The DG discretization converges optimally with order $O\left(h^{p+1}\right)$
in the $L_{2}\left(\Omega\right)$ norm, i.e., $\left\Vert y-y_{\mathrm{exact}}\right\Vert _{L_{2}\left(\Omega\right)}$,
when no interior faces are present in the domain, but it converges
sub-optimally with order $O\left(h^{p+\nicefrac{1}{2}}\right)$ in
the $L_{2}\left(\Omega\right)$ norm due to the numerical flux at
interior faces. Furthermore, since the discretization is adjoint consistent,
we expect super-optimal convergence of the order $2p+1$ for a functional
output~\citep{Har13}. In the context of ODE integration, we are
interested in the behavior of the error of the terminal condition,
i.e., $\left|y\left(t_{f}\right)-y_{\mathrm{exact}}\left(t_{f}\right)\right|$
and can therefore expect the error to exhibit $O\left(h^{2p+1}\right)$
super-convergence~\citep{Bau95,Har13}. Finally, DGODE is stiffly
A-stable for all polynomial degrees $p$~\citep{Del81,Bau95}.

\subsubsection{$hp$-adaptivity \label{subsec:dg-ode-hp-adaptivity}}

We adaptively control both the time step, $h$, and the degree of
the polynomial approximation, $p$, to ensure efficient integration
of the nonlinear system and that the solution achieved is within a
desired accuracy. The the norm of the local error estimate
\begin{equation}
\mathrm{err}_{h}=\left\Vert \frac{\mathcal{R}_{i}\left(y\right)}{\epsilon_{\mathrm{abs}}+\epsilon_{\mathrm{rel}}\left|y_{i}\right|}\right\Vert \label{eq:dg-ode-error}
\end{equation}
is used to define the nonlinear convergence criteria, which is given
as
\begin{equation}
\mathrm{err}_{h}<1\label{eq:dg-ode-convergence-criteria}
\end{equation}
for $i=1,\ldots,n$, where $\epsilon_{\mathrm{abs}}=10^{-6}$ and
$\epsilon_{\mathrm{rel}}=10^{-12}$ are user specified absolute and
relative tolerances. If~(\ref{eq:dg-ode-convergence-criteria}) is
not satisfied within $n=5$ iterations, the time step is reduced by
a factor of ten. Otherwise, we estimate the accuracy of the solution
and adapt the polynomial degree of the approximation accordingly.

In order to determine if adaptation of the polynomial degree is necessary,
we apply a residual based a posteriori error estimation~\citep{Har13}.
This approach has proved to be simple and robust. The the residual
based error estimate is defined in terms of the interior cell residual
and the interface residual:
\begin{eqnarray}
\mathcal{R}_{i}^{p}\left(y\right) & = & \sum_{\kappa\in\mathcal{T}}\left(\nabla\cdot\mathcal{F}\left(y\right),\varphi_{i}\right)_{\kappa}-\sum_{\kappa\in\mathcal{T}}\left(\mathcal{S}\left(y\right),\varphi_{i}\right)_{\kappa},\label{eq:dg-ode-residual-p-adaptation-cell}\\
r_{i}^{p}\left(y\right) & = & \sum_{\epsilon\in\mathcal{E_{\partial}}}\left(h_{\partial}\left(y^{+},n\right),\varphi_{i}^{+}\right)_{\mathcal{E}_{\partial}}+\sum_{\epsilon\in\mathcal{E_{I}}}\left(h\left(y^{+},y^{-},n\right),\left\llbracket \varphi_{i}\right\rrbracket \right)_{\mathcal{E_{I}}}.\label{eq:dg-ode-residual-p-adaptation-boundary}
\end{eqnarray}
In the case of the single-element, $\Omega=\kappa=\left(t_{0},t_{f}\right)$,
spectral discretization, Equation~(\ref{eq:dg-ode-residual-p-adaptation-boundary})
reduces to
\begin{equation}
r_{i}^{p}\left(y\right)=\left(\left(y_{0}-y^{+}\right),\varphi_{i}^{+}\right)_{\mathcal{E}_{0}},\label{eq:dg-ode-residual-p-adaptation-boundary-one-element}
\end{equation}
where the only interface contribution is due to the inflow boundary
condition.

The norm of the local error estimate for $p$-refinement is
\begin{equation}
\mathrm{err}_{p}=\left\Vert \mathcal{R}_{i}^{p}\left(y\right)\right\Vert _{L^{2}\left(\Omega\right)}+\left\Vert r_{i}^{p}\left(y\right)\right\Vert _{L^{2}\left(\Omega\right)},\label{eq:dg-ode-error-p}
\end{equation}
and the accuracy criteria is
\begin{equation}
\mathrm{err}_{p}<\epsilon_{p},\label{eq:dg-ode-convergence-criteria-p}
\end{equation}
where $\epsilon_{p}=10^{-6}$ is a user specified tolerance.

If the nonlinear solver converges, i.e. ~(\ref{eq:dg-ode-convergence-criteria})
is satisfied within $n=5$ iterations, and~(\ref{eq:dg-ode-convergence-criteria-p})
is satisfied, the solution is updated and a new time-step is determined
using Gustafsson's method~\citep{Hai96}. Otherwise, the polynomial
degree of the approximation is refined until a user specified maximum
polynomial degree is reached. Algorithm~\ref{alg:DGODE} summarizes
the $hp$-adaptive DGODE method.
\begin{algorithm}[H]
\caption{\label{alg:DGODE}DGODE: $hp$-adaptive discontinuous Galerkin method
for ordinary differential equation integration.}
\begin{algorithmic}[1]
\While{$t<t_f$}
\State Initialize the one-dimensional space-time solution $y$ by extruding the initial condition $y_0$.
\For{$k=1,\cdots,n$}
\State Solve $\mathcal{R}'\left[ y^k \right]\delta y=-\mathcal{R}\left(y^k \right)$ for $\delta y$.
\State Increment the solution $y^{k+1} \mathrel{+}= \delta y$.
\If{$\left\Vert \frac{\mathcal{R}\left( y^k \right)}{\epsilon_{\mathrm{abs}} + \epsilon_{\mathrm{rel}}\left| y^k\right|} \right\Vert < 1$}
\State Converged return;
\EndIf
\EndFor
\If{converged}
\If{$\left( \left\Vert \left( \nabla y - S\left( y\right), v \right)_{\kappa} \right\Vert < \epsilon_p \right)$ or  $\left( p\mathrel{=}= p_{\mathrm{max}} \right)$}
\State Restrict the space-time solution $y_0=y\left( t+\delta t \right)$.
\State increment $t\mathrel{+}=\delta t$.
\State Compute a new time step $\delta t$.
\Else
\State $p$-refine the space-time solution $y$ and goto 3.
\EndIf
\Else
\State reduce the time step $\delta t$.
\EndIf
\EndWhile
\end{algorithmic}
\end{algorithm}

\subsubsection{Convergence of the ordinary differential equation integration under
$hp$-refinement \label{subsec:dg-ode-convergence}}

We apply the DGODE discretization to Equation~(\ref{eq:ode-strong-form})
and study convergence under $hp$-refinement. In this case, we consider
the initial value problem defined for the piecewise smooth, $\mathbb{R}^{1}$-valued,
function $y$ as

\begin{align}
y'-\lambda y=0 & \textup{ in }\left(0,1\right),\label{eq:ode-strong-example}\\
y\left(0\right)-1=0 & \textup{ on }0,\label{eq:ode-initial-condition-example}
\end{align}
where $\lambda=-1$.

\begin{figure}[H]
\subfloat[\label{fig:dg-ode-h-convergence}Convergence with respect to grid
$\left(h\right)$ refinement. The coarsest grid consisted of one linear
line cell and the finest grid consisted of $64$ linear line cells.
The DGODE discretization converges with the expected super-optimal
($2p+1$) order for a functional output.]{\begin{centering}
\includegraphics[width=0.45\linewidth]{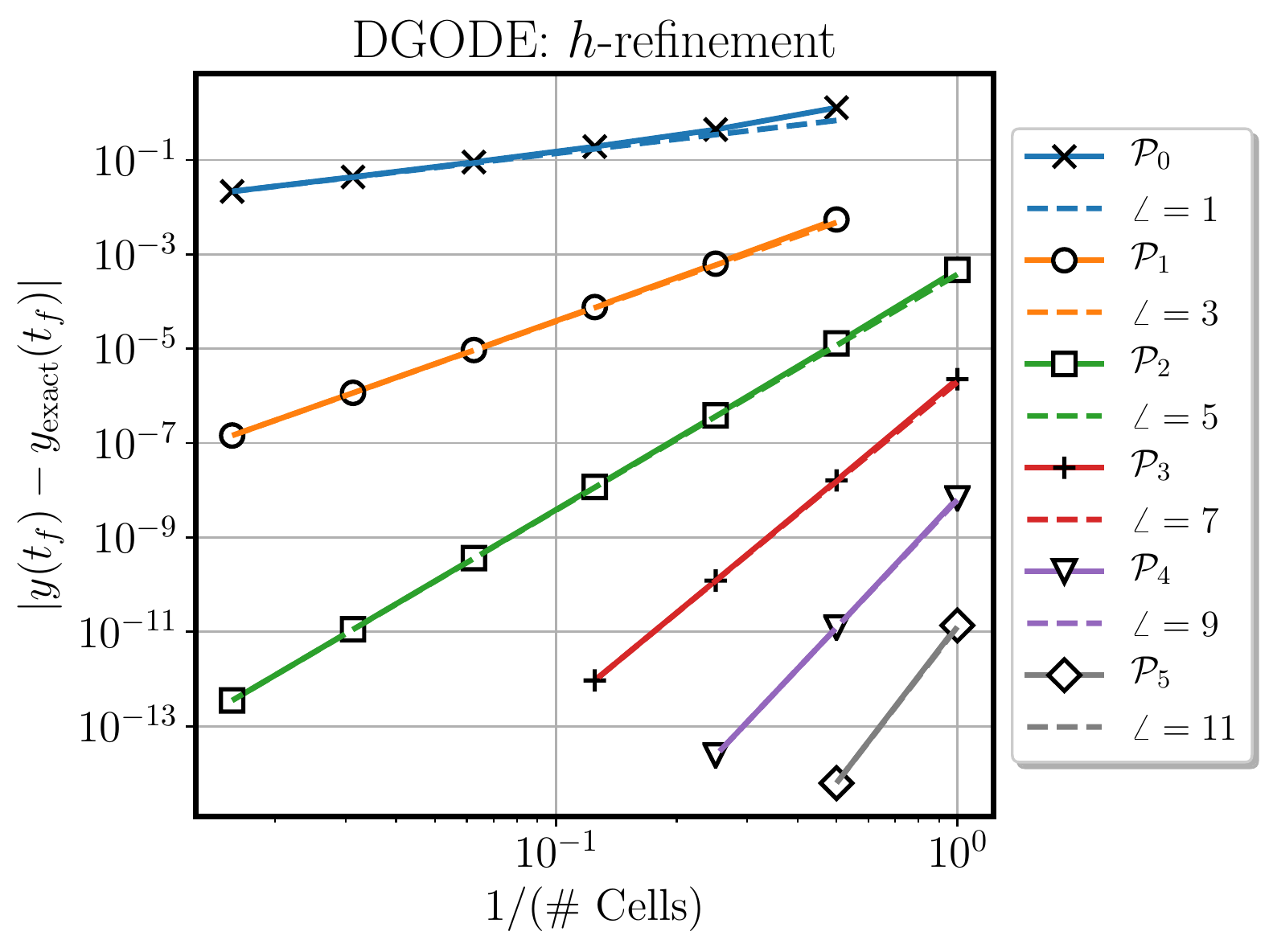}
\par\end{centering}
}\hfill{}
\begin{centering}
\subfloat[\label{fig:dg-ode-p-convergence}Convergence with respect to polynomial
$\left(p\right)$ refinement for grids consisting of $2$, $4$, and
$8$ linear line cells.]{\begin{centering}
\includegraphics[width=0.45\linewidth]{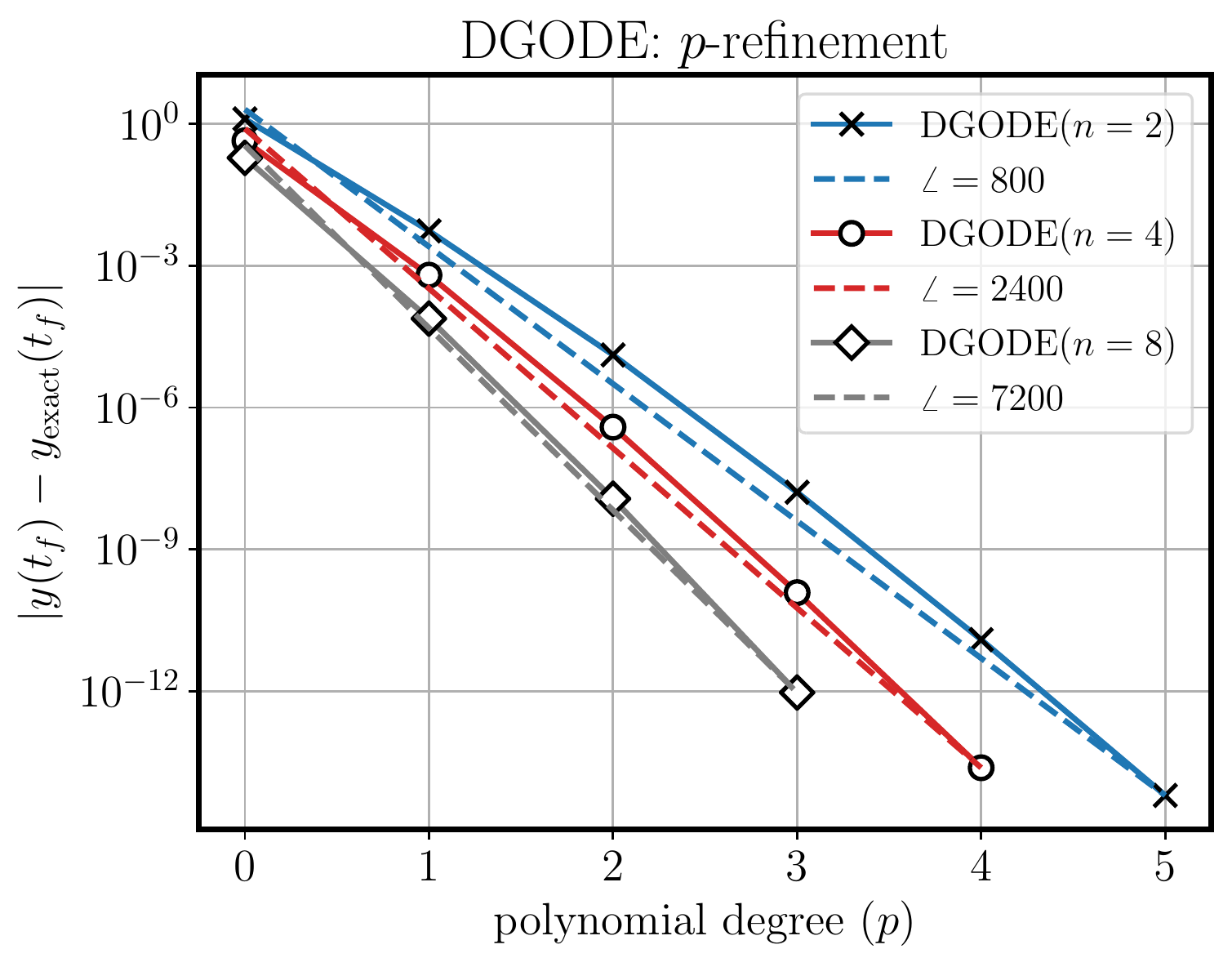}
\par\end{centering}
}
\par\end{centering}
\caption{\label{fig:dg-ode-convergence}Convergence plots for the terminal
condition, $\left|y\left(t_{f}\right)-y_{\mathrm{exact}}\left(t_{f}\right)\right|$,
with respect to both grid $\left(h\right)$ and polynomial $\left(p\right)$
refinement. The exact solution at $t_{f}=1$, given by Equation~(\ref{eq:ode-exact}),
is $\exp\left(-1\right)$.}
\end{figure}

The exact solution is given as

\begin{equation}
y\left(t\right)=\exp\left(\lambda t\right).\label{eq:ode-exact}
\end{equation}
Since the DG discretization is adjoint consistent, we expect super-optimal
convergence of the order $2p+1$ for a functional output. In the context
of ODE integration, the functional output we are most concerned with
is the $L_{\infty}$ error of the solution at the final time, i.e.,
the terminal condition: $\left|y\left(t_{f}\right)-y_{\mathrm{exact}}\left(t_{f}\right)\right|$.
Figure~\ref{fig:dg-ode-convergence} presents convergence plots for
the terminal condition corresponding to the exact solution with respect
to both grid, $h$, and polynomial, $p$, refinement. Figure~\ref{fig:dg-ode-h-convergence}
presents the convergence of the terminal condition under grid refinement
for polynomial degrees $p=0,\cdots,5$. The coarsest grid consisted
of one linear line cell and the finest grid consisted of $64$ linear
line cells. The DGODE discretization converges at the expected super-optimal
rate of $2p+1$.

Polynomial refinement is also an effective strategy for increasing
the accuracy of the approximation. Figure~\ref{fig:dg-ode-p-convergence}
presents the convergence of the terminal condition under polynomial
refinement for grids consisting of $2$, $4$, and $8$ linear line
cells. For each grid resolution, the error associated with the terminal
condition decreases exponentially. 

\subsubsection{Ordinary differential equation integration of a homogeneous reactor\label{subsec:homogeneous-reactor}}

\begin{figure}[H]
\subfloat[\label{fig:homogeneous-reactor-GRI-3.0-major-speicies}The mass fractions
of species with peak value $2\times10^{-2}<Y_{i}$ and adapted polynomial
degree, $p$(solid black line), computed with DGODE for the homogeneous
reactor corresponding to the GRI-3.0 mechanism ($\Diamond$) compared
with the mass fractions computed using Cantera (solid colored lines).]{\begin{centering}
\includegraphics[width=0.45\linewidth]{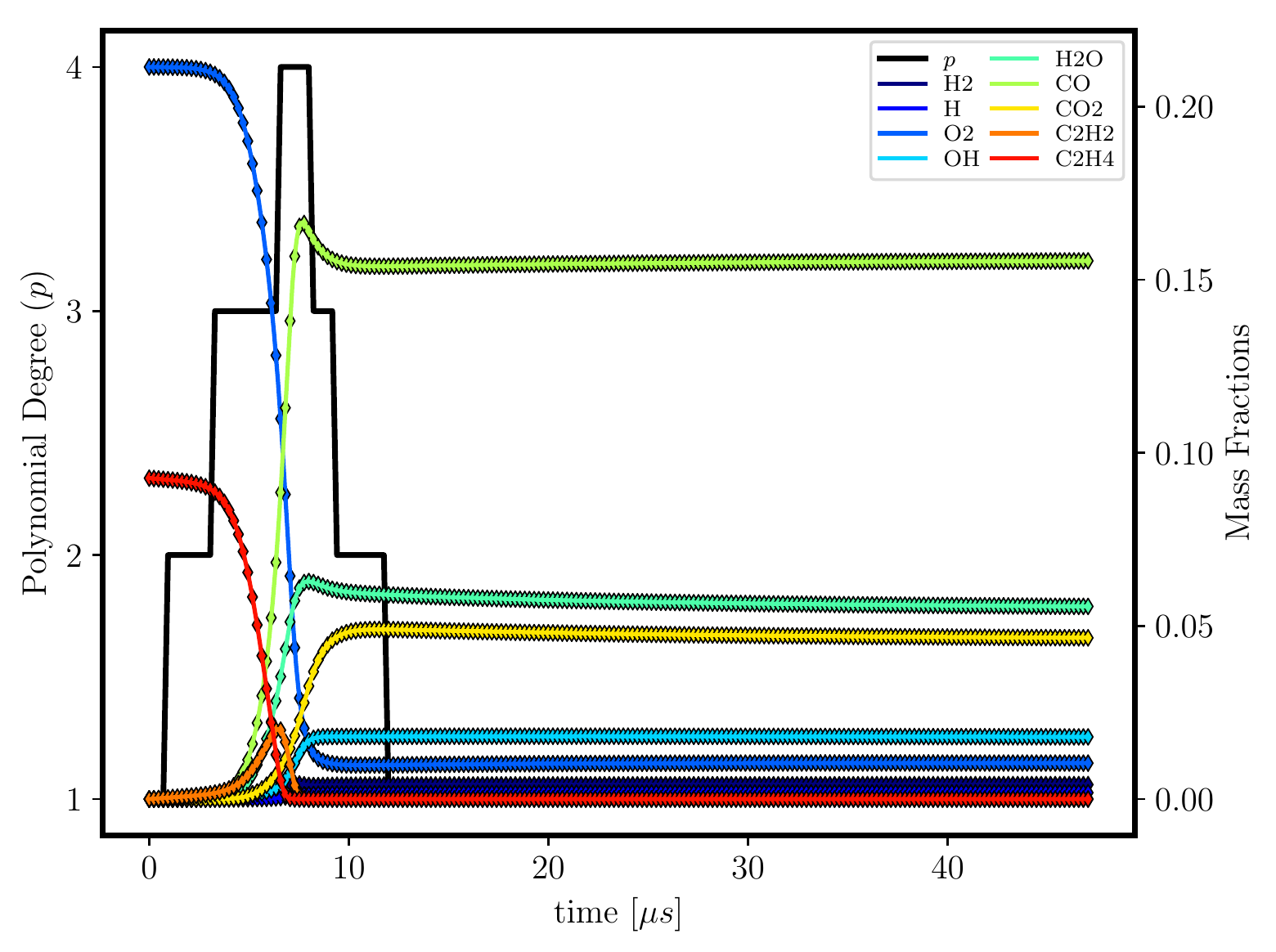}
\par\end{centering}
}\hfill{}\subfloat[\label{fig:homogeneous-reactor-GRI-3.0-speicies-minor}The mass fractions
of species $2\times10^{-5}<Y_{i}<2\times10^{-2}$ and adapted polynomial
degree, $p$(solid black line), computed with DGODE for the homogeneous
reactor corresponding to the GRI-3.0 mechanism ($\Diamond$) compared
with the mass fractions computed using Cantera (solid colored lines).]{\begin{centering}
\includegraphics[width=0.45\linewidth]{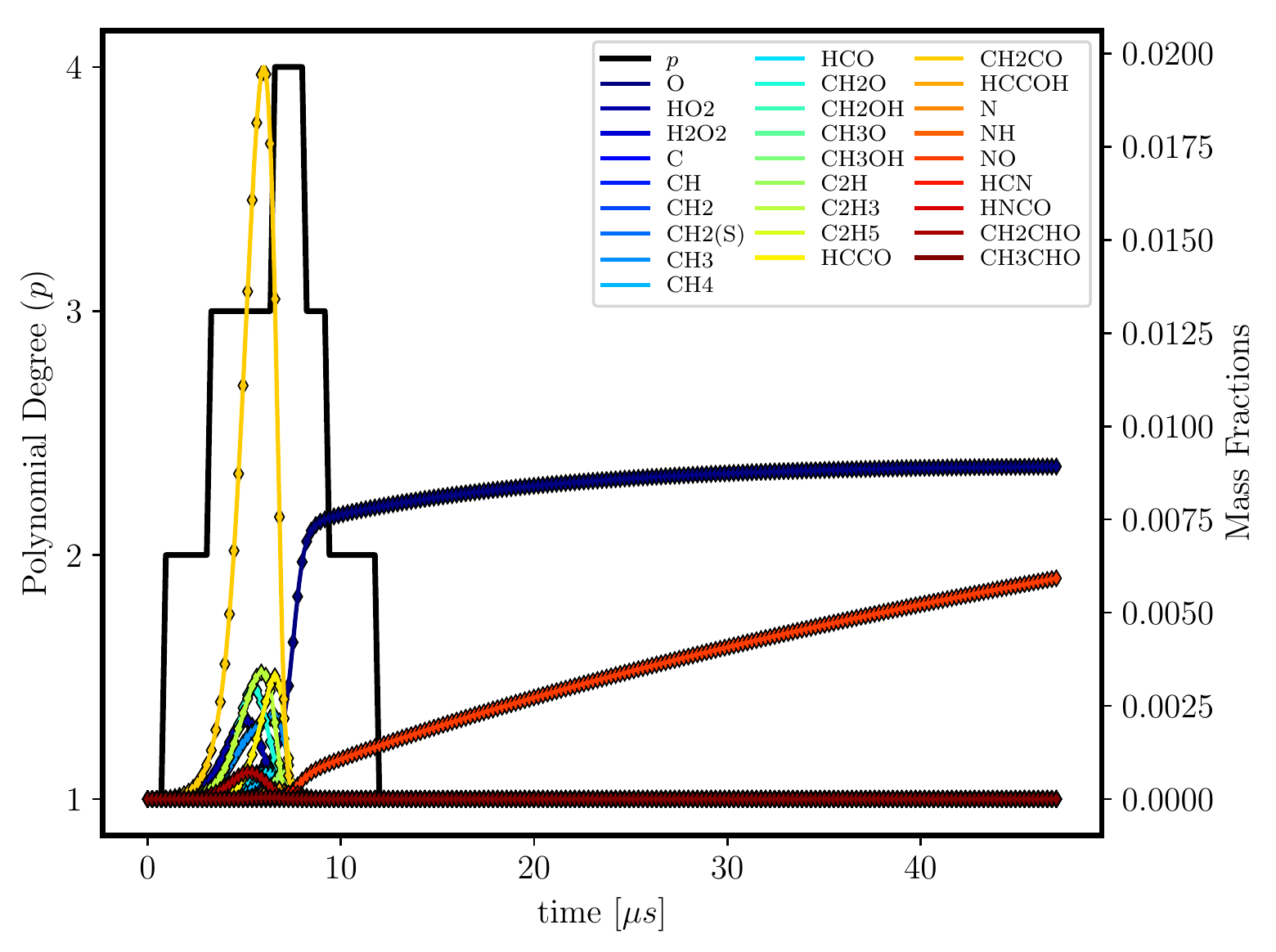}
\par\end{centering}
}\hfill{}\subfloat[\label{fig:homogeneous-reactor-GRI-3.0-minor-minor-speicies}The mass
fractions of species $Y_{i}<2\times10^{-5}$ and adapted polynomial
degree, $p$(solid black line), computed with DGODE for the homogeneous
reactor corresponding to the GRI-3.0 mechanism ($\Diamond$) compared
with the mass fractions computed using Cantera (solid colored lines).]{\begin{centering}
\includegraphics[width=0.45\linewidth]{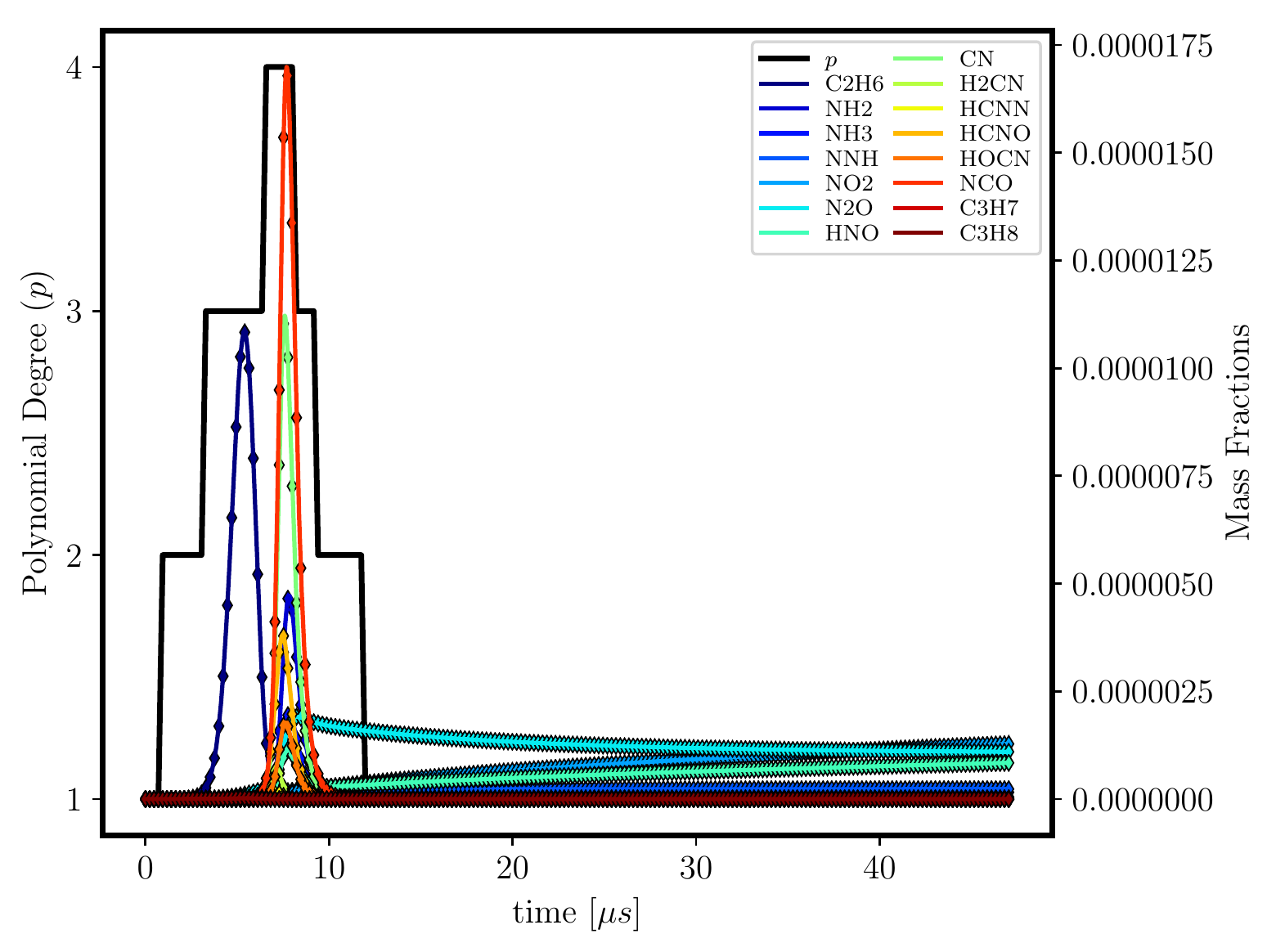}
\par\end{centering}
}\hfill{}\subfloat[\label{fig:homogeneous-reactor-GRI-3.0-temperature}The temperature
($\Diamond$) and adapted polynomial degree, $p$(solid black line),
computed with DGODE for the homogeneous reactor corresponding to the
GRI-3.0 mechanism compared with the temperature profile computed using
Cantera (solid green line).]{\begin{centering}
\includegraphics[width=0.45\linewidth]{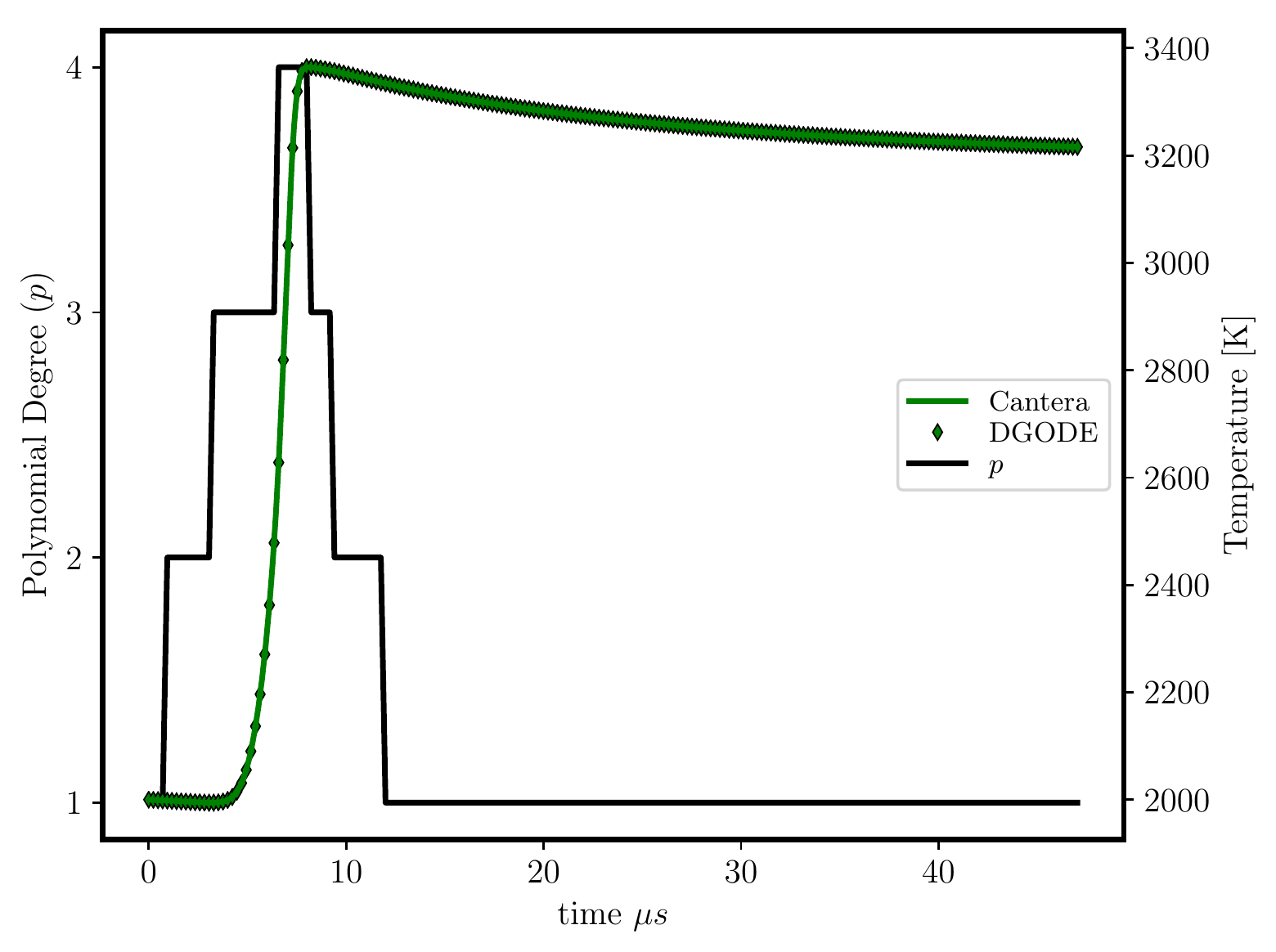}
\par\end{centering}
}

\caption{\label{fig:homogeneous-reactor-GRI-3.0-speicies}The DGODE solution
for mass fractions, temperature, and adapted polynomial degree for
homogeneous reactor corresponding to the GRI-3.0 mechanism. DGODE
automatically adapts the polynomial degree to accurately resolve the
disparate time-scales associated with the various species}
\end{figure}

We solve the one-dimensional system of ODEs corresponding to a homogeneous
reactor with detailed kinetics using the GRI-3.0 mechanism ~\citep{GRI30}
to demonstrate the ability of DGODE to employ $hp$-adaptivity to
ensure accurate and efficient integration of the stiff chemical system.
The mechanism consists of $53$ species and $325$ reactions. We initialized
the solution with an equivalence ratio of one, $\phi=1$, for ethylene-air
and an initial pressure, $p_{0}=1\,\mathrm{atm}$, and an initial
temperature, $T_{0}=2000\,\mathrm{K}$. Figures~\ref{fig:homogeneous-reactor-GRI-3.0-major-speicies}-\ref{fig:homogeneous-reactor-GRI-3.0-temperature}
show the results of the species mass fractions and temperature as
a function of time as well as the adapted polynomial degree compared
to results using Cantera~\citep{cantera}. The adapted polynomial
degree is the greatest in the region of highest chemical activity.
Almost all species are created and destroyed in regions where $p>1$.
As time progresses, all species reach an equilibrium with the exception
of $NO$, which is slowly created in regions of hot products. 

\begin{figure}[H]
\begin{centering}
\includegraphics[width=0.6\linewidth]{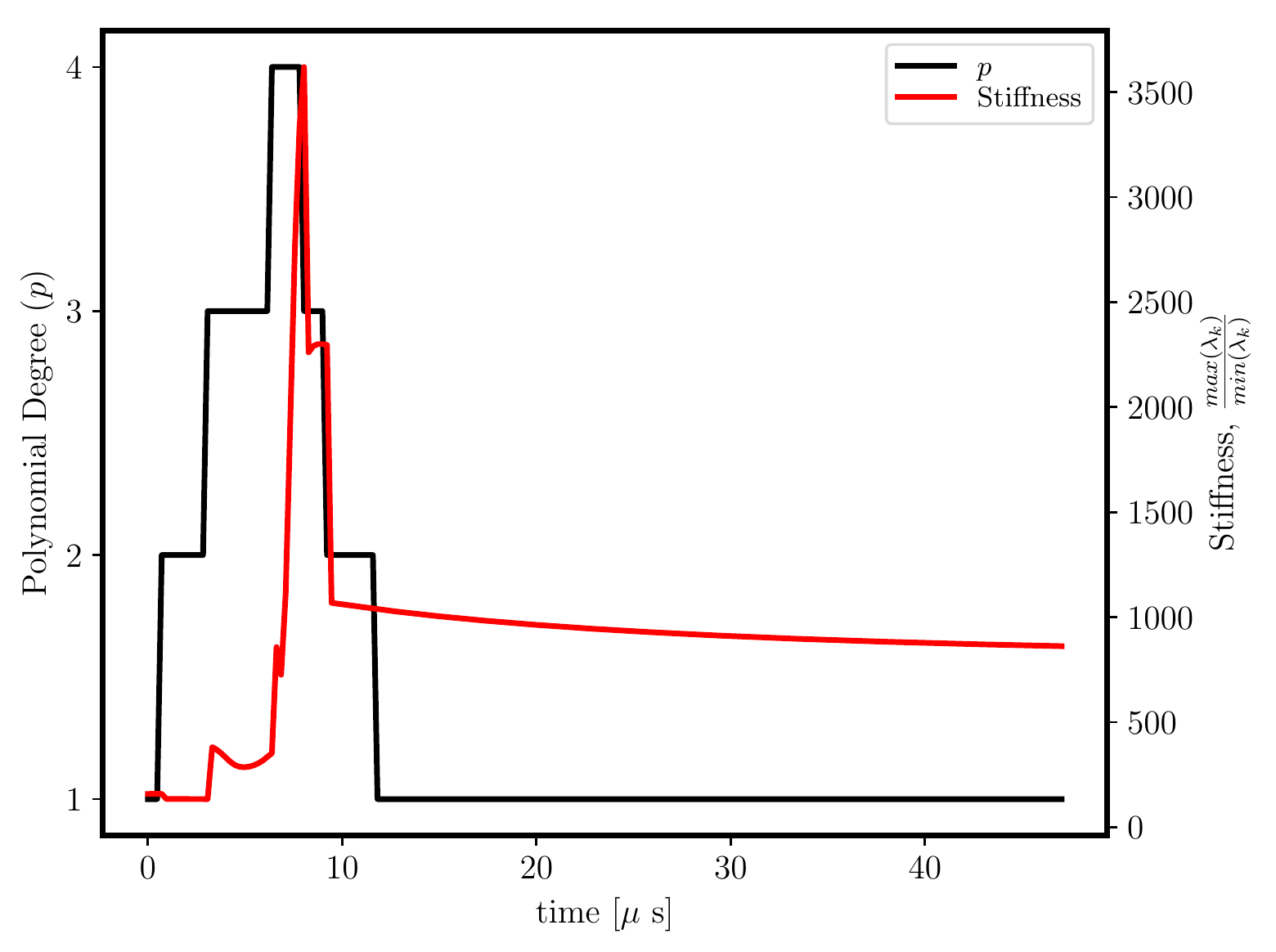}
\par\end{centering}
\caption{\label{fig:homogeneous-reactor-GRI-3.0-stiffness}Stiffness ratio
of the analytical Jacobian for the chemical source term corresponding
to the GRI-3.0 mechanism compared to the adapted polynomial degree
computed using DGODE for the homogeneous reactor. DGODE automatically
adapts the polynomial degree to accurately integrate the potentially
stiff chemical source term. High-order integration is only used in
the stiffest regions.}
\end{figure}

The stiffness associated with the analytical Jacobian of the chemical
source term is defined as the ratio of the absolute value of the the
real part of the largest and smallest eigenvalues,

\begin{equation}
\frac{\left|Re\left(\lambda_{\mathrm{max}}\right)\right|}{\left|Re\left(\lambda_{\mathrm{min}}\right)\right|},\label{eq:stiffness-ratio}
\end{equation}
where $\lambda_{\mathrm{max}}$ and $\lambda_{\mathrm{min}}$ are
the largest and smallest eigenvalues respectively of the source term
Jacobian, $\mathcal{S}'$, see~\ref{sec:Source-term-Jacobian}. Figure~\ref{fig:homogeneous-reactor-GRI-3.0-stiffness}
shows the stiffness from Equation~(\ref{eq:stiffness-ratio}) as
a function of time for the aforementioned homogeneous reactor as well
as the adapted polynomial degree, $p$. The adapted polynomial degree
is at a maximum in regions where the stiffness is greatest, indicating
that DGODE is efficiently integrating the source term using high-order
approximations when necessary.

\begin{figure}[H]
\subfloat[\label{fig:detonation-1d-temperature}The temperature profile compared
to the polynomial degree of the ODE integration for the one-dimensional
$C_{2}H_{4}$ and air detonation wave.]{\begin{centering}
\includegraphics[width=0.45\linewidth]{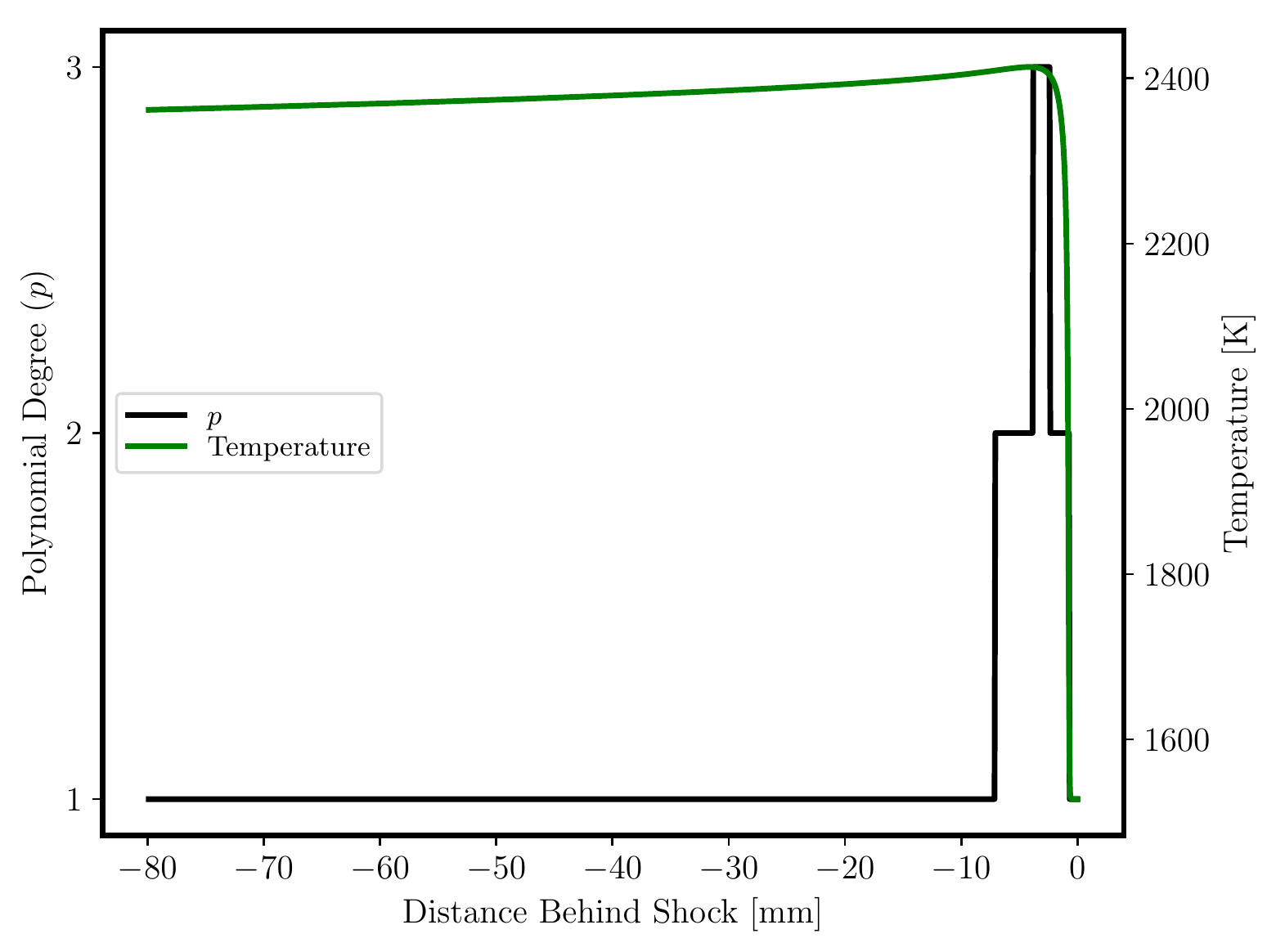}
\par\end{centering}
}\hfill{}
\begin{centering}
\subfloat[\label{fig:detonation-1d-stiffness}Stiffness associated with the
analytical Jacobian of the chemical source term corresponding to the
GRI-3.0 mechanism compared to the polynomial degree of the ODE integration.]{\begin{centering}
\includegraphics[width=0.45\linewidth]{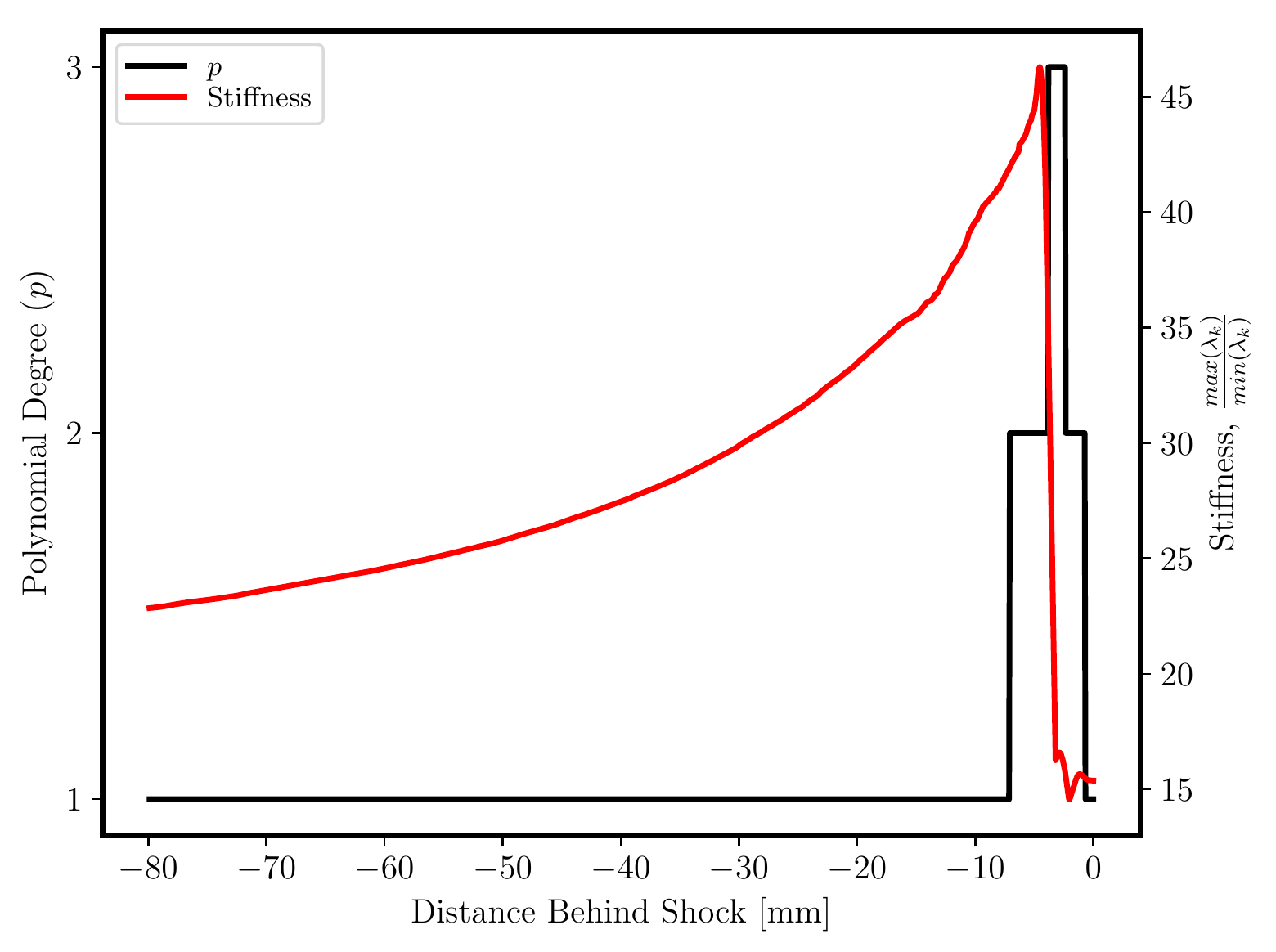}
\par\end{centering}
}
\par\end{centering}
\caption{\label{fig:detonation-1d}The temperature profile and stiffness, measured
as the ratio of the largest and smallest eigenvalue of the chemical
source term Jacobian corresponding to the GRI-3.0 mechanism compared
to the polynomial degree of the ODE integration for the one-dimensional
$C_{2}H_{4}$ and air detonation wave. DGODE automatically adapts
the polynomial degree to accurately integrate the potentially stiff
chemical source term. High-order integration is only used in the stiffest
regions in space.}
\end{figure}

We test the effectiveness of local polynomial adaptivity in the case
of a temporally evolving and spatially varying problem with stiff
chemistry by analyzing a one dimensional detonation wave propagating
through a premixed medium of ethylene-air at equivalence ratio of
one, $\phi=1$. The mixture ahead of the detonation is at a pressure
of $p=101325$ Pa and a temperature of $T=300$ K. The detonation
wave is moving from left to right. To the left of the shock, a region
of high temperature and pressure causes the mixture to react and sustain
the detonation. In a stable scenario, this detonation velocity is
steady in time, and the structures can be numerically predicted. For
this example, the profiles for species and the thermodynamic state
in the reacting region to left of the shock were obtained via the
Shock and Detonation Toolbox~\citep{sdtoolbox}. Figures~\ref{fig:detonation-1d-temperature}
and~\ref{fig:detonation-1d-stiffness} show the temperature and the
stiffness as well as the adapted polynomial degree for a time step
of $\Delta t=10^{-8}$ s using DGODE. As expected, the polynomial
degree of the approximation adapts to the stiffness of the chemical
system, which sharply increases to the left of the detonation front,
where heat release is the greatest, and decays as the distance from
the detonation front increases. 

The local adaptation of the polynomial order is also demonstrated
for time dependent results in combination with our method for simulating
multi-component fluid dynamics corresponding to hydrogen detonations
in Sections~\ref{subsec:One-dimensional-detonation-wave} and~\ref{subsec:Two-dimensional-detonation-wave}.
Finally, the polynomial degree of the local ODE approximation can
be used to accurately, and unambiguously, identify regions of stiff
chemical reactions. This is potentially a useful indicator for load
balancing across multiple processors, cf.~\citep{Ora98,Cuo13}.

\section{Discrete pressure equilibrium at material interfaces\label{sec:discrete-pressure-equilibrium}}

Various methods have been used to suppress unwanted pressure oscillations
in smooth regions of multi-component flows~\citep{Abg01,Bil03,Bil11,Hou11,Lv15}.
Jenny et al. summarized the conditions where pressure oscillations
do not exist in~\citep{Jen97}, which were restated in the Section~\ref{sec:Background}
of this work. Their conclusions also apply to the formulation presented
in this work, as detailed in~\ref{sec:Discontinuities}. Therefore,
we do not expect the conservative DG formulation to generate pressure
oscillations via numerical mixing of species concentrations at the
same temperature. Discontinuities in temperature will generate pressure
oscillations if the species internal energies are nonlinear with respect
to temperature. However, the magnitude of these oscillations is reduced
as the solution is better resolved and the oscillations do not cause
the numerical simulation to fail. In this section we study several
test cases in order to verify the results of~\ref{sec:Discontinuities},
which were derived in the context of simplified numerics.

To study the generation of unphysical pressure oscillations for the
formulation presented in Section~\ref{sec:Formulation}, we test
three different methods for evaluating the thermodynamic quantities
required to evaluate the convective and numerical fluxes. These methods
are listed in Table~\ref{tab:flux_evaluation}, where basic steps
for evolving the state temporally are listed. The first column corresponds
to the formulation with the exact thermodynamics considered in this
work and the second column corresponds to a formulation with frozen
thermodynamics where $\bar{\gamma}$ and $\bar{c}_{p}$ are held constant
throughout each time step. The formulation with exact thermodynamics
does not evaluate $\bar{c}_{p}$ in order to calculate the convective
flux, or numerical flux, however the speed of sound~(\ref{eq:speed-of-sound})
is required for the evaluation of the numerical flux. Instead the
nonlinear equation~(\ref{eq:newton_eq}) for temperature is solved
and pressure is then calculated via~(\ref{eq:EOS-1}). For the formulation
with frozen thermodynamics, we consider two methods for evaluating
the specific heat at constant pressure: the mean value $\bar{c}_{p}$
given by~(\ref{eq:cp_eff_enthalpy}) and directly evaluating $c_{p}$
given by~(\ref{eq:mixture_average_cp_polynomial}) based on the NASA
polynomials expressions.
\begin{table}[H]
\caption{\label{tab:flux_evaluation}Procedure for temporally evolving the
state.}

\centering{}%
\begin{tabular}{|c|>{\centering}p{1.75in}|>{\raggedright}p{1.5in}|>{\raggedright}p{1.5in}|}
\hline 
Step & Formulation with Exact Thermodynamics & \multicolumn{2}{>{\centering}p{3in}|}{Formulation with Frozen Thermodynamics}\tabularnewline
\hline 
\hline 
1 & Calculate pressure from~(\ref{eq:EOS-1}) using the temperature obtained
by solving ~(\ref{eq:internal_energy_conserved_state}). & \multicolumn{2}{>{\centering}p{3in}|}{Calculate pressure from~(\ref{eq:gamma_m_one_equivalent}) based
on frozen $\bar{\gamma}$.}\tabularnewline
\hline 
2 & Update state using $h_{HLLC}\left(y^{+},y^{-},n\right)$ and $\mathcal{F}_{k}^{c}\left(y\right)$ & \multicolumn{2}{>{\centering}p{3in}|}{Update state using $h_{HLLC}\left(y^{+},y^{-},n\right)$ and $\mathcal{F}_{k}^{c}\left(y\right)$.}\tabularnewline
\hline 
3 &  & Calculate and freeze the new $\bar{c}_{p}$ from~ (\ref{eq:cp_eff_enthalpy})
using the updated state and temperature, which is calculated from~(\ref{eq:EOS-1})
where the pressure is calculated via~(\ref{eq:gamma_m_one_equivalent})
using the frozen $\bar{\gamma}$ .  & Calculate and freeze the new $\bar{c}_{p}$ as equivalent to $c_{p}$
from ~(\ref{eq:mixture_average_cp_polynomial}) using the updated
state and temperature, which is calculated from~(\ref{eq:EOS-1})
where the pressure is calculated via~(\ref{eq:gamma_m_one_equivalent})
using the frozen $\bar{\gamma}$ . \tabularnewline
\hline 
4 &  & \multicolumn{2}{>{\centering}p{3in}|}{Calculate and freeze new $\bar{\gamma}$ from $\bar{c}_{p}$ in Step
3 and $R$ from the updated state. }\tabularnewline
\hline 
\end{tabular}
\end{table}

If $\bar{\gamma}$ is frozen as presented in the second column of
Table~\ref{tab:flux_evaluation}, then pressure oscillations will
occur as derived in~\citep{Kar94,Abg96}. This is problematic for
smooth regions of reacting multi-component flows, as it requires additional
methods for suppressing the unphysical pressure oscillations, even
in regions of the flow where temperature is continuous. Although,
we do not make use of it otherwise, the formulation with frozen thermodynamics
is considered in this section to demonstrate that it is potential
source of unphysical pressure oscillations. %

In Sections~\ref{subsec:species-discontinuity-constant-temperature}-\ref{subsec:p0_oscillations}
we consider the advection of material discontinuities for different
species and temperature profiles where the concentrations are comprised
of two fictitious species for reproducibility purposes. The initial
conditions all have the form
\begin{eqnarray}
\left(v,T,p,Y_{1},Y_{2}\right) & = & \begin{cases}
\left(10\textrm{ m/s},\:T,\:1\textrm{ atm},\:0,\:1\right) & \text{if }0.025<x<0.075\\
\left(10\textrm{ m/s},\:T,\:1\textrm{ atm},\:1,\:0\right) & \text{otherwise}
\end{cases},\label{eq:material-interface-initialization}
\end{eqnarray}
where $T$ is problem specific.

In Section~\ref{subsec:thermal-bubble} we revisit the case of a
slowly moving one dimensional hydrogen and oxygen thermal bubble previously
considered by~\citep{Lv15} where it was reported that conservative
schemes would generate unphysical pressure oscillations. For the cases
considered in this section, numerical instabilities are not suppressed
via artificial viscosity, limiting, or filtering in order to emphasize
the effect of numerical mixing on pressure in the absence of any additional
stabilization.

\subsection{Species discontinuities at constant temperature\label{subsec:species-discontinuity-constant-temperature}}

In this Section, and Sections~\ref{subsec:species-discontinuity-temperature-discontinuity}
and~\ref{subsec:p0_oscillations} as well, we approximate exact solutions
corresponding to material interfaces by solving the non-reacting,
inviscid formulation of Equations~(\ref{eq:conservation-law-strong-form})-(\ref{eq:conservation-law-flux-boundary-condition})
using $\mathrm{DG}(p=1)$, $\mathrm{DG}(p=2)$, and $\mathrm{DG}(p=3)$,
without artificial viscosity or limiting. For ease of reproducibility,
we constructed two fictitious species, $n_{s}=2$, with different
molecular weights, $W_{1}=20$ and $W_{2}=70$. Here we used a nonlinear
function for internal energy, with $n_{p}=3$, $a_{1k}=\{0,2.08\times10^{4},83.1,-2.77\times10^{-2}\}$
and $a_{2k}=\{0,3.33\times10^{4},83.1,-2.22\times10^{-2}\}$, of the
form $u_{i}=\sum_{k=0}^{n_{p}}a_{ik}T^{k}$ where $u_{i}$ is the
units J/kg with temperature, $T$, in K. The enthalpy of each species
is therefore $h_{i}=\sum_{k=0}^{n_{p}}a_{ik}T^{k}+R^{0}T$ and the
specific heat at constant pressure of each species is $c_{p,i}=\sum_{k=0}^{n_{p}}ka_{ik}T^{k-1}+R^{o}$.
For each test case the domain is $0.1$ m, with grid spacing of $h=0.002$
m. All reported results are using the formulation with exact thermodynamics
unless otherwise specified.

\begin{figure}[H]
\subfloat[\label{fig:species_discontinuity_constT_nl_p1}$\mathrm{DG}\left(p=1\right)$]{\includegraphics[width=0.32\columnwidth]{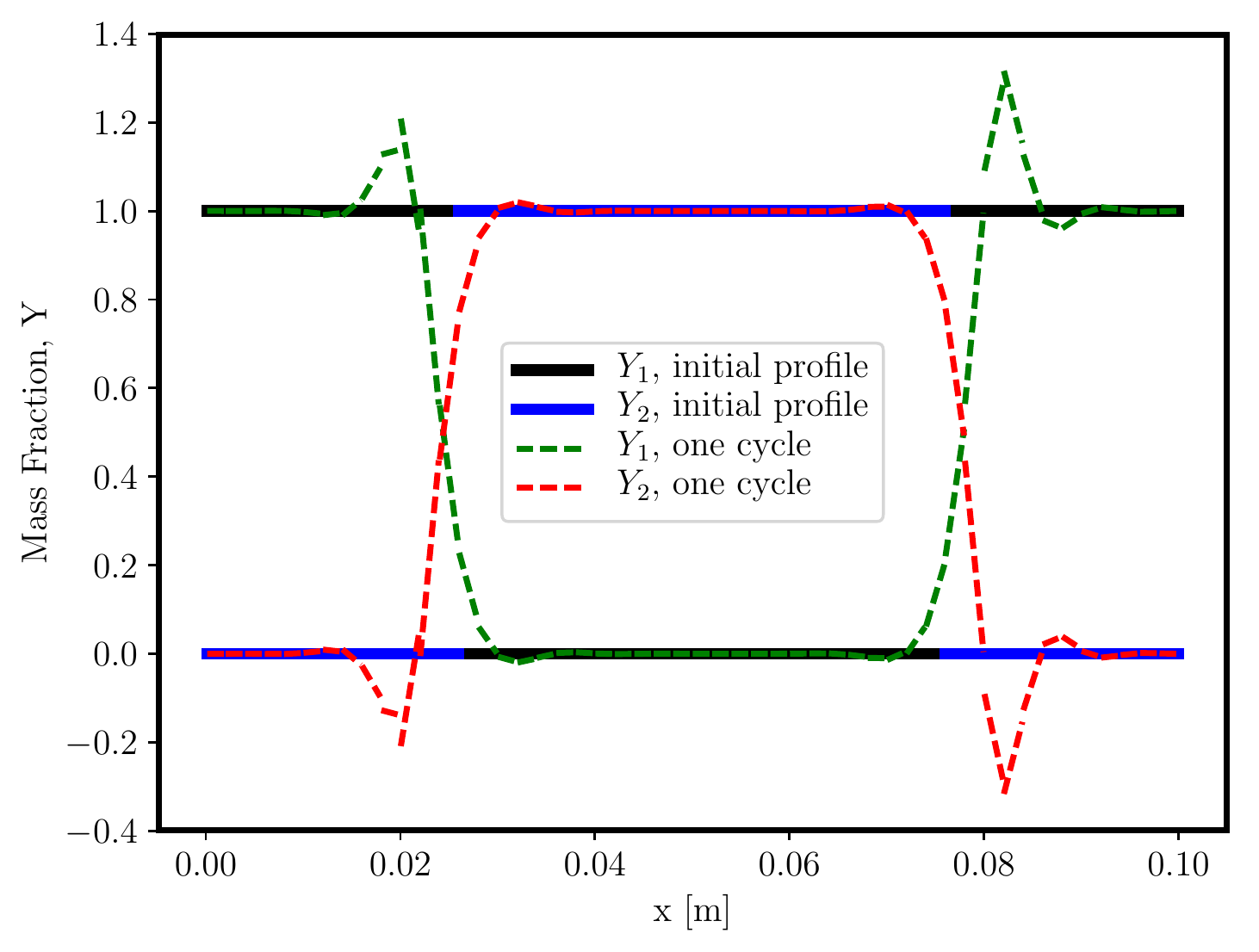}

}\hfill{}\subfloat[\label{fig:species_discontinuity_constT_nl_p2}$\mathrm{DG}\left(p=2\right)$]{\includegraphics[width=0.32\textwidth]{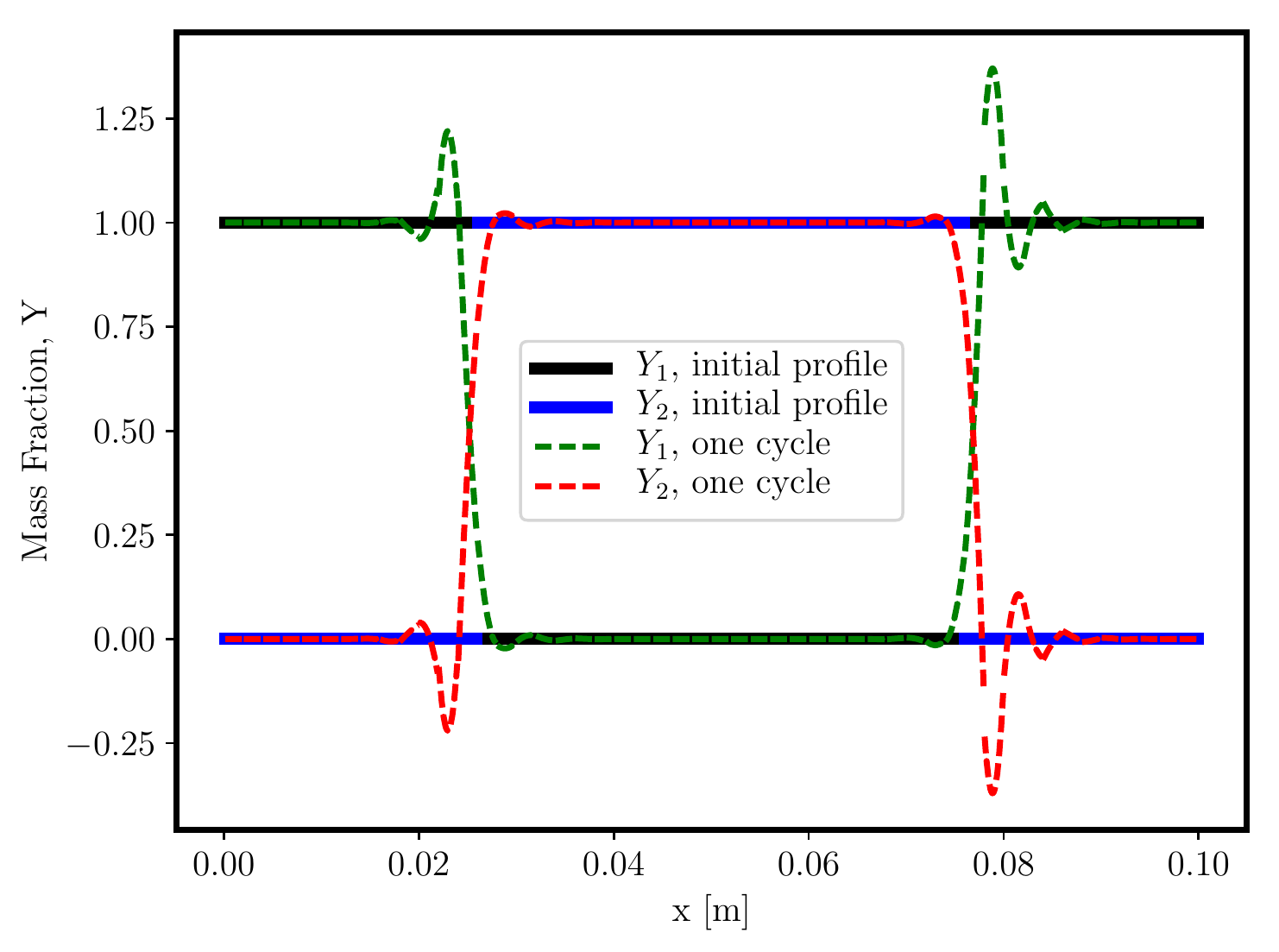}

}\hfill{}\subfloat[\label{fig:species_discontinuity_constT_nl_p3}$\mathrm{DG}\left(p=3\right)$]{\includegraphics[width=0.32\textwidth]{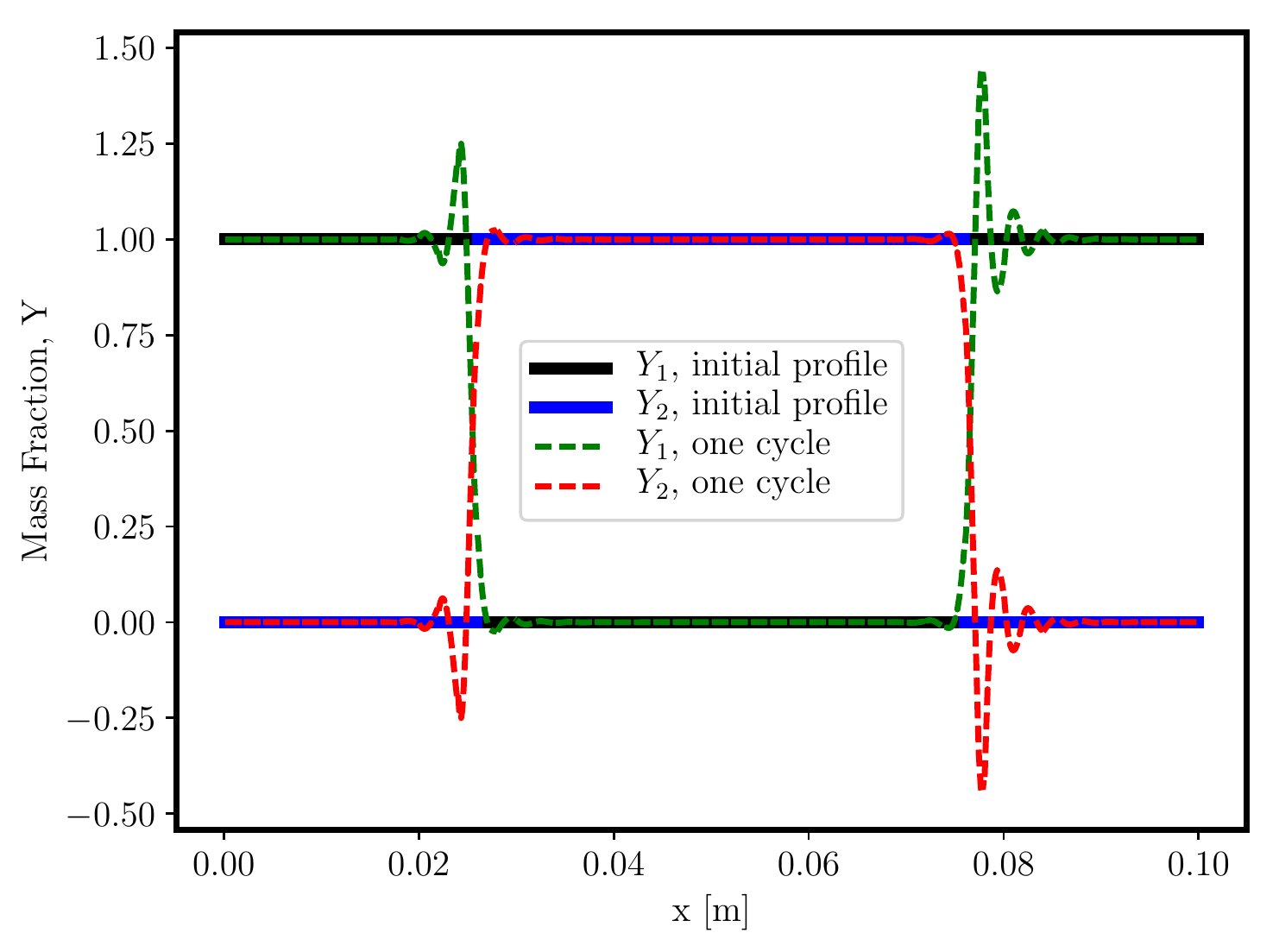}

}\caption{\label{fig:species_discontinuity_constT-1}The mass fractions profiles
of the DG solutions for $p=1,2,3$ corresponding to the species discontinuity~(\ref{eq:material-interface-initialization})
at a constant temperature, $T=300\textrm{ K}$, test case after one
cycle through the domain where the thermodynamics were evaluated exactly.
The formulation behaves exactly like its single-component counterpart,
discontinuous interfaces generate numerical instabilities that need
to be suppressed in practical simulations with artificial viscosity,
as in Section~\ref{subsec:helium_bubble}, or physical diffusion
as in Section~\ref{subsec:hydrogen-air-shear-layer}, but they do
not lead to unphysical pressure oscillations, as shown in Figure~\ref{fig:species_discontinuity_constT_p},
when the thermodynamics are evaluated exactly.}
\end{figure}

In this first test case, we consider a species discontinuity at a
constant temperature, $T=300\textrm{ K}$, given by~(\ref{eq:material-interface-initialization}).
The test problem is run on a periodic domain for one full cycle with
$\mathrm{CFL}=0.1$. Figures~\ref{fig:species_discontinuity_constT_nl_p1},~\ref{fig:species_discontinuity_constT_nl_p2},
and~\ref{fig:species_discontinuity_constT_nl_p3} show the species
mass fractions for the $\mathrm{DG}(p=1)$, $\mathrm{DG}(p=2)$, and
$\mathrm{DG}(p=3)$ solutions. All three solutions present numerical
overshoots and mixing of the species mass fractions across the discontinuities.
Some of these numerical instabilities cause the mass fractions to
be greater than one or less than zero, and, as expected, the higher
order solutions are more oscillatory for the species mass fractions.
In practical simulations, these instabilities would be suppressed
via limiting or artificial viscosity, however these approaches would
also suppress pressure oscillations, the generation of which these
test cases were created to study.

\begin{figure}[H]
\subfloat[\label{fig:species_discontinuity_constT_pressure}Pressure]{\includegraphics[width=0.45\columnwidth]{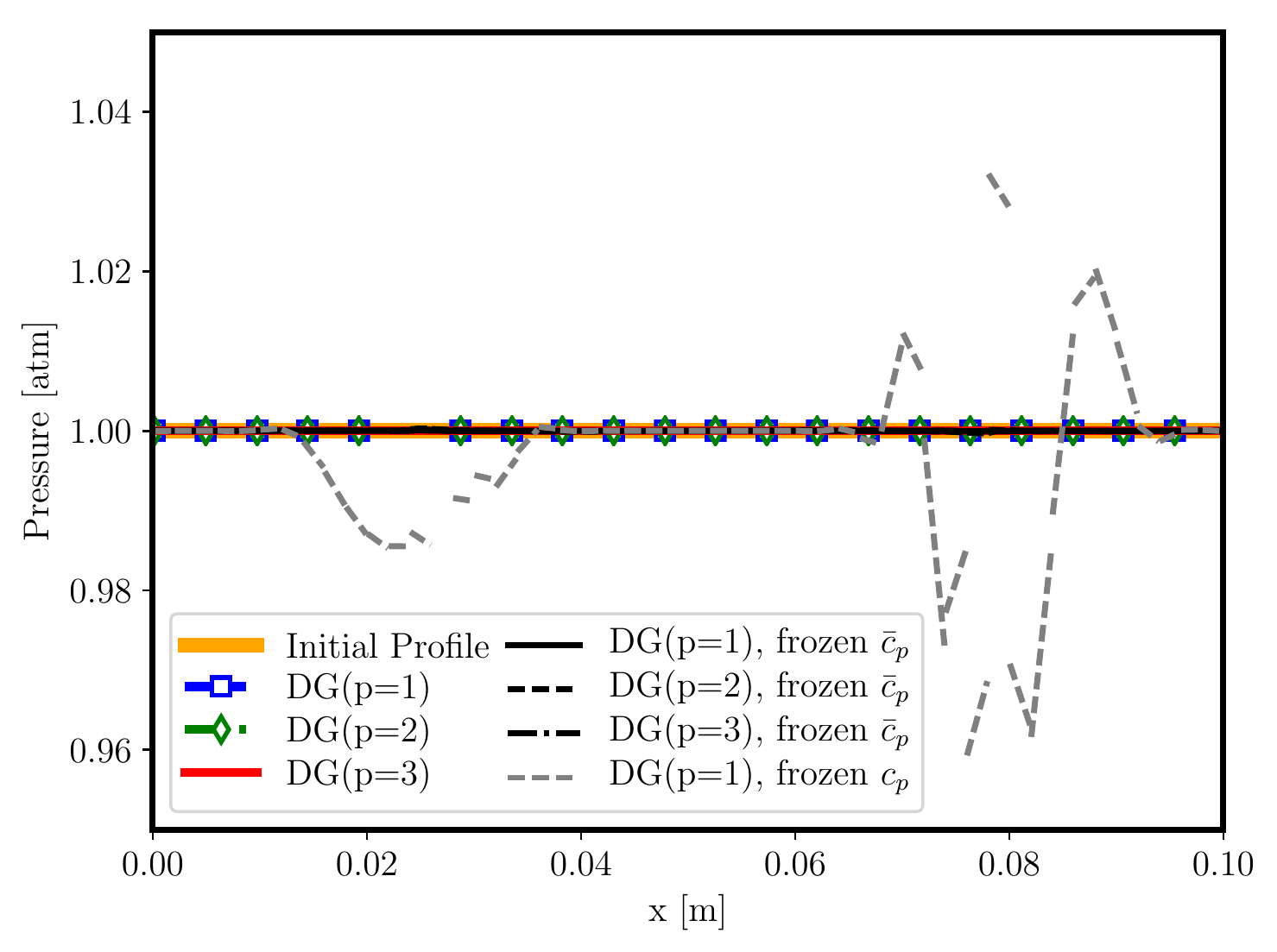}

}\hfill{}\subfloat[\label{fig:species_discontinuity_constT_pressure_zoomed}Pressure
zoomed view]{\includegraphics[width=0.45\textwidth]{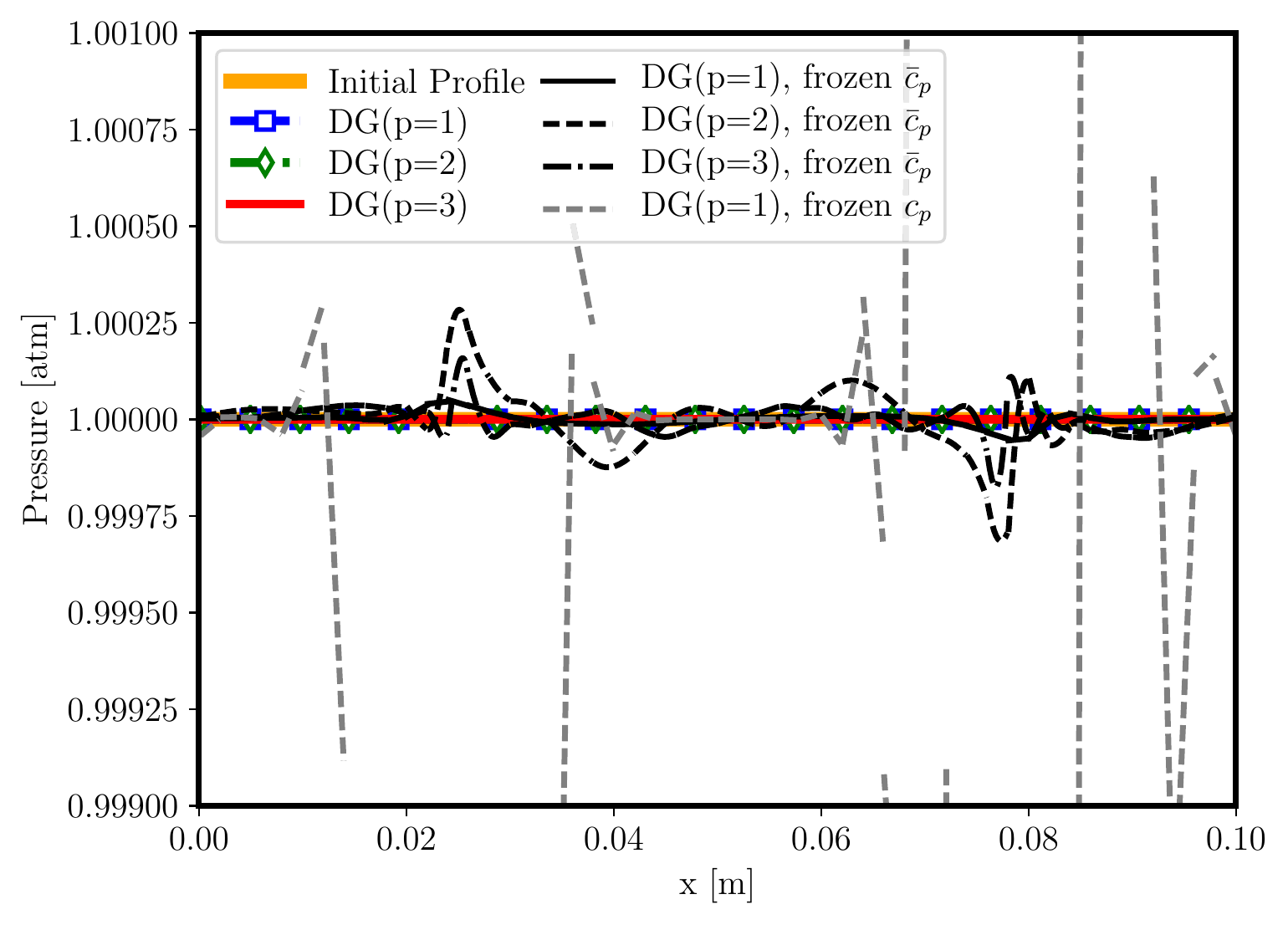}

}\caption{\label{fig:species_discontinuity_constT_p}The pressure profiles corresponding
to the $\mathrm{DG}(p=1)$, $\mathrm{DG}(p=2)$, and $\mathrm{DG}(p=3)$
solutions after one cycle for species discontinuity~(\ref{eq:material-interface-initialization})
at a constant temperature, $T=300\textrm{ K}$, test case. The formulation
behaves exactly like its single-component counterpart, discontinuous
interfaces generate numerical instabilities, as shown in Figure~\ref{fig:species_discontinuity_constT-1},
that need to be suppressed in practical simulations with artificial
viscosity, as in Section~\ref{subsec:helium_bubble}, or physical
diffusion as in Section~\ref{subsec:hydrogen-air-shear-layer}, but
they do not lead to unphysical pressure oscillations when the thermodynamics
are evaluated exactly, as shown above.}
\end{figure}

Figure~\ref{fig:species_discontinuity_constT_p} shows the pressure
after one cycle for the formulation with exact thermodynamics and
for the formulation using frozen $\bar{c}_{p}$ and frozen $c_{p}$.
The frozen $c_{p}$ solution has pressure oscillations that are on
the order of one percent error. Small perturbations of pressure caused
by freezing $\bar{c}_{p}$ are shown in Figure~\ref{fig:species_discontinuity_constT_pressure_zoomed}.
The error caused by the frozen $\bar{c}_{p}$ formulation is reduced
as the approximation order is increased from $\mathrm{DG}(p=1)$ to
$\mathrm{DG}(p=3)$. In contrast, the pressure for each solution using
the exact thermodynamics maintains a flat profile despite numerical
instabilities generated at the discontinuous interface and the error
never exceeds $10^{-7}$ atm, which is consistent with the analysis
in~\ref{sec:Discontinuities} and~\citep{Jen97}.

\begin{figure}[H]
\subfloat[\label{fig:species_discontinuity_constT_pressure_cfl}Pressure at
$\mathrm{CFL}$ 0.1 and $\mathrm{CFL}$ 0.3 conditions with frozen
$\bar{c_{p}}$ formulation]{\begin{centering}
\includegraphics[width=0.45\columnwidth]{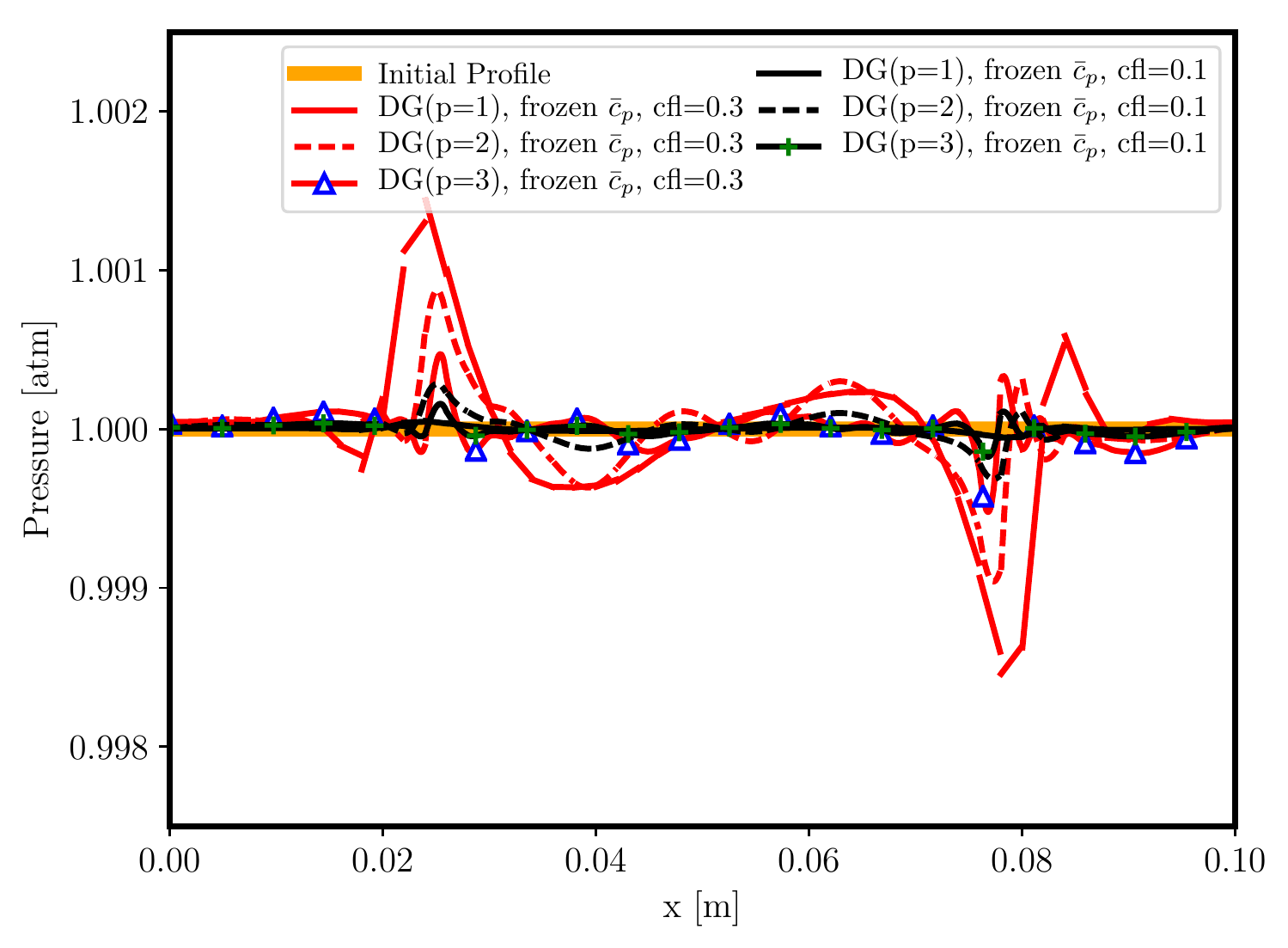}
\par\end{centering}
}\hfill{}\subfloat[\label{fig:species_discontinuity_constT_pressure_cycle}Pressure after
100 cycles for $\mathrm{DG}(p=1)$ for the formulation with exact
thermodynamics and the formulation with frozen $\bar{c_{p}}$ .]{\begin{centering}
\includegraphics[width=0.45\columnwidth]{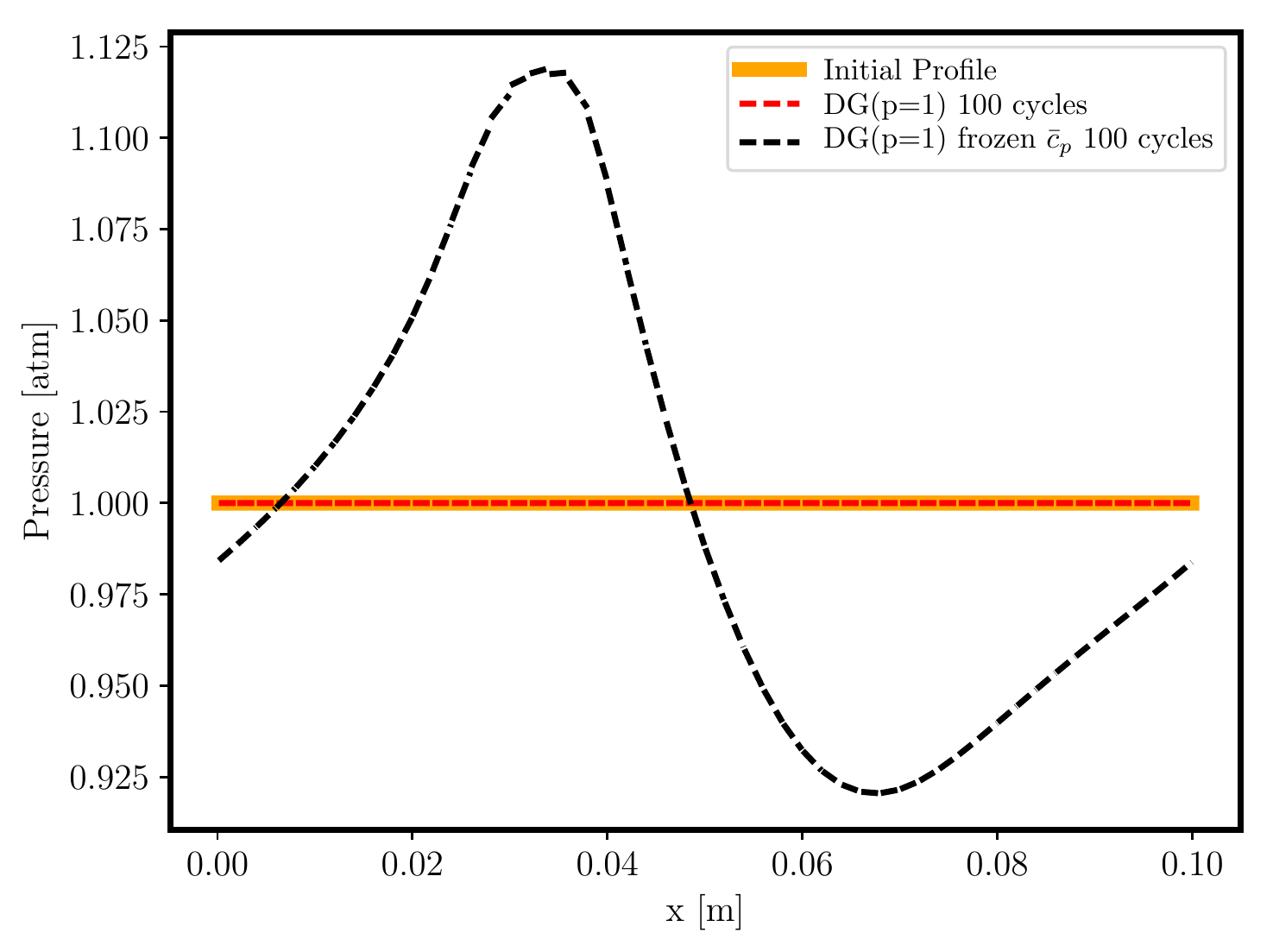}
\par\end{centering}
}
\centering{}\caption{\label{fig:species_discontinuity_constT_cfl_cycles}Pressure for $\mathrm{DG}(p=1)$,
$\mathrm{DG}(p=2)$, and $\mathrm{DG}(p=3)$ solutions for the advection
of a species discontinuity~(\ref{eq:material-interface-initialization})
at a constant temperature, $T=300\textrm{ K}$, after one cycle using
$\mathrm{CFL}$ or 0.1 and $\mathrm{CFL}$ of 0.3 with frozen $\bar{c}_{p}$
and pressure after 100 cycles using the formulation with exact thermodynamics
and the formulation with frozen $\bar{c}_{p}$ for $\mathrm{DG}(p=1)$.
The solutions were computed without additional stabilization, e.g.,
artificial viscosity, limiting, or filtering, in order to emphasize
the effect of the thermodynamic formulation on the generation of unphysical
pressure oscillations.}
\end{figure}

Figure~\ref{fig:species_discontinuity_constT_pressure_cfl} shows
the solutions for $\mathrm{DG}(p=1)$, $\mathrm{DG}(p=2)$, and $\mathrm{DG}(p=3)$
for the formulation with frozen $\bar{c}_{p}$ at $\mathrm{CFL}$
of $0.1$ and $0.3$. The larger time steps exacerbate the instability
introduced by freezing $\bar{c}_{p}$ to exceed $0.001$ atm, which
is still an order of magnitude less than the error of the frozen $c_{p}$
solutions. The solutions for the formulation with exact thermodynamics
is unaffected by the time step and therefore the corresponding results
are not shown.

Additionally, Figure~\ref{fig:species_discontinuity_constT_pressure_cycle}
shows the pressure profile from the $\mathrm{DG}(p=1)$ solution for
the formulation with frozen $\bar{c}_{p}$ and the formulation with
exact thermodynamics after 100 cycles, i.e., $t=1$ s. The pressure
oscillations for the 100 cycle solution using the formulation with
frozen $\bar{c}_{p}$ grow in time to be on the order of the one cycle
with frozen $c_{p}$. The pressure solution for the formulation with
exact thermodynamics remains constant after 100 cycles. 

For brevity, the temperature solutions are not presented. The formulation
with frozen $c_{p}$ temperature solution fluctuated on the order
of $1$ K whereas the formulation with frozen $\bar{c}_{p}$ solution
fluctuates less than $0.025$ K. These fluctuations were located at
the species interfaces. The temperature solution corresponding to
the formulation using exact thermodynamics remains flat and does not
deviate by more than $10^{-4}$ K from the exact value of $300$ K.
\subsection{Species discontinuities with temperature discontinuities\label{subsec:species-discontinuity-temperature-discontinuity}}

\begin{figure}[H]
\subfloat[\label{fig:species_discontinuity_discT_p} Pressure solutions after
one cycle.]{\includegraphics[width=0.45\columnwidth]{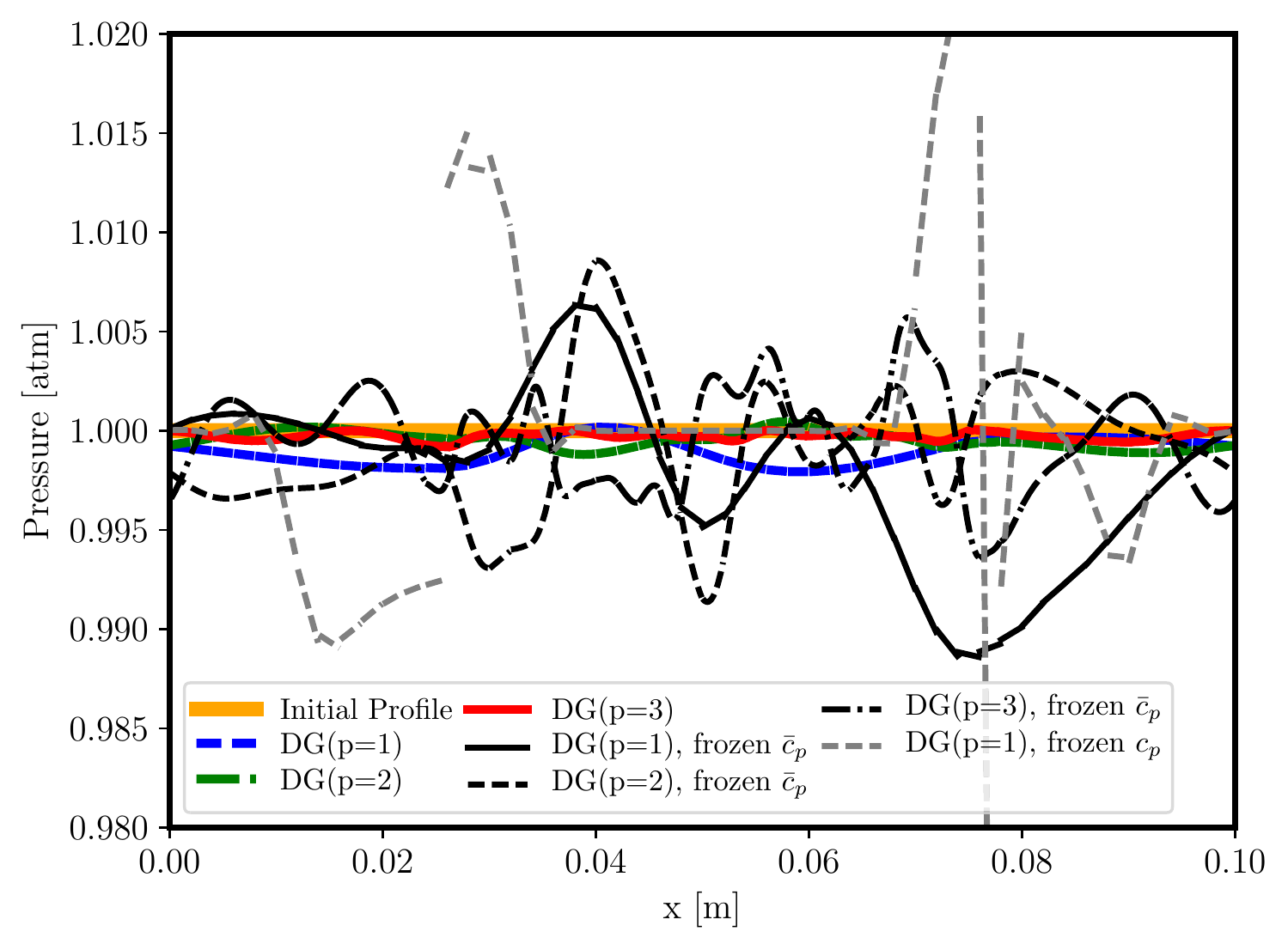}

}\hfill{}\subfloat[\label{fig:species_discontinuity_discT_T}Temperature solutions after
one cycle.]{\includegraphics[width=0.45\textwidth]{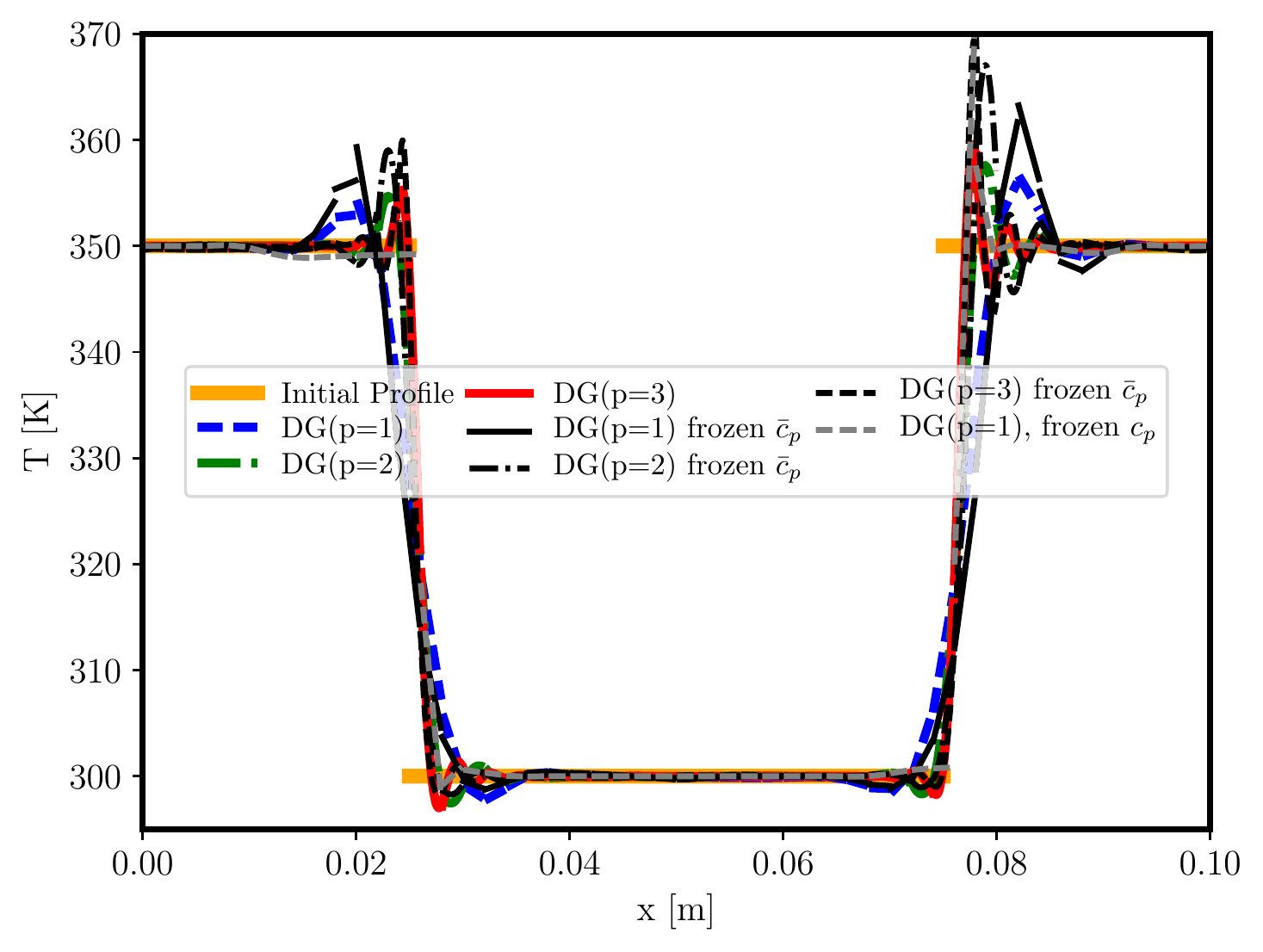}

}\caption{\label{fig:species_discontinuity_discT_p_T}The pressure and temperature
profiles of the $\mathrm{DG}(p=1)$, $\mathrm{DG}(p=2)$, and $\mathrm{DG}(p=3)$
solutions after one cycle for the advection of both discontinuous
species and temperature profiles. The initial condition is given by~(\ref{eq:material-interface-initialization})
and the piece-wise constant temperature profile is given by~(\ref{eq:discontinuous-species-discontinuous-temperature}).
The frozen $c_{p}$ solution at 100 steps is shown in grey. The solutions
were computed without additional stabilization, e.g., artificial viscosity,
limiting, or filtering, in order to emphasize the effect of the thermodynamic
formulation on the generation of unphysical pressure oscillations.}
\end{figure}

We consider the same discontinuous profile as in Section~\ref{subsec:species-discontinuity-constant-temperature}
and introduce a temperature discontinuity defined as

\begin{eqnarray}
T & = & \begin{cases}
300\textrm{ K} & \text{if }0.025<x<0.075\\
350\textrm{ K} & \text{otherwise}
\end{cases}.\label{eq:discontinuous-species-discontinuous-temperature}
\end{eqnarray}
Similar to the previous test case, the three solutions present numerical
overshoots and mixing of the species mass fractions across the discontinuities
but are not shown graphically for brevity. Figures~\ref{fig:species_discontinuity_discT_p}
and \ref{fig:species_discontinuity_discT_T} show the pressure and
temperature, respectively, after one cycle for the formulation using
exact and frozen thermodynamics. The formulation with frozen $c_{p}$
fails before one complete cycle and is shown after 100 time steps
as a dashed grey line. The pressure for the solutions using the formulation
with exact thermodynamics causes pressure oscillations that are an
order of magnitude less than the frozen $\bar{c_{p}}$ simulations.
The oscillations in the formulation with exact thermodynamics are
due to the numerical mixing of species across a temperature discontinuity
and are expected based on the discussion in~\ref{sec:Discontinuities}
and~\citep{Jen97}. Both the formulation with exact thermodynamics
and formulation with frozen $\bar{c}_{p}$ produce overshoots and
undershoots at the temperature discontinuities. The temperature oscillations
associated with the formulation using frozen $\bar{c}_{p}$ are larger
than the oscillations corresponding the formulation with exact thermodynamics.

\begin{figure}[H]
\subfloat[\label{fig:species_discontinuity_discT_100_T}Temperature solutions
after 100 cycles.]{\includegraphics[width=0.45\textwidth]{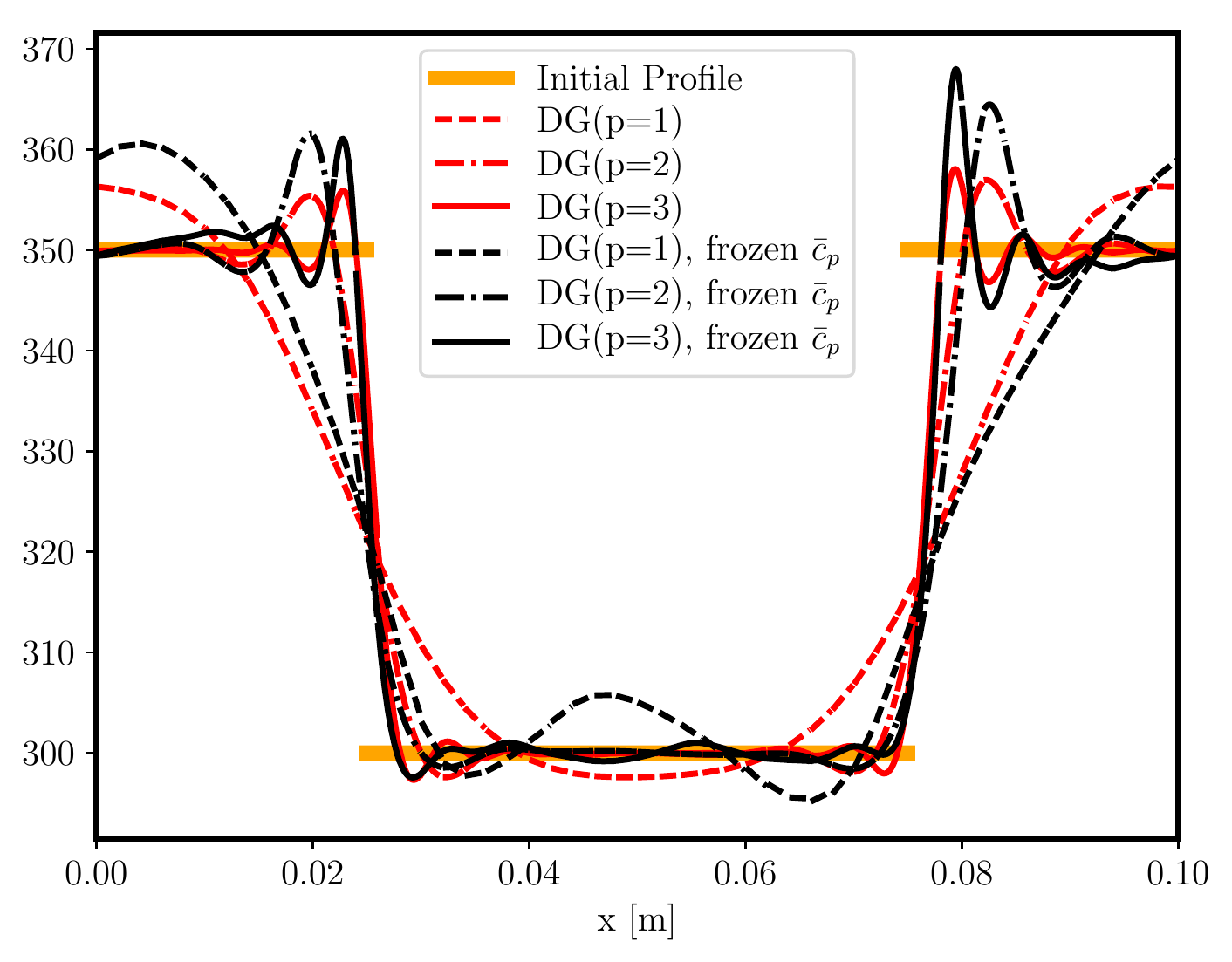}

}\hfill{}\subfloat[\label{fig:species_discontinuity_discT_100_p}Pressure solutions after
100 cycles.]{\includegraphics[width=0.45\columnwidth]{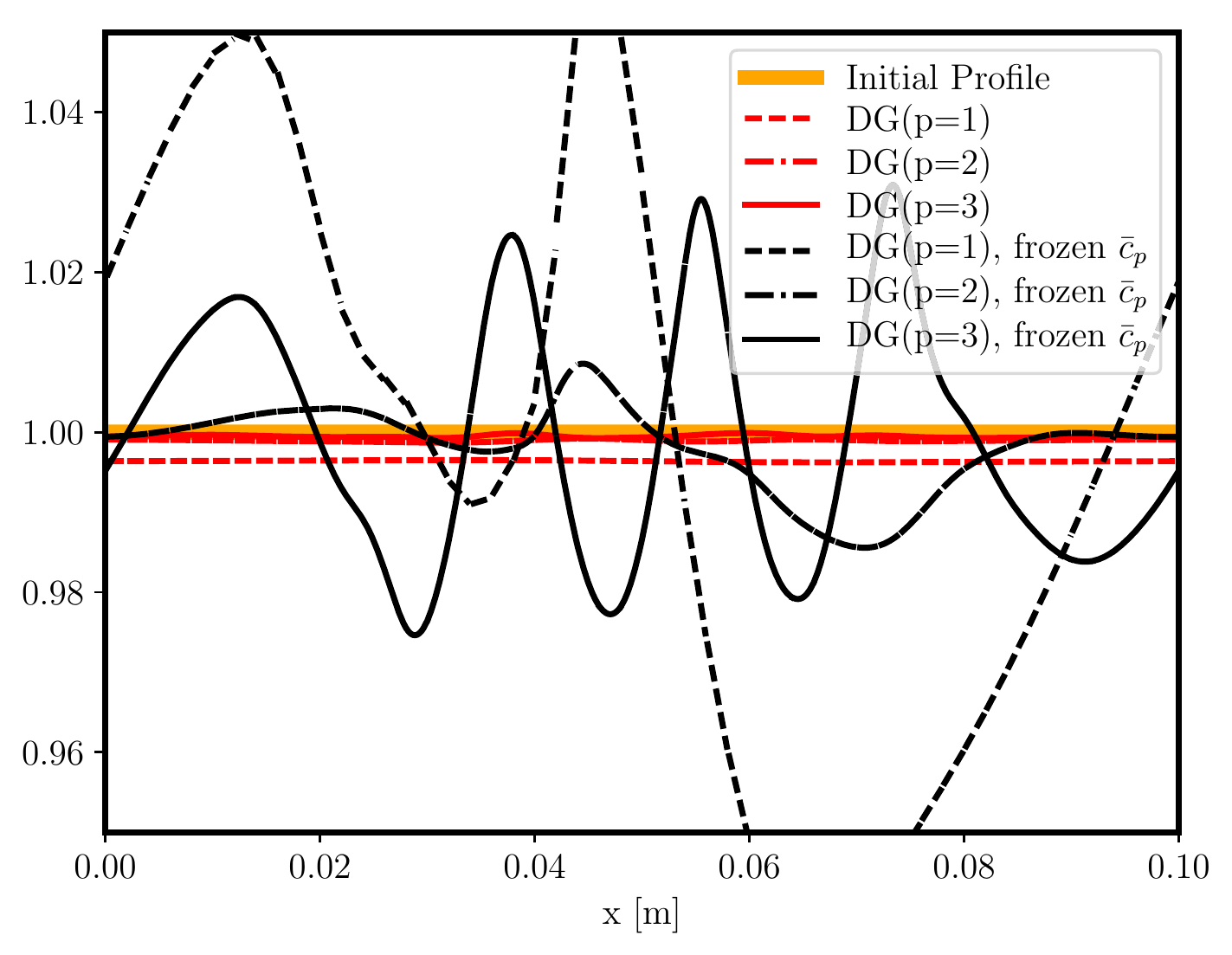}

}\caption{\label{fig:species_discontinuity_discT_p_T-1}The pressure and temperature
profiles for $\mathrm{DG}(p=1)$, $\mathrm{DG}(p=2)$, and $\mathrm{DG}(p=3)$
solutions after 100 cycles for the advection of both discontinuous
species and temperature profiles. The initial condition is given by~(\ref{eq:material-interface-initialization})
and the piece-wise constant temperature profile is given by~(\ref{eq:discontinuous-species-discontinuous-temperature}).
The solutions were computed without additional stabilization, e.g.,
artificial viscosity, limiting, or filtering, in order to emphasize
the effect of the thermodynamic formulation on the generation of unphysical
pressure oscillations.}
\end{figure}

Figures~\ref{fig:species_discontinuity_discT_100_T} and~\ref{fig:species_discontinuity_discT_100_p}
present the temperature and pressure solutions, respectively, for
the formulation with exact thermodynamics and the formulation with
frozen $\bar{c}_{p}$ after 100 cycles, i.e. $t=1$ s. The temperature
profiles become more diffuse with the larger number of cycles (see
Figure~\ref{fig:species_discontinuity_discT_T} for the comparison
of one cycle). The pressure oscillations for the formulation with
exact thermodynamics cause the ambient pressure to fall below 1 atm.
This departure from ambient was improved by increasing the approximation
order from $\mathrm{DG}(p=1)$ to $\mathrm{DG}(p=3)$. The pressure
oscillations for the 100 cycle solution using the formulation with
frozen $\bar{c}_{p}$ grew in time regardless of approximation order
as shown in Figure~\ref{fig:species_discontinuity_discT_p} for the
one cycle solution for frozen $\bar{c}_{p}$. Furthermore, the oscillations
associated with the frozen $\bar{c}_{p}$ formulation are an order
of magnitude larger than the oscillations associated with the formulation
using exact thermodynamics.

\subsection{Discrete pressure equilibrium $\mathrm{DG}(p=0)$\label{subsec:p0_oscillations}}

It is important to note that the $\mathrm{DG}(p=0)$ method is equivalent
to low order finite volume methods. For $\mathrm{DG}(p>0)$ discrete
approximation can accurately represent a continuous solution, however,
for $\mathrm{DG}(p=0)$ the discrete approximation is inherently discontinuous.
These discontinuities cause unphysical pressure oscillations for species
that are nonlinear with respect to temperature as shown in Section~\ref{subsec:species-discontinuity-temperature-discontinuity}.
As such, we expect the method to generate unphysical pressure oscillations
for smooth profiles of temperature. 

\begin{figure}[H]
\subfloat[\label{fig:species_discontinuity_varT_temperature}Temperature profiles
for $\mathrm{DG}(p=0)$ and $\mathrm{DG}(p=1)$ solutions.]{\includegraphics[width=0.45\columnwidth]{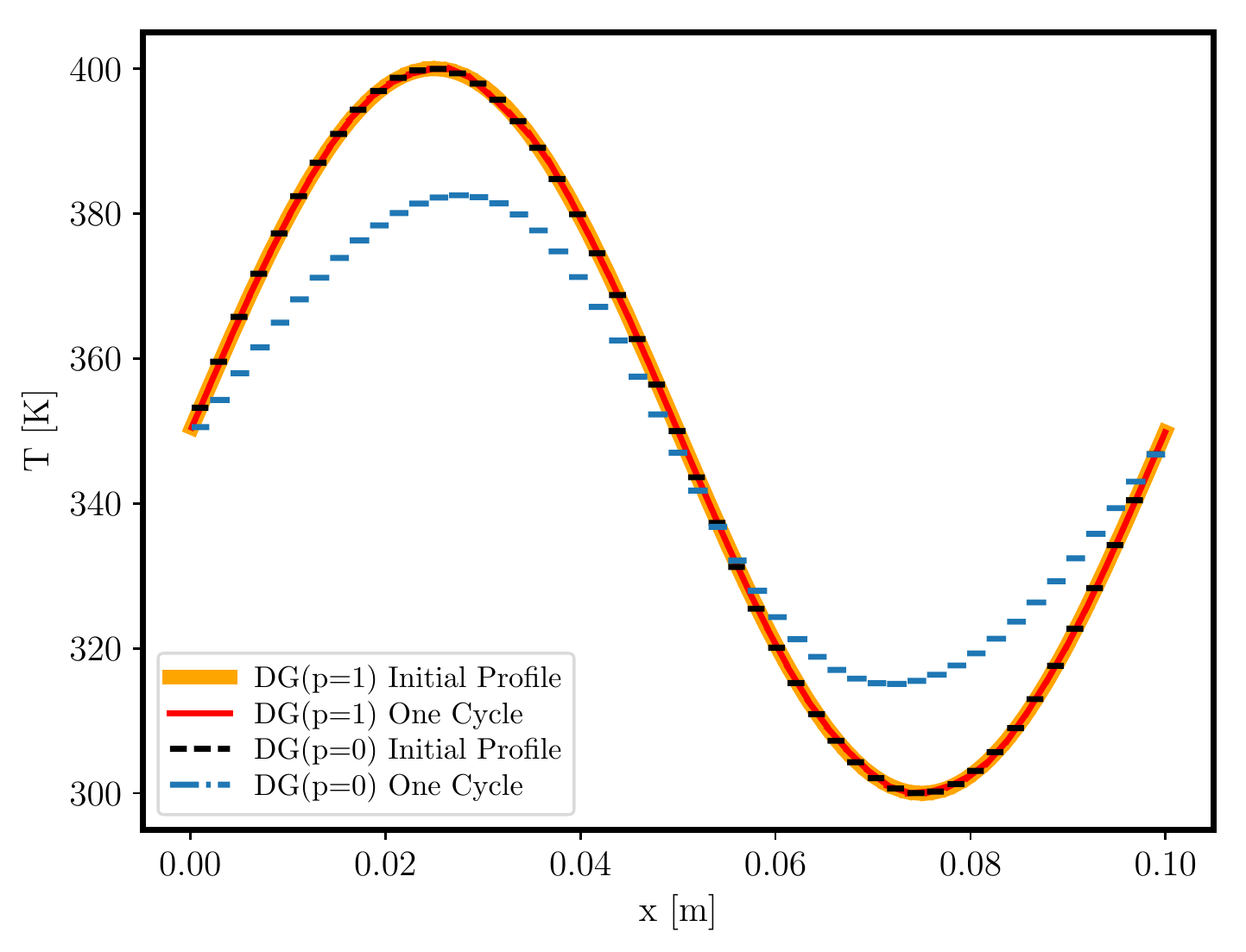}

}\hfill{}\subfloat[\label{fig:species_discontinuity_varT_pressure}Pressure profiles
for $\mathrm{DG}(p=0)$ and $\mathrm{DG}(p=1)$ solutions.]{\includegraphics[width=0.45\textwidth]{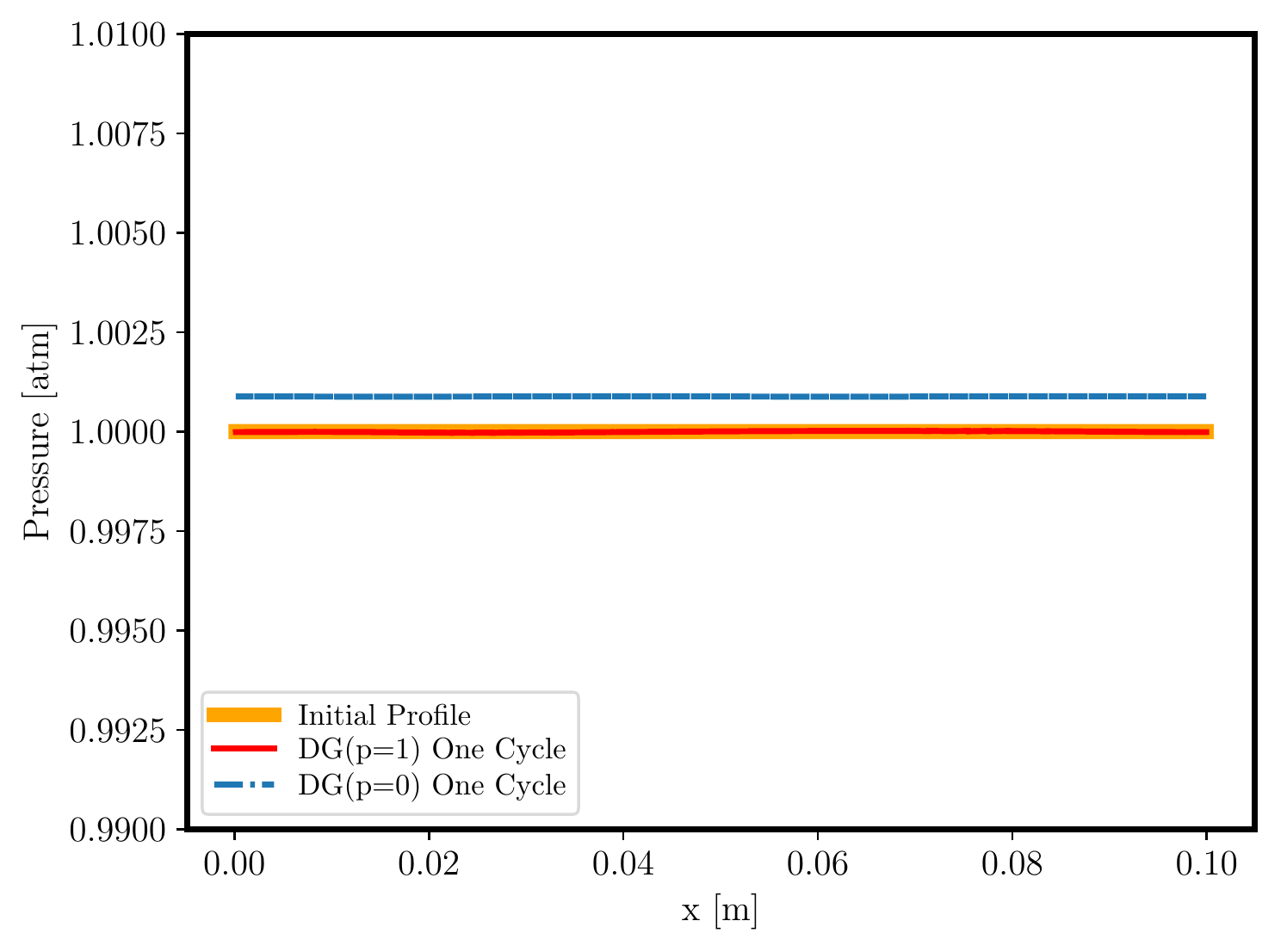}

}\caption{\label{fig:species_discontinuity_varT_pnT}The temperature and pressure
profiles of the $\mathrm{DG}(p=0)$ and $\mathrm{DG}(p=1)$ solutions
after one cycle for the advection of a species discontinuity~(\ref{eq:material-interface-initialization})and
a continuous temperate profile, $T=350+50\sin\left(20\pi x\right)\textrm{ K}$.}
\end{figure}

Again, we consider the discontinuous species profile given by~(\ref{eq:material-interface-initialization})
and introduce a continuous variation in temperature, defined as

\begin{eqnarray}
T & = & 350+50\sin\left(20\pi x\right)\textrm{ K}.\label{eq:discontinuous-species-continuous-temperature}
\end{eqnarray}
Figure~\ref{fig:species_discontinuity_varT_temperature} and~\ref{fig:species_discontinuity_varT_pressure}
show the initial and final profiles of temperature and pressure, respectively,
for solutions using $\mathrm{DG}(p=1)$ and $\mathrm{DG}(p=0)$. The
initial temperature profile for $\mathrm{DG}(p=1)$ is represented
smoothly, whereas the initial profile using $\mathrm{DG}(p=0)$ is
piecewise constant. The initial discontinuities between adjacent elements
in the $\mathrm{DG}(p=0)$ solution lead to small pressure oscillations.
After one cycle the pressure errors have dispersed and the overall
pressure has diverged by $0.1$ \% for $\mathrm{DG}(p=0)$. As expected
the $\mathrm{DG}(p=1)$ pressure does not diverge from the expected
constant solution after one cycle and is consistent with the initial
profile.

Although not the focus of this manuscript, it is worth noting that
discontinuous discrete solution corresponding to $\mathrm{DG}(p=0)$
generates unphysical pressure oscillations even when the formulation
with exact thermodynamics is used. However, it is also important to
note that the solution remains stable even in the presence of unphysical
pressure oscillations. Finally, these issues are not present for $\mathrm{DG}(p>0)$
solutions as temperature and pressure equilibrium are maintained in
well resolved regions of the flow.

\subsection{Thermal bubble \label{subsec:thermal-bubble}}

\begin{figure}[H]
\subfloat[Pressure solution after one cycle. \label{fig:tb_1D_lcp_p}]{\includegraphics[width=0.45\columnwidth]{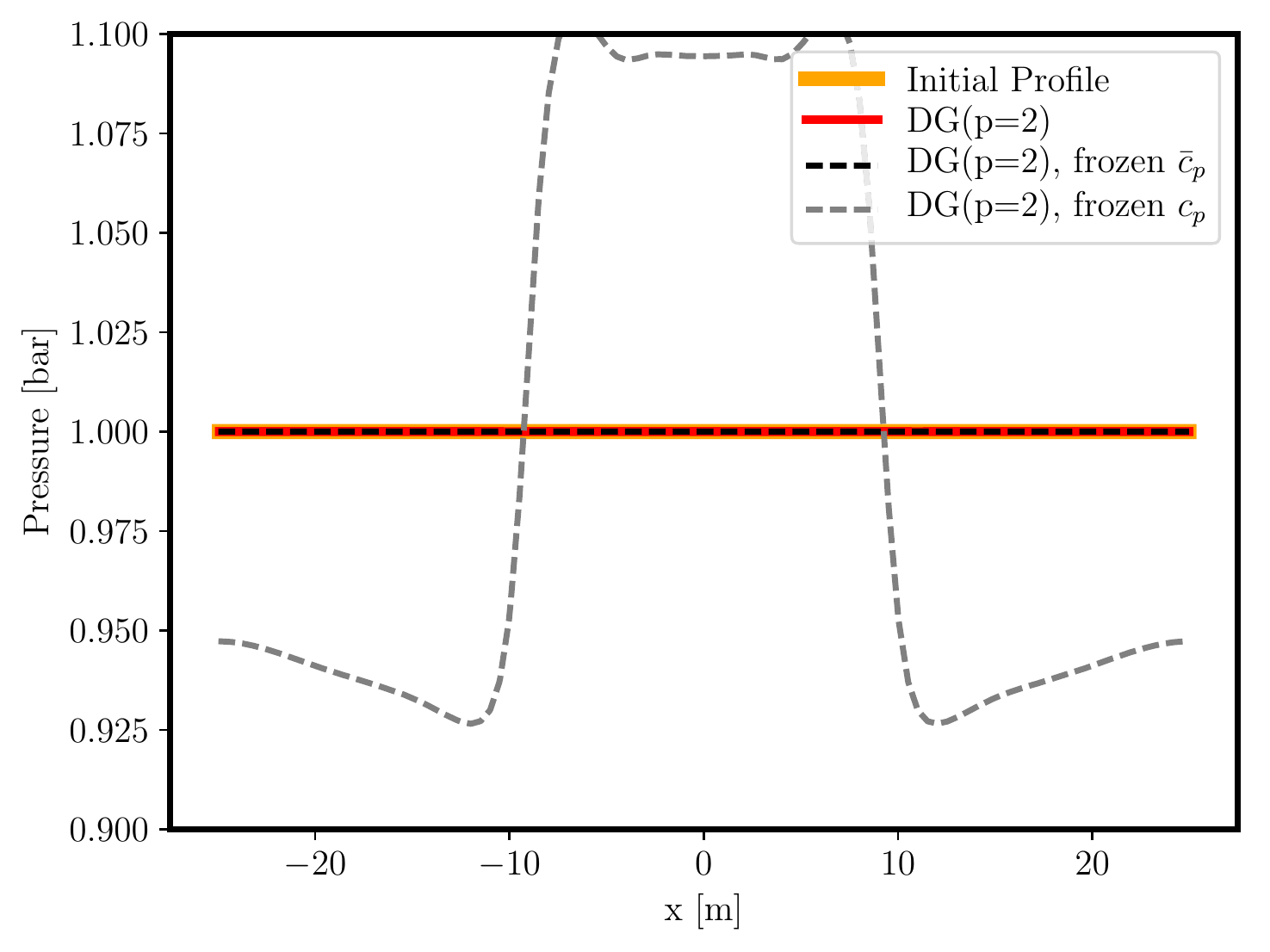}

}\hfill{}\subfloat[\label{fig:tb_1D_lcp_T}The pressure profile of the $\mathrm{DG}\left(p=1\right)$
and $\mathrm{DG}\left(p=2\right)$ solutions with $h=0.5$ m at $t=50$
s (one cycle).]{\includegraphics[width=0.45\columnwidth]{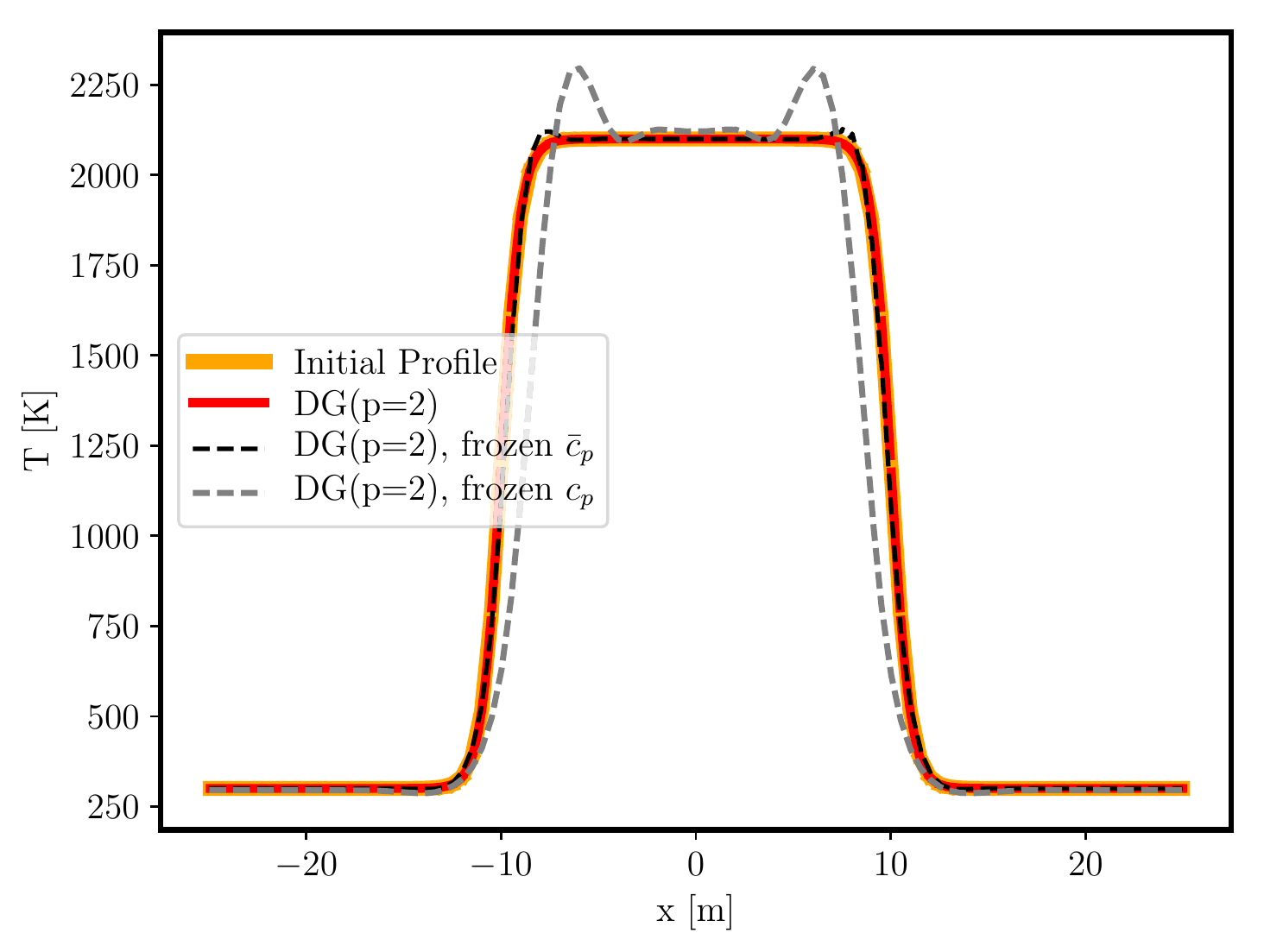}

}%

\caption{\label{fig:tb_1D_lcp}The temperature and pressure profiles of the
$\mathrm{DG}\left(p=2\right)$ solutions for advection of a thermal
bubble~(\ref{eq:thermal-bubble})after one cycle. The solutions were
computed without additional stabilization, e.g., artificial viscosity,
limiting, or filtering, in order to emphasize the effect of the thermodynamic
formulation on the generation of unphysical pressure oscillations.}
\end{figure}

Here we present the one dimensional thermal bubble test case previously
presented by~\citep{Lv15}. For this test case, a periodic $50$
m domain, $(-25,25)$ m, with grid spacing $h=0.5$~m, and the following
initial conditions

\begin{eqnarray}
v & = & 1\textrm{ m/s},\label{eq:thermal-bubble}\\
Y_{H_{2}} & = & \frac{1}{2}\left[1-\tanh\left(|x|-10\right)\right],\nonumber \\
Y_{O_{2}} & = & 1-Y_{H_{2}},\nonumber \\
T & = & 1200-900\tanh\left(|x|-10\right)\textrm{ K},\nonumber \\
p & = & 1\textrm{ bar}.\nonumber 
\end{eqnarray}
The test case is run for 1 cycle, $t=50$ s, using $\mathrm{DG}(p=2)$
with $\mathrm{CFL}=0.1$ and the inviscid, non-reacting formulation
of Equations~(\ref{eq:conservation-law-strong-form})-(\ref{eq:conservation-law-flux-boundary-condition}).
No artificial viscosity or limiting is used in this test case. The
mesh resolution was too coarse to stably compute a $\mathrm{DG}(p=1)$
solution without artificial viscosity, limiting, or filtering. Like
the previous test cases, the analytical solution after one cycle is
the same as the initial profile. Figures~\ref{fig:tb_1D_lcp_p} and~\ref{fig:tb_1D_lcp_T}
show the results for pressure and temperature, respectively. The pressure
is constant throughout for both the formulation with exact thermodynamics
and the formulation with frozen $\bar{c}_{p}$ having variations on
the order of $10^{-5}\,$ atm. The pressure for the formulation with
frozen $c_{p}$ fluctuates on the order of $10$\% of the expected
ambient pressure. Previous work reported that without the double flux
method the pressure fluctuated throughout the solution by $3\%$ of
the expected ambient pressure for frozen thermodynamic formulations~\citep{Lv15}.
These issues do not occur in the solutions corresponding to the formulation
presented in this work when the thermodynamic quantities are computed
exactly and or frozen and $\bar{c}_{p}$ is correctly evaluated.

\begin{figure}[H]
\subfloat[\label{fig:thermal_bubble_Y_10_cycles}The hydrogen mass fraction
profile of the $\mathrm{DG}\left(p=2\right)$ solution after each
cycle for a total of ten cycles using the formulation with exact thermodynamics..]{\includegraphics[width=0.45\columnwidth]{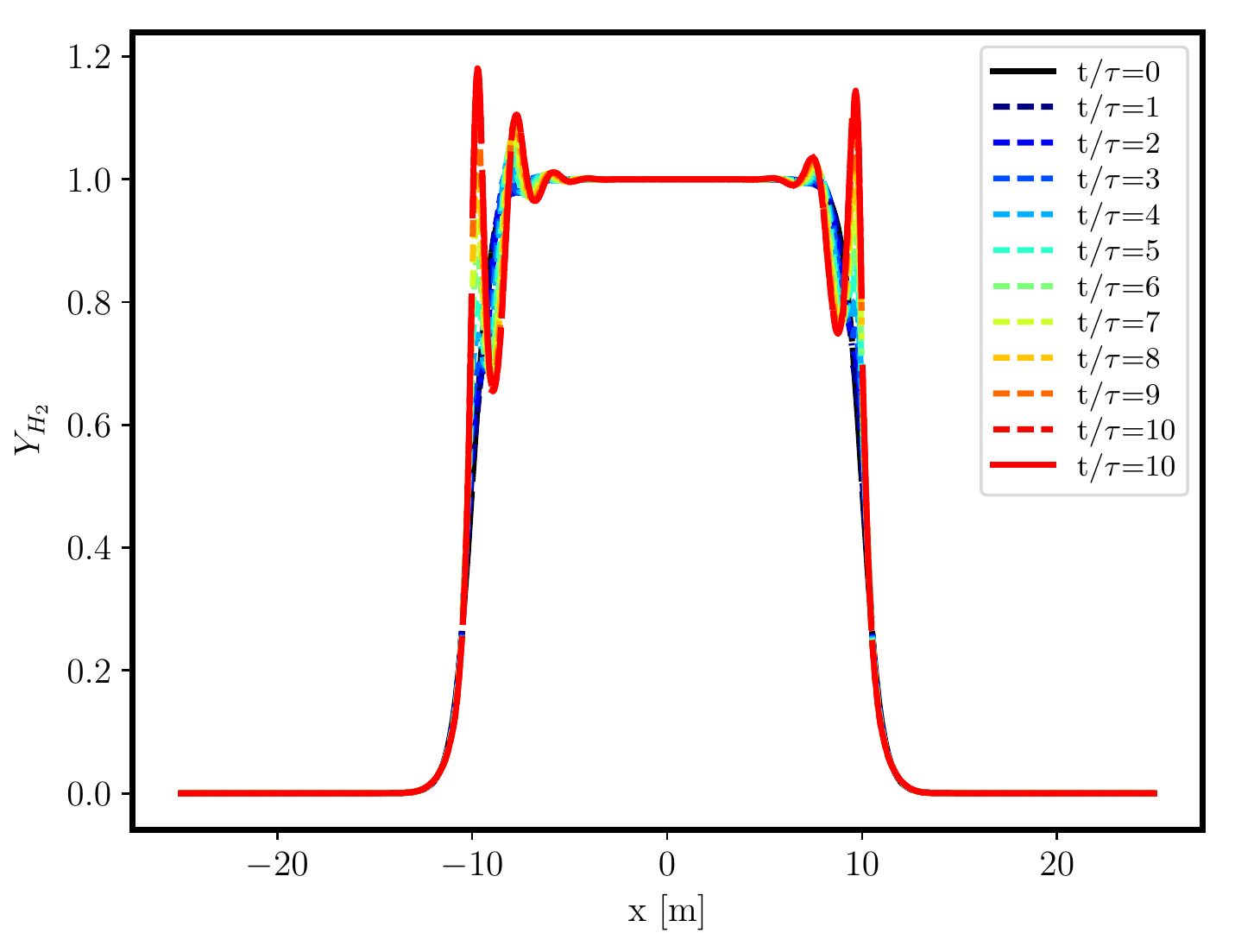}

}\hfill{}\subfloat[\label{fig:thermal_bubble_delta_p_10_cycles}The temporal evolution
of the deviation of the pressure from the initial pressure for the
$\mathrm{DG}\left(p=2\right)$ solutions.]{\includegraphics[width=0.45\columnwidth]{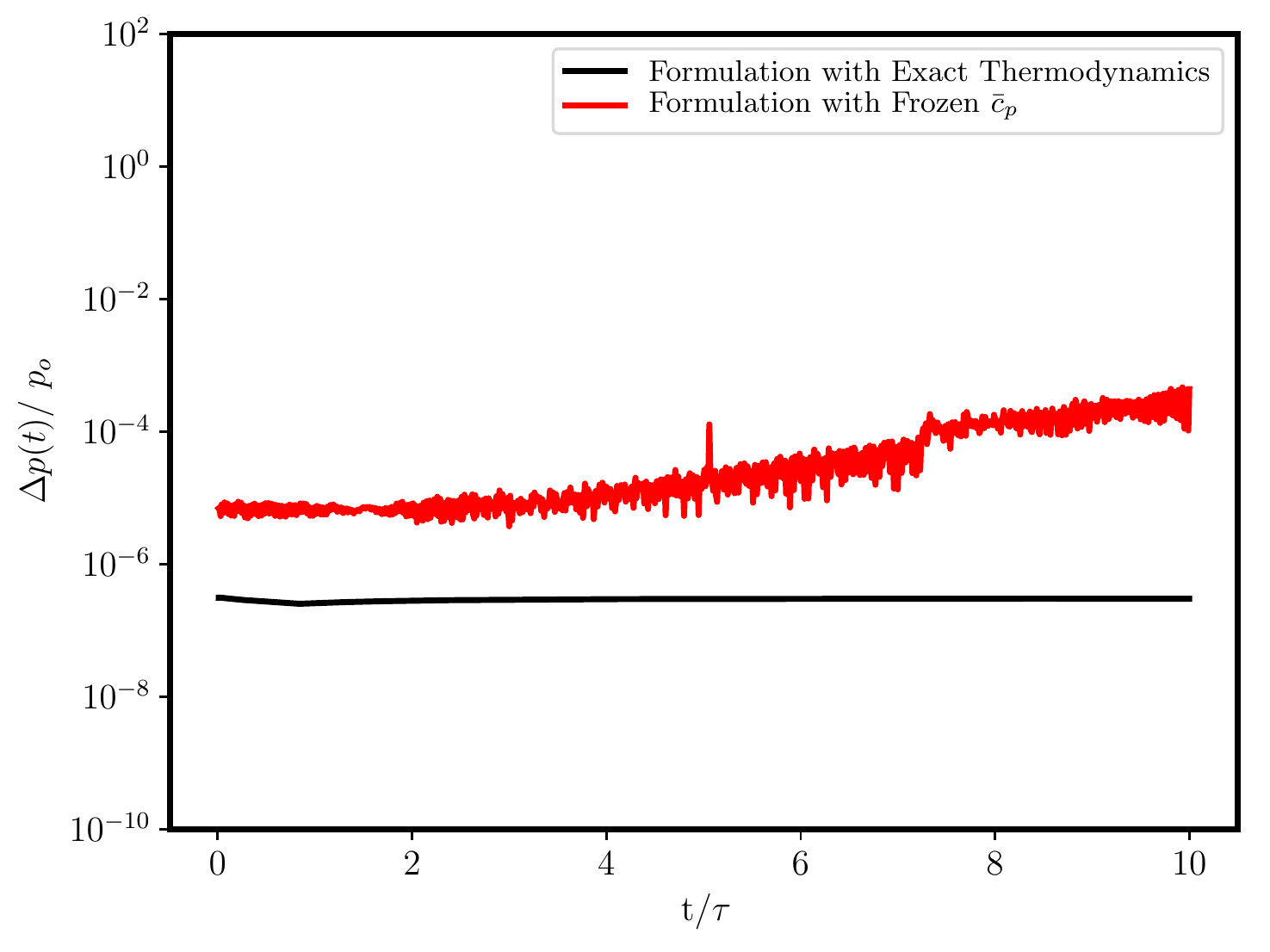}

}

\caption{\label{fig:thermal_bubble_10_cycles}The hydrogen mass fraction profile
and the temporal evolution of the deviation of the pressure from the
initial pressure for the $\mathrm{DG}\left(p=2\right)$ solution after
ten cycles for advection of a thermal bubble~(\ref{eq:thermal-bubble}).
The solutions were computed without additional stabilization, e.g.,
artificial viscosity, limiting, or filtering, in order to emphasize
the effect of the thermodynamic formulation on the generation of unphysical
pressure oscillations.}
\end{figure}

The formulation with frozen $\bar{c}_{p}$ does not generate large
pressure fluctuations after one cycle. This is due to the the fact
that the concentrations are represented as a smooth continuous profile,
which results in less numerical mixing than the discontinuous interfaces
present in the previous test cases. However, after $10$ cycles, numerical
mixing does occur. Figure~\ref{fig:thermal_bubble_Y_10_cycles} presents
the mass fraction profiles of $H_{2}$ for all $10$ cycles using
the formulation with exact thermodynamics. Without additional stabilization,
the species concentrations eventually oscillate in regions of the
flow with steep gradients. Figure~\ref{fig:thermal_bubble_delta_p_10_cycles}
presents the deviation from initial pressure, $\Delta p=p-p_{0}$,
normalized by the initial pressure, $p_{0}=1$ bar as a function of
physical time, $t$, which itself is normalized by the cycle time,
$\tau=50$ s. The error in the formulation with frozen $\bar{c}_{p}$
has errors that grow to be two orders of magnitude greater than the
formulation with exact thermodynamics, which does not grow despite
the numerical mixing after $10$ cycles. 

\subsubsection{Convergence under grid refinement}

As stated in Section~\ref{subsec:Discretization}, the expected order
of accuracy for the RK2+DG method presented in this work is $\mathcal{O}\left(h^{p+1}+\tau^{2}\right)$.
Since we wish to verify that the spatial discretization convergences
under gird refinement with the optimal rate, we seek to minimize the
error due to the temporal discretization by restricting the time-step
using a $\mathrm{CFL}=0.1$ and integrated temporally via SSP-RK3,
as opposed to the SSP-RK2 method we apply throughout this work. Therefore,
we expect the order of accuracy of the discretization, under these
circumstances, to be approximately $\mathcal{O}\left(h^{p+1}\right)$.
\begin{figure}[H]
\subfloat[\label{fig:RKDG-convergence-state}Convergence under grid, $h$, refinement
of the normalized state $L^{2}$ error $\left\Vert \hat{y}-\hat{y}_{\mathrm{exact}}\right\Vert _{L_{2}\left(\Omega\right)}$
with respect to the exact solution.]{\begin{centering}
\includegraphics[width=0.45\linewidth]{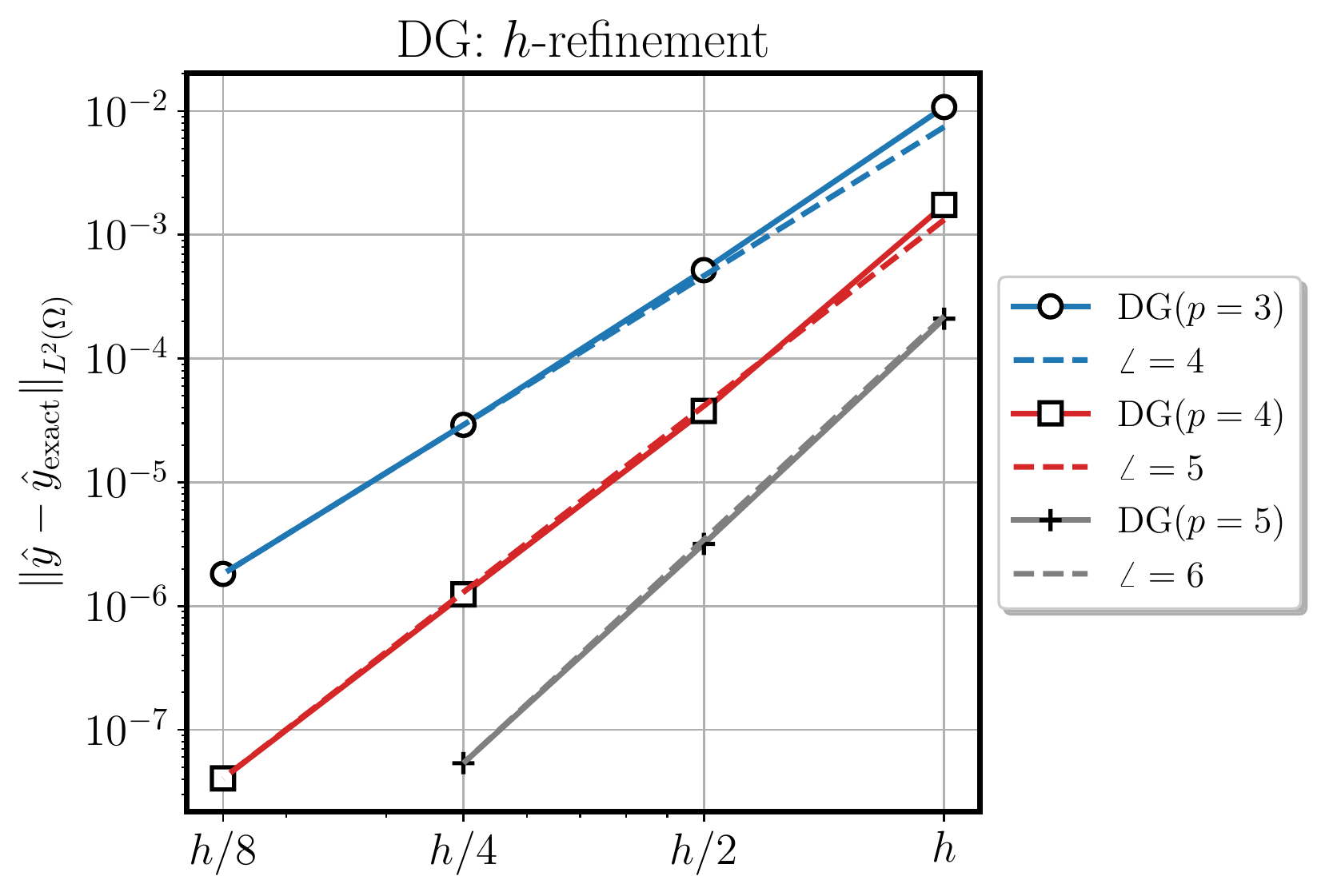}
\par\end{centering}
}\hfill{}
\begin{centering}
\subfloat[\label{fig:RKDG-convergence-temperature}Convergence with respect
to grid, $h$, refinement of the normalized temperature $L^{2}$ error
$\footnotesize\left\Vert \hat{T}-\hat{T}_{\mathrm{exact}}\right\Vert _{L_{2}\left(\Omega\right)}$
with respect to the exact temperature profile compared to the error
corresponding to the $L^{2}$ projection of the exact temperature
profile, which provides an upper bound on the attainable accuracy
since it minimizes the error in the $L^{2}$ norm.]{\begin{centering}
\includegraphics[width=0.45\linewidth]{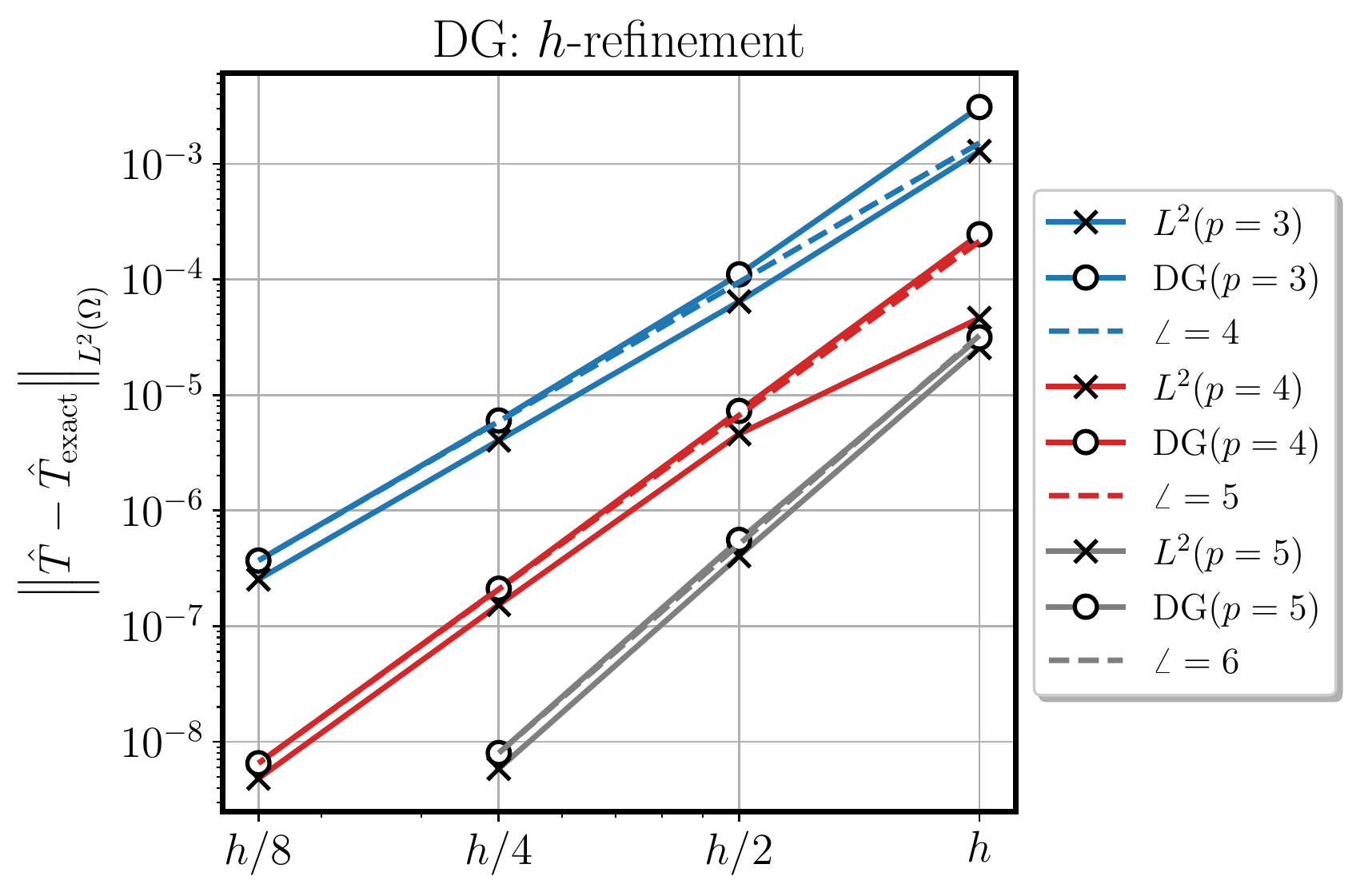}
\par\end{centering}
}
\par\end{centering}
\caption{\label{fig:RKDG-convergence}Convergence under grid refinement for
the advection of a thermal bubble at time $t=5\,\mathrm{s}$ on a
grid consisting linear line elements with $h=1\,\mathrm{m}$, where
the initial condition is given by~(\ref{eq:thermal-bubble}). The
RKDG spatial discretization converges with the expected optimal ($p+1$)
rate, thus confirming that the multi-component formulation behaves
exactly like its single-component counterpart, achieving high-order
accuracy in smooth regions of the flow.}
\end{figure}

The expected order of convergence is verified for the advection of
a thermal bubble at time $t=5\,\mathrm{s}$, corresponding to one
tenth of a cycle, on coarse grid, $h=1\,\mathrm{m}$, consisting of
linear line elements, where the initial condition given by~(\ref{eq:thermal-bubble}).
We compute the $L_{2}$ error in terms of normalized values, defined
via the following relationships
\begin{equation}
T=\hat{T}\cdot T_{\mathrm{ref}},\quad\rho=\hat{\rho}\cdot\rho_{\mathrm{ref}},\quad v_{i}=\hat{v}_{i}\cdot\sqrt{p_{\mathrm{ref}}/\rho_{\mathrm{ref}}},\quad\rho e_{t}=\hat{\rho e_{t}}\cdot p_{\mathrm{ref}},\quad C_{i}=\hat{C}_{i}\cdot p_{\mathrm{ref}}/\left(R^{0}\cdot T_{\mathrm{ref}}\right),\label{eq:thermal-bubble-normalization}
\end{equation}
where $T_{\mathrm{ref}}=1000\,\mathrm{K}$, $\rho_{\mathrm{ref}}=1\,\mathrm{kg\cdot}\mathrm{m}^{-3}$,
and $p_{\mathrm{ref}}=101325\,\mathrm{Pa}$ are reference values.
Figures~\ref{fig:RKDG-convergence-state} and~\ref{fig:RKDG-convergence-temperature}
present the convergence results corresponding to the normalized conserved
state and the normalized temperature, respectively. The solutions
corresponding to $\mathrm{DG}\left(p=3\right)$, $\mathrm{DG}\left(p=4\right)$,
and $\mathrm{DG}\left(p=5\right)$ were refined until the error associated
with the conserved state was on the order of the time-step at which
point error associated with the temporal integration began to pollute
the results. Both the conserved state and the nonlinear temperature
converge at the expected rate for $\mathrm{DG}\left(p=3\right)$,
$\mathrm{DG}\left(p=4\right)$, and $\mathrm{DG}\left(p=5\right)$
as shown in Figure~\ref{fig:RKDG-convergence-state} and~\ref{fig:RKDG-convergence-temperature}.
Thus confirming that, in the case of convergence under grid refinement,
the multi-component formulation behaves exactly like its single-component
counterpart, achieving high-order accuracy in smooth regions of the
flow.

In order to access the accuracy of the formulation in in relation
to other methods, Figure~\ref{fig:RKDG-convergence-temperature}
also compares the normalized temperature $L^{2}$ error $\left\Vert \hat{T}-\hat{T}_{\mathrm{exact}}\right\Vert _{L_{2}\left(\Omega\right)}$
to the error corresponding to the $L^{2}$ projection of the exact
temperature profile, which minimizes the error in the $L^{2}$ norm,
thus providing an upper bound on the attainable accuracy. As solution
is better resolved, i.e., smaller $h$ or larger $p$, the DG approximation
approaches the $L^{2}$ projection in terms of accuracy. Finally,
there is zero temporally integration error associated with the $L^{2}$
projection of the exact solution, which might account for the small
differences still present in errors associated with the most well
resolved approximations as the temporally integration error was limiting
but did not vanish in the case of the DG approximations.

\subsection{Concluding remarks regarding discrete pressure equilibrium at material
interfaces\label{subsec:Concluding-remarks-pressure-equilibrium}}

In the previous Sections we presented several test cases to assess
when pressure oscillations would be generated for multi-component
non-reacting flows without physical diffusion. Additional stabilization,
e.g., artificial viscosity, limiting, or filtering, was not applied
in any of the test cases to avoid the unintentional suppression of
pressure oscillations. The results confirm the analysis presented
in~\ref{sec:Discontinuities} and are summarized below.

In the case of frozen thermodynamics, unphysical pressure oscillations
are generated at material interfaces and in smooth regions where mixing
occurs, requiring additional stabilization, or nonconservative methods.
The magnitude of the oscillations depends on the method with which
$\bar{c}_{p}$ is evaluated. Furthermore, these oscillations are dependent
on size of time step and grow as the solution evolves, even in smooth
regions of the flow, as shown in Figure~\ref{fig:thermal_bubble_delta_p_10_cycles}.

In contrast, if the thermodynamics are evaluated exactly, pressure
oscillations are not generated when the temperature is continuous,
regardless of species mixing at material interfaces. However, pressure
oscillations are generated at temperature discontinuities but do not
cause the solution to diverge, as the oscillations do not grow as
the solution evolves. Finally, pressure oscillations are not generated
in regions of sharp continuous profiles of species and temperature,
as described in Section~\ref{subsec:thermal-bubble}.

Therefore, we expect the $\text{DG}(p>0)$ formulation for the multi-component
reacting Navier-Stokes flows described in this work to behave no differently
than its counterpart for single-component Navier-Stokes flows, namely:
high order accuracy is achieved in smooth regions of the flow and
additional stabilization of the form~(\ref{eq:artifical-viscosity})
is only required in the presence of discontinuities.{\footnotesize{}}%
{\footnotesize\par}

\section{Results\label{sec:results}}

In order to demonstrate the practicality of the conservative DG method
described in this work, we apply it to both reacting and non-reacting
multi-component flows. We have selected challenging problems in one,
two, and three dimensions that showcase the ability of the discretization
to stably compute solutions to problems with both discontinuous interfaces
and sharp, but smooth, gradients, while maintaining conservation without
generating unphysical pressure oscillations. The first two test cases,
presented in Section~\ref{subsec:shock_tube_non_reacting} and Section~\ref{subsec:helium_bubble},
approximate the interaction of shocks with multi-component non-reacting
gases for cases previously reported in literature.

The final four test cases, presented in Sections~\ref{subsec:One-dimensional-detonation-wave}-\ref{subsec:hydrogen-air-shear-layer},
consider chemically reacting multi-component flows. The ODEs describing
the chemical reactions are all solved using DGODE. The first two test
cases, presented in Section~\ref{subsec:One-dimensional-detonation-wave}
and Section~\ref{subsec:Two-dimensional-detonation-wave}, explore
sustained detonations formed via an overdriven initialization. These
cases require artificial viscosity to stabilize the large pressure
and temperature discontinuity present in the region of the detonation.

The last two test cases, presented in Section~\ref{subsec:One-dimensional-premixed-flame}
and Section~\ref{subsec:hydrogen-air-shear-layer}, correspond to
full multi-component reacting Navier-Stokes solutions where the thermodynamics
and transport properties were given by the tran.dat file provided
by USC Mech II~\citep{uscmech}. The chemically reacting shear layer
cases did not require artificial viscosity to run stably. 

\subsection{One-dimensional non-reacting multi-component shock tube\label{subsec:shock_tube_non_reacting}}

\begin{figure}[H]
\subfloat[\label{fig:shock-tube-1d-primitive-variables}The mass fraction profiles
for $He$ and $N_{2}$, as well as the pressure, $p$, normalized
by 10~atm, at time $t=300\,\mu\mathrm{s}$. The initialization is
given by~(\ref{eq:shock-tube-initialization}). The estimated shock
speed of~711.9 m/s is in close agreement with the exact value 712~m/s~\citep{Hou11},
as is expected for a fully conservative discretization.]{\begin{centering}
\includegraphics[width=0.45\linewidth]{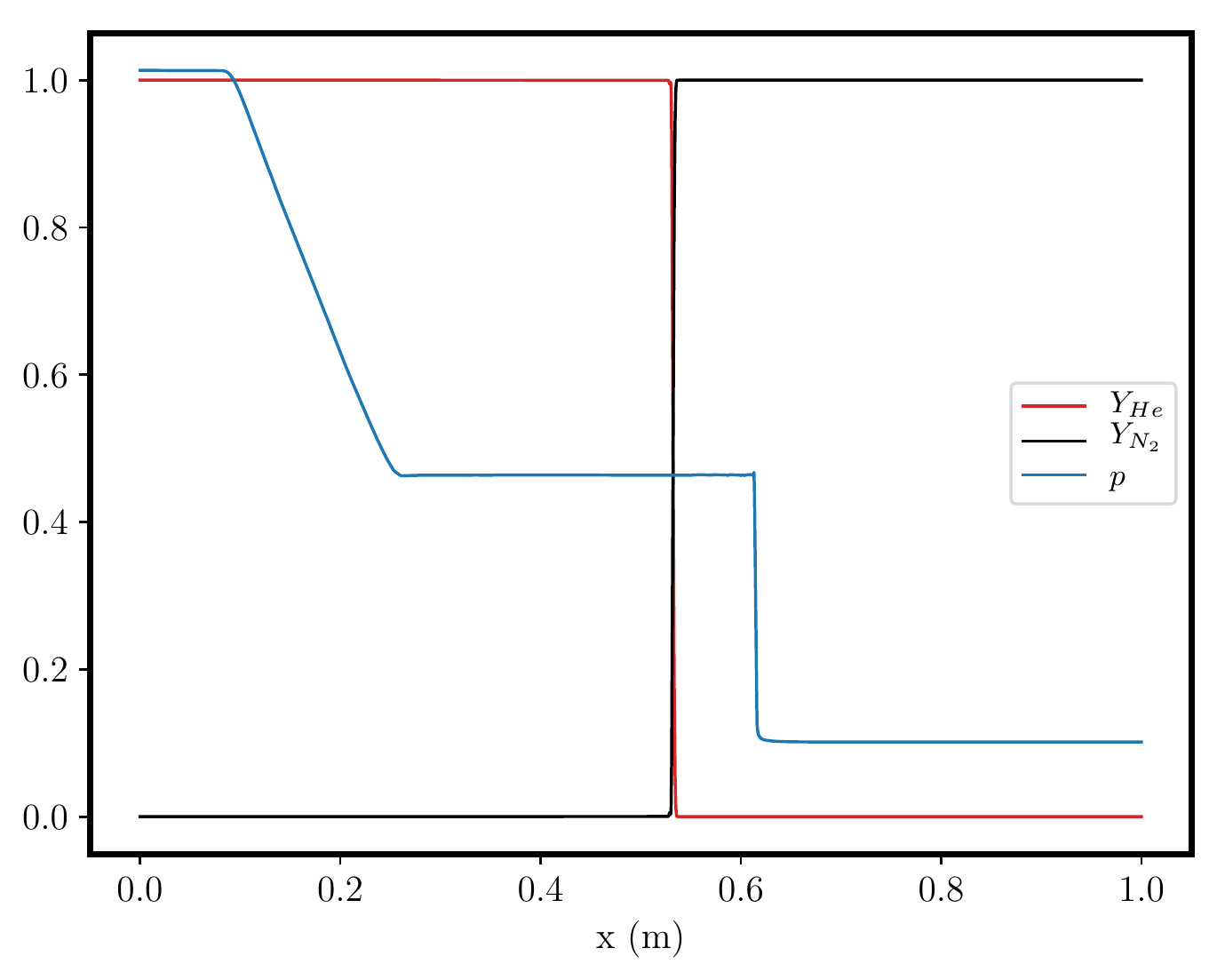}
\par\end{centering}
}\hfill{}
\begin{centering}
\subfloat[\label{fig:shock-tube-1d-energy-conservation}Percent loss of conservation
quantities, $\left(\int_{\Omega}\left(\phi\right)_{0}-\int_{\Omega}\left(\phi\right)_{t}\right)/\int_{\Omega}\left(\phi\right)_{0}\times100$,
for the one-dimensional multi-component shock tube computed as a function
of physical time computed using the initial integrated quantity, $\int_{\Omega}\left(\phi\right)_{0}d\Omega$,
and current integrated quantity, $\int_{\Omega}\left(\phi\right)_{t}d\Omega$,
where $\phi=\rho e_{t}$ for total energy and $\phi=\rho$ for density.
The conservation error is on the order of $10^{-14}$~\%. ]{\begin{centering}
\includegraphics[width=0.45\linewidth]{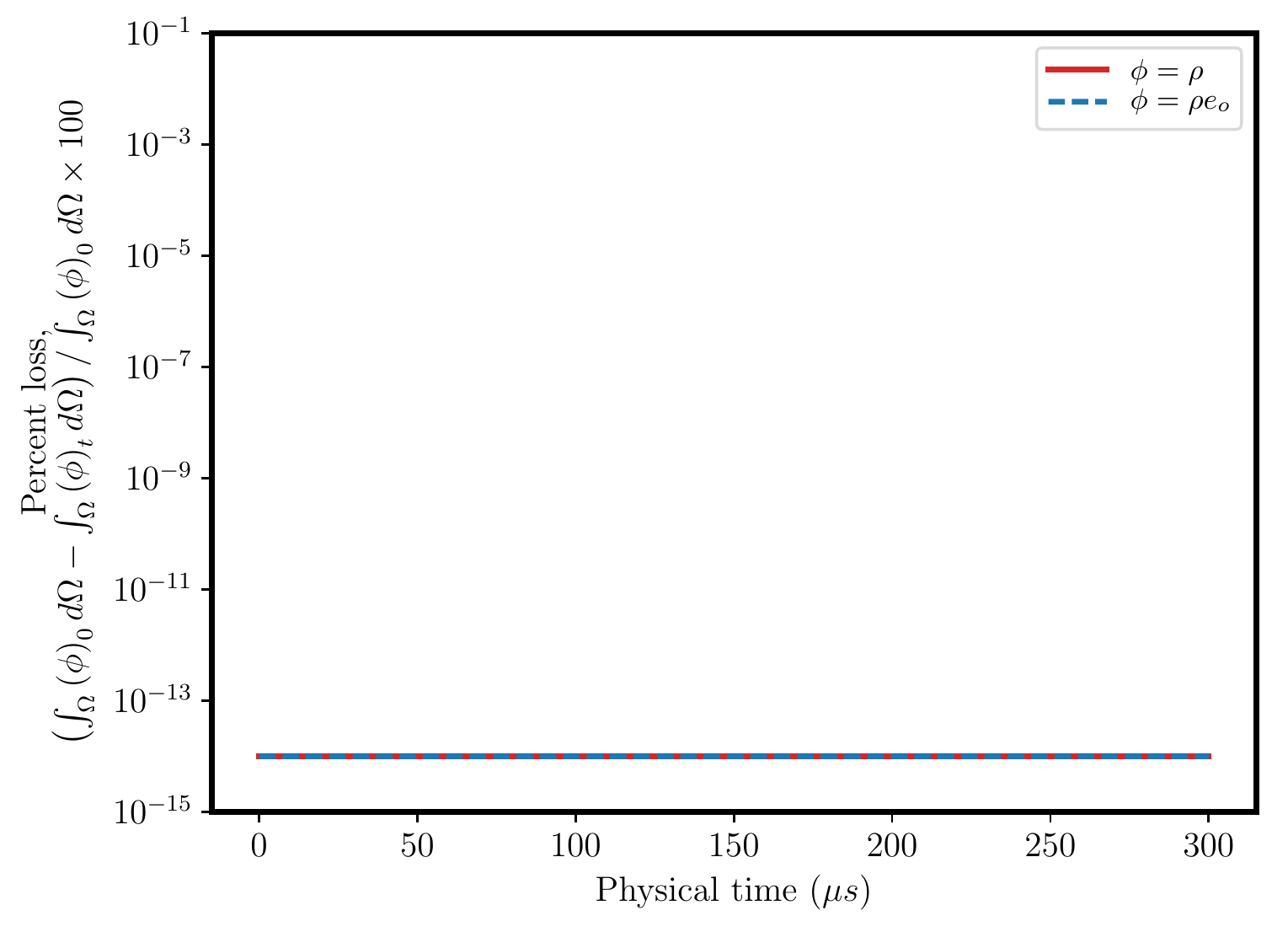}
\par\end{centering}
}
\par\end{centering}
\caption{\label{fig:shock-tube-1d}Solution and conservation error for the
one-dimensional multi-component shock tube computed using $\mathrm{DG}\left(p=2\right)$
on a grid consisting of linear line elements with $h=0.0005$~m.
The initialization is given by~(\ref{eq:shock-tube-initialization}).}
\end{figure}

This case was first presented by Houim and Kuo~\citep{Hou11}. A
$1$ m long one-dimensional domain, $(0,1)$ m, with walls at both
ends, and grid spacing $h=0.0005$ m is used with the following initial
conditions

\begin{eqnarray}
\left(v_{1},T,p,Y_{N_{2}},Y_{He}\right) & = & \begin{cases}
\left(0\text{ m/s},\:300\text{ K},\:\phantom{1}1\text{ atm},\:1,\:0\right) & x>0.4\\
\left(0\text{ m/s},\:300\text{ K},\:10\text{ atm},\:0,\:1\right) & x<0.4
\end{cases}.\label{eq:shock-tube-initialization}
\end{eqnarray}
Figure~\ref{fig:shock-tube-1d-energy-conservation} presents the
mass fraction profiles for $He$ and $N_{2}$, as well as the pressure,
$p$, profile normalized by $10$~atm at time $t=300\,\mu s$ computed
using $\mathrm{DG}\left(p=2\right)$ on 2000 linear line elements.
The shock front is maintained without spurious pressure oscillations.
The exact shock speed calculated by Houim and Kuo~\citep{Hou11}
was 712~m/s and our solution gave a shock speed of 711.9~m/s. Figure~\ref{fig:shock-tube-1d-energy-conservation}
shows the energy conservation percent loss and density conservation
percent loss as a function of physical time which stays at the expected
machine precision of $10^{-14}\%$ assuring that this formulation
is indeed conservative.

\subsection{Two-dimensional non-reacting air shock with helium bubble interaction\label{subsec:helium_bubble}}

\begin{figure}[H]
\begin{centering}
\includegraphics[width=0.9\columnwidth]{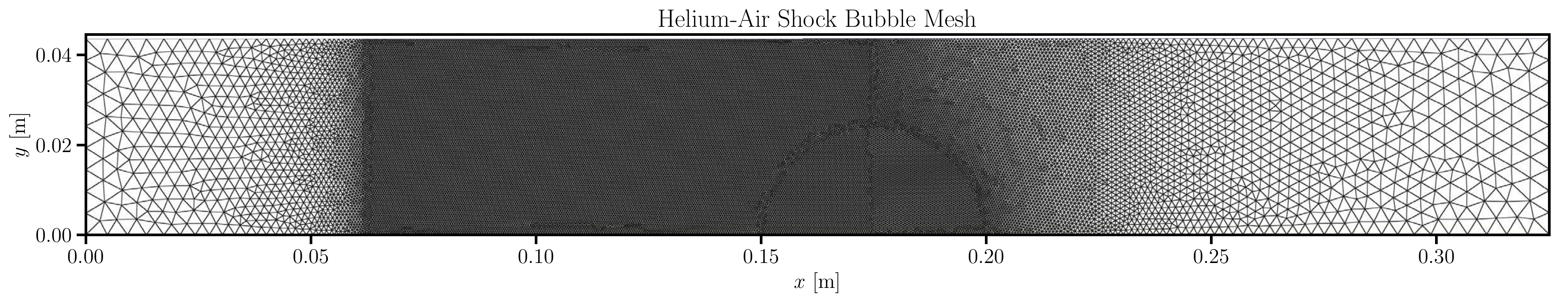}
\par\end{centering}
\caption{\label{fig:shock_bubble_mesh}Unstructured mesh for Helium-air shock
bubble test case.}
\end{figure}

This two-dimensional test case is modeled after an experiment of an
air shock wave traveling though a suspended helium bubble that was
performed by Hass and Sturtevant~\citep{Haa87}. Although this configuration
is naturally three-dimensional, it has been used previously as a two-dimensional
simulation validation case~\citep{Qui96,Mar03,Joh06,Lv15}. Here
we present the same two-dimensional case as previously simulated on
a rectangular domain, $\text{\ensuremath{\Omega}}=\left(0,0.325\right)\,\mathrm{\text{m}}\times\left(0,0.0455\right)\,\mathrm{\text{m}}$,
that has initial conditions with a normal shock located at $x=0.225$
m and a helium bubble to the left of the shock,

\begin{eqnarray}
\left(v_{1},v_{2}\right) & = & \begin{cases}
\left(M_{2}c_{2}-M_{1}c_{1},0\right) & x\geq0.225\\
\left(0,0\right)\text{ m/s} & x<0.225
\end{cases},\nonumber \\
Y_{N_{2}} & = & \begin{cases}
0.785 & \sqrt{(x-0.175)^{2}+y^{2}}\geq0.025\text{ m}\\
0 & \sqrt{(x-0.175)^{2}+y^{2}}<0.025\text{ m}
\end{cases}\nonumber \\
Y_{O_{2}} & = & \begin{cases}
0.215 & \sqrt{(x-0.175)^{2}+y^{2}}\geq0.025\text{ m}\\
0 & \sqrt{(x-0.175)^{2}+y^{2}}<0.025\text{ m}
\end{cases},\label{eq:shock-helium-bubble-interaction-initialization}\\
Y_{He} & = & \begin{cases}
0 & \sqrt{(x-0.175)^{2}+y^{2}}\geq0.025\text{ m}\\
1 & \sqrt{(x-0.175)^{2}+y^{2}}<0.025\text{ m}
\end{cases},\nonumber \\
T & = & \begin{cases}
300\frac{T_{2}}{T_{1}}\text{ K} & x\geq0.225\\
300\text{ K} & x<0.225
\end{cases},\nonumber \\
p & = & \begin{cases}
\frac{p_{2}}{p_{1}}\text{ bar} & x\geq0.225\\
1\text{ bar} & x<0.225
\end{cases}.\nonumber 
\end{eqnarray}
where $c=\sqrt{\gamma RT}$ and the normal shock ratios, $T_{2}/T_{1}$
and $p_{2}/p_{1}$, and post shock Mach number, $M_{2}$, can be calculated
from the isentropic flow relations for air $\gamma=1.4$ which yields,
for a $M_{1}=1.22$ normal shock, $T_{2}/T_{1}=1.14054133$, $p_{2}/p_{1}=1.56979999$,
and $M_{2}=0.82998648$. The shifting of velocity by the pre-shock
velocity condition gives an inflow boundary condition on the right
hand side. For this configuration we used the unstructured linear
triangular grid shown in Figure~\ref{fig:shock_bubble_mesh}. The
grid is built so that it is refined where the shock and bubble interact.
The grid has a target grid spacing of $h=0.005$~m on the left and
right boundaries and refines to $h=0.0001$~m where the shock and
bubble interact. We also constructed the mesh so that the helium bubble
is initially grid aligned. %

\begin{figure}[H]
\subfloat[\label{fig:shock-bubble-p1}Density profile using $\mathrm{DG}\left(p=1\right)$
for shock and helium bubble interaction at different times.]{\begin{centering}
\includegraphics[width=0.45\linewidth]{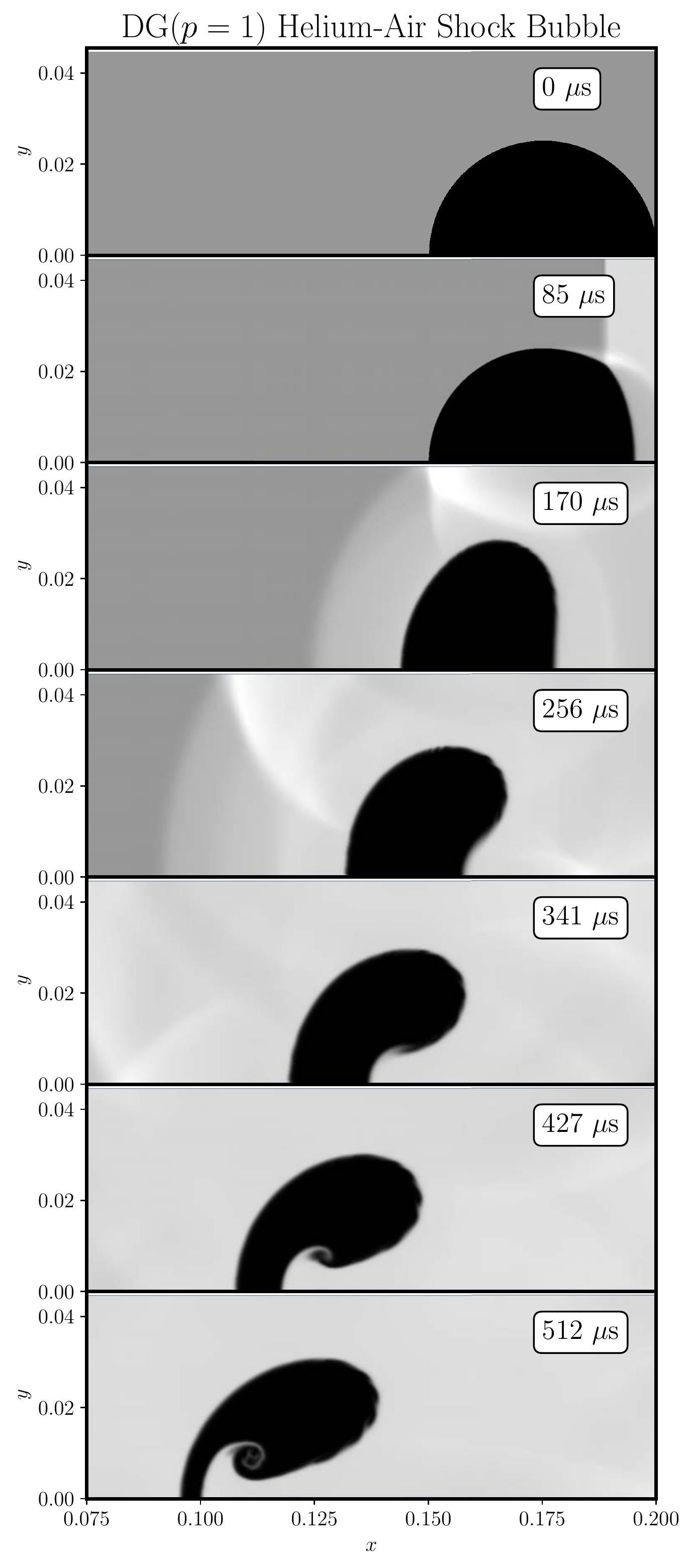}
\par\end{centering}
}\hfill{}\subfloat[\label{fig:shock-bubble-p2}Density profile using $\mathrm{DG}\left(p=2\right)$
for shock and helium bubble interaction at different times.]{\begin{centering}
\includegraphics[width=0.45\linewidth]{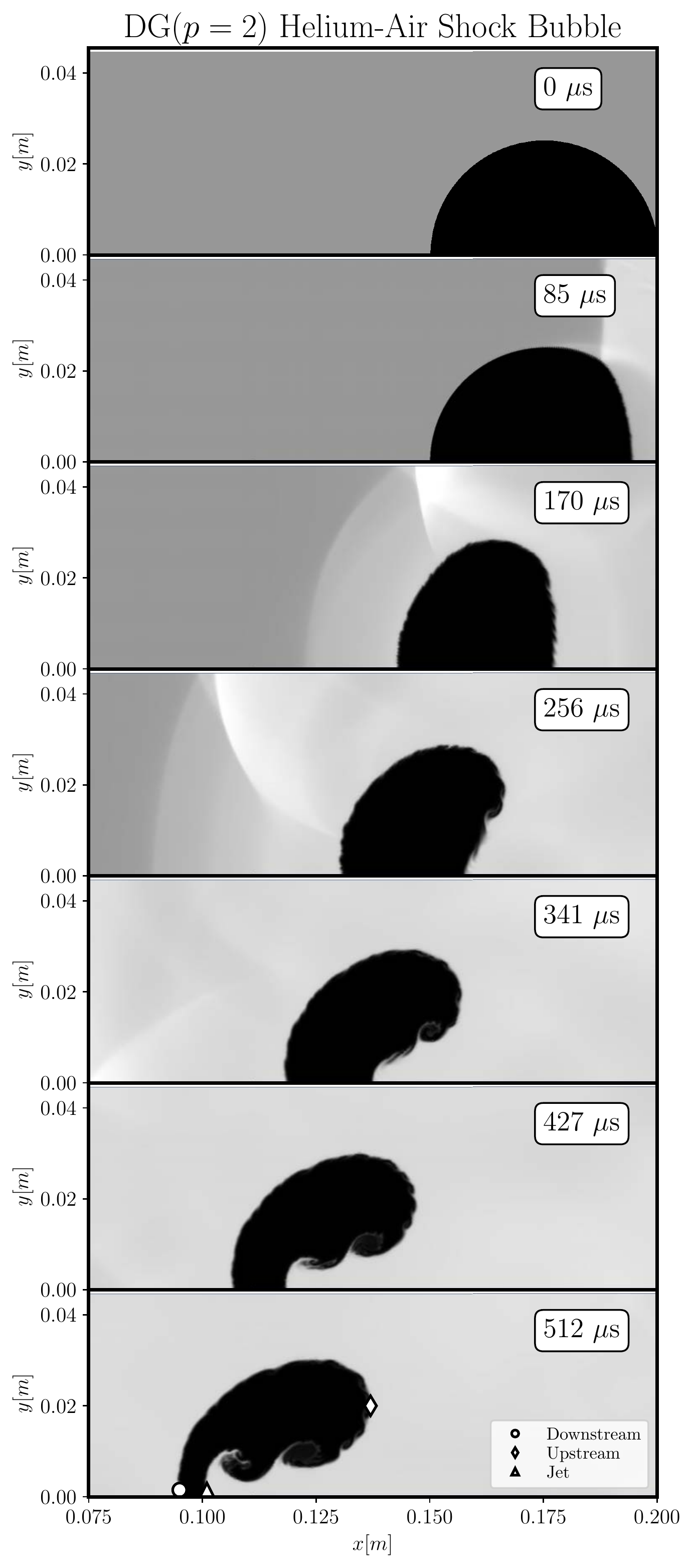}
\par\end{centering}
}

\caption{\label{fig:shock-bubble-time}Density profile snapshots for $\mathrm{DG}\left(p=1\right)$
and $\mathrm{DG}\left(p=2\right)$ at different times. The initialization
is given by~(\ref{eq:shock-helium-bubble-interaction-initialization}).
Unphysical pressure oscillations at the material interface were not
observed as pressure and temperature discontinuities were stabilized
with the addition of artificial viscosity, resulting in sharp, yet
continuous, profiles, thus confirming the results presented in Section~\ref{sec:discrete-pressure-equilibrium}.}
\end{figure}

The evolution of the helium bubble as it interacts with the shock
is shown for $\mathrm{DG}\left(p=1\right)$ and $\mathrm{DG}\left(p=2\right)$
in Figures~\ref{fig:shock-bubble-p1} and~\ref{fig:shock-bubble-p2},
respectively. The shock wave collides with the helium bubble and causes
compression and expansion as well as mixing of the helium into the
surrounding air. The $\mathrm{DG}\left(p=2\right)$ solution more
accurately resolves this interaction of the pressure waves than the
$\mathrm{DG}\left(p=1\right)$ solution. Additionally, the mixing
of the helium on the left and right sides of the bubble are more resolved
with the $\mathrm{DG}\left(p=2\right)$ solution. For these simulations
we did not notice any spurious pressure oscillations at the material
interfaces, as is expected based on the results presented in Section~\ref{sec:discrete-pressure-equilibrium}
as the shock waves are stabilized with the addition of artificial
viscosity, resulting in sharp, yet smooth, profiles in temperature
and pressure.

\begin{figure}[H]
\begin{centering}
\includegraphics[width=0.9\columnwidth]{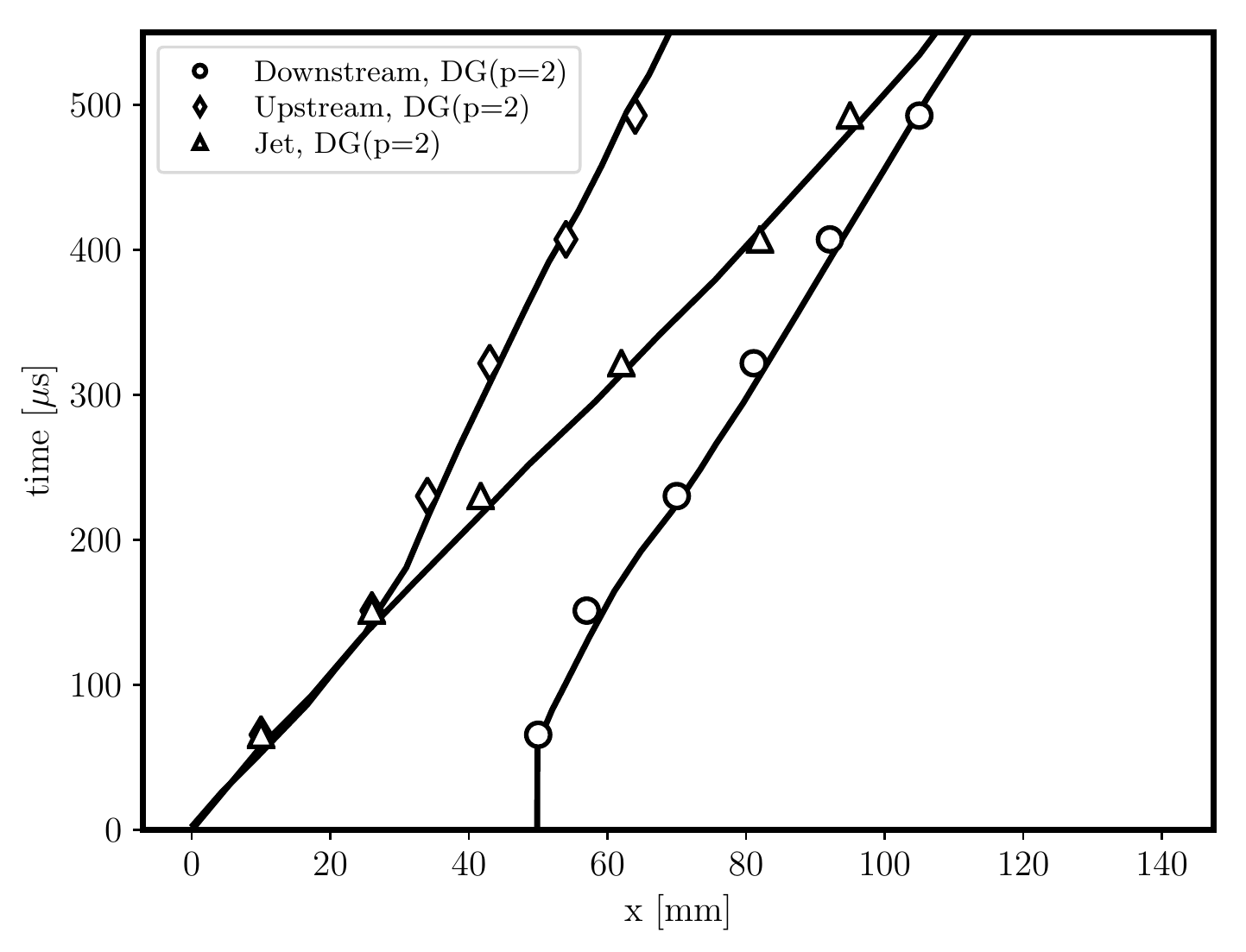}
\par\end{centering}
\caption{\label{fig:front_locations}Front Locations for the $\mathrm{DG}\left(p=2\right)$
solution (symbols) and accompanying data from \citep{Ter09} (solid
lines).}
\end{figure}

We have labeled three points as downstream, jet, and upstream for
the solution at time $t=512$ $\mu\mathrm{s}$ in Figure~\ref{fig:shock-bubble-p2}.
The downstream point is the left-most location of the helium bubble,
the jet point is the right-most location of the helium bubble at $y=0$~m,
and the upstream point is the right-most location of the helium bubble
in the entire domain. The impact of the shock on the helium bubble
causes these locations to move as a function of time, where the jet
and downstream location eventually merge. Figure~\ref{fig:front_locations}
shows the trajectory of these points for the $\mathrm{DG}\left(p=2\right)$
solution and agrees well with the accompanying solid lines from \citep{Ter09}.

\subsection{One-dimensional detonation wave\label{subsec:One-dimensional-detonation-wave}}

Here we present a one-dimensional hydrogen-oxygen detonation wave
diluted in Argon. A $0.45$ m long one-dimensional domain, $(0,0.45)$
m, with walls at both ends and grid spacing, $h=9\times10^{-5}$ m
is used with the following initial conditions

\begin{equation}
\begin{array}{cccc}
\qquad\qquad\qquad\qquad v_{1} & = & 0\text{ m/s},\\
\quad\quad\quad\;X_{Ar}:X_{O_{2}}:X_{H_{2}} & = & 7:1:2 & \text{ }x>0.025\text{ m},\\
X_{Ar}:X_{O_{2}}:X_{H_{2}}:X_{OH} & = & 7:1:2:0.01 & \text{ }0.015\text{ m}<x<0.025\text{ m,}\\
\quad\quad\;X_{Ar}:X_{H_{2}O}:X_{OH} & = & 8:2:0.01 & x<0.015\text{ m},\\
\qquad\qquad\qquad\qquad p & = & \begin{cases}
\expnumber{5.50}5 & \text{ Pa}\\
\expnumber{6.67}3 & \text{ Pa}
\end{cases} & \begin{array}{c}
x<0.015\text{ m}\\
x>0.015\text{ m}
\end{array},\\
\qquad\qquad\qquad\qquad T & = & \begin{cases}
298 & \text{ K}\\
350 & \text{ K}\\
3500\text{\hspace{1em}\hspace{1em}} & \text{ K}
\end{cases} & \begin{array}{c}
\text{ }x>0.025\text{ m}\\
\text{ }0.015\text{ m}<x<0.025\text{ m}\\
x<0.015\text{ m}
\end{array}.
\end{array}\label{eq:detonation-1d-initialization}
\end{equation}
The region $0.015$~m~$<x<0.025$~m has a small amount of $OH$-radical
with the premixed fuel to encourage reactivity as the large shock
moves to the right from the region with the initial driver pressure,
$x<0.015$~m. The detailed chemical kinetics are described by the
reaction mechanism of Westbrook~\citep{Wes82}. This test case is
similar to the cases presented by~\citep{Hou11} and~\citep{Lv15}
with the difference being in the initialization. Regardless of the
initialization, the final detonation will correspond to the Chapman-Jouget
detonation solution for a mixture with the mole fraction ratios $X_{Ar}:X_{O_{2}}:X_{H_{2}}=7:1:2$.

\begin{figure}[H]
\begin{centering}
\includegraphics[width=0.8\columnwidth]{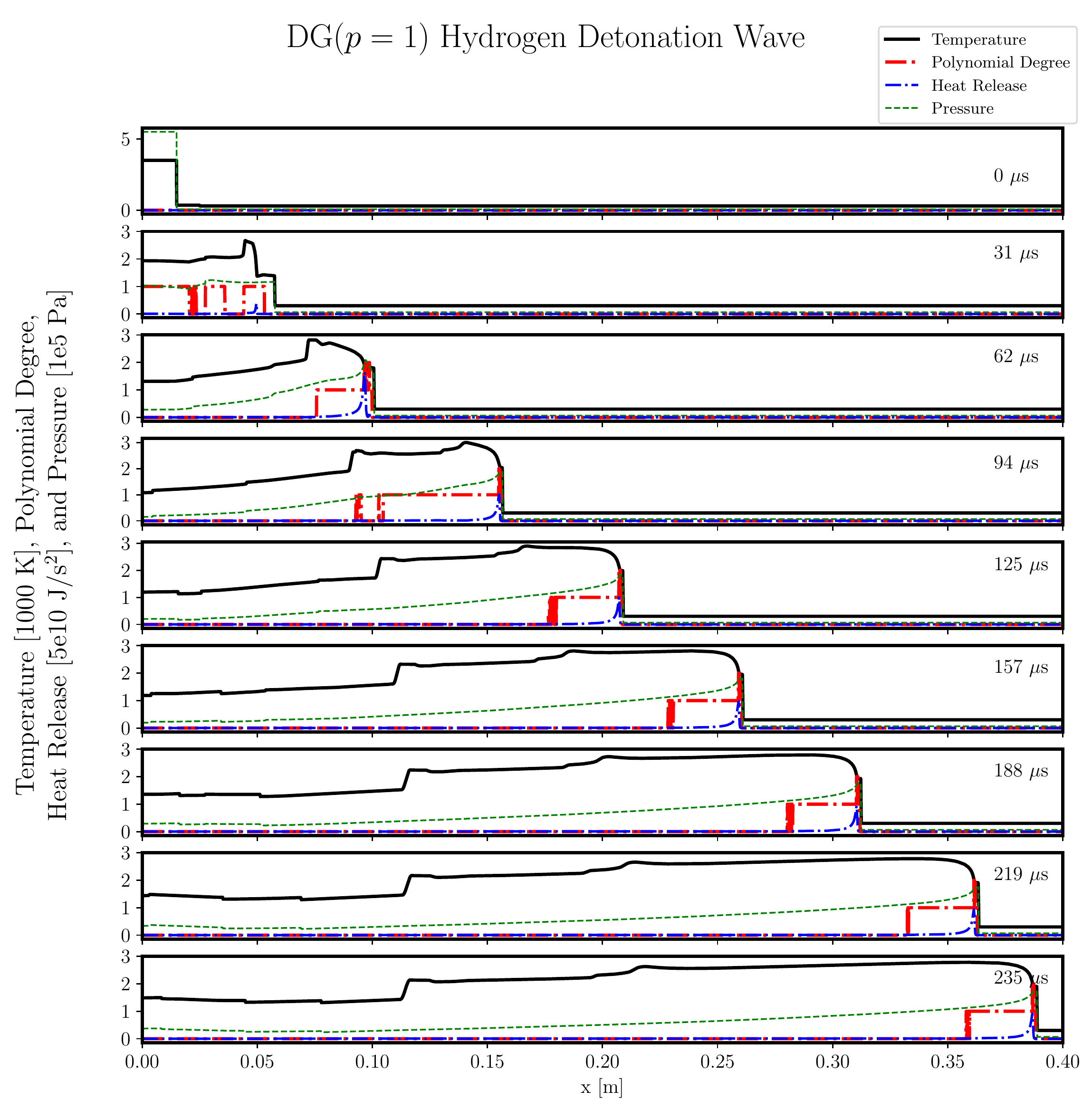}
\par\end{centering}
\caption{\label{fig:traveling_detonation}One-dimensional hydrogen detonation
test case temperature and pressure evolution from a high pressure
and high temperature initialization to a steady traveling detonation
front. Heat release,$-\sum_{k=1}^{n_{s}}\omega_{k}W_{k}h_{k}^{0}$
, and DGODE polynomial degree are also shown to indicate areas of
chemical activity and stiffness. The discontinuous solution at the
detonation front is stabilized with the addition of artificial viscosity.
The initialization is given by~(\ref{eq:detonation-1d-initialization}).}
\end{figure}

Figure~\ref{fig:traveling_detonation} shows the temperature, adapted
polynomial degree of the DGODE approximation, heat release, $-\sum_{k=1}^{n_{s}}\omega_{k}W_{k}h_{k}^{0}$,
and pressure as function of distance as the detonation evolves in
time. Initially a shock travels to the right with a reaction zone
lagging behind the high pressure front, as indicated by the heat release
at solution times $31$~$\mu\mathrm{s}$ and $62$~$\mu\mathrm{s}$.
As the reactivity increases the largest heat release reaches the shock
front, at solution time $94$~$\mu\mathrm{s}$, and a steady detonation
front is achieved and maintained, as shown for solution times greater
than $125$~$\mu\mathrm{s}$.

For the steady detonation, the polynomial degree of the DGODE approximation
is greatest at the detonation front, reaching a value of $p=2$, and
relaxes to $p=0$ as the distance from the detonation increases. At
start-up, the polynomial degree of the DGODE approximation is $p=1$
in the high temperature regions where the initialization is only water
and hydroxyl. This is due to the large initial temperature, $T=3500$~K,
in the driver zone that causes backwards reactions. As the temperature
in this region decreases due to the backwards moving expansion, the
polynomial degree of the DGODE approximation decreases accordingly,
as can be seen with solutions at time $31$~$\mu\mathrm{s}$ and
$62$~$\mu\mathrm{s}$. A spike in the polynomial degree of the DGODE
approximation exists to the left of the detonation wave for the solution
at time $94$~$\mu\mathrm{s}$. This location has a weaker discontinuity
than the detonation but is strong enough to offset the reacted zone
away from chemical equilibrium to require high-order chemistry integration.

\begin{figure}[H]
\begin{centering}
\subfloat[Velocity of the leading shock front as function of time, the front
location is defined as the right most location where the jump in temperature
from $300$~K was larger than $10$~K and a second order finite
differencing scheme was used to calculate the instantaneous velocity.\label{fig:front_velocity}]{\begin{centering}
\includegraphics[width=0.45\linewidth]{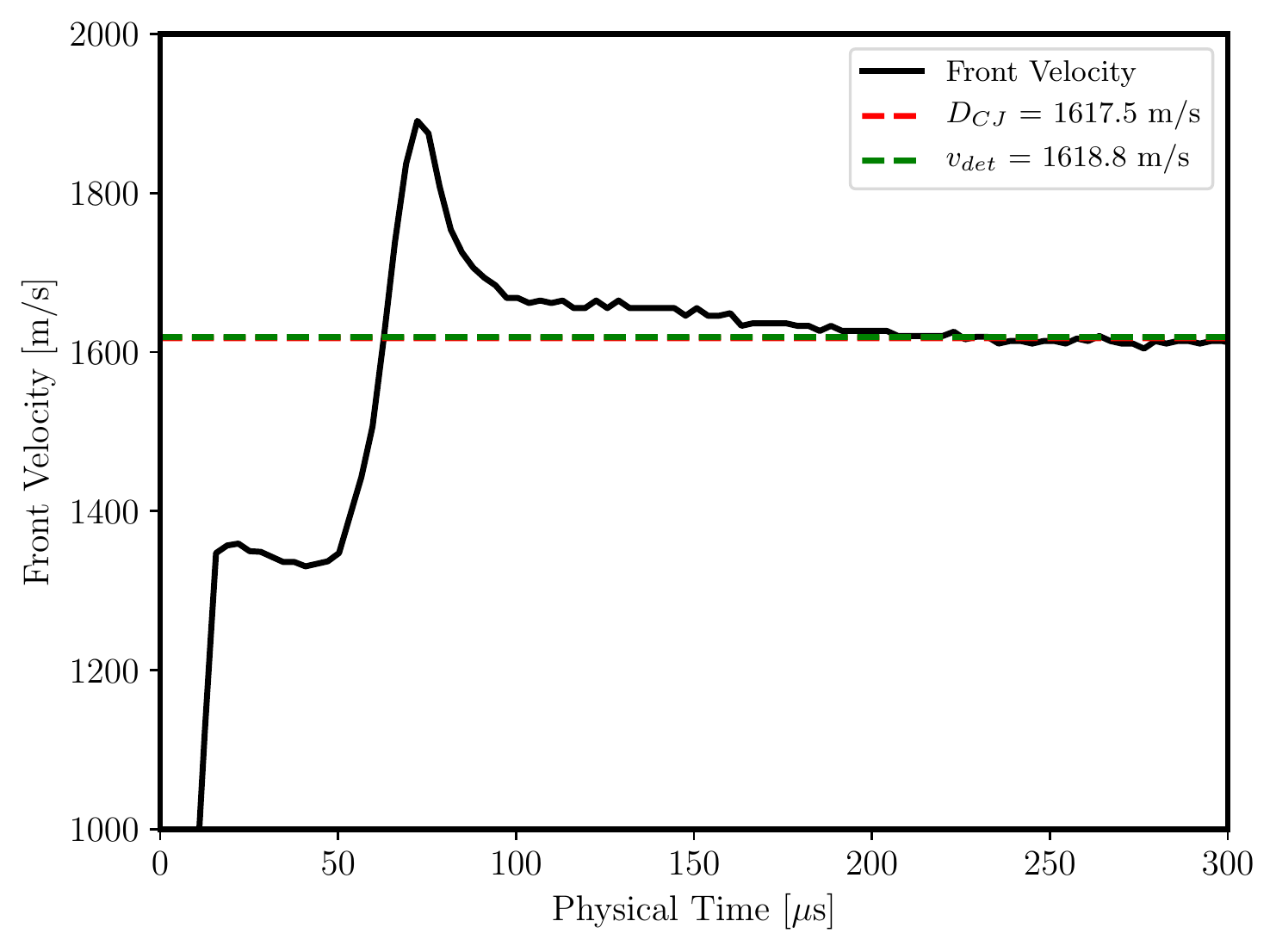}
\par\end{centering}
}\hfill{}\subfloat[The calculated front location and the approximated linear propagation
using the calculated velocity of $1618.8$~m/s using linear regression
with $r^{2}=0.999994$. \label{fig:front_velocity_position}]{\begin{centering}
\includegraphics[width=0.45\linewidth]{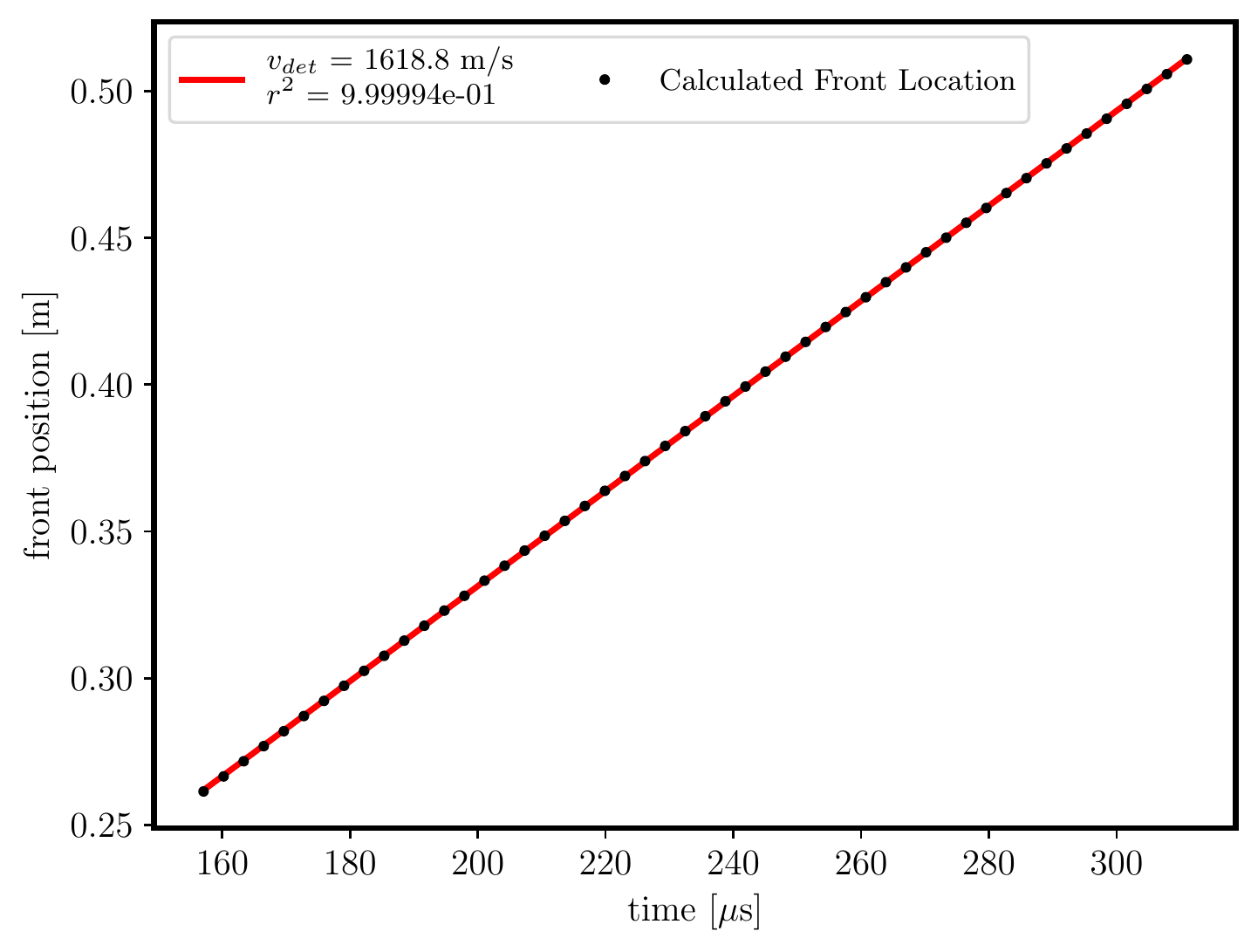}
\par\end{centering}
}
\par\end{centering}
\caption{\label{fig:detonation_velocity}Front location and velocity as a function
of physical time.}
\end{figure}

\begin{figure}[H]
\subfloat[\label{fig:compare_to_SDtoolbox_TP}Comparison of the temperature
and pressure profiles corresponding to $\mathrm{DG}\left(p=1\right)$
(solid red lines) and the solution computed via the Shock Detonation
Toolbox (dashed black lines) for the one-dimensional hydrogen detonation
described in Section~\ref{subsec:One-dimensional-detonation-wave}.]{\begin{centering}
\includegraphics[width=0.45\linewidth]{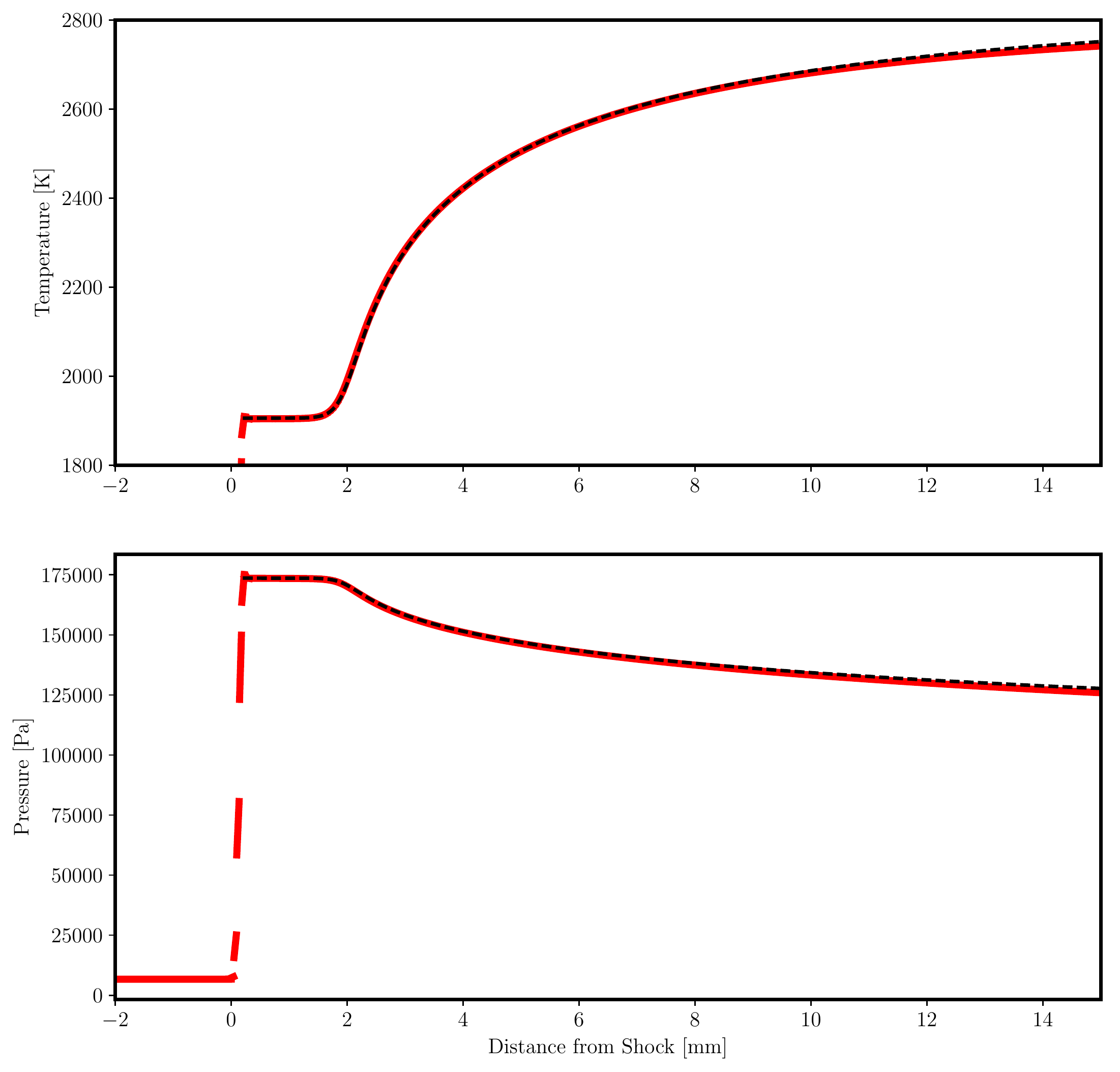}
\par\end{centering}
}\hfill{}
\begin{centering}
\subfloat[\label{fig:compare_to_SDtoolbox_Y}Comparison between Shock Detonation
Toolbox (dashed black lines) and current simulations (solid colored
lines) for species mass fractions.]{\begin{centering}
\includegraphics[width=0.45\linewidth]{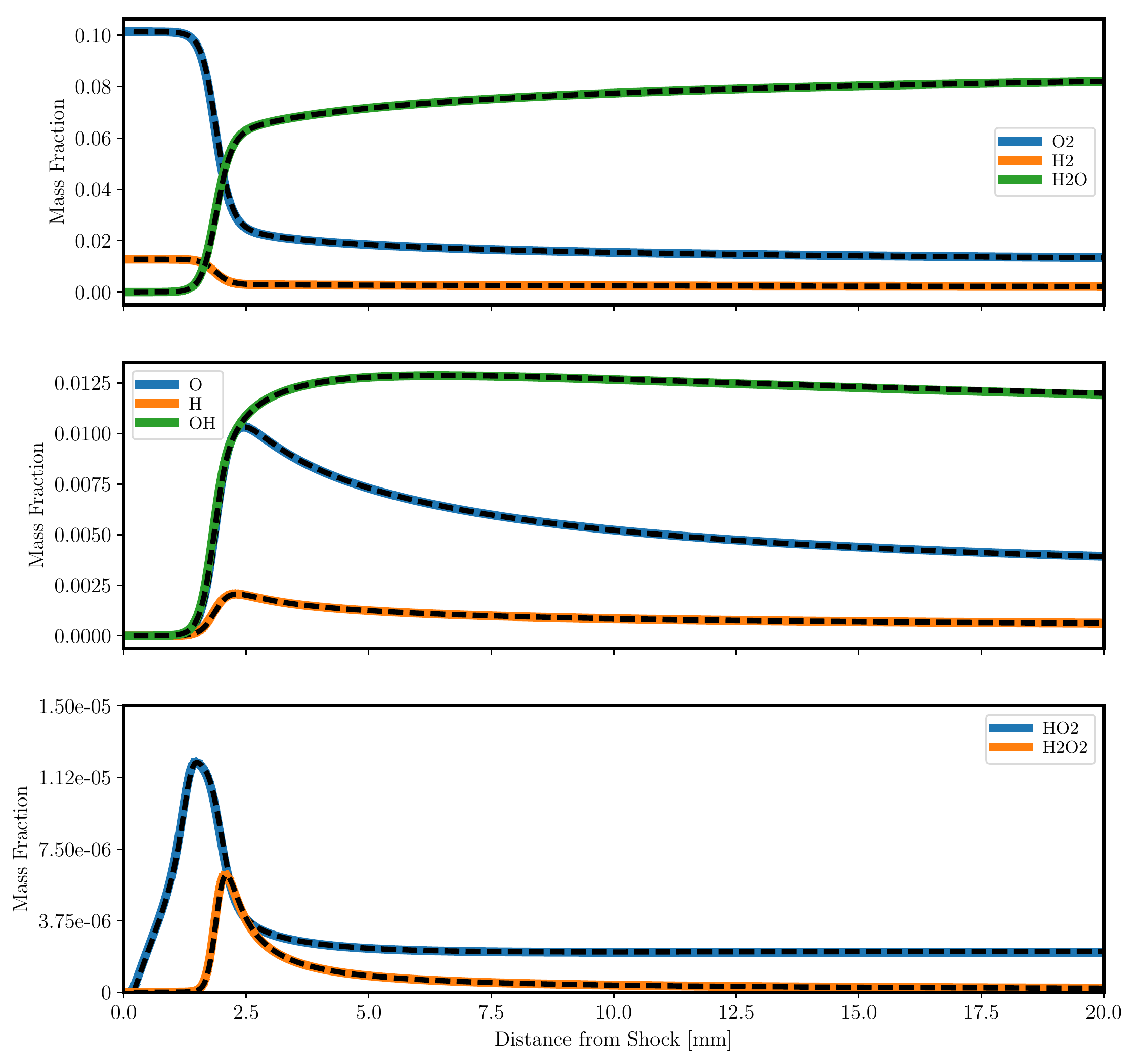}
\par\end{centering}
}
\par\end{centering}
\caption{\label{fig:compare_to_SDToolbox}Comparison of the $\mathrm{DG}\left(p=1\right)$
solution for $h=9\times10^{-5}$~m and the solution computed via
the Shock Detonation Toolbox at at time $t=235$~$\mu\mathrm{s}$
for the one-dimensional hydrogen detonation described in Section~\ref{subsec:One-dimensional-detonation-wave}.
The discontinuous solution at the detonation front is stabilized with
the addition of artificial viscosity.}
\end{figure}

Figure~\ref{fig:front_velocity} shows the velocity of the leading
shock front as function of time. We define the front location as the
right most location where the jump in temperature from $300$~K was
larger than $10$~K and a second order finite differencing scheme
was used to calculate the instantaneous velocity. Initially, the front
moves as a detached oblique shock traveling through the unreacted
medium, as seen in solution time $40$-$80$~$\mu\mathrm{s}$. Eventually
the reaction zone behind the front begins to influence the shock and
an overdriven detonation event occurs. The detonation velocity, $v_{det}$,
falls to the steady value of $1618.8$~m/s which was calculated using
linear regression with $r^{2}=0.999994$. The calculated front location
and the approximated linear propagation using the computed detonation
velocity are shown in Figure~\ref{fig:front_locations}. The Chapman-Jouget
detonation velocity, $v_{det}=D_{CJ}=1617.5$~m/s, as well as the
post shock profiles for temperature, pressure, and species concentrations
can be calculated using Shock and Detonation Toolbox~\citep{sdtoolbox}.
Figures~\ref{fig:compare_to_SDtoolbox_TP} and~\ref{fig:compare_to_SDtoolbox_Y}
show the solution at time $t=235$~$\mu\mathrm{s}$ in direct comparison
to the Shock Detonation Toolbox computed results (dashed black lines)
for temperature, pressure, and species mass fractions. Additionally,
Table~\ref{tab:post_shock_values} summarizes the post shock values
and detonation velocity corresponding to the current simulation and
compares them with the values calculated using the Shock Detonation
Toolbox and the values reported by Houim and Kuo~\citep{Hou11} as
well as Lv and Ihme~\citep{Lv15}. The results presented in this
work agree well with all past reported solutions and the Shock Detonation
Toolbox in particular.
\begin{table}[H]
\caption{\label{tab:post_shock_values}Table of post-shock values for one-dimensional
detonation wave case.}

\centering{}%
\begin{tabular}{|c|c|c|c|}
\hline 
Reference & $p_{vN}$ {[}kPa{]} & $T_{vN}$ {[}K{]} & $v_{det}$ {[}m/s{]}\tabularnewline
\hline 
\hline 
Shock and Detonation Toolbox~\citep{sdtoolbox} & 173.6 & 1904.7 & 1617.5\tabularnewline
\hline 
Current Simulations & 173.5 & 1904.7 & 1618.8\tabularnewline
\hline 
Houim and Kuo~\citep{Hou11} & 173.9 & 1915.2 & 1619.8\tabularnewline
\hline 
Lv and Ihme~\citep{Lv15} & 179.4 & 1926.0 & 1634.6\tabularnewline
\hline 
\end{tabular}
\end{table}

Spurious pressure oscillations were not generated during the simulation
of this test case, indicating the formulation can stably compute detonations.
For this test case, the integrated total energy, $\int_{\Omega}\left(\rho e_{t}\right)_{0}d\Omega$,
and the density, $\int_{\Omega}\left(\rho\right)_{0}d\Omega$, are
conserved as the domain has walls at both the left and right boundaries.
Figure~\ref{fig:traveling_detonation} shows the percent loss of
the integrated total energy, $\left(\int_{\Omega}\left(\rho e\right)_{0}d\Omega-\int_{\Omega}\left(\rho e\right)_{t}d\Omega\right)/\int_{\Omega}\left(\rho e\right)_{0}d\Omega\times100$,
as well as the percent loss of the integrated density, $\left(\int_{\Omega}\left(\rho\right)d\Omega-\int_{\Omega}\left(\rho\right)_{t}d\Omega\right)/\int_{\Omega}\left(\rho\right)_{0}d\Omega\times100$,
as a function of time. The error remains on the order of $\expnumber 1{-13}$\%
for the entire simulation which is expected for a conservative method.
\begin{figure}[H]
\begin{centering}
\includegraphics[width=0.6\columnwidth]{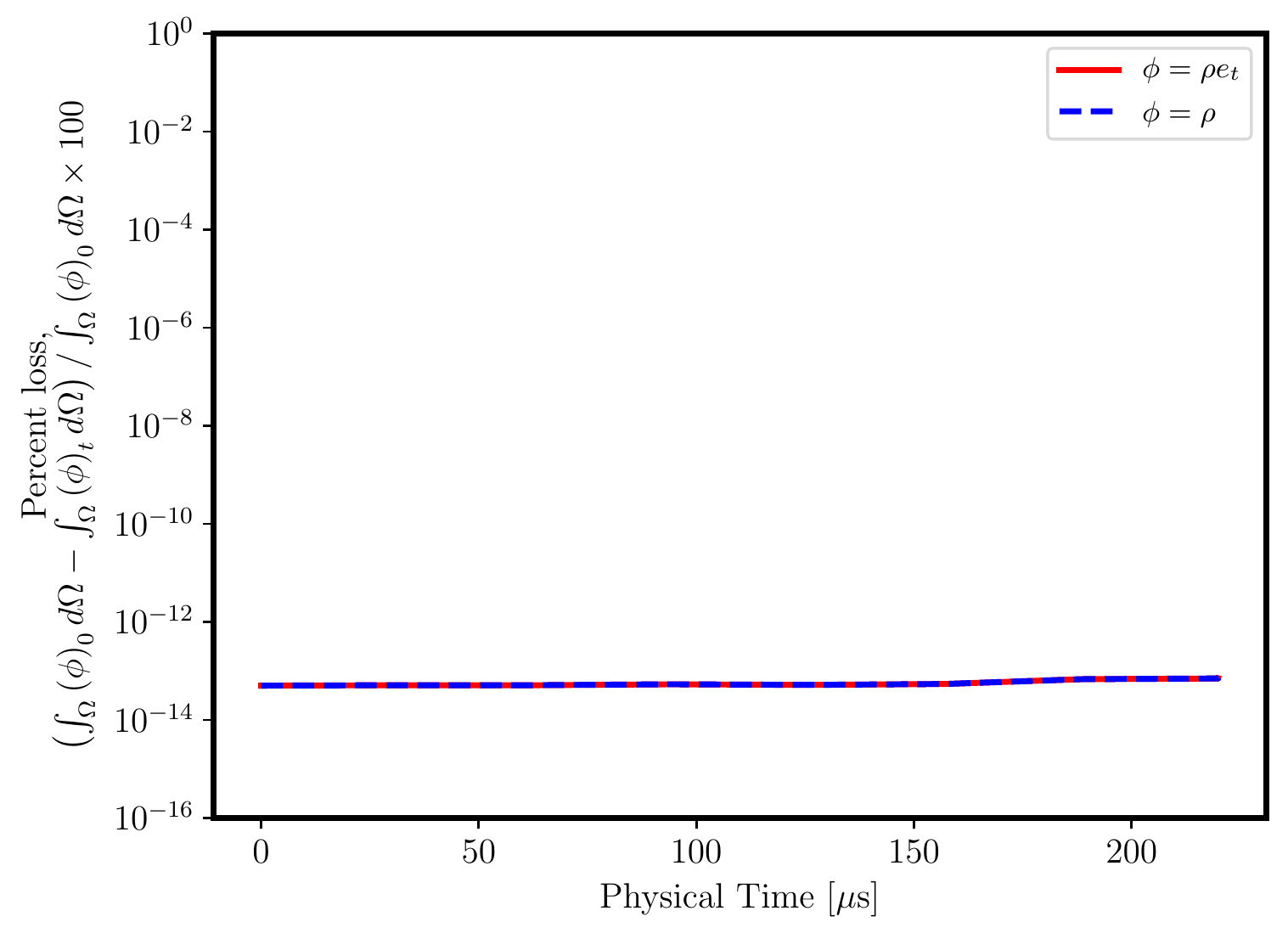}
\par\end{centering}
\caption{\label{fig:one_dimensional_detonation_energy_conservation}Percent
loss of conservation quantities, $\left(\int_{\Omega}\left(\phi\right)_{0}-\int_{\Omega}\left(\phi\right)_{t}\right)/\int_{\Omega}\left(\phi\right)_{0}\times100$,
for the one-dimensional detonation computed using $\mathrm{DG}\left(p=1\right)$
for $h=9\times10^{-5}$~m as a function of physical time computed
using the initial integrated quantity, $\int_{\Omega}\left(\phi\right)_{0}d\Omega$,
and current integrated quantity, $\int_{\Omega}\left(\phi\right)_{t}d\Omega$,
where $\phi=\rho e_{t}$ for total energy and $\phi=\rho$ for density.}
\end{figure}

\subsection{Two-dimensional detonation wave\label{subsec:Two-dimensional-detonation-wave}}

\begin{figure}[H]
\begin{centering}
\includegraphics[width=0.9\columnwidth]{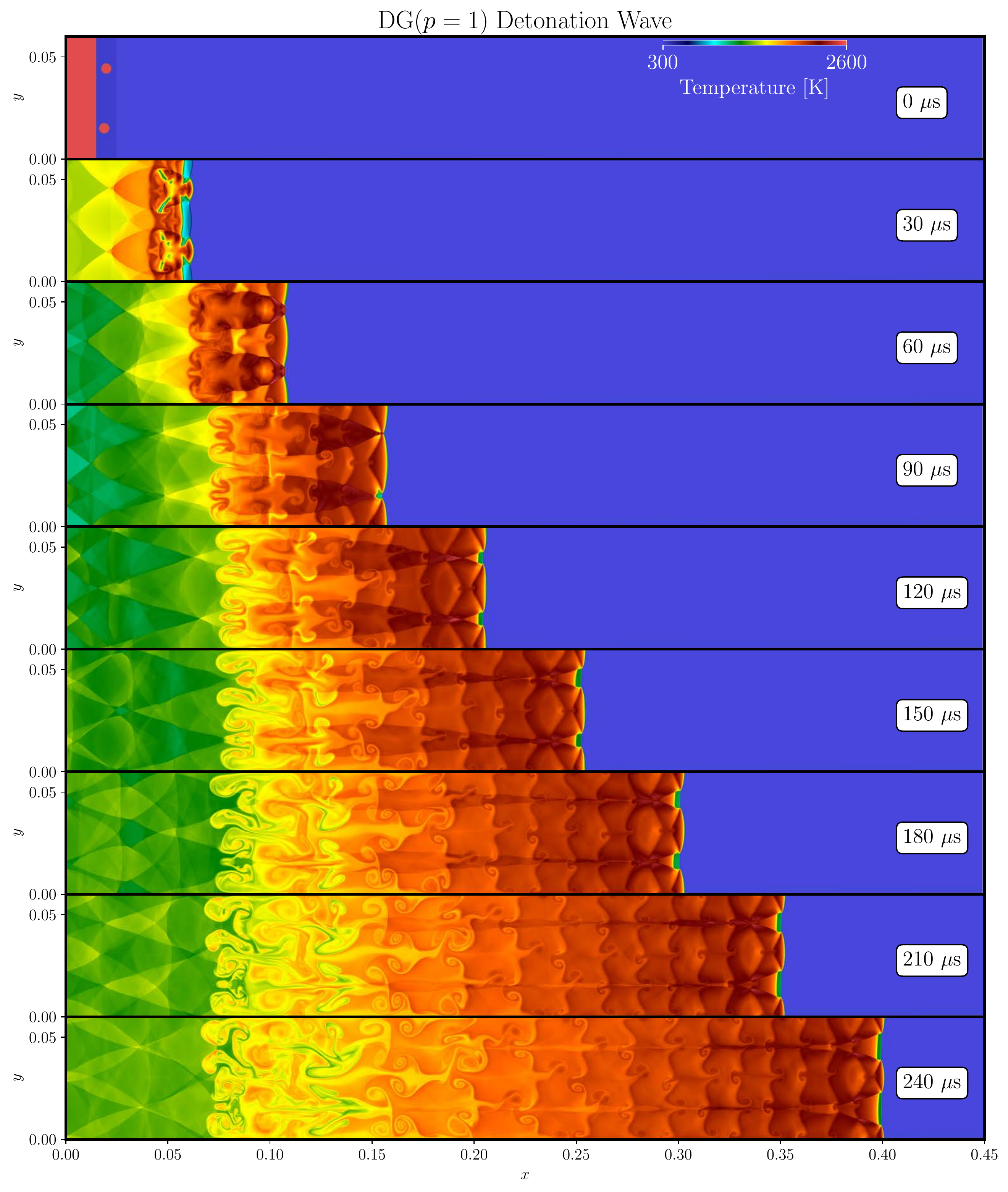}
\par\end{centering}
\caption{\label{fig:two-dimesiona-detonation-wave_temperature}Temperature
contours for two-dimensional detonation wave. The discontinuous solution
at the detonation front is stabilized with the addition of artificial
viscosity. The corresponding one-dimensional initialization given
by~(\ref{eq:shock-helium-bubble-interaction-initialization}) is
augmented with two additional circular two additional high pressure
and high temperature regions~(\ref{eq:detonation-1d-initialization}). }
\end{figure}

Here we present the two-dimensional extension of the test case presented
in Section~\ref{subsec:One-dimensional-detonation-wave} with a $0.45$~m
long domain that has a channel height of $0.06$~m , $\text{\ensuremath{\Omega}}=\left(0,0.45\right)\mathrm{m}\times\left(0,0.06\right)\mathrm{m}$,
with a uniform quadrilateral mesh with grid spacing $h=9\times10^{-5}$~m.
Each quadrilateral element was split into two triangular elements.
The results presented here used a $\text{DG}\left(p=1\right)$ approximation.
The top and bottom boundaries are simulated as walls. The initial
conditions are the same as Section~\ref{subsec:One-dimensional-detonation-wave}
with the exception of two additional high pressure and high temperature
regions, located within the circles $\sqrt{\left(x-0.019\right)^{2}+\left(y-0.015\right)^{2}}=0.0025$
m and $\sqrt{\left(x-0.010\right)^{2}+\left(y-0.044\right)^{2}}=0.0025$
m, with conditions

\begin{equation}
\begin{array}{ccc}
\left(v_{1},v_{2}\right) & = & \left(0,0\right)\text{ m/s},\\
X_{Ar}:X_{H_{2}O}:X_{OH} & = & 8:2:0.01,\\
p & = & \expnumber{5.5}5\text{ Pa},\\
T & = & 3500\text{ K}.
\end{array}\label{eq:detonantion-2d-initialization}
\end{equation}
These circular regions were added to perturb the detonation and develop
cellular structures and they where aligned with the unstructured grid
interfaces at initialization as shown in Figure~(\ref{fig:two-dimesiona-detonation-wave_temperature}).
This test case is similar to one performed by Oran et al.~\citep{Ora98}
with the exception that they initialized their solution from a one-dimensional
detonation and their domain was $0.10$~m longer, $\text{\ensuremath{\Omega}}=\left(0,0.55\right)\mathrm{m}\times\left(0,0.06\right)\mathrm{m}$.
Their simulations revealed two detonation cells in the vertical direction
with each cell computed to be $\left(0.055\right)\mathrm{m}\times\left(0.03\right)\mathrm{m}$.
Their results were consistent with experiments performed by Lefebvre
et al.~\citep{Lef98}. Houim and Kuo~\citep{Hou11}, as well as
Lv and Ihme~\citep{Lv15}, ran similar test cases but the domain
was half the height of the one considered in this work. Houim and
Kuo~\citep{Hou11} shifted their simulation to be in the detonation
frame of reference whereas Lv and Ihme~\citep{Lv15} ran the simulation
in the laboratory frame of reference. In both cases, the solutions
were initialized from the corresponding one-dimensional detonation
solutions. Their simulations revealed one detonation cell, which is
is consistent with domain height considered in their respective works.

Figure~(\ref{fig:two-dimesiona-detonation-wave_temperature}) presents
the temperature solutions as the detonation front progresses through
the simulation domain. As the initial shock progresses into the domain,
it collides with the two circular high pressure and high temperature
regions before $t=30$~$\mu\mathrm{s}$, the detonation is then established
and the perturbations lead to transverse waves traveling in the vertical
directions that reflect off the top and bottom walls. To the left
of the detonation, complex, two-dimensional structures develop, including
Kelvin--Helmholtz instabilities as well as the triple point that
connects the Mach stems and incident shock.

\begin{figure}[H]
\begin{centering}
\includegraphics[width=0.9\columnwidth]{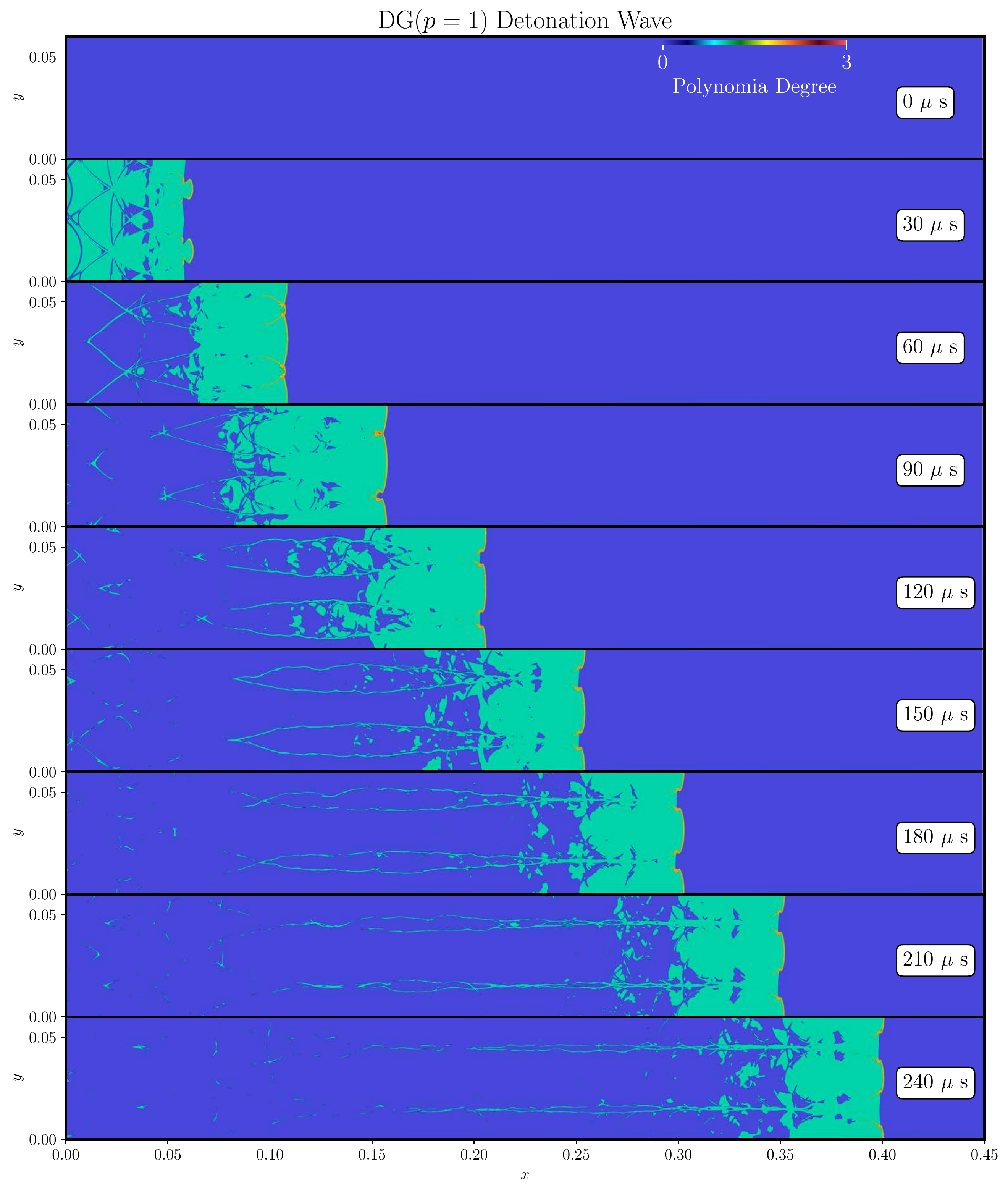}
\par\end{centering}
\caption{\label{fig:two-dimensional-detonation-wave_ode_degree}The adapted
polynomial degree of the DGODE approximation for two-dimensional detonation
wave.}
\end{figure}

Figure~(\ref{fig:two-dimensional-detonation-wave_ode_degree}) shows
the adapted polynomial degree for the DGODE chemistry solve at each
degree of freedom as the detonation propagates through the simulation
domain. The adapted polynomial degree is largest in the vicinity of
the detonation, reaching $p=3$ in the regions behind the detonation
front and at the triple points. The majority of the adapted polynomial
degree reduces to $p=0$ downstream of the shock. Thin regions of
$p=1$ exist downstream of the shock along the vertically traveling
transverse waves. This indicates that the traveling waves are strong
enough to push the chemistry out of equilibrium and require high-order
integration to solve the chemistry. These thin regions of higher order
are similar to the spike in the adapted polynomial degree as shown
in Section~\ref{subsec:One-dimensional-detonation-wave}'s one-dimensional
detonation results at time $t=94$~$\mu\mathrm{s}$, see Figure~\ref{fig:traveling_detonation}.

\begin{figure}[H]
\begin{centering}
\includegraphics[width=0.9\columnwidth]{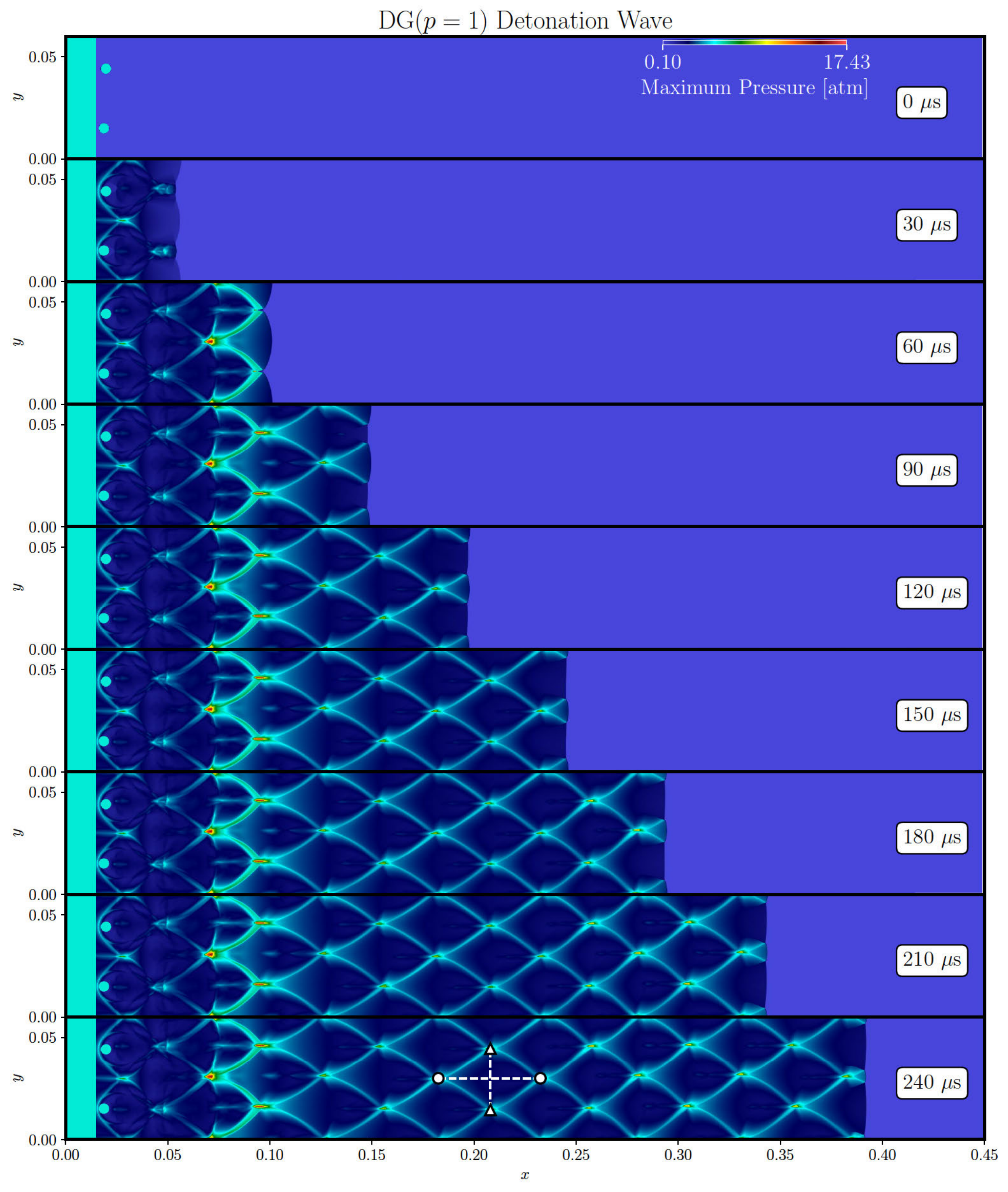}
\par\end{centering}
\caption{\label{fig:two-dimensional-detonation-wave_pmax}Maximum pressure,
$\max(p,p_{max})$, revealing two distinct detonation cellular structure,
cf.~\citep{Ora98} and~\citep{Lef98}.}
\end{figure}

Figure~\ref{fig:two-dimensional-detonation-wave_pmax} shows the
maximum pressure experienced at each location as the solution evolved,
$max(p,p_{max})$. This metric is a good indicator of the detonation
cell structures and shows two distinct detonation cells in the vertical
direction. A large burst of pressure exists between $x=0.04$~m and
$x=0.1$~m from the over driven detonation as the shock transitions
to a detonation in the early stages of development. The last solution,
at $t=240$~$\mu\mathrm{s}$, displays two dashed white lines that
indicated the measured horizontal and vertical size of a detonation
cell. For this solution the detonation cell was computed to be $\left(0.055\right)\mathrm{m}\times\left(0.03\right)\mathrm{m}$
which is consistent with experimental work~\citep{Lef98} and previous
numerical results~\citep{Ora98}. 

\begin{figure}[H]
\begin{centering}
\includegraphics[width=0.6\columnwidth]{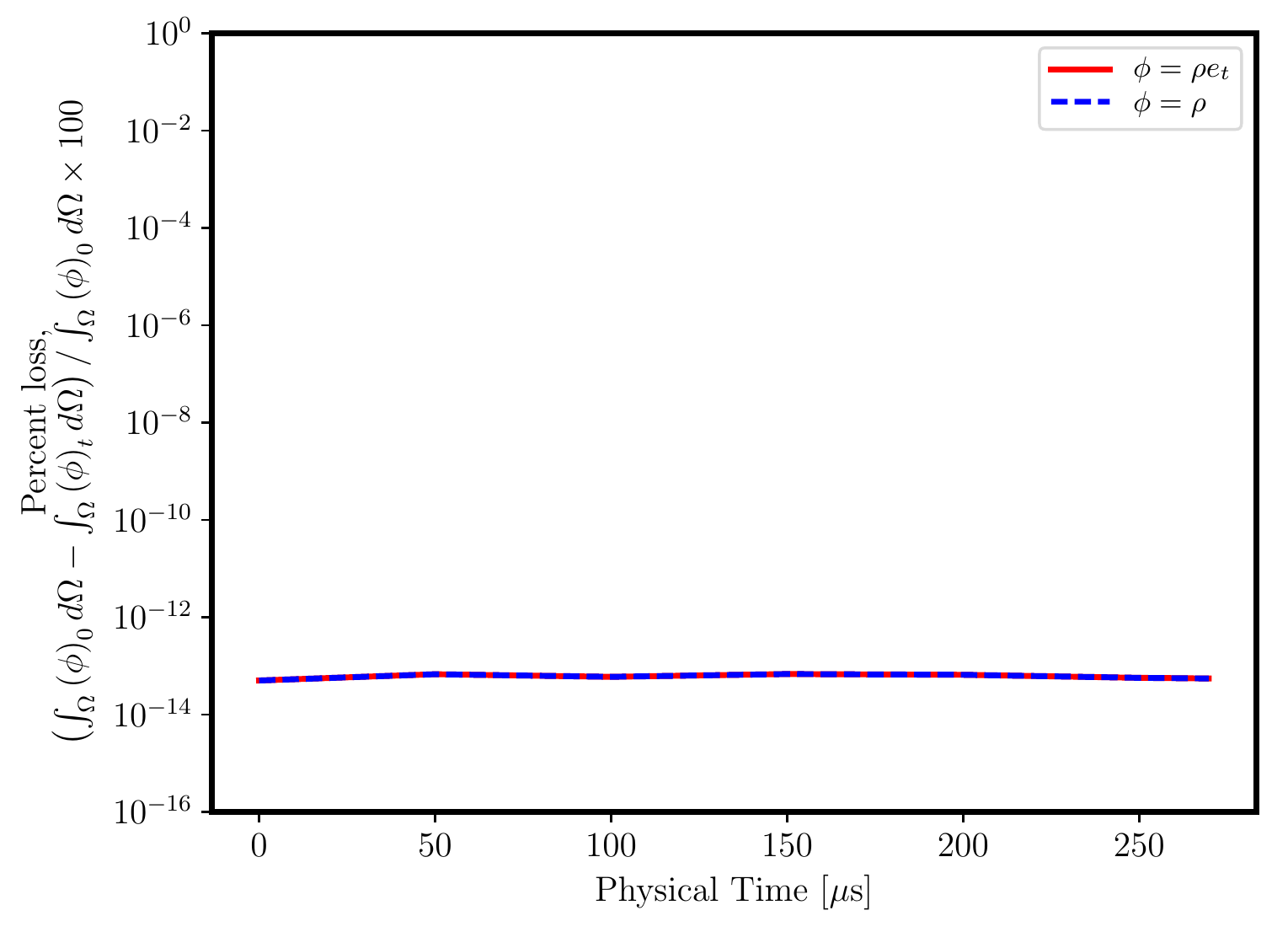}
\par\end{centering}
\caption{\label{fig:one_dimensional_detonation_energy_conservation_2d}Percent
loss of conservation quantities, $\left(\int_{\Omega}\left(\phi\right)_{0}-\int_{\Omega}\left(\phi\right)_{t}\right)/\int_{\Omega}\left(\phi\right)_{0}\times100$,
for the two-dimensional detonation computed computed using the initial
integrated quantity, $\int_{\Omega}\left(\phi\right)_{0}d\Omega$,
and current integrated quantity, $\int_{\Omega}\left(\phi\right)_{t}d\Omega$,
where $\phi=\rho e_{t}$ for total energy and $\phi=\rho$ for density.}
\end{figure}

No spurious pressure oscillations were noticed for the entire simulation
of this test case, indicating that the formulation presented in this
work is stable for two-dimensional detonations. For this test case,
the integrated total energy, $\int_{\Omega}\left(\rho e_{t}\right)_{0}d\Omega$,
and the density, $\int_{\Omega}\left(\rho\right)_{0}d\Omega$, are
conserved as the domain has walls at all boundaries. Figure~\ref{fig:one_dimensional_detonation_energy_conservation_2d}
shows the percent loss of the integrated total energy, $\left(\int_{\Omega}\left(\rho e\right)_{0}d\Omega-\int_{\Omega}\left(\rho e\right)_{t}d\Omega\right)/\int_{\Omega}\left(\rho e\right)_{0}d\Omega\times100$,
as well as the percent loss of the integrated density, $\left(\int_{\Omega}\left(\rho\right)d\Omega-\int_{\Omega}\left(\rho\right)_{t}d\Omega\right)/\int_{\Omega}\left(\rho\right)_{0}d\Omega\times100$,
as a function of time. The error remains on the order of $10^{-13}$~\%
for the entire simulation which is on the order of the expected machine
error and assures that the method remains conservative.

\subsection{One-dimensional premixed flame\label{subsec:One-dimensional-premixed-flame}}

In this section we consider a fully compressible multi-component reacting
Navier-Stokes flow corresponding to a one-dimensional premixed flame
using the Hydrogen-Air chemistry from~\citep{Oco04}. The chosen
domain is $6$ cm in length and discretized with uniform linear line
elements with $h=5\times10^{-5}\,$m. The initialization is given
by
\begin{equation}
\begin{array}{cccc}
\left(p,T,X_{O_{2}},X_{H_{2}},X_{N_{2}}\right) & = & \left(1\,\mathrm{\text{bar}},300\,\mathrm{K},0.170,0.188,0.642\right) & x<5.8\,\mathrm{\text{cm}}\\
y & = & \hat{y} & x>5.8\,\mathrm{\text{cm}}
\end{array},\label{eq:premixed-flame-initialization}
\end{equation}
where $\hat{y}$ corresponds to the fully reacted state from Cantera's
homogeneous constant pressure reactor simulation calculated from the
unreacted conditions specified at $x<5.8$ cm ~\citep{cantera}.
The solution at the initially discontinuous interface immediately
diffuses to form a continuous, smooth, solution. No additional stabilization,
e.g., artificial viscosity, limiting or filtering was needed for the
duration of the simulation.

The right hand side boundary is a fixed pressure outflow characteristic
boundary condition with $p_{\infty}=1$ bar. The left hand side boundary
is a characteristic farfield wave boundary condition with $T_{\infty}=300$
K and $v_{\infty}=0$ m/s. This boundary allows any pressure waves
caused by the initialization to exit the domain. The initialization
contains a temperature and species discontinuity which gives rise
to a pressure wave that leaves the system as the reaction front diffuses
into the unreacted region. Eventually, the solution converges to a
propagating flame traveling at a unique velocity.
\begin{table}[H]
\caption{\label{tab:1D-flame-speeds}Computed flame speeds corresponding to
the one-dimensional premixed flame described in Section~\ref{subsec:One-dimensional-premixed-flame}. }

\centering{}%
\begin{tabular}{|c|c|}
\hline 
Method & Flame Speed {[}m/s{]}\tabularnewline
\hline 
\hline 
Cantera & 0.643\tabularnewline
\hline 
$\mathrm{DG}\left(p=1\right)$ & 0.641\tabularnewline
\hline 
$\mathrm{DG}\left(p=2\right)$ & 0.643\tabularnewline
\hline 
\end{tabular}
\end{table}

\begin{figure}[H]
\subfloat[\label{fig:flame_T_p1}The temperature profile correspond to the $\mathrm{DG}\left(p=1\right)$
solution (grey lines) and the Cantera solution (solid lines) on a
uniform $h=5\times10^{-5}\,$m grid. Both the $\mathrm{DG}\left(p=1\right)$
and Cantera solutions have been shifted so that $T=400$ K at $x=0$.]{\includegraphics[width=0.45\columnwidth]{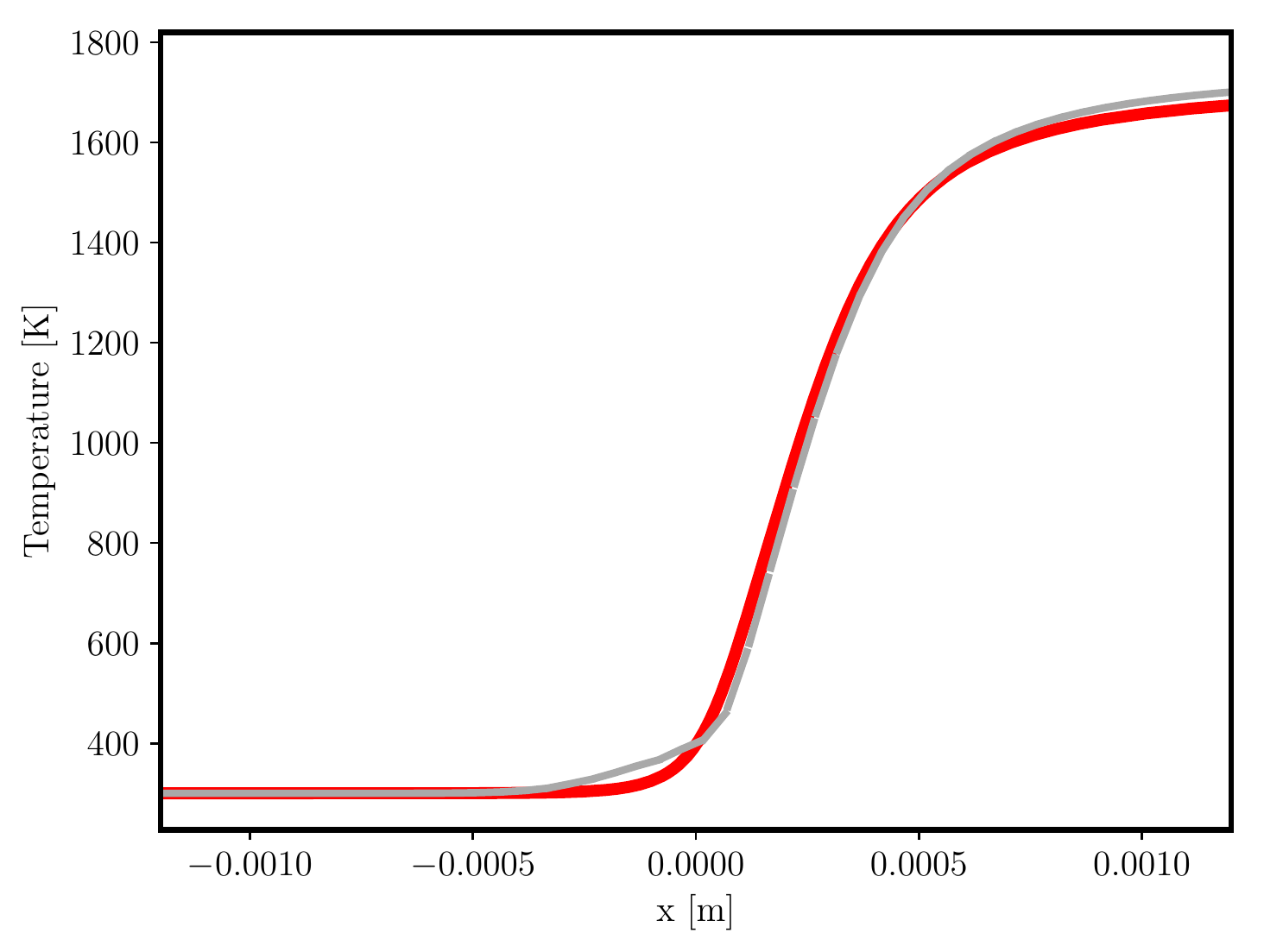}}\hfill{}\subfloat[\label{fig:flame_minor_p1}Minor species profiles correspond to the
$\mathrm{DG}\left(p=1\right)$ solution (grey lines) and the Cantera
solution (solid lines) on a uniform $h=5\times10^{-5}\,$m grid. Both
the $\mathrm{DG}\left(p=1\right)$ and Cantera solutions have been
shifted so that $T=400$ K at $x=0$. ]{\includegraphics[width=0.45\columnwidth]{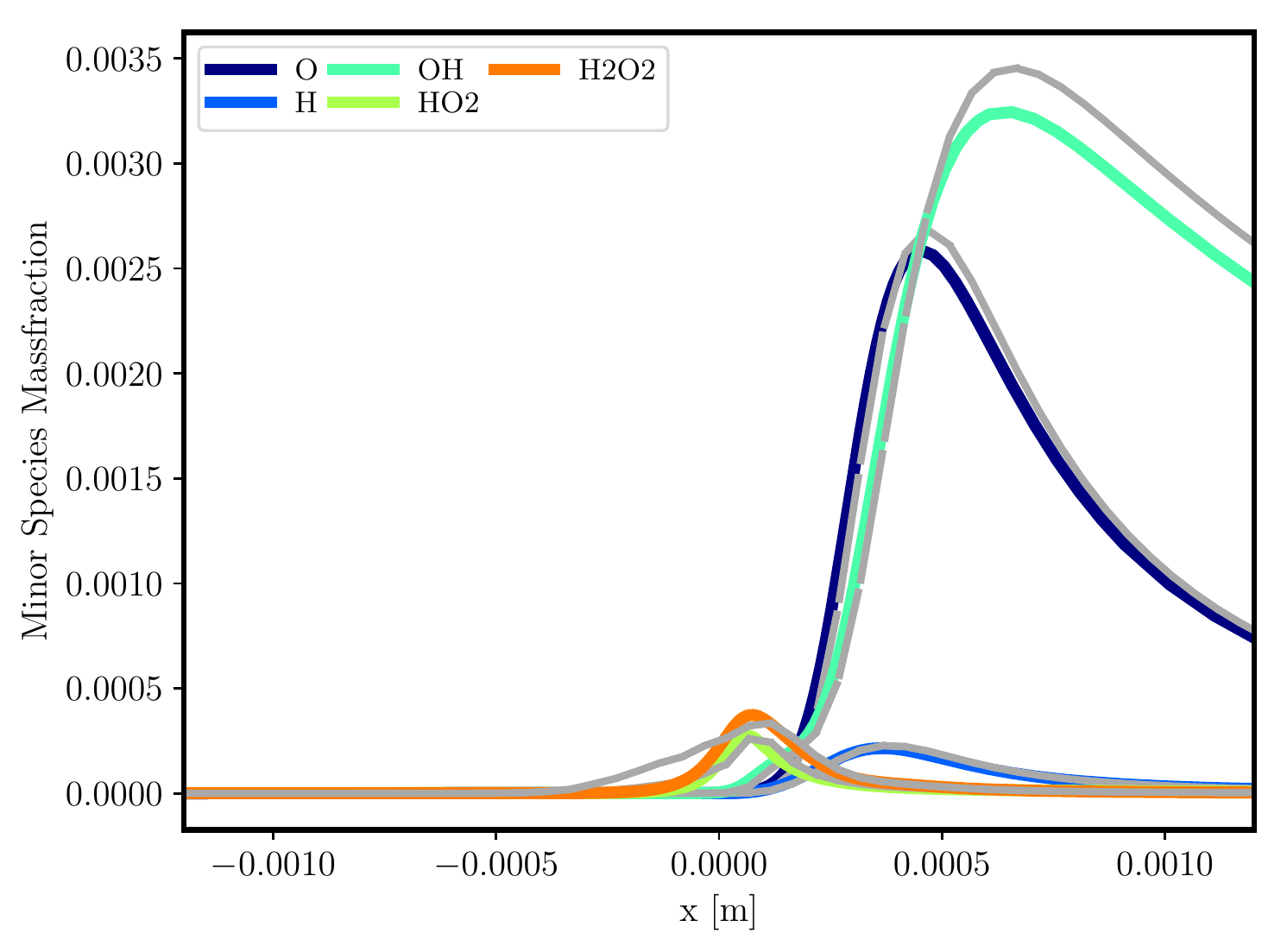}}\hfill{}\subfloat[\label{fig:flame_major_p1}Major species profiles correspond to the
$\mathrm{DG}\left(p=1\right)$ solution (grey lines) and the Cantera
solution (solid lines) on a uniform $h=5\times10^{-5}\,$m grid. Both
the $\mathrm{DG}\left(p=1\right)$ and Cantera solutions have been
shifted so that $T=400$ K at $x=0$.]{\includegraphics[width=0.45\columnwidth]{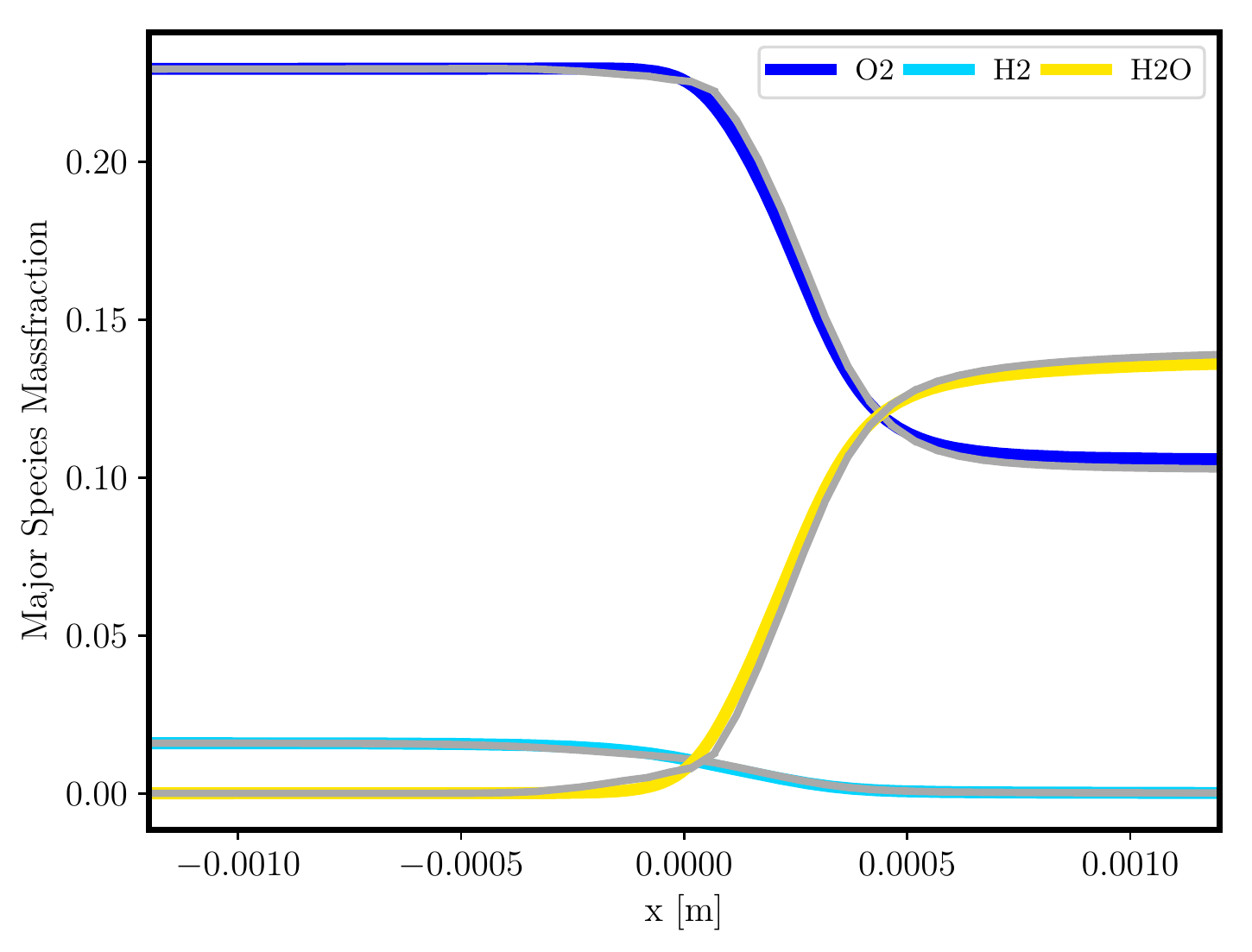}}\hfill{}\subfloat[\label{fig:flame_p_p1}The pressure profile corresponding to the $\mathrm{DG}\left(p=1\right)$
solution (grey line) for the entire simulation domain. For reference
the the pressure profile corresponding to a constant 1 bar solution
is included (solid red line). The initialization is given by~(\ref{eq:premixed-flame-initialization}).]{\includegraphics[width=0.45\columnwidth]{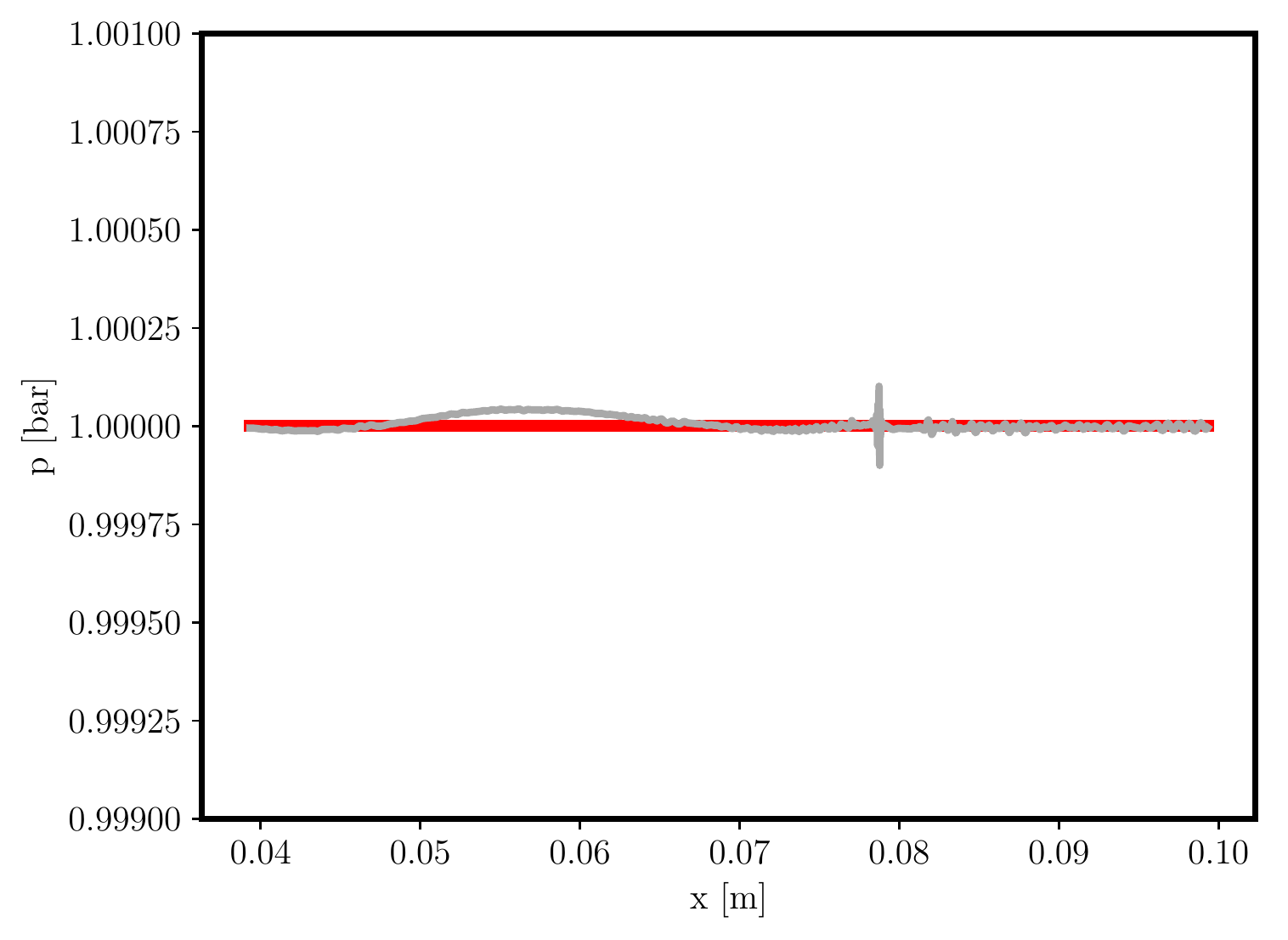}}

\caption{\label{fig:flame_p1}The profiles corresponding to the $\mathrm{DG}\left(p=1\right)$
solution for the one-dimensional premixed flame, described in Section~\ref{subsec:One-dimensional-premixed-flame},
computed on a uniform grid with on a uniform $h=5\times10^{-5}\,$m
grid. The initialization is given by~(\ref{eq:detonation-1d-initialization}).
The flame speed of 0.641 m/s corresponding to the $\mathrm{DG}\left(p=1\right)$
solution compares well with the flame speed corresponding to the Cantera
solution of 0.643 m/s. This smooth solution of a multi-component chemically
reacting Navier-Stokes flow was computed without the use of additional
stabilization, e.g., artificial viscosity, limiting or filtering.
The pressure oscillations visible in Figure~\ref{fig:flame_p_p1}
are not present in the corresponding $\mathrm{DG}\left(p=2\right)$
solution shown in Figure~\ref{fig:flame_p_p2}, indicating the $\mathrm{DG}\left(p=1\right)$
solution is under-resolved.}
\end{figure}

\begin{figure}[H]
\subfloat[\label{fig:flame_T_p2}The temperature profile correspond to the $\mathrm{DG}\left(p=2\right)$
solution (grey lines) and the Cantera solution (solid lines) on a
uniform $h=5\times10^{-5}\,$m grid. Both the $\mathrm{DG}\left(p=2\right)$
and Cantera solutions have been shifted so that $T=400$ K at $x=0$.]{\includegraphics[width=0.45\columnwidth]{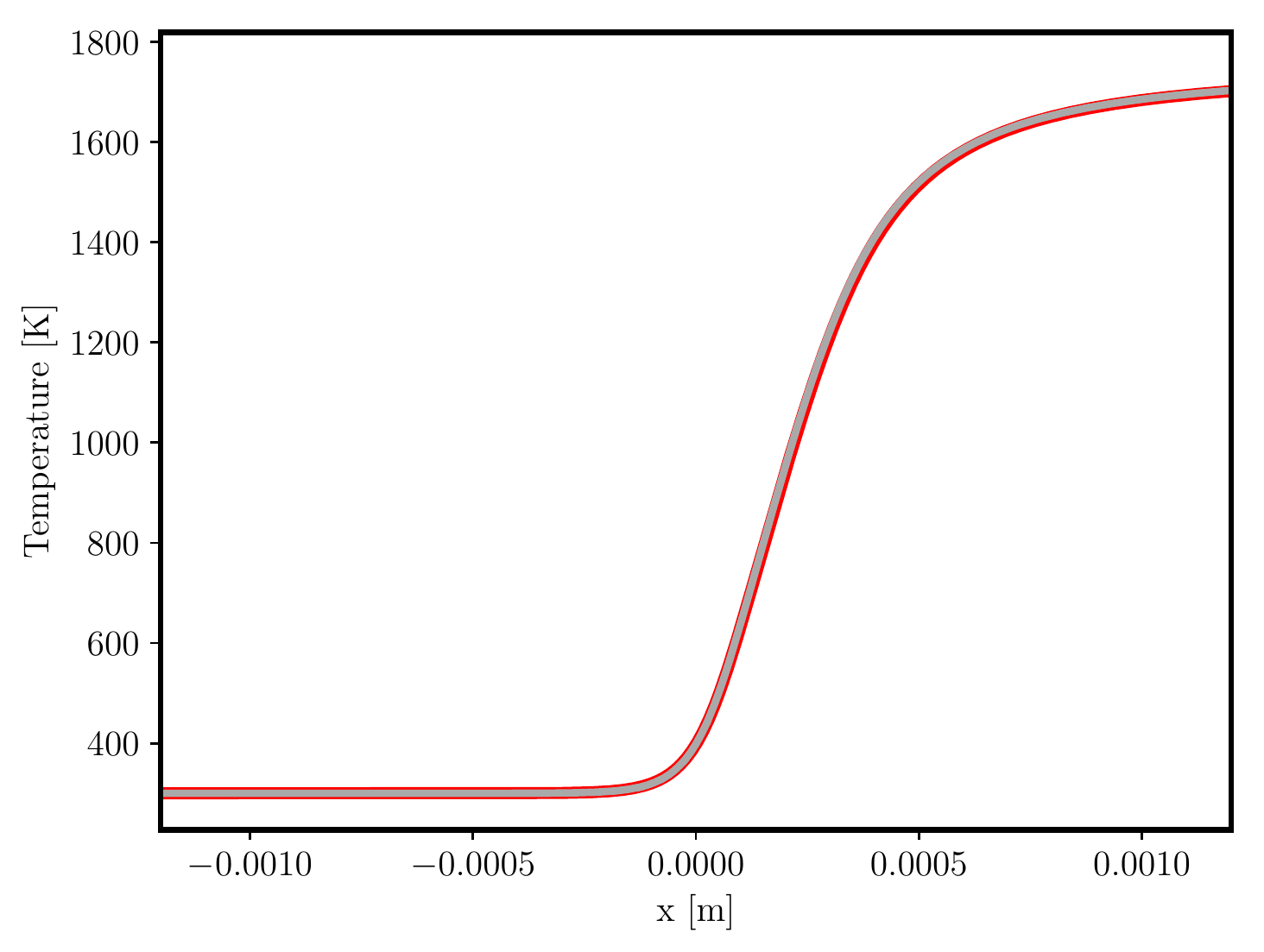}}\hfill{}\subfloat[\label{fig:flame_minor_p2}Minor species profiles correspond to the
$\mathrm{DG}\left(p=2\right)$ solution (grey lines) and the Cantera
solution (solid lines) on a uniform $h=5\times10^{-5}\,$m grid. Both
the $\mathrm{DG}\left(p=2\right)$ and Cantera solutions have been
shifted so that $T=400$ K at $x=0$.]{\includegraphics[width=0.45\columnwidth]{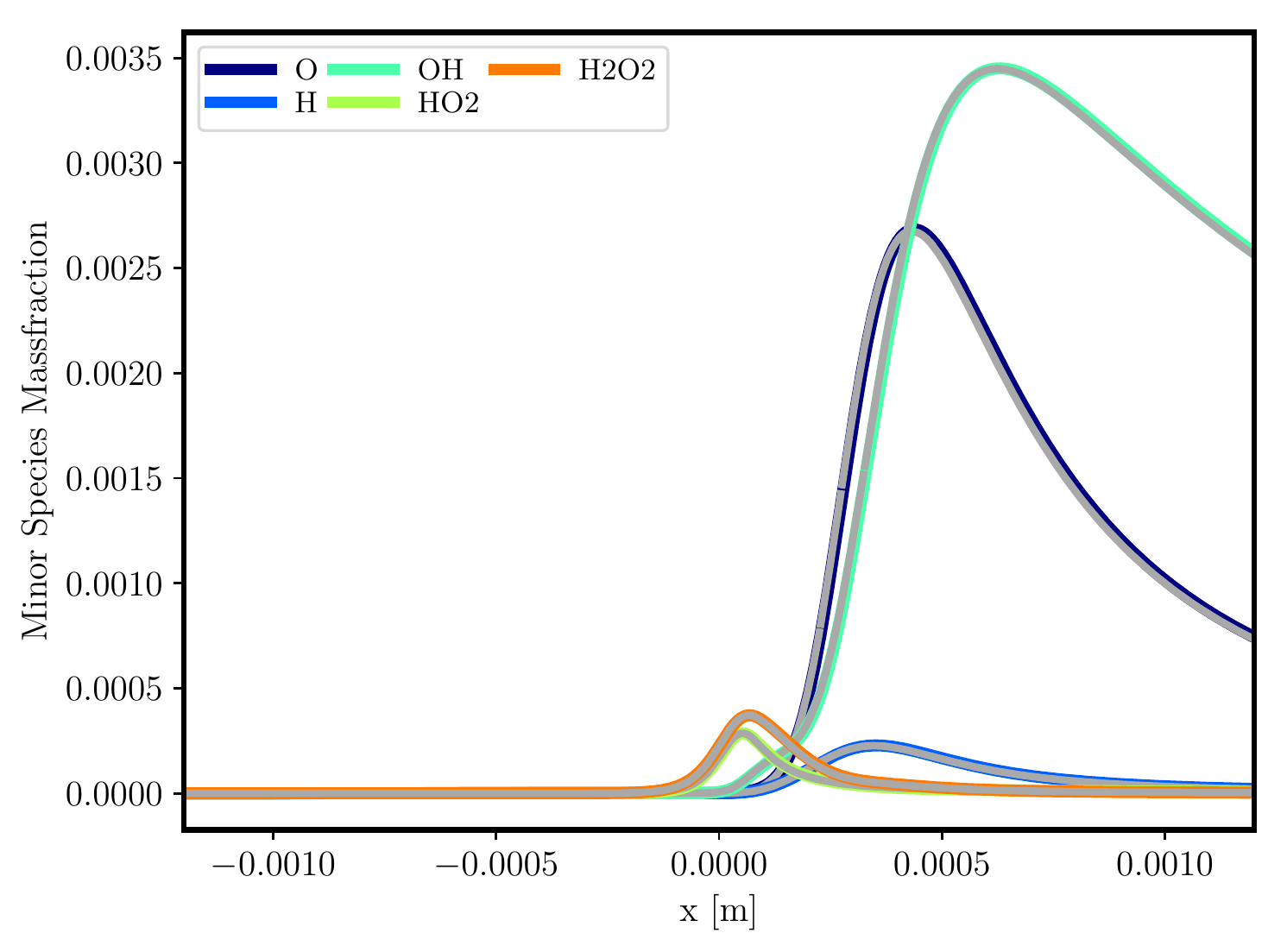}}\hfill{}\subfloat[\label{fig:flame_major_p2}Major species profiles correspond to the
$\mathrm{DG}\left(p=2\right)$ solution (grey lines) and the Cantera
solution (solid lines) on a uniform $h=5\times10^{-5}\,$m grid. Both
the $\mathrm{DG}\left(p=2\right)$ and Cantera solutions have been
shifted so that $T=400$ K at $x=0$.]{\includegraphics[width=0.45\columnwidth]{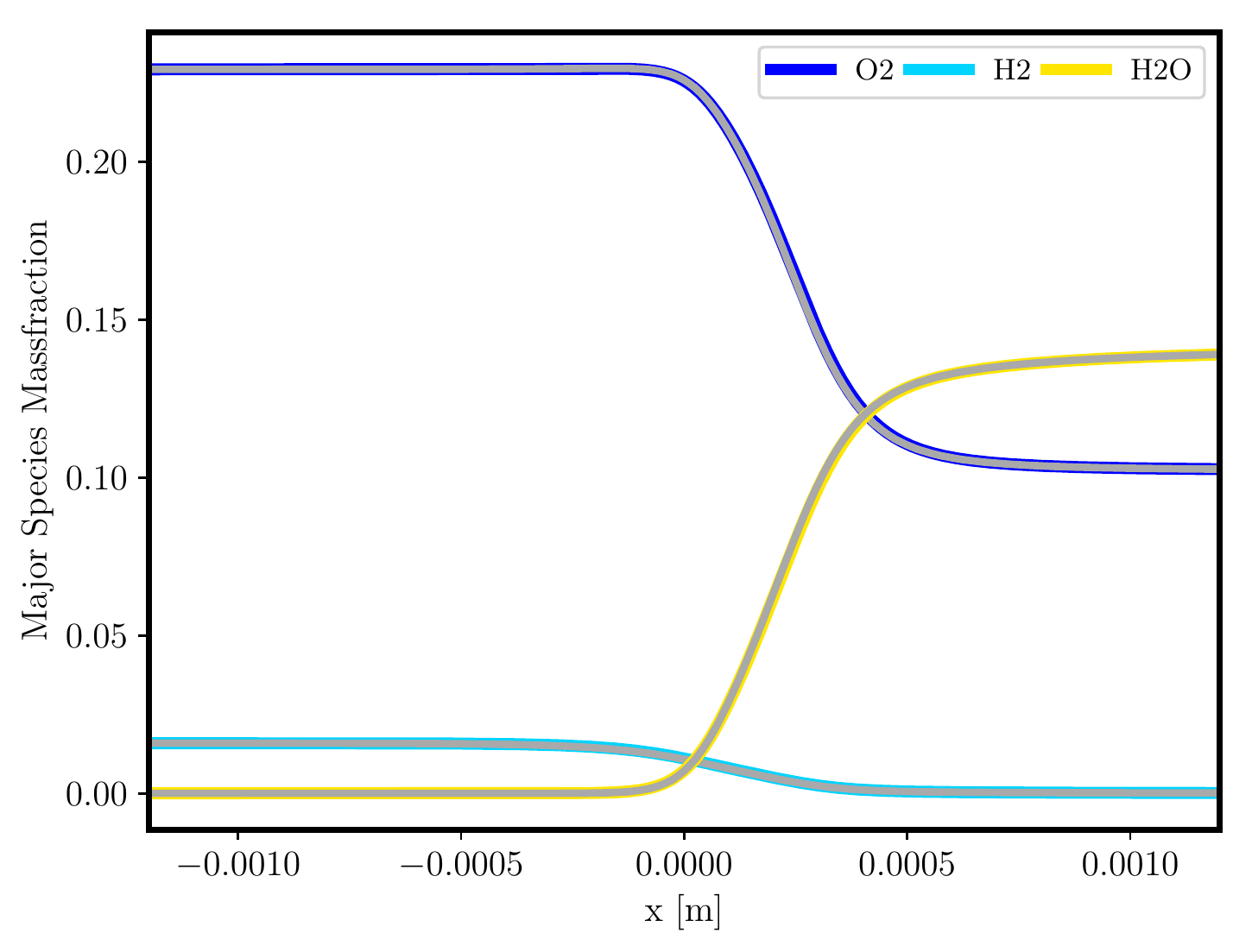}}\hfill{}\subfloat[\label{fig:flame_p_p2}The pressure profile corresponding to the $\mathrm{DG}\left(p=2\right)$
solution (grey line) for the entire simulation domain. For reference
the the pressure profile corresponding to a constant 1 bar solution
is included (solid red line). ]{\includegraphics[width=0.45\columnwidth]{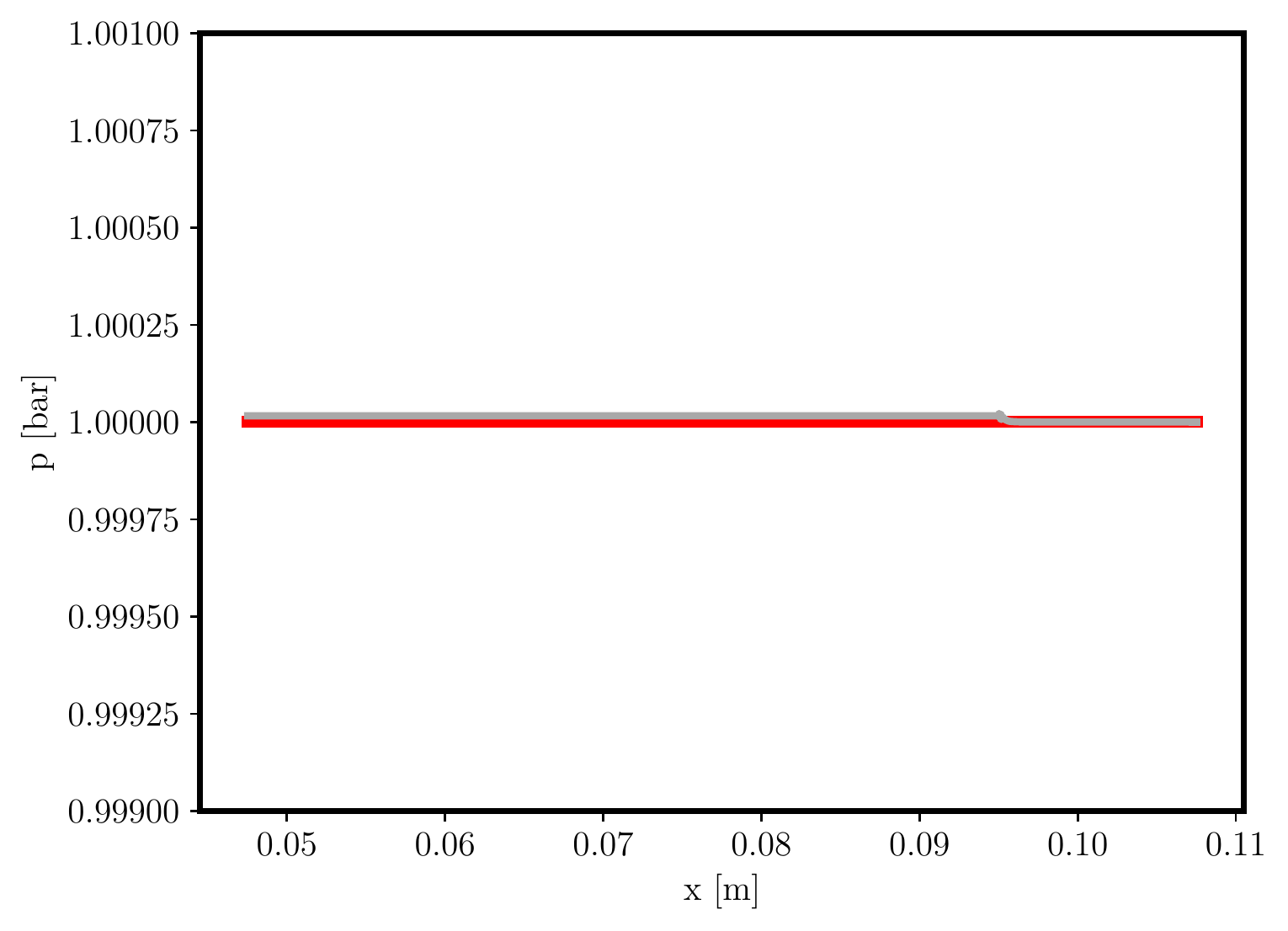}}

\caption{\label{fig:flame_p2}The profiles corresponding to the $\mathrm{DG}\left(p=2\right)$
solution for the one-dimensional premixed flame, described in Section~\ref{subsec:One-dimensional-premixed-flame},
computed on a uniform grid with on a uniform $h=5\times10^{-5}\,$m
grid. The initialization is given by~(\ref{eq:detonation-1d-initialization}).
The flame speed of 0.643 m/s corresponding to the $\mathrm{DG}\left(p=2\right)$
solution compares well with the flame speed corresponding to the Cantera
solution of 0.643 m/s. This smooth solution of a multi-component chemically
reacting Navier-Stokes flow was computed without the use of additional
stabilization, e.g., artificial viscosity, limiting or filtering.}
\end{figure}

Figures~\ref{fig:flame_p1} and~\ref{fig:flame_p2} present the
$\mathrm{DG}\left(p=1\right)$ and $\mathrm{DG}\left(p=2\right)$
solutions, respectively. The temperature and species mass fraction
profiles are compared to the Cantera flame solution with $h=10^{-5}\,$
m and the profiles are shifted so that $T=400$ K at $x=0$ m. The
$\mathrm{DG}\left(p=1\right)$ solutions reach the correct reacted
state but cannot fully resolve the flame structures in the $-0.0005<x<0$
region. The $\mathrm{DG}\left(p=2\right)$ solution overcomes these
errors and is in good agreement with the Cantera solution. Despite
the under resolved profiles in the $\mathrm{DG}\left(p=1\right)$
solution, the flame speeds corresponding to both solutions compare
well to the flame speed calculated in Cantera as presented in Table~\ref{tab:1D-flame-speeds}.
The Cantera flame speed, given as the inflow velocity for the constant
mass flow-rate, is 0.643 m/s. We considered the flame front in the
unsteady $\mathrm{DG}\left(p=1\right)$ and $\mathrm{DG}\left(p=2\right)$
solutions to be the location corresponding to $T=1000$ K. We tracked
this location and computed a steady velocity of 0.641 m/s and 0.643
m/s for the $\mathrm{DG}\left(p=1\right)$ and $\mathrm{DG}\left(p=2\right)$
solutions, respectively.

Figures~\ref{fig:flame_p_p1} and~\ref{fig:flame_p_p2} show the
pressure through the entire computational domain for $\mathrm{DG}\left(p=1\right)$
and $\mathrm{DG}\left(p=2\right)$, respectively. For the $\mathrm{DG}\left(p=1\right)$
solution, there are small oscillations, on the order of $0.25$\%
of the ambient pressure. These oscillations are not present for the
$\mathrm{DG}\left(p=2\right)$ solution, where only a slight variation
is seen through the flame front but is constant on both sides of the
flame within $0.1$\% of the desired ambient pressure. The lack of
pressure oscillations in the higher order solution indicate that the
$\mathrm{DG}\left(p=1\right)$ solution is under-resolved.

\subsection{Hydrogen-Air chemically reacting shear layer\label{subsec:hydrogen-air-shear-layer}}

\begin{figure}[H]
\begin{centering}
\includegraphics[width=0.8\columnwidth]{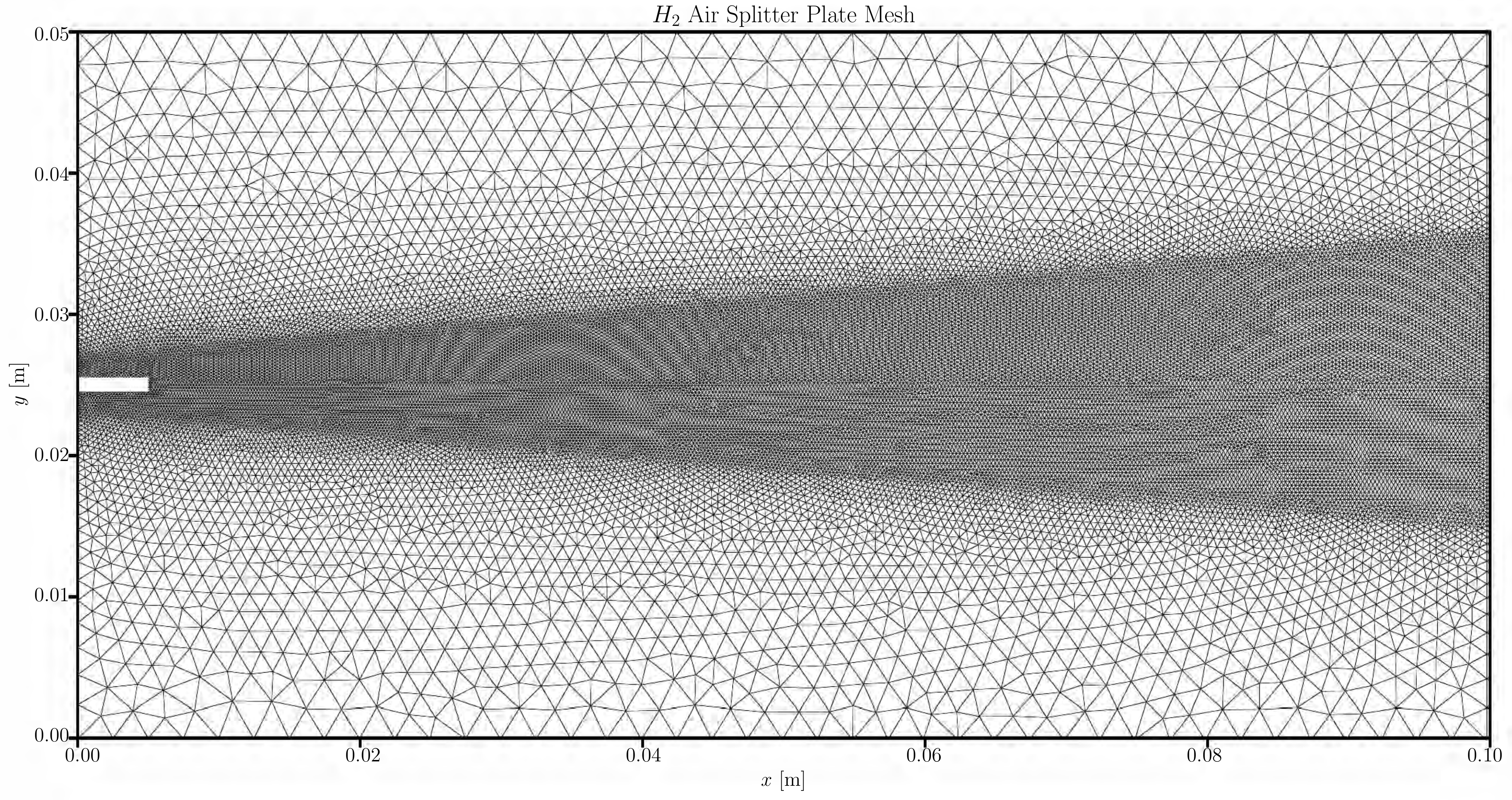}
\par\end{centering}
\caption{\label{fig:mesh}Splitter plate mesh comprised of linear triangular
elements with increased resolution mixing region. This mesh was generated
using Gmsh~\citep{Geu09}. }
\end{figure}

We use the viscous reacting formulation of Equations~(\ref{eq:conservation-law-strong-form})-
(\ref{eq:conservation-law-flux-boundary-condition}) with detailed
chemical kinetics and transport to approximate the solution of a chemically
reacting shear layer using the Hydrogen-Air chemistry extracted from
the GRI Mech II mechanism~\citep{Oco04}. This test case was designed
to test the stability and robustness of the formulation presented
in this work and verify that smooth multidimensional multi-component
reacting flows can be computed without artificial viscosity, limiting,
or filtering. The entire simulation domain is depicted in Figure~\ref{fig:mesh}.
The bounding domain is $0.10$ m in length and $0.05$ m in height,
$\text{\ensuremath{\Omega}}=\left(0,0.10\right)\mathrm{m}\times\left(0,0.05\right)\mathrm{m}$.
A splitter plate is used to separate the incoming streams, anchor
the flame, and to test slip and adiabatic wall boundary conditions
described in Section~\ref{subsec:Discretization}. The plate is $0.001$
m thick and $0.005$ m long and is centered at $y=0.025$ m. The flow
is initialized with the following conditions 

\begin{equation}
\begin{array}{ccc}
\left(M,T,p,Y_{O_{2}},Y_{N_{2}},Y_{H_{2}}\right) & = & \begin{cases}
\left(0\text{.8},\:2000\text{ K},\:101325\text{ Pa},\:0.21,\:0.79,\:0\right) & y<0.025\text{ m}\\
\left(0\text{.8},\:2000\text{ K},\:101325\text{ Pa},\:\phantom{.21}0,\:\phantom{.79}0,\:1\right) & y>0.025\text{ m}
\end{cases},\end{array}\label{eq:2D_splitter_plate}
\end{equation}
where $M$ is the freestream Mach number.

The upstream half of the plate, $0\,\mathrm{m}<x<0.0025\,\mathrm{m}$,
is specified as a slip wall, whereas the downstream half of the plate,
$0.0025\,\mathrm{m}<x<0.005\,\mathrm{m}$, is an adiabatic no-slip
wall. The remaining boundaries, at $x=0\,\mathrm{m}$, $x=0.1\,\mathrm{m}$,
$y=0\,\mathrm{m}$ , and $y=0.05\,\mathrm{m}$, are specified as non-reflecting
characteristic boundary conditions where the freestream state is given
by~(\ref{eq:2D_splitter_plate}). The characteristic conditions downstream
of the plate at $x=0.1\,\mathrm{m}$ allow for pockets of supersonic
flow to exit the domain if the heat release from reactions is large
enough to transition the flow to supersonic.

Initial simulations were run to determine the regions of refinement
required to adequately capture the reacting mixing layer. Figure~\ref{fig:mesh}
presents the final grid corresponding to the entire domain, which
was constructed using Gmsh~\citep{Geu09}. The finest resolution
in the mixing region corresponds to a mesh size of $200\times10^{-6}\,\mathrm{m}$
and transitions to $500\times10^{-6}\,\mathrm{m}$ at $x=0.1$ m.
This region begins at a thickness of $0.0035\,\mathrm{m}$, $0.02325\,\mathrm{m}<y<0.02675\,\mathrm{m}$,
at $x=0\,\mathrm{m}$, and grows to a thickness of $0.015\,\mathrm{m}$,
$0.0155\,\mathrm{m}<y<0.0355\,\mathrm{m}$, at $x=0.1\,\mathrm{m}$.

\begin{figure}[H]
\subfloat[\label{fig:splitter_plate_time_p1} Evolution of temperature profile
using $\mathrm{DG}\left(p=1\right)$ for H2-air splitter plate. ]{\begin{centering}
\includegraphics[width=0.45\linewidth]{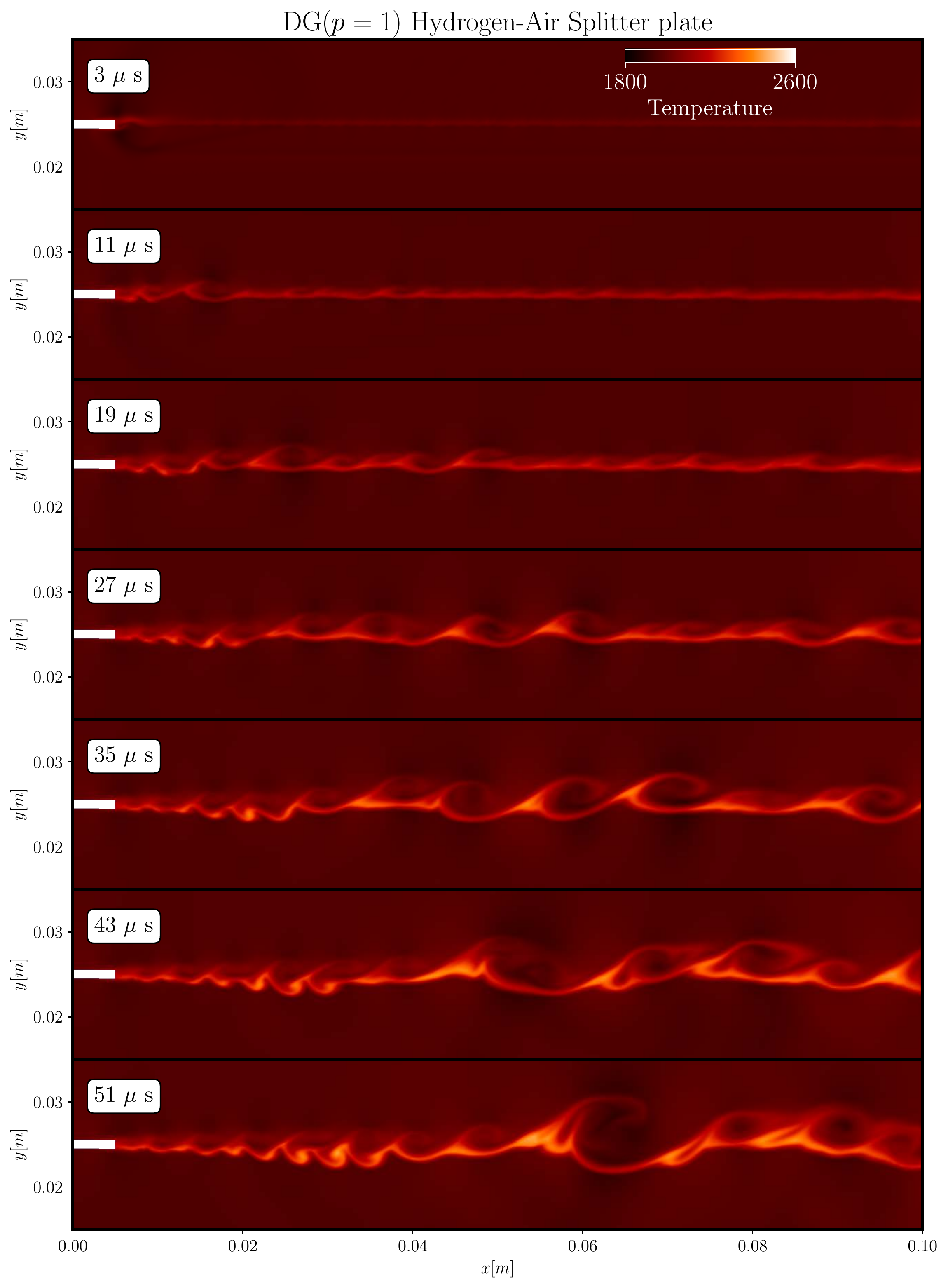}
\par\end{centering}
}\hfill{}\subfloat[\label{fig:splitter_plate_time_p2}Evolution of temperature profile
using $\mathrm{DG}\left(p=2\right)$ for H2-air splitter plate.]{\begin{centering}
\includegraphics[width=0.45\linewidth]{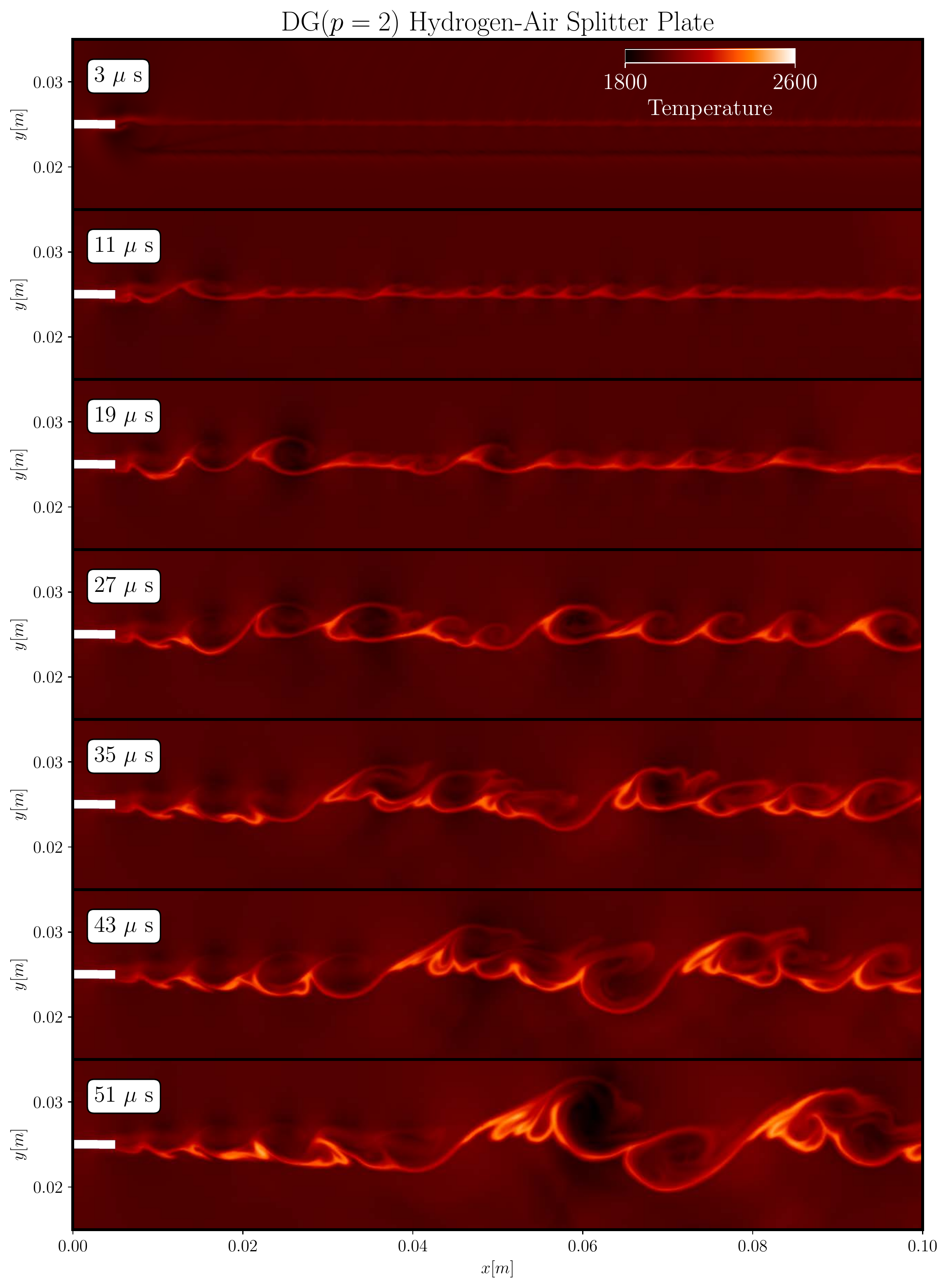}
\par\end{centering}
}

\caption{\label{fig:splitter_plate_time} Evolution of temperature profile
corresponding to the $\mathrm{DG}\left(p=1\right)$ and $\mathrm{DG}\left(p=2\right)$
solutions at different times. The initialization is given by~(\ref{eq:2D_splitter_plate}).
This smooth solution of a multi-component chemically reacting Navier-Stokes
flow was computed without the use of additional stabilization, e.g.,
artificial viscosity, limiting or filtering.}
\end{figure}

\begin{figure}[H]
\centering{}\includegraphics[width=0.8\columnwidth]{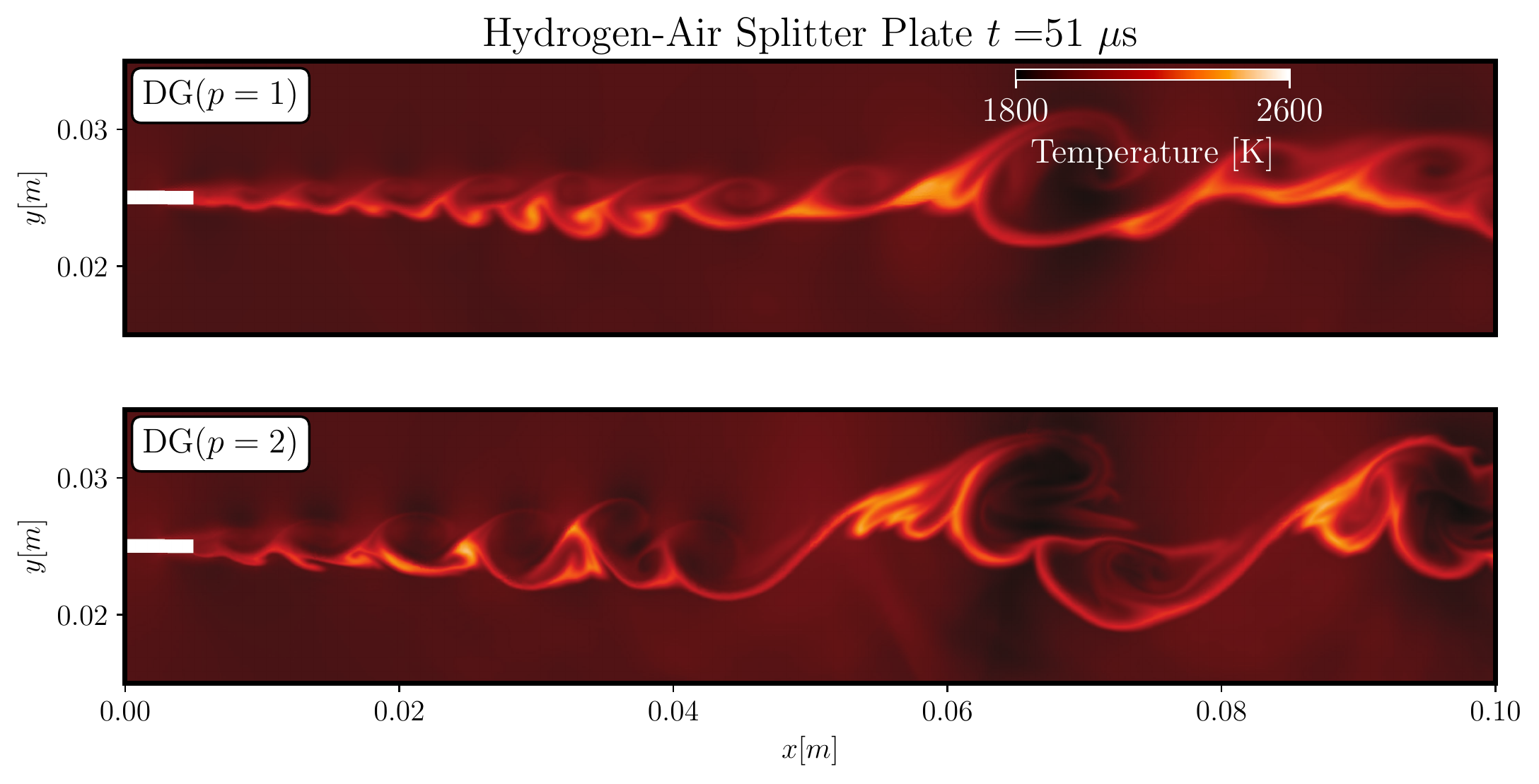}\caption{Comparison of splitter plate temperature solution for $\mathrm{DG}\left(p=1\right)$
and $\mathrm{DG}\left(p=2\right)$ at $t=51$ $\mu$s. \label{fig:splitter_plate_temperature_compare}}
\end{figure}

\begin{figure}[H]
\subfloat[\label{fig:H2O_splitter_upstream} Upstream structures of $H_{2}O$
for $\mathrm{DG}\left(p=1\right)$ and $\mathrm{DG}\left(p=2\right)$
at $t=51$ $\mu$s.]{\begin{centering}
\includegraphics[width=0.45\linewidth]{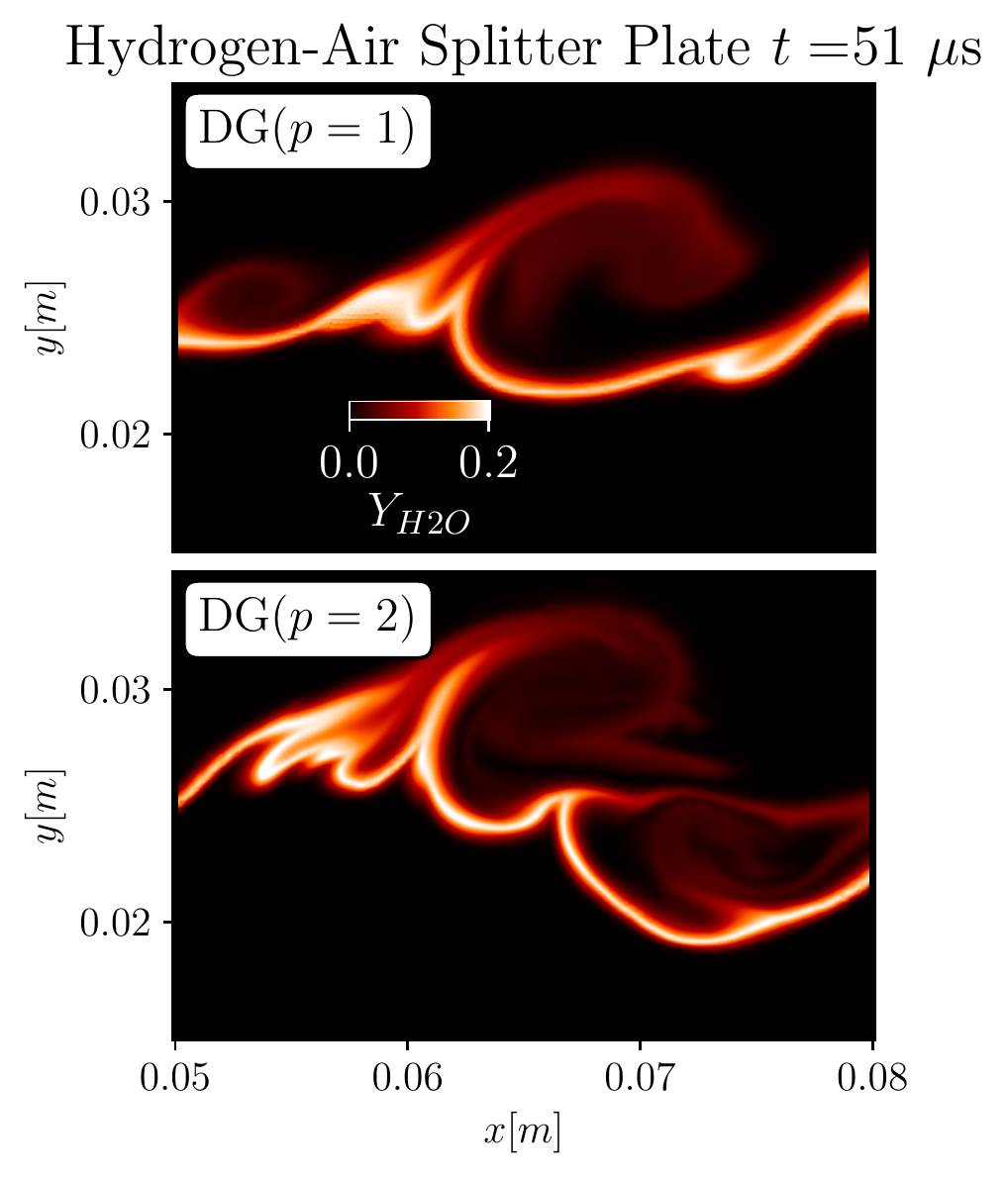}
\par\end{centering}
}\hfill{}\subfloat[\label{fig:H2O_splitter_downstream}Downstream structures of $H_{2}O$
for $\mathrm{DG}\left(p=1\right)$ and $\mathrm{DG}\left(p=2\right)$
at $t=51$ $\mu$s..]{\begin{centering}
\includegraphics[width=0.45\linewidth]{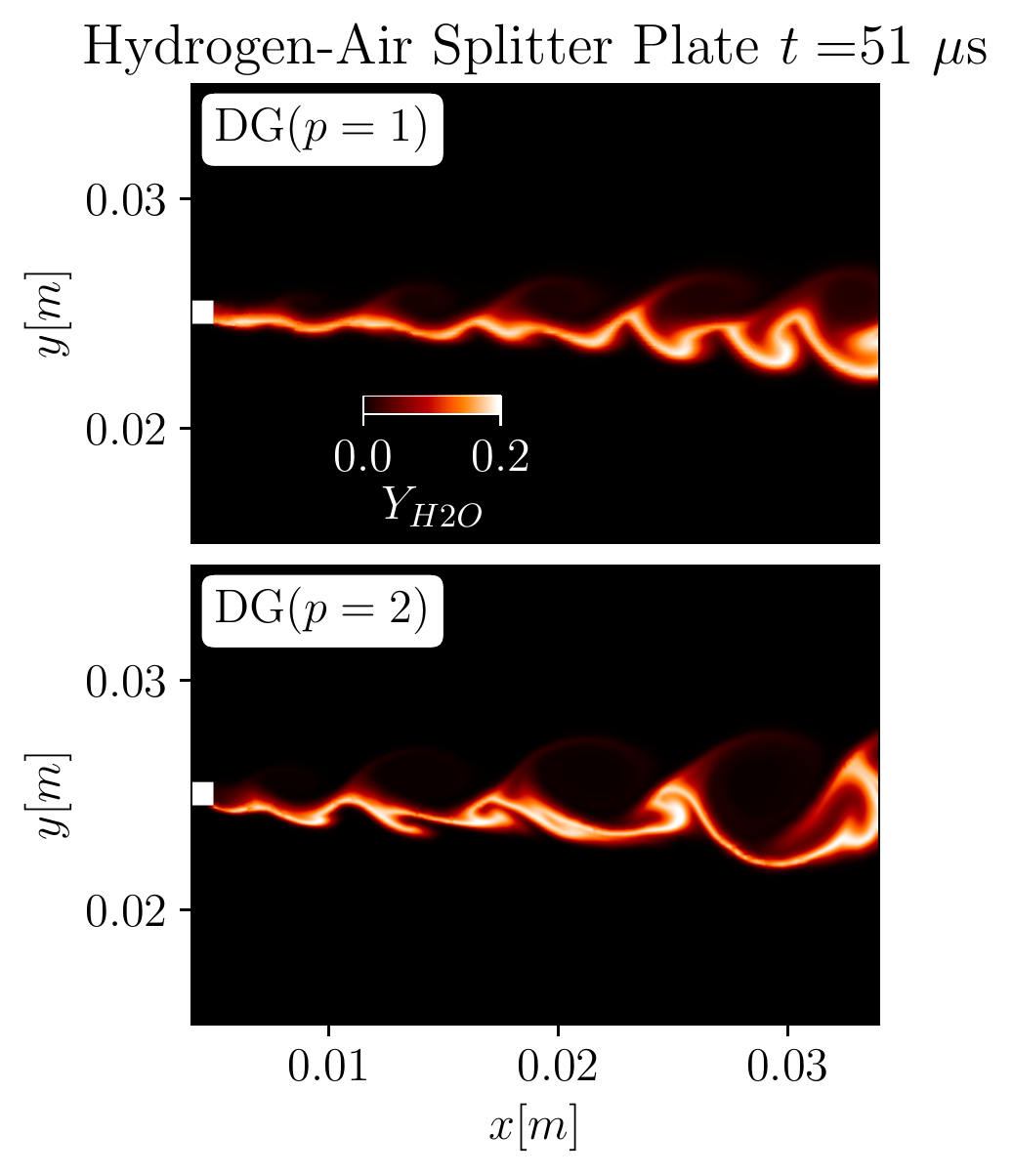}
\par\end{centering}
}

\caption{\label{fig:H2O_splitter} Comparison of splitter plate $H_{2}O$ mass
fraction solution for $\mathrm{DG}\left(p=1\right)$ and $\mathrm{DG}\left(p=2\right)$
at $t=51$ $\mu$s. The initialization is given by~(\ref{eq:2D_splitter_plate}).
This smooth solution of a multi-component chemically reacting Navier-Stokes
flow was computed without the use of additional stabilization, e.g.,
artificial viscosity, limiting or filtering.}
\end{figure}

Figures~\ref{fig:splitter_plate_time_p1} and~\ref{fig:splitter_plate_time_p2}
show the temporal evolution for $\mathrm{DG}(p=1)$ and $\mathrm{DG}(p=2)$
temperature solutions, respectively. The initial interface between
H2 and air diffuses and reactions occur along the mid-line. As the
reactions progress, flow features develop that create mixing structures
that grow in the downstream direction. As the $\mathrm{DG}(p=1)$
and $\mathrm{DG}(p=2)$ solutions evolve in time, they begin to diverge
from each other. Figure~\ref{fig:splitter_plate_temperature_compare}
shows a direct comparison of the solutions at $t=51$ $\mu$s. The
profiles for the $\mathrm{DG}(p=2)$ solution are thinner and less
diffuse than the $\mathrm{DG}(p=1)$ solution which contributes to
the difference in the downstream features. Figures~\ref{fig:H2O_splitter_upstream}
and~\ref{fig:H2O_splitter_downstream} show the upstream and down
stream $Y_{H_{2}O}$ profiles, respectively, for both the $\mathrm{DG}(p=1)$
and $\mathrm{DG}(p=2)$ at $t=51$ $\mu$s. The $\mathrm{DG}(p=2)$
solution resolves a sharper flame attachment to the splitter wall
than the $\mathrm{DG}(p=1)$ solution. In the downstream region a
large structure is apparent in both solutions, with the $\mathrm{DG}(p=2)$
solution resolving more fine scale features than the $\mathrm{DG}(p=1)$
solution.

We extended this two-dimensional shear layer problem to three dimensions
by extruding the mesh in the $z$-direction. To save on computational
cost, the domain was changed to $0.03$ m in height but maintained
a length of $0.10$ m, $\text{\ensuremath{\Omega}}=\left(0,0.10\right)\mathrm{m}\times\left(0,0.03\right)\mathrm{m}$.
The splitter plate is the same size but is centered at $y=0.015$
m. The mixing region resolution and size is the same as the two dimensional
problem but shifted to center the new plate location. The mesh was
constructed using Gmsh and was extruded in the $z$-direction at a
cellular width of $200\times10^{-6}$ m for a total width of $0.0028$
m. The boundary conditions in the resulting $xz$-planes and $yz$-planes
were the same as the two dimensional splitter plate shifted to be
centered at $y=0.015$ m for the left hand side boundary with the
exception of a spatiotemporal variation introduced to the momentum
at the $x=0$ boundaries,

\begin{equation}
\begin{array}{cccc}
M & = & 0.8,\\
v_{1} & = & Mc+\frac{M}{10}c\sin\left({\frac{\pi}{w}z}\right)\sin\left({\frac{2\pi}{\tau}t}\right) & \text{ m/s, }\\
v_{2} & = & 0 & \text{ m/s, }\\
v_{3} & = & \frac{M}{5}c+\frac{M}{50}c\sin\left({\frac{\pi}{w}z}\right)\sin\left({\frac{2\pi}{\tau}t}\right) & \text{ m/s, }
\end{array}\label{eq:spatiotemporal_variation}
\end{equation}
where $c=\sqrt{\gamma RT}$ with $\gamma$ and $R$ calculated from
the mixture properties at the boundary. This inflow boundary condition
was used to encourage three-dimensional breakup and unsteadiness in
the $z$-direction. Periodicity is enforced between the resulting
$xy$-planes at $z=0$ m and $z=0.0028$ m. The initialization is
the same as in two dimensional case, which is given by~(\ref{eq:2D_splitter_plate})
at time $t=0\,\mathrm{s}$. The $\mathrm{DG}(p=1)$ approximation
was used for the first $100$ $\mu$s of the simulation, at which
point the approximate solution was interpolated to the space corresponding
to the $\mathrm{DG}(p=2)$ approximation. No artificial viscosity,
limiting, or filtering was required throughout the entire simulation
as the physical diffusion present in the solution provided adequate
stabilization.

\begin{figure}[H]
\centering{}\includegraphics[width=0.8\columnwidth]{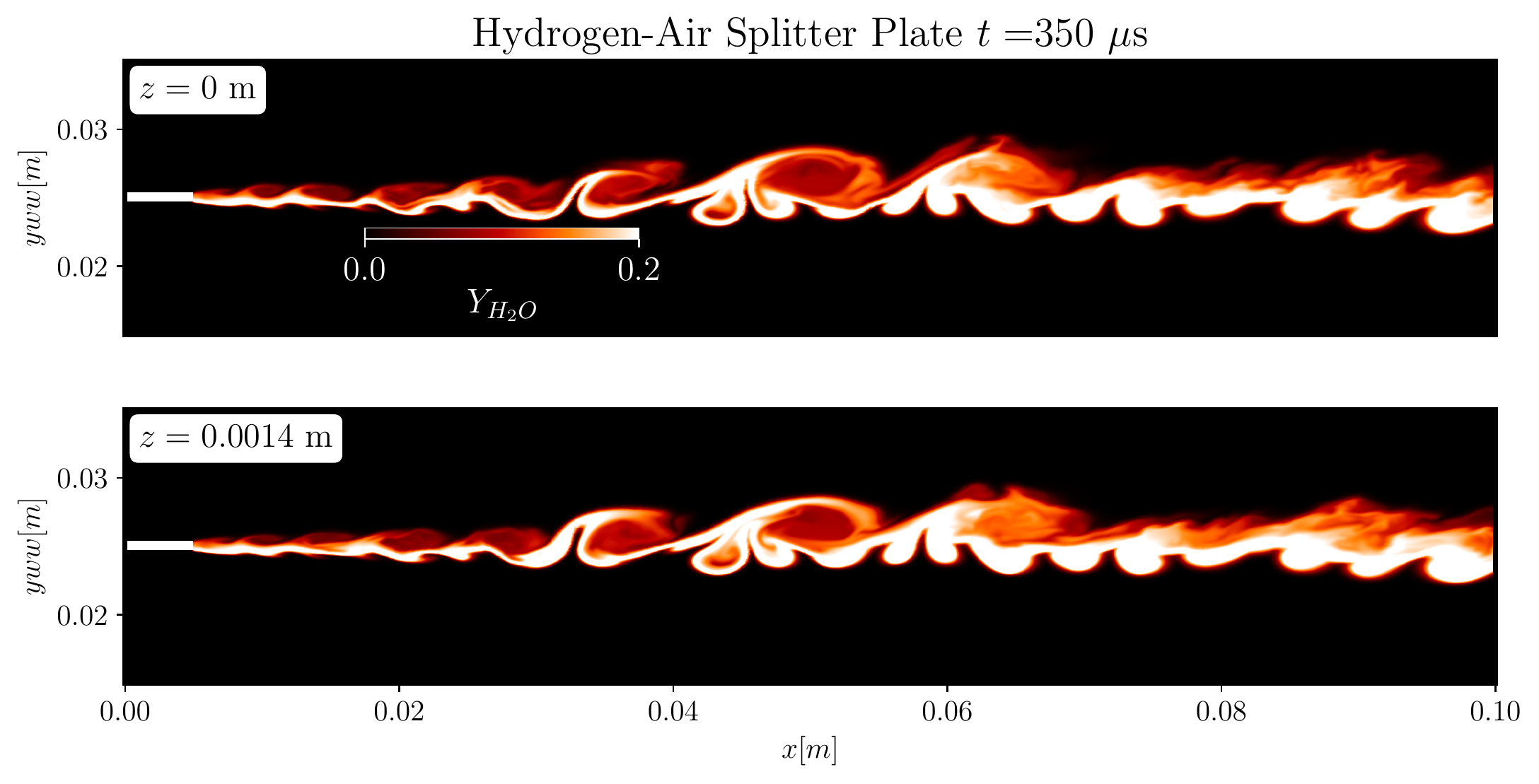}\caption{\label{fig:splitter_plate_3D_h2o_compare}Three dimensional $\mathrm{DG}\left(p=2\right)$
splitter plate $Y_{H_{2}O}$ xy-plane solutions at two different z-locations
at $t=350$ $\mu$s. This smooth solution of a multi-component chemically
reacting Navier-Stokes flow was computed without the use of additional
stabilization, e.g., artificial viscosity, limiting or filtering.}
\end{figure}

\begin{figure}[H]
\centering{}\includegraphics[width=0.8\columnwidth]{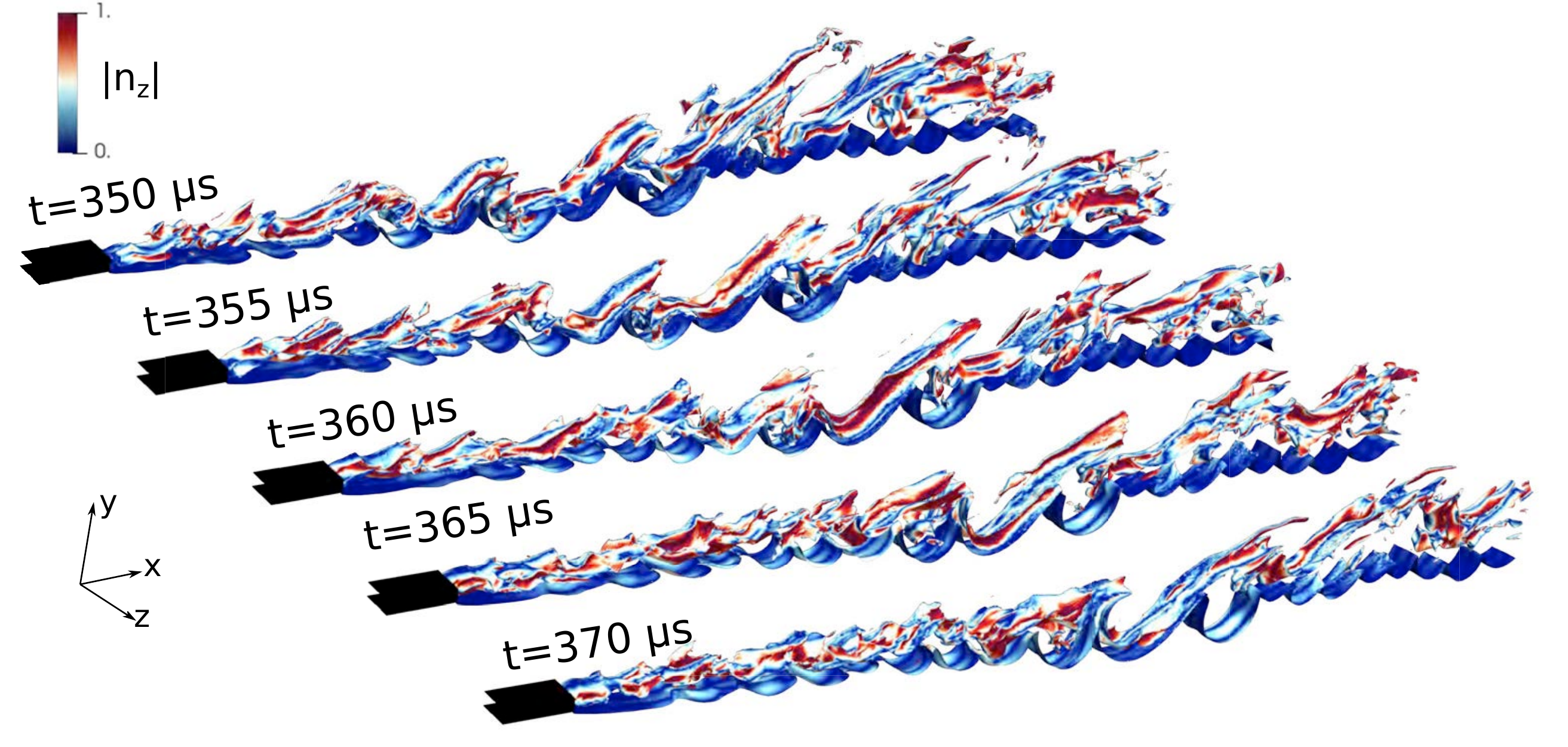}\caption{\label{fig:splitter_plate_3D_iso}Three dimensional $\mathrm{DG}\left(p=2\right)$
splitter plate $T=2200$ K isocontour solution from $t=350$ to $t=370$
$\mu$s. This smooth solution of a multi-component chemically reacting
Navier-Stokes flow was computed without the use of additional stabilization,
e.g., artificial viscosity, limiting or filtering.}
\end{figure}

Figure~\ref{fig:splitter_plate_3D_h2o_compare} shows the $Y_{H_{2}O}$
solution for $xy$ -planes located at $z=0$ m and $z=0.0014$ m.
The variation of $Y_{H_{2}O}$ in the z-direction is greater downstream
of $x=0.04$ m. This is seen particularly in the solution for $y>0$,
where the flow-field inertia is not as great since the species has
less mass and is therefore more susceptible to mixing due to the inflow
perturbations. Figure~\ref{fig:splitter_plate_3D_iso} shows the
three dimensional isocontours for $T=2200$ K between times $t=350$
and $t=370$ $\mu$s. The splitter plate is shown as a black surfaces,
where as the isocontours are colored by the absolute value of the
$z$-component of the isocontour surface normal, labeled as $|n_{z}|$.
Red and white surfaces represent areas where the flame is wrinkled
in the $z$-direction, whereas blue surfaces represent areas of the
flame that are planar in the $z$-direction, $|n_{z}|\rightarrow0$.
These isocontours confirm that the flame becomes less uniform in the
$z$-direction in the region downstream of $x=0.04$ m, particularly
in the top half of the domain, $y>0.015$ m.

These results indicate that the conservative DG formulation presented
in this work is capable of computing smooth multidimensional, multi-component,
chemically reacting Navier-Stokes flows without the need for artificial
viscosity, limiting, filtering. Furthermore, since the formulation
maintains discrete temperature and pressure equilibrium it is capable
of incorporating the viscous and diffusive operators without caveats
or creating additional concerns. As expected, the formulation behaves
exactly like its single-component counterpart for the case of viscous
multi-component reacting flows.

\section{Conclusion}

We have presented a a fully conservative discontinuous Galerkin finite
element method for the multi-component chemically reacting Navier-Stokes
equations that retains the desirable properties of DG, namely discrete
conservation and high-order accuracy in smooth regions of the flow.
In contrast to previous DG implementations for multi-component chemically
reacting flows, the formulation presented in this work does not freeze
the thermodynamics; instead a nonlinear relationship for temperature
is solved which ensures consistency between the internal energy of
the conserved state and the internal energy defined by a mixture averaged
polynomial expression based on temperature. Furthermore, the discrete
solution is represented in terms of a nodal basis with coefficients
defined on the element interfaces so that temperature and pressure
equilibrium between adjacent elements is maintained and the resulting
weak form is evaluated in terms of basis coefficients that are designed
to maintain pressure equilibrium. As such, unphysical pressure oscillations
are not generated at material interfaces if the temperature is continuous.
Furthermore, if the temperature is discontinuous, the unphysical pressure
oscillations remain small and do not lead to numerical failure. Other
numerical instabilities do arise at discontinuous interfaces, e.g.,
shocks and detonations, but these are not specific to multi-component
flows and they are suppressed in this work by the addition of a residual
based artificial viscosity.

A coupled $hp$-adaptive ODE solver, DGODE, has been developed for
the integration of the stiff chemical source term. The solver adapts
the local time-step, $h$, as well as the polynomial degree, $p$,
corresponding to the local approximation in order to efficiently integrate
the system of ODEs. The solver was shown to successfully integrate
the stiff source term corresponding to the one- and two-dimensional
$H_{2}$-$O_{2}$-$Ar$ detonation waves, where the local polynomial
degree adapted to the local stiffness of the chemical system. 

The generation of unphysical pressure oscillations was studied in
the context of the formulation described in this work. We found that
pressure oscillations generated for formulations based on frozen thermodynamics
did not reach the magnitudes previously reported in literature~\citep{Abg01,Bil03,Bil11,Hou11,Lv15}.
However, when it was assumed that the mean specific heat at constant
pressure was the same as the NASA polynomial specific heat at constant
pressure, $\bar{c_{p}}=c_{p}$, the oscillations did reach the same
levels as previously reported, indicating that the severity of the
pressure oscillations is directly related to the method in which the
thermodynamic quantities are evaluated. Regardless, frozen thermodynamic
formulations generate pressure oscillations that grow in time as the
solution evolves. In contrast, evaluating the thermodynamics quantities
exactly does not lead to pressure oscillations at a material interface
if the temperature is continuous. We therefore concluded that the
conservative DG discretization presented in this work can stably compute
solutions to smooth multi-component chemically reacting Navier-Stokes
flows without additional stabilization. 

The discretization was applied to several multi-component non-reacting
and reacting test cases including: a one-dimensional $He$ and $N_{2}$
shock-tube, a two dimension shock interaction with a helium bubble
suspended in air, one-dimensional $H_{2}$-$O_{2}$-$Ar$ detonation
wave, and two-dimensional $H_{2}$-$O_{2}$-$Ar$ detonation. We confirmed
that additional stabilization is not required for well-resolved smooth
regions of the flow. Where applicable, i.e., for wall-bounded problems,
we computed the integrated percent loss of the conserved quantities
at each time-step. The value was shown to be on the order of machine
error while remaining constant for the duration of the simulation.
The solutions for the one- and two-dimensional $H_{2}$-$O_{2}$-$Ar$
detonation simulations were compared to results previously reported
in the literature. The profile of the one-dimensional solution in
the region of the detonation was shown to be in close agreement with
the profile predicted by the detonation toolbox and two-dimensional
solution was shown to reproduce the correct cellular detonation structure.

We also presented a one-dimensional premixed hydrogen-air flame and
two- and three-dimensional reacting shear layer with a splitter plate.
The $H_{2}$-air deflagration flame speed compared well to the speed
computed via Cantera for both the $\mathrm{DG}(p=1)$ and $\mathrm{DG}(p=2)$
solutions. Discrepancies between the Cantera solution and the $\mathrm{DG}(p=1)$
solution were not observed in the $\mathrm{DG}(p=2)$ solution, indicating
the $\mathrm{DG}(p=2)$ solution better resolved the flame. The two-
and three-dimensional shear layer simulations confirmed, in a practical
setting, that the conservative DG formulation does not require additional
stabilization for smooth multi-component chemically reacting Navier-Stokes
flows.

\section*{Acknowledgements}

This work is sponsored by the Office of Naval Research through the
Naval Research Laboratory 6.1 Computational Physics Task Area.

\section*{}

\bibliographystyle{elsarticle-num}
\bibliography{citations}

\appendix

\section{Jacobian of the chemically reacting source term\label{sec:Source-term-Jacobian}}

For brevity we do not present details on the differentiation of the
complex reactions found in large chemical mechanisms and instead refer
the reader to ~\citep{Nie17} and ~\citep{Per12}. Therefore, in
the following we assume the derivatives of chemical source term, $\omega_{i}$,
with respect to concentrations and temperature, $\frac{\partial\omega_{i}}{\partial C_{j}}$
and $\frac{\partial\omega_{i}}{\partial T}$ are known for all $i$
and $j$.

The perturbation of the source term is found via the chain rule

\begin{equation}
\mathcal{\delta S}=\mathcal{S}'\delta y=\frac{\partial\mathcal{S}}{\partial T}T'\delta y+\sum_{j=1}^{N_{s}}\frac{\partial\mathcal{S}}{\partial C_{j}}C'_{j}\delta y.\label{eq:source_term_perturbation}
\end{equation}
The first term, $\frac{\partial\mathcal{S}}{\partial T}T'\delta y$,
relies on temperature and its implicit relationship to the conserved
state, whereas the second term, $\frac{\partial\mathcal{S}}{\partial C_{j}}C_{j}'\delta y$,
can be extracted from the species elements corresponding to the perturbation,

\begin{equation}
C_{j}'\delta y=\left(\begin{array}{c}
0,\dots,0,\delta_{j,1},\dots\delta_{j,N_{s}}\end{array}\right)\left(\begin{matrix}\delta\rho v_{1}\\
\vdots\\
\delta\rho v_{d}\\
\delta\rho e_{t}\\
\delta C_{1}\\
\vdots\\
\delta C_{N_{s}}
\end{matrix}\right)\begin{cases}
\delta_{j,i}=1 & i=j\\
\delta_{j,i}=0 & \text{otherwise}
\end{cases}=\delta C_{j}.\label{eq:species_jacobian}
\end{equation}
The temperature jacobian, $T'$, in Equation~\ref{eq:source_term_perturbation}
is extracted by first applying the chain rule to the definition total
energy, 
\begin{equation}
\rho e_{t}=\rho u+\frac{1}{2}\sum_{k=1}^{d}\rho v_{k}v_{k}\label{eq:total_energy}
\end{equation}
and grouping elements of the state perturbation together. Since temperature
is implicitly related to total energy through the internal energy,
$\rho u=\sum_{j=1}^{n_{s}}W_{j}C_{j}\sum_{k=1}^{n_{p}}a_{jk}T^{k}$,
we extract the temperature perturbation by using the derivatives of
internal energy, $\frac{\partial\rho u}{\partial T}=\sum_{j=1}^{n_{s}}W_{j}C_{j}\sum_{k=1}^{n_{p}}ka_{jk}T^{k-1}$
and $\frac{\partial\rho u}{\partial C_{j}}=W_{j}\sum_{k=1}^{n_{p}}a_{jk}T^{k}$,
combined with the perturbations of the kinetic energy, 

\begin{eqnarray}
\delta\rho e_{t} & = & \frac{\partial\rho u}{\partial T}\delta T+\sum_{j=1}^{n_{s}}\frac{\partial\rho u}{\partial C_{j}}\delta C_{j}+\frac{1}{2}\left(\sum_{k=1}^{d}\left(\delta\rho v_{k}\right)v_{k}+\sum_{k=1}^{d}\rho v_{k}\delta v_{k}\right).\label{eq:energy_perturbation_1}
\end{eqnarray}
The internal energy derivatives are can be evaluated if the polynomial
fits for internal energy have been defined such that they are differentiable.
The perturbations of total energy, $\rho e_{t}$, species, $C_{k}$,
and momentum, $\delta\rho v_{i}$, all come from the perturbed state.
In order to move forward, we need the velocity perturbation which
comes from the following definition,

\begin{eqnarray}
v_{k} & = & \frac{\rho v_{k}}{\rho}\nonumber \\
\delta v_{k} & = & \frac{\delta\rho v_{k}}{\rho}-\frac{\rho v_{k}}{\rho^{2}}\delta\rho,\label{eq:velocity_perturbation}
\end{eqnarray}
 and the perturbation of density,

\begin{eqnarray}
\rho & = & \sum_{i=1}^{n_{s}}W_{i}C_{i}\nonumber \\
\delta\rho & = & \sum_{i=1}^{n_{s}}W_{i}\delta C_{i}.\label{eq:density_perturbation}
\end{eqnarray}
By substituting~\ref{eq:velocity_perturbation} and~\ref{eq:density_perturbation}
into~\ref{eq:energy_perturbation_1} we arrive at a total energy
perturbation dependent only on the state and temperature perturbations,

\begin{eqnarray*}
\delta\rho e_{t} & = & \frac{\partial\rho u}{\partial T}\delta T+\sum_{j=1}^{n_{s}}\frac{\partial\rho u}{\partial C_{j}}\delta C_{j}+\sum_{k=1}^{d}\left(\delta\rho v_{k}\right)v_{k}-\frac{1}{2\rho}\sum_{k=1}^{d}v_{k}v_{k}\sum_{j=1}^{n_{s}}W_{j}\delta C_{j}.
\end{eqnarray*}
The temperature perturbation is then solved for,

\begin{eqnarray}
\delta T & =T'\delta y=-\frac{\partial\rho u}{\partial T}^{-1} & \left(\sum_{k=i}^{n_{s}}\frac{\partial\rho u\left(C_{j,}T\right)}{\partial C_{i}}\delta C_{i}+\sum_{k=1}^{d}\left(\delta\rho v_{k}\right)v_{k}\right.\nonumber \\
 &  & \left.-\frac{1}{2\rho}\sum_{k=1}^{d}v_{k}v_{k}\sum_{j=1}^{n_{s}}W_{j}\delta C_{j}-\delta\rho e_{t})\right)
\end{eqnarray}
and the temperature jacobian is extracted by grouping like terms,

\begin{eqnarray}
\delta T & =T'\delta y=-\frac{\partial\rho u}{\partial T}^{-1} & \left(\begin{array}{c}
v_{1},\dots,v_{d},-1,\frac{\partial\rho u}{\partial C_{1}}-\frac{1}{2\rho}\sum_{k=1}^{d}v_{k}v_{k}W_{1},\dots,\frac{\partial\rho u}{\partial C_{n_{s}}}-\frac{1}{2\rho}\sum_{k=1}^{d}v_{k}v_{k}W_{n_{s}}\end{array}\right)\left(\begin{matrix}\delta\rho v_{1}\\
\vdots\\
\delta\rho v_{d}\\
\delta\rho e_{t}\\
\delta C_{1}\\
\vdots\\
\delta C_{n_{s}}
\end{matrix}\right)\nonumber \\
 & T' & =-\frac{\partial\rho u}{\partial T}^{-1}\left(\begin{array}{c}
v_{1},\dots,v_{d},-1,\alpha_{1},\dots,\alpha_{n_{s}}\end{array}\right),\label{eq:temperatuer_jacobian}
\end{eqnarray}
where

\begin{equation}
\alpha_{i}=\frac{\partial\rho u}{\partial C_{i}}-\frac{1}{2\rho}\sum_{k=1}^{d}v_{k}v_{k}W_{i}.\label{eq:alpha}
\end{equation}
We now use~\ref{eq:temperatuer_jacobian} and~\ref{eq:species_jacobian}
together to form the source term jacobian,

\begin{eqnarray}
\mathcal{S}'= & \frac{d\mathcal{S}}{dT}T'+\sum_{j=1}^{n_{s}}\frac{d\mathcal{S}}{dC_{j}}C_{j}'= & -\frac{\partial\rho u}{\partial T}^{-1}\left(\begin{matrix}0\\
\vdots\\
0\\
0\\
\frac{d\omega_{1}}{dT}\\
\vdots\\
\frac{d\omega_{n_{s}}}{dT}
\end{matrix}\right)\left(\begin{array}{c}
v_{1},\dots,v_{d},-1,\alpha_{1},\dots,\alpha_{n_{s}}\end{array}\right)\label{eq:src_vec_1}\\
 &  & +\sum_{j=1}^{n_{s}}\left(\begin{matrix}0\\
\vdots\\
0\\
\frac{\partial\omega_{1}}{\partial C_{j}}\\
\vdots\\
\frac{\partial\omega_{n_{s}}}{\partial C_{j}}
\end{matrix}\right)\left(\begin{array}{c}
0,\dots,0,0,\delta_{j,1},\dots\delta_{j,n_{s}}\end{array}\right)\begin{cases}
\delta_{j,i}=1 & i=j\\
\delta_{j,i}=0 & \text{otherwise}
\end{cases}
\end{eqnarray}
which gives

\begin{equation}
\mathcal{S}'=\left(\begin{array}{ccccccc}
0 & \dots & 0 & 0 & 0 & \dots & 0\\
\vdots & \ddots & \vdots & \vdots & \vdots & \ddots & \vdots\\
0 & \dots & 0 & 0 & 0 & \dots & 0\\
0 & \dots & 0 & 0 & 0 & \dots & 0\\
-\beta_{1}v_{1} & \dots & -\beta_{1}v_{d} & \beta_{1} & \chi_{11} & \dots & \chi_{n_{s}1}\\
\vdots & \ddots & \vdots & \vdots & \vdots & \ddots & \vdots\\
-\beta_{n_{s}}v_{1} & \dots & -\beta_{n_{s}}v_{d} & \beta_{N_{s}} & \chi_{1n_{s}} & \dots & \chi_{n_{s}n_{s}}
\end{array}\right)\label{eq:source_term_jacobian}
\end{equation}
where $\beta_{i}=\frac{\partial\rho u}{\partial T}^{-1}\frac{d\omega_{i}}{dT}$
, \textbf{$\chi_{ij}=-\alpha_{i}\beta_{j}+\frac{\partial\omega_{j}}{\partial C_{i}}$},
and finally

\begin{equation}
\delta\mathcal{S}=\left(\begin{array}{ccccccc}
0 & \dots & 0 & 0 & 0 & \dots & 0\\
\vdots & \ddots & \vdots & \vdots & \vdots & \ddots & \vdots\\
0 & \dots & 0 & 0 & 0 & \dots & 0\\
0 & \dots & 0 & 0 & 0 & \dots & 0\\
-\beta_{1}v_{1} & \dots & -\beta_{1}v_{d} & \beta_{1} & \chi_{11} & \dots & \chi_{n_{s}1}\\
\vdots & \ddots & \vdots & \vdots & \vdots & \ddots & \vdots\\
-\beta_{n_{s}}v_{1} & \dots & -\beta_{n_{s}}v_{d} & \beta_{N_{s}} & \chi_{1n_{s}} & \dots & \chi_{n_{s}n_{s}}
\end{array}\right)\left(\begin{matrix}\delta\rho v_{1}\\
\vdots\\
\delta\rho v_{n}\\
\delta\rho e_{t}\\
\delta C_{1}\\
\vdots\\
\delta C_{N_{s}}
\end{matrix}\right)=\left(\begin{matrix}0\\
\vdots\\
0\\
0\\
\frac{\partial\omega_{1}}{\partial T}\delta T+\sum_{i=1}^{n_{s}}\frac{\partial\omega_{1}}{\partial C_{k}}\delta C_{k}\\
\vdots\\
\frac{\partial\omega_{N_{s}}}{\partial T}\delta T+\sum_{i=1}^{n_{s}}\frac{\partial\omega_{N_{s}}}{\partial C_{i}}\delta C_{i}
\end{matrix}\right)\label{eq:source_perturbation}
\end{equation}

\section{Non-reflecting boundary conditions\label{sec:Non-reflective-Inflow-Outflow}}

The Riemann invariants for the calorically perfect Euler equations
with boundary normal, $n=\left(n_{1},\dots,n_{d}\right)$, and normal
velocity $v_{n}=\sum_{k=1}^{d}v_{k}\cdot n_{k}$ are specified in
Kuzmin et al.~\citep{Kuz12} 
\begin{equation}
w=\left(v_{n}-\frac{2c}{\bar{\gamma}-1},\bar{c_{v}}\textup{log}\left(\frac{p}{\rho^{\bar{\gamma}}}\right),v_{\xi},v_{\eta},v_{n}+\frac{2c}{\bar{\gamma}-1}\right)^{T},\label{eq:invariants}
\end{equation}
where $v_{\xi}$ and $v_{\eta}$ are the velocity components tangential
to the boundary surface and $c=\sqrt{\gamma RT}$ is the speed of
sound where $R$ is calculated from Equation~(\ref{eq:R_mix}). The
specific heat at constant volume is $\bar{c}_{v}=\bar{c}_{p}-R$.
These invariants have the corresponding eigenvalues
\begin{equation}
\Lambda=\textup{diag}\left\{ v_{n}-c,v_{n},v_{n},v_{n},v_{n}+c\right\} \label{eq:eigenvectors}
\end{equation}
 which determine what is specified at a subsonic inflow and outflow
boundary conditions.

For a subsonic outlet $v_{n}>0$ and $M<1$ so $\Lambda_{1}=v_{n}-c$
is negative and requires $w_{1}$ to be specified from the prescribed
state, $y_{\infty}$, and $w_{2}$-$w_{5}$ come from the interior
state, $y^{+}$,

\begin{eqnarray}
w_{1} & = & v_{n,\infty}-\frac{2c_{\infty}}{\bar{\gamma}_{\infty}-1}\label{eq:w1_subout}\\
w_{2} & = & \bar{c_{v}}^{+}\textup{log}\left(\frac{p^{+}}{\left(\rho^{\bar{\gamma}}\right)^{+}}\right)\label{eq:w2_subout}\\
w_{3} & = & v_{\xi}^{+}\label{eq:w3_subout}\\
w_{4} & = & v_{\eta}^{+}\label{eq:w4_subout}\\
w_{5} & = & v_{n}^{+}+\frac{2c^{+}}{\bar{\gamma}^{+}-1}.\label{eq:w6_subout}
\end{eqnarray}
The species mass fractions and mole fractions are assumed to be from
the interior state which gives $c_{\infty}=\sqrt{\gamma^{+}R^{+}T_{\infty}}$
where $T_{\infty}$ is specified but the gas properties come from
the interior. It is often inconvenient to specify temperature and
velocity, which is required for $w_{1}$ at an outflow, unless the
boundary is specified far enough away to not interfere with the flow.
For a downstream outflow condition, the pressure, $p_{\infty}$, can
be specified and the temperature, $T_{\infty}$, can be calculated
from a constant entropy process assuming the mole fractions are constant.
Here entropy is 

\begin{equation}
S=\sum_{i=1}^{n_{s}}X_{i}\sum_{k=0}^{n_{p}}s_{ik}T^{k}-X_{i}R\log X_{i}-X_{i}R\ln{\frac{p}{p_{atm}}}\label{eq:entropy}
\end{equation}
where $s_{ik}$ are the polynomial coefficients for the species specific
entropies. The constant entropy process temperature, $T_{\partial}$,
is computed such that the following is satisfied to machine precision
for a given initial temperature guess:

\begin{equation}
\delta T=\frac{\sum_{i=1}^{n_{s}}X_{i}\sum_{k=0}^{n_{p}}s_{ik}\left(T^{+}\right)^{k}+R\ln{\frac{p_{\partial}}{p^{+}}}-\sum_{i=1}^{n_{s}}X_{i}\sum_{k=0}^{n_{p}}s_{ik}T_{\partial}^{k}}{\frac{\partial S}{\partial T}},\label{eq:entropy_decrement}
\end{equation}
where $\delta T$ is the temperature decrement corresponding to Newton's
method and

\begin{equation}
\frac{\partial S}{\partial T}=\sum_{i=1}^{n_{s}}X_{i}\sum_{k=1}^{K}ks_{ik}T_{\partial}^{k-1}.\label{eq:entropy_derivative}
\end{equation}
With $T_{\partial}$ known, $\bar{\gamma}_{\partial}$ and $\rho_{\partial}$
can be calculated and the incoming characteristic can be specified,

\begin{equation}
w_{1}=w_{5}-\frac{4}{\bar{\gamma_{\infty}}-1}\sqrt{\frac{\bar{\gamma_{\infty}}p_{\infty}}{\rho_{\infty}}}.\label{eq:w5_recalc}
\end{equation}
 The thermodynamic state for the boundary is then given by

\begin{eqnarray}
\rho^{*} & = & \left(\frac{c_{\infty}^{2}}{\bar{\gamma}^{+}}e^{-\frac{w_{2}}{\bar{c_{v}}^{+}}}\right)^{\frac{1}{\bar{\gamma}^{+}-1}}\label{eq:rho_star}\\
c^{*} & = & \left(\gamma^{+}-1\right)\left(w_{5}-w_{1}\right)\\
p^{*} & = & \frac{\rho^{*}\left(c^{*}\right)^{2}}{\bar{\gamma}^{+}}\\
T^{*} & = & \frac{p^{*}}{\rho^{*}\bar{R}^{+}}\\
v^{*} & = & \frac{w_{1}+w_{5}}{2}\left(n_{1},\dots,n_{d}\right)+v^{+}-v_{n}\left(n_{1},\dots,n_{d}\right)\\
Y_{i}^{*} & = & Y_{i}^{+}.
\end{eqnarray}

Likewise, the subsonic inflow requires $w_{1}-w_{4}$ to be specified,

\begin{eqnarray}
w_{1} & = & v_{n}-\frac{2c_{\infty}}{\bar{\gamma}_{\infty}-1},\\
w_{2} & = & c_{v,\partial}\textup{log}\left(\frac{p_{\infty}}{\left(\rho^{\bar{\gamma}}\right)_{\infty}}\right),\\
w_{3} & = & v_{\xi,\infty,}\\
w_{4} & = & v_{n,\infty},
\end{eqnarray}
where all the gas properties come from the specified exterior state.
The last invariant comes from the interior, $w_{5}=v_{n}^{+}+\frac{2c^{+}}{\bar{\gamma}^{+}-1}$.
This gives the thermodynamic state at the boundary as

\begin{eqnarray}
\rho^{*} & = & \rho_{\infty}\\
c^{*} & = & \left(\gamma_{\infty}-1\right)\frac{\left(w_{5}-w_{1}\right)}{4}\\
p^{*} & = & \frac{\rho^{*}\left(c^{*}\right)^{2}}{\bar{\gamma_{\infty}}}\\
T^{*} & = & \frac{p^{*}}{\rho^{*}\bar{R}_{\infty}}\\
v^{*} & = & \frac{w_{1}+w_{5}}{2}\left(n_{1},\dots,n_{d}\right)+v_{\infty}-v_{n}\left(n_{1},\dots,n_{d}\right)\\
Y_{i}^{*} & = & Y_{i,\infty}
\end{eqnarray}
The estimates for the thermodynamic state are used to construct the
boundary value,

\begin{equation}
y^{*}\left(y^{+},y_{\infty},n^{+}\right)=\left(\rho^{*}v_{1}^{*},\ldots,\rho^{*}v_{n}^{*},\rho e_{t},\frac{\rho^{*}Y_{1}^{*}}{W_{i}},\ldots,\rho Y_{n_{s}}^{*}\right),
\end{equation}
where 

\begin{equation}
\rho e_{t}=\rho u+\frac{1}{2}\sum_{k=1}^{d}\rho^{*}v_{k}^{*}v_{k}^{*}
\end{equation}
with 

\begin{equation}
\rho u=\sum_{i=1}^{n_{s}}W_{i}C_{i}\sum_{k=0}^{n_{p}}a_{ik}\left(T^{*}\right)^{k}.
\end{equation}

\section{Material discontinuities in multi-component flows\label{sec:Discontinuities}}

A material discontinuity is defined as a discontinuity across which
there is no mass flow. The velocity and pressure are constant across
the discontinuity but other material quantities are not. In this appendix
we analyze problems involving material interfaces for the formulation
presented in this work by considering the non-reacting inviscid formulation
of Equations~(\ref{eq:conservation-law-strong-form})-(\ref{eq:conservation-law-flux-boundary-condition})
where $\mathcal{F}^{v}\left(y,\nabla y\right)=\left(0,\dots,0\right)$
in Equation~(\ref{eq:flux_function}) and $\mathcal{S}\left(y\right)=\left(0,\ldots,0\right)$
in Equation~(\ref{eq:conservation-law-strong-form}). A discontinuous
solution, in one dimension satisfies, the inviscid form of Equations~(\ref{eq:conservation-law-strong-form})-(\ref{eq:conservation-law-flux-boundary-condition})
if the jump in the flux is equal to the product of the jump in the
state and the material interface velocity~\citep{Mad84},

\begin{equation}
\mathcal{F}\left(y_{r}\right)-\mathcal{F}\left(y_{l}\right)=v_{s}\left(y_{r}-y_{l}\right),\label{eq:jump_condition_solution}
\end{equation}
where $y_{r}$ is the state on the right of the discontinuity, $y_{l}$
is the state on the left of the discontinuity, and $v_{s}$ is the
material velocity normal to the interface.

\begin{figure}[H]
\subfloat[\label{fig:interface_continuous}The speed of the material interface
is the slope of the trajectory $v_{s}=\bar{v}=\angle$.]{\begin{centering}
\includegraphics[width=0.45\columnwidth]{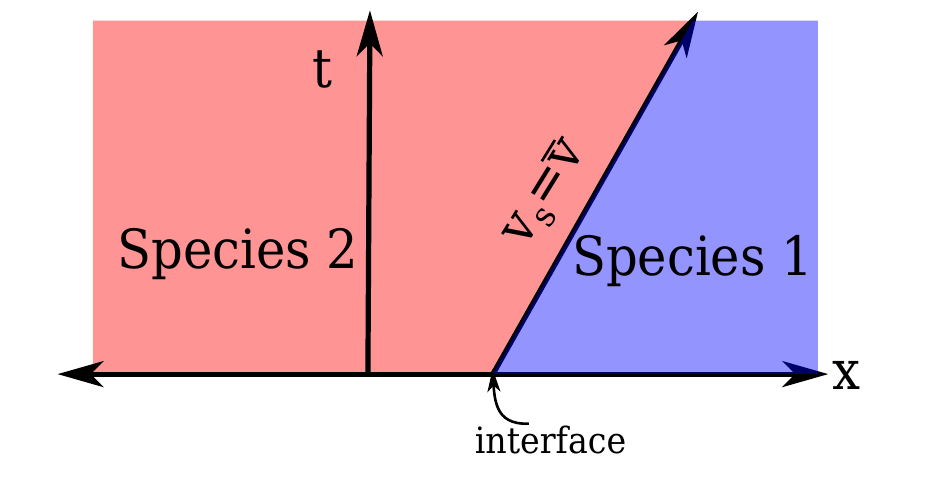}
\par\end{centering}
}\hfill{}\subfloat[\label{fig:interface_numerical}Numerical diffusion of material interface
at discrete times.]{\begin{centering}
\includegraphics[width=0.45\columnwidth]{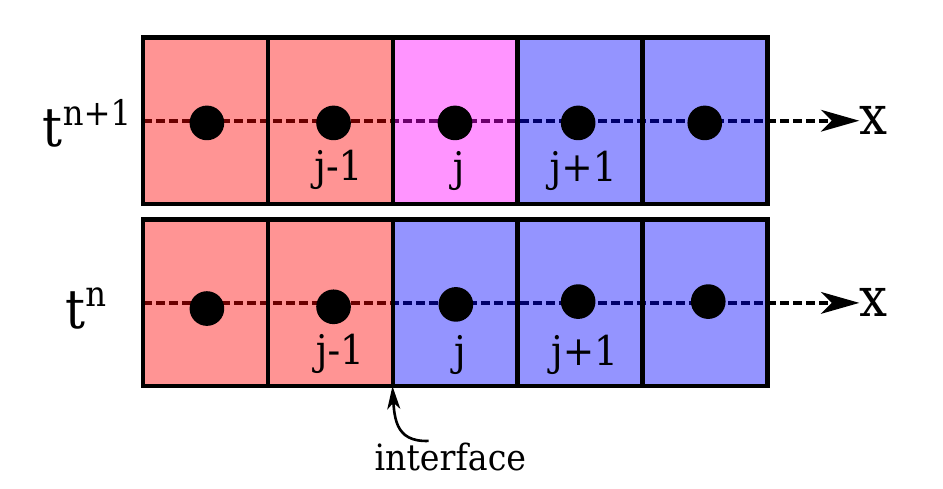}
\par\end{centering}
}
\centering{}\caption{\label{fig:interface_diagram}Diagrams of moving material interfaces.}
\end{figure}

Below we introduce a material discontinuity by considering a one-dimensional
two species discontinuity at $x_{j}$ where the velocity and pressure
are constant and the temperature is discontinuous,

\begin{eqnarray*}
v & = & \bar{v},\\
C_{1} & = & \begin{cases}
C_{1}^{0} & \text{if }x>x_{j}\\
0 & \text{otherwise}
\end{cases},\\
C_{2} & = & \begin{cases}
0 & \text{if }x<x_{j}\\
C_{2}^{0} & \text{otherwise}
\end{cases},\\
T & = & \begin{cases}
\eta\bar{T} & \text{if }x<x_{j}\\
\bar{T} & \text{otherwise}
\end{cases},\\
p & = & \bar{p.}
\end{eqnarray*}
The species with index $i=1$, species 1, has molecular weight $W_{1}$
and the species with index $i=2$, species 2, has molecular weight
$W_{2}$. The initial fluid state from Equation~(\ref{eq:reacting-navier-stokes-state})
is therefore

\begin{equation}
y\left(x,t=0\right)=\begin{cases}
\left(W_{1}C_{1}^{0}\bar{v},\frac{1}{2}W_{1}C_{1}^{0}\bar{v}^{2}+W_{1}C_{1}^{0}\sum_{k=0}^{n_{p}}a_{1k}\bar{T}^{k},C_{1}^{0},0\right) & x>x_{j}\\
\left(W_{2}C_{2}^{0}\bar{v},\frac{1}{2}W_{2}C_{2}^{0}\bar{v}^{2}+W_{2}C_{2}^{0}\sum_{k=0}^{n_{p}}a_{2k}\left(\eta\bar{T}\right)^{k},0,C_{2}^{0}\right) & \text{otherwise}
\end{cases}.\label{eq:material_states}
\end{equation}
Substituting the fluid state from Equation~(\ref{eq:material_states})
in Equation~(\ref{eq:jump_condition_solution}) we arrive at the
following condition

\begin{eqnarray}
W_{1}C_{1}^{0}\bar{v}^{2}-W_{2}C_{2}^{0}\bar{v}^{2} & = & v_{s}\left(W_{1}C_{1}^{0}v-W_{2}C_{2}^{0}v\right),\label{eq:interface_momentum_sub}\\
\left(\frac{1}{2}W_{1}C_{1}^{0}v^{2}+C_{1}^{0}W_{1}\sum_{k=0}^{n_{p}}a_{1k}T^{k}+\bar{p}\right)\bar{v}\nonumber \\
-\left(\frac{1}{2}W_{2}C_{2}^{0}v^{2}+C_{2}^{0}W_{2}\sum_{k=0}^{n_{p}}a_{2k}\left(\eta T\right)^{k}+\bar{p}\right)\bar{v} & = & v_{s}\left(\frac{1}{2}W_{1}C_{1}^{0}v^{2}+W_{1}C_{1}^{0}\sum_{k=0}^{n_{p}}a_{1k}T^{k}\right.\nonumber \\
 &  & \left.-\frac{1}{2}W_{2}C_{2}^{0}v^{2}-W_{2}C_{2}^{0}\sum_{k=0}^{n_{p}}a_{2k}\left(\eta T\right)^{k}\right),\label{eq:interface_energy_sub}\\
C_{1}^{0}\bar{v} & = & v_{s}\left(C_{1}^{0}\right),\label{eq:interface_species_1_sub}\\
-C_{2}^{0}\bar{v} & = & v_{s}\left(-C_{2}^{0}\right).\label{eq:interface_species_2_sub}
\end{eqnarray}
Therefore a material discontinuity where velocity and pressure are
constant and the temperature is discontinuous satisfies Equations~(\ref{eq:conservation-law-strong-form})-(\ref{eq:conservation-law-flux-boundary-condition})
with $v_{s}=\bar{v}$. A diagram of the space-time solution is shown
in Figure~\ref{fig:interface_continuous}. 

We now present the effect of a linear discretization on the same two
species discontinuity. Using the notation from Abgrall and Karni~\citep{Abg01},
the inviscid non-reacting conservation equations can be written as

\begin{align}
\delta\left(\rho v\right)+\nu\Delta\left(\rho v^{2}+p\right)= & 0,\label{eq:linear_momentum}\\
\delta\left(\rho e_{t}\right)+\nu\Delta\left(\left(\rho e_{t}+p\right)v\right)= & 0,\label{eq:linear_energy}\\
\delta\left(C_{i}\right)+\nu\Delta\left(C_{i}v\right)= & 0\text{ for }i=1\dots n_{s},\label{eq:linear_species}
\end{align}
where the inviscid form of Equation~(\ref{eq:conservation-strong-primal})
has been linearized with respect to space and time. Here, $\delta\left(\right)=\left(\right)_{j}^{n+1}-\left(\right)_{j}^{n}$
denotes the temporal change of the state, $\Delta\left(\right)$ denotes
spatial variation across the interface $\Delta\left(\right)=\left(\right)_{j}^{n}-\left(\right)_{j-1}^{n}$,
and $\nu=\frac{\Delta t}{\Delta x}$ where $\Delta t$ is the chosen
time step and $\Delta x$ is the spatial distance across the interface.
As in the ~\citep{Abg01} and ~\citep{Joh12}, $\Delta\left(\right)=\left(\right)_{j}^{n}-\left(\right)_{j-1}^{n}$
is assumed to have the following properties: $\Delta\left(ab+c\right)=a\left(b\right)_{j}^{n}-a\left(b\right)_{j-1}^{n}+\left(c\right)_{j}^{n}-\left(c\right)_{j-1}^{n}$
for \textbf{$a$} constant, and $b$ and $c$ not constant. We are
deriving in the context of first order approximations without reconstruction,
however, it should be noted that these properties do not always apply
to all schemes. The material interface is initially between two nodes,
$j$ and $j-1$, as depicted at time $t^{n}$ in Figure~\ref{fig:interface_numerical}.
Specifically, the initial flow state at $t^{n}$ is

\begin{eqnarray}
y_{j}^{n} & = & \left(W_{1}C_{1}^{0}\bar{v},\frac{1}{2}W_{1}C_{1}^{0}\bar{v}^{2}+W_{1}C_{1}^{0}\sum_{k=0}^{n_{p}}a_{1k}\bar{T}^{k},C_{1}^{0},0\right),\label{eq:initial_state_numerical_1}\\
y_{j-1}^{n} & = & \left(W_{2}C_{2}^{0}\bar{v},\frac{1}{2}W_{2}C_{2}^{0}\bar{v}^{2}+W_{2}C_{2}^{0}\sum_{k=0}^{n_{p}}a_{2k}\left(\eta\bar{T}\right)^{k},0,C_{2}^{0}\right).\label{eq:initial_state_numerical_2}
\end{eqnarray}
For simplification purposes, we define the initial concentration of
species 2 in terms of the initial concentration of species 1 through
the constant initial pressure conditions, $p_{j}^{n}=p_{j-1}^{n}=\bar{p}$,
and the equation of state, Equation~(\ref{eq:EOS-1}),

\begin{equation}
R^{o}\bar{T}C_{1}^{0}=R^{o}\eta\bar{T}C_{2}^{0}\rightarrow C_{2}^{0}=\frac{C_{1}^{0}}{\eta}.\label{eq:species_relationship}
\end{equation}
The species conservation, Equation~(\ref{eq:linear_species}), gives
the concentrations at $t^{n+1}$ in terms of the initial species 1
concentration,

\begin{eqnarray}
C_{1,j}^{n+1} & = & C_{1}^{0}-\nu\bar{v}C_{1}^{0},\label{eq:change_species_1}\\
C_{2,j}^{n+1} & = & \nu\bar{v}C_{2}^{0}=\nu\bar{v}\frac{C_{1}^{0}}{\eta}.\label{eq:change_species_2}
\end{eqnarray}
Equations~(\ref{eq:change_species_1})~and~(\ref{eq:change_species_2})
show that there is numerical mixing of the species at time $t^{n+1}$
and node $j$, as depicted in Figure~\ref{fig:interface_numerical}.
This is a departure from the exact solution that satisfies the interface
condition, depicted in Figure~\ref{fig:interface_continuous}, and
we continue in this section by examining the effect that the numerical
mixing of the species concentrations has on the stability of the material
interface.

Substituting the concentrations from Equations~(\ref{eq:change_species_1})~and~(\ref{eq:change_species_2})
into Equation~(\ref{eq:density_definition}) gives the density at
$t^{n+1}$,

\begin{equation}
\rho_{j}^{n+1}=\left(W_{1}+\nu\bar{v}\left(\frac{W_{2}}{\eta}-W_{1}\right)\right)C_{1}^{0}.\label{eq:change_density}
\end{equation}
Using Equations~(\ref{eq:change_species_1})-(\ref{eq:change_density})
and Equation~(\ref{eq:density_definition}) for the initial density
and substituting into Equation~(\ref{eq:linear_momentum}) reveals
that the velocity remains constant,

\begin{eqnarray}
\left(W_{1}+\nu\bar{v}\left(\frac{W_{2}}{\eta}-W_{1}\right)\right)C_{1}^{0}v^{n+1}-W_{1}C_{1}^{0}\bar{v}+\nu\bar{v}^{2}\left(W_{1}C_{1}^{0}-W_{2}\frac{C_{1}^{0}}{\eta}\right) & = & 0\rightarrow v_{j}^{n+1}=\bar{v}.\label{eq:constant_velocity}
\end{eqnarray}

Using Equations~(\ref{eq:change_species_1})-(\ref{eq:constant_velocity})
we consider the change in total energy to analyze the stability of
the material interface. We derive a relationship for kinetic energy
by multiplying Equation~(\ref{eq:linear_momentum}) by $\frac{1}{2}\bar{v}$,

\begin{eqnarray}
\left(\frac{1}{2}\rho_{j}^{n+1}\bar{v}^{2}-\frac{1}{2}\rho_{j}^{n}\bar{v}^{2}+\nu\bar{v}\left(\frac{1}{2}\rho_{j}^{n}\bar{v}^{2}-\frac{1}{2}\rho_{j-1}^{n}\bar{v}^{2}\right)\right) & = & 0,\label{eq:kinetic_energy_eliminate}
\end{eqnarray}
 and we derive a relationship for pressure by noting $pv$ is constant
across the interface at $t^{n}$,

\begin{eqnarray}
\Delta\left(pv\right) & = & 0.\label{eq:pressure_differential_eliminate}
\end{eqnarray}
Combining Equation~(\ref{eq:kinetic_energy_eliminate}) and Equation~(\ref{eq:pressure_differential_eliminate})
with Equation~(\ref{eq:linear_energy}) we remove the kinetic energy,
contained in $\rho e_{t}=\rho v^{2}/2+\rho u$, and the pressure term
to yield a linear relationship for the internal energy across the
interface,

\begin{eqnarray}
\delta\rho u+\nu v\Delta\left(\rho u\right) & = & 0.\label{eq:linear_internal_energy}
\end{eqnarray}
We substitute Equation~(\ref{eq:internal_energy_polynomial}), Equation~(\ref{eq:species_relationship}),
and Equations~(\ref{eq:change_species_1})-(\ref{eq:change_density})
in Equation~(\ref{eq:linear_internal_energy}) and arrive at $n_{p}$
expressions for the temperature at time $t^{n+1}$ by collecting like
terms,

\begin{eqnarray}
T_{j}^{n+1} & = & \bar{T}\frac{\left(W_{1}a_{11}-\nu\bar{v}\left(W_{1}a_{11}-W_{2}a_{21}\right)\right)}{W_{1}\left(1-\nu\bar{v}\right)a_{11}+\frac{W_{2}}{\eta}\left(\nu\bar{v}\right)a_{21}}\nonumber \\
\vdots & \vdots & \vdots\nonumber \\
\left(T_{j}^{n+1}\right)^{n_{p}} & = & \left(\bar{T}\right)^{n_{p}}\frac{\left(W_{1}a_{1n_{p}}-\nu\bar{v}\left(W_{1}a_{1N_{p}}-W_{2}a_{2N_{p}}\eta^{n_{p}-1}\right)\right)}{W_{1}\left(1-\nu\bar{v}\right)a_{1n_{p}}+\frac{W_{2}}{\eta}\left(\nu\bar{v}\right)a_{2n_{p}}}.\label{eq:temperature_relationships}
\end{eqnarray}
Finally, the change in pressure is given as

\begin{eqnarray}
p_{j}^{n+1}-p_{j}^{n} & = & R^{o}T_{j}^{n+1}\left(C_{1}^{0}-\nu\bar{v}C_{1}^{0}+\nu\bar{v}\frac{C_{1}^{0}}{\eta}\right)-R^{o}\bar{T}C_{1}^{0}.\label{eq:pressure_change}
\end{eqnarray}
From analyzing Equations~(\ref{eq:temperature_relationships})~and~(\ref{eq:pressure_change})
we come to the same conclusions to those of Jenny et al~\citep{Jen97},
that pressure oscillations, $p_{j}^{n+1}-p_{j}^{n}\ne0$, do not exist
if one of the following conditions is true
\begin{enumerate}
\item The temperature is continuous, $\eta=1$.
\item The contact discontinuity remains grid aligned, $\nu\bar{v}=1$.
\item The contact discontinuity is stationary, $\bar{v}=0$.
\item The internal energies are linear, $n_{p}=1$, with respect to temperature
\emph{and} the species are the same across the interface, i.e., molecular
weights are constant across the interface, $W_{1}=W_{2}$, and the
internal energies are the same across the interface, $a_{1k}=a_{2k}$.
\end{enumerate}
For condition (1), the numerical mixing of species concentrations,
Equations~(\ref{eq:change_species_1})~and~(\ref{eq:change_species_2}),
inside the cell does not cause a pressure oscillation as both species
are at the same temperature despite having different internal energies. 

When $\eta\ne1$ the temperature is discontinuous and stabilization,
e.g., artificial viscosity, would be required if (2)-(4) were not
satisfied. Satisfaction of condition (2) would requires an interface
fitting method~\citep{Zah18,Cor18} that dynamically fits a priori
unknown discontinuities and is therefore beyond the scope of this
manuscript. Condition (3) is a trivial case. Condition (4) applies
to ideal gases that have a linear relationship between temperature
and internal energy and are assumed to be the same species in all
regions of the flow. 

Applying the same linearization to Equation~(\ref{eq:gamma_m_one_equivalent})
we can arrive at a similar relationship for the internal energy based
on $\bar{\ensuremath{\gamma}}$ and $p$,

\begin{eqnarray}
\frac{p_{j}^{n+1}}{\bar{\gamma}_{j}^{n+1}-1} & = & \frac{\bar{p}}{\bar{\gamma_{1}}-1}-\nu v\left(\frac{\bar{p}}{\bar{\gamma_{1}}-1}-\frac{\bar{p}}{\bar{\gamma}_{2}-1}\right),\label{eq:pressure_gamma_relationship}
\end{eqnarray}
where $\bar{\gamma}_{1}$ and $\bar{\gamma}_{2}$ are the known specific
heat ratios of the right and left hand side based on $C_{1}^{0}$
at temperature $\bar{T}$ and $C_{2}^{0}$ at temperature $\eta\bar{T}$,
respectively. The equivalent process for this formulation would be
to use the definition of pressure and $\bar{\gamma}$ in terms of
known concentrations, $C_{1}^{n+1}$, $C_{2}^{n+1}$, $C_{i}^{0}$,
and $C_{2}^{0}$, and temperatures, $\eta\bar{T}$ and $\bar{T}$,
to solve for $T^{n+1}$. This results in similar nonlinear relationships
for temperature but instead from the $h_{i}$ polynomials. It follows
that the same stability properties found for formulation presented
in this work also apply when the formulation is written in terms of
specific heats that are computed exactly, i.e., not frozen. However,
when $\bar{\gamma}$ is frozen and the temperature is not solved for
exactly, pressure oscillations occur according to the analysis of~\citep{Abg01}.
\end{document}